\documentclass[11pt]{article}
\pdfoutput=1
\usepackage{jheppub}
\usepackage{amsmath}
\usepackage{bm}
\usepackage{slashed}
\usepackage{graphicx}
\usepackage{mathrsfs}

\newlength{\dummysp}
\newcommand{\beq}{\begin{eqnarray}}
\newcommand{\eeq}{\end{eqnarray}}

\newcommand{\gappeq}{\mathrel{\rlap {\raise.5ex\hbox{$>$}}
{\lower.5ex\hbox{$\sim$}}}}
\newcommand{\lappeq}{\mathrel{\rlap{\raise.5ex\hbox{$<$}}
{\lower.5ex\hbox{$\sim$}}}}

\newcommand{\ben}{\begin{enumerate}}
\newcommand{\een}{\end{enumerate}}

\newcommand{\bit}{\begin{itemize}}
\newcommand{\eit}{\end{itemize}}

\def\[{\left [}
\def\]{\right ]}
\def\({\left (}
\def\){\right )}

\def\({\left (}
\def\){\right )}
 
\title{Gaugino condensates in ${\cal N}=1$ super-Yang--Mills theory on Eguchi–Hanson space}
   
 \author{Mohamed M. Anber}
  
\affiliation{Centre for Particle Theory, Department of Mathematical Sciences, Durham University, South Road, Durham DH1 3LE, UK}

\emailAdd{mohamed.anber@durham.ac.uk}  

\abstract{

{\flushleft{W}}e study gaugino condensates in ${\cal N}=1$ $SU(N)$ super-Yang--Mills theory on Eguchi--Hanson space using instanton methods. Fractional topological charge and the spectral asymmetry of the asymptotic $\mathbb{RP}^3$ boundary modify the instanton calculus. A charge-$\frac12$ instanton carries only two adjoint gaugino zero modes and can therefore contribute directly to the bilinear. Because the Eguchi--Hanson path integral prepares a boundary state on $\mathbb{RP}^3$, the observables are boundary-sector-projected amplitudes, and the chiral phases of individual sectors are charges rather than anomalies. For $N\equiv2\pmod 4$, the mixed chiral-center anomaly is nevertheless realized on the $\mathbb{RP}^3$ Hilbert space, with the boundary torsion replacing the 't Hooft flux. Supersymmetric Ward identities ensure that chiral correlators are independent of insertion points and the Eguchi--Hanson resolution parameter. For the $SU(2)$ instanton, with central holonomy, the collective-coordinate integral gives $\langle{\rm tr}(\lambda\lambda)\rangle=16\pi^2\Lambda^3$ in a chiral vacuum, reproducing the flat-space value. For a single $SU(2)$-block embedding in $SU(N>2)$, the noncentral holonomy requires a continuum determinant for renormalization-group invariance, so zero modes alone do not determine the bilinear. We also study a charge-$1$ $SU(4)$ two-block saddle with central holonomy, $16$ bosonic moduli, and $8$ adjoint zero modes. Its collective-coordinate integral gives $\langle\prod_{i=1}^4{\rm tr}(\lambda\lambda)(x_i)\rangle=\frac16(16\pi^2\Lambda^3)^4$. A charge-$1$ trivial-holonomy sector, an embedded $SU(2)$ instanton, contributes at the same order, so clustering constrains only the sum of the two sectors, and the factor $1/6$ alone does not determine the condensate.
}

\begin{document}

\maketitle

\flushbottom

\section{Introduction}

Understanding dynamical chiral-symmetry breaking in strongly coupled gauge theories remains a central challenge in quantum field theory, and placing such theories on curved backgrounds introduces further conceptual and technical difficulties. Supersymmetric theories provide a particularly useful setting in which to address this problem, since holomorphy, nonrenormalization theorems, and exact Ward identities can render otherwise inaccessible quantities calculable. The paradigmatic example is pure ${\cal N}=1$ $SU(N)$ super-Yang--Mills theory. The theory possesses a discrete chiral symmetry $\mathbb Z_{2N}^{\chi}$, which is expected to break spontaneously to $\mathbb Z_2$, giving rise to $N$ supersymmetric vacua. The corresponding order parameter is the gaugino bilinear $\operatorname{tr}_{\mathbf N}(\lambda\lambda)$.

The determination of the gaugino condensate has a long history; see, for
example, Ref.~\cite{Shifman:2012zz} and references therein. Early attempts
were based on direct BPST-instanton calculus \cite{Belavin:1975fg} on $\mathbb R^4$ \cite{Novikov:1983ee,Rossi:1983bu}. A charge-one
$SU(N)$ instanton possesses $2N$ adjoint gaugino zero modes and therefore
cannot contribute directly to the bilinear. Instead, it contributes to the
$N$-point chiral correlator
$\left\langle\prod_{i=1}^{N}\operatorname{tr}_{\mathbf N}
(\lambda\lambda)(x_i)\right\rangle$. Supersymmetric Ward identities were
used to argue that this correlator is independent of the insertion points,
after which cluster decomposition was invoked to extract the one-point
condensate. The difficulty is that the BPST size modulus extends to
arbitrarily large values. The collective-coordinate integral therefore
probes instantons of size $\Lambda^{-1}$, the inverse strong scale, and larger, where the theory is
strongly coupled, and the Gaussian approximation is no longer reliable. The
ensuing disagreement with the exact condensate became known as the
strong-coupling-instanton puzzle.

A controlled calculation can instead be performed by deforming the theory.
One introduces $N-1$ fundamental flavors and moves to a region of the
scalar moduli space in which the gauge group is completely Higgsed at a
scale parametrically larger than $\Lambda$. Instanton calculus is then
genuinely weakly coupled. Holomorphy permits the masses and scalar
expectation values to be varied continuously, allowing the matter
multiplets to be decoupled and pure super-Yang--Mills theory to be
recovered \cite{Finnell:1995dr}. This weak-coupling instanton method avoids the uncontrolled
large-instanton region and yields the exact condensate.

Further semiclassical realizations become available upon compactification.
On $\mathbb R^3\times S^1$, with periodic boundary conditions for the
gaugino and a sufficiently small circle, the theory abelianizes, and the
relevant semiclassical configurations are the fundamental
monopole-instantons, including the Kaluza--Klein monopole \cite{Kraan:1998pm,Lee:1998bb}. Each carries two
gaugino zero modes and can therefore contribute directly to the bilinear condensate \cite{Davies:1999uw,Davies:2000nw}.
A complementary construction is available on a small four-torus
$\mathbb T^4$ with suitable 't~Hooft twists \cite{tHooft:1979rtg,tHooft:1981sps,tHooft:1981nnx,GarciaPerez:2000aiw,Gonzalez-Arroyo:2019wpu}. The twists support fractional
and multi-fractional instantons of charge $k/N$, whose $2k$ gaugino zero
modes saturate the corresponding $k$-bilinear correlators \cite{Anber:2023sjn,Anber:2025yub}. After the
 path integral is normalized by summing over the
gauge-inequivalent zero-action configurations, the resulting correlators
obey cluster decomposition and reproduce the exact four-dimensional
answer~\cite{Anber:2022qsz,Anber:2024mco}. Localization techniques can also be used to compute the condensates on $S^3\times S^1$ \cite{Terashima:2014yua,Hayashi:2026srq}.

The holomorphic calculation on $\mathbb R^4$ and the controlled
semiclassical calculations on $\mathbb R^3\times S^1$ and $\mathbb T^4$
give 
\begin{equation}
 \left\langle
  \operatorname{tr}_{\mathbf N}
  (\lambda\lambda)
 \right\rangle_k
 =
 16\pi^2\Lambda^3
 \exp\left(\frac{2\pi ik}{N}\right),
 \qquad
 k=0,\ldots,N-1,
 \label{eq:intro-exact-gaugino-condensate}
\end{equation}
at vanishing vacuum angle $\theta$. For general $\theta$, the phase is replaced by
$\exp[i(\theta+2\pi k)/N]$. The agreement among these conceptually distinct
methods provides a stringent benchmark for any proposed calculation of the
condensate on a different spacetime.

In this work, we investigate the gaugino condensate on Eguchi--Hanson (EH)
space, the simplest noncompact asymptotically locally Euclidean (ALE)
gravitational instanton~\cite{Eguchi:1978xp,Eguchi:1978gw}. The EH space is a spin manifold and its Ricci-flat hyperk\"ahler
metric admits one chirality of the rigid supersymmetry \cite{Hawking:1978ghb}, and hence, there are only two preserved supercharges on this manifold.
It contains a distinguished exceptional two-sphere
$E\simeq S^2$ (the bolt) and approaches
$S^3/\mathbb Z_2\simeq\mathbb{RP}^3$ asymptotically. These geometric
features allow EH space to support globally well-defined, liftable
$SU(N)$ gauge bundles with fractional topological charge \cite{Bianchi:1996zj}
\begin{equation}
 Q_{(N)}=\frac12.
 \label{eq:intro-EH-half-charge}
\end{equation}

The fractional charge has a direct geometric origin.  The exceptional sphere has self-intersection
$2$, with the sign reversed upon reversing the orientation of
EH. Its intersection lattice is therefore the $A_1$ root lattice, which
has determinant two and is not self-dual. The cohomology class dual to the
bolt is governed by the inverse intersection form and consequently has
half-integral squared norm. A gauge flux supported on this class can thus
give a half-integral contribution to
$\int_{\rm EH}\operatorname{tr}(F\wedge F)/(8\pi^2)$. The same determinant-two
mismatch appears at infinity through the order-two torsion-one cycle of $\mathbb{RP}^3$: 
$H_1(\mathbb{RP}^3,\mathbb Z)\simeq H^2(\mathbb{RP}^3,\mathbb Z)\simeq\mathbb Z_2$. The self-intersection of
the exceptional sphere therefore provides the local geometric origin of
the fractional charge, while the noncompact boundary permits the
corresponding nonunimodular intersection lattice.

EH space provides a particularly interesting setting for strongly coupled
gauge dynamics for several related reasons:

\begin{itemize}

\item Dynamical chiral-symmetry breaking on curved backgrounds remains considerably less understood than its flat-space counterpart. Curvature, topology, and asymptotic boundary conditions may affect the infrared spectrum and vacuum selection, even when supersymmetry protects certain local observables. EH space provides a useful setting in which to investigate these questions, and the gaugino condensate offers a concrete probe of whether the familiar breaking pattern $\mathbb Z_{2N}^{\chi}\rightarrow\mathbb Z_2$  persists on this background.

 \item Although the EH gravitational instanton has been known for nearly
 half a century, the physical meaning of quantum field theory on this
 noncompact background remains comparatively unexplored. A notable early
 exception is the work of 't~Hooft~\cite{tHooft:1988wxy}, who interpreted
 the EH instanton in Hilbert-space language. This viewpoint was subsequently
 developed into an interpretation of the EH path integral as an amplitude
 connecting an initial entangled configuration to the
 vacuum, which was investigated in the context of gauging the $\mathbb Z_6^{[1]}$ one-form symmetry of the Standard Model~\cite{Anber:2025gvb}. More recently, it was shown explicitly in
 free Maxwell theory that the path integral on ALE spaces naturally
 prepares a state in the Hilbert space that lives on its asymptotic boundary, rather
 than defining a single scalar partition function~\cite{Anber:2026pfg}.
 Computing fermion condensates on EH space and comparing them with the
 corresponding results on $\mathbb R^4$, $\mathbb R^3\times S^1$, and
 $\mathbb T^4$ may therefore provide further insight into the physical
 interpretation of quantum field theory on this background.

\item
It was argued in~\cite{Anber:2025dxy} that EH space can probe anomalies that may not be visible on conventional closed manifolds. For a theory with global symmetry $G_1\times G_2$, the path-integral measure was found to acquire a nontrivial phase under a $G_2$ transformation in the presence of a background gauge field for $G_1$. This phase was interpreted as a mixed 't~Hooft anomaly between $G_1$ and $G_2$. However, as emphasized recently in~\cite{Anber:2026pfg}, the path integral on EH space prepares a state in the Hilbert space associated with its asymptotic boundary, rather than an ordinary scalar partition function. The interpretation of these phases should therefore be revisited within this boundary-state framework. Their origin may involve an interplay between the cohomology class generated by the exceptional sphere, the torsion of the asymptotic $\mathbb{RP}^3$ boundary, and the spectral asymmetry measured by the boundary $\eta$-invariant. Studying fermion condensation on EH space may further clarify how these anomalous responses are realized in the infrared, including the possible roles of spontaneous chiral-symmetry breaking and the structure of the boundary state\footnote{The index computations of Ref.~\cite{Anber:2025dxy} remain correct, but the interpretation of the resulting phases as 't~Hooft anomalies overstated their meaning. As shown in Sec.~\ref{sec:EH-anomalies-APS}, the Fujikawa phase of a boundary sector is a chiral charge of the EH-prepared state, measured relative to the Dirac sea of that sector, rather than an anomaly. A genuine anomaly arises only when a symmetry acting on the boundary Hilbert space fails to commute with the conserved chiral operator; for the center symmetry this happens for $N\equiv2\bmod4$. Conclusions drawn there from the EH phases alone, in particular constraints on infrared realizations, should be reconsidered in this light.}.
 
 \item Supersymmetric field theories in curved space provide a setting in which exact Ward identities and holomorphy can be combined with nontrivial geometry and topology; see, e.g., \cite{Konishi:1988mb,Festuccia:2011ws,Dumitrescu:2012ha,Dumitrescu:2012at,Closset:2013vra,Closset:2014uda,Komargodski:2010rb}. The gaugino condensate is a particularly natural observable in this context: it is a chiral quantity whose dependence on the insertion points, the complexified coupling, and the EH resolution parameter is strongly constrained by the preserved supersymmetry. At the same time, its value is sensitive to nonperturbative gauge dynamics and asymptotic boundary conditions. Computing it on EH space therefore offers a concrete test of both the power and the limitations of supersymmetric protection on a curved, noncompact background.

 \item The minimal $SU(N)$ instanton on EH carries
 $Q_{(N)}=1/2$ but supports precisely two normalizable adjoint-fermion zero
 modes. It can therefore contribute directly to the gaugino bilinear. This
 is in sharp contrast with the charge-one BPST instanton on $\mathbb R^4$,
 which possesses $2N$ adjoint zero modes and cannot directly saturate a
 bilinear insertion. EH space consequently raises the intriguing
 possibility of obtaining, through direct instanton calculus, the gaugino
 condensate that is inaccessible to the conventional one-instanton
 calculation on $\mathbb R^4$.

\end{itemize}

The reduction to two zero modes is itself a global effect. The local gauge
contribution to the adjoint index is
$2T_{\rm adj}Q_{(N)}=2NQ_{(N)}=N$, which is already smaller than the $2N$
zero modes of a charge-one BPST instanton but is not the complete answer.
Since EH space is noncompact, the relevant index theorem is the
Atiyah--Patodi--Singer (APS) theorem \cite{Atiyah:1976jg,Atiyah:1976qjr,Atiyah:1975jf}. In addition to the gauge contribution, the
index receives a bulk gravitational term and an $\eta$-invariant boundary correction
determined by the spectral asymmetry of the adjoint Dirac operator on
$\mathbb{RP}^3$. For the $Q_{(N)}=1/2$ connection and its
associated boundary holonomy, these contributions combine as
${\cal I}_{\rm adj}^{\rm EH}
=2NQ_{(N)}+{\cal I}_{\rm grav}
-\eta_{\rm adj}(\mathbb{RP}^3)
=N+(2-N)=2$. The number of zero modes is therefore not determined by the
fractional topological charge alone. It follows from the interplay among
the gauge bundle, the gravitational Pontryagin density, and the
$\eta$-invariant of the boundary Dirac operator. The asymptotic boundary is
thus an essential part of the mechanism that reduces the adjoint index to
two~\cite{Anber:2025dxy}. Yet, these zero-mode profiles are localized near the bolt and do not escape to the asymptotic boundary. 

The same background probes the discrete chiral symmetry. Under its
generator, $\lambda\longrightarrow e^{i\pi/N}\lambda$, the fermionic
measure in the two-zero-mode sector acquires the phase\\
${\cal D}\lambda\longrightarrow
\exp(-2\pi i/N){\cal D}\lambda$. Because the connection defines a liftable $SU(N)$ bundle and the path integral prepares a boundary state, this phase is not by itself an anomaly; we explain below and in Sec.~\ref{sec:EH-anomalies-APS} that it is a chiral charge, and how the mixed anomaly between $\mathbb Z_{2N}^{\chi}$ and the center symmetry is nevertheless realized on EH space without any 't~Hooft flux.
The gaugino bilinear transforms with the compensating phase,
$\operatorname{tr}_{\mathbf N}(\lambda\lambda)\longrightarrow
\exp(2\pi i/N)\operatorname{tr}_{\mathbf N}(\lambda\lambda)$, and
simultaneously absorbs the two Grassmann zero modes. The corresponding
instanton-induced amplitude is therefore invariant.

This microscopic selection rule raises a natural infrared question. If a
gaugino condensate forms and breaks
$\mathbb Z_{2N}^{\chi}\longrightarrow\mathbb Z_2$, the broken symmetry
permutes the resulting $N$ chiral vacua, and their condensates differ by
the phases appearing in \eqref{eq:intro-exact-gaugino-condensate}. It is
therefore natural to ask whether this vacuum structure provides the
infrared realization of the chiral response detected
on EH space. The answer is not automatic: because the EH path integral
prepares a boundary state, one must distinguish its sector-resolved
amplitudes from expectation values in the four-dimensional chiral vacua
and determine how the chiral symmetry acts on the boundary-state
components. 

Our aim is to carry out these calculations directly. We construct the relevant
self-dual $SU(N)$ backgrounds, determine their bosonic and fermionic
collective coordinates, and evaluate the instanton-induced chiral
correlators. Particular attention is paid to the role of the asymptotic boundary state and the extent to which
the resulting sector amplitudes can be interpreted as ordinary vacuum
condensates. 

\paragraph{Summary of result.} 
Because EH is noncompact, an instanton sector is characterized not only by
its bulk topological data but also by the framing, or equivalently by the
conjugacy class of the limiting flat connection on the asymptotic boundary
$Y_\infty\simeq\mathbb{RP}^3$. For an $SU(N)$ bundle the first Chern class
vanishes, while the relevant bulk invariant is the second Chern
number $Q_{(N)}$; for a $U(N)$ bundle both the first and second Chern classes
may be nontrivial. The asymptotic data are encoded in the holonomy
$h_c^{(\ell)}$ around the generator $c$ of
$\pi_1(\mathbb{RP}^3)=\mathbb Z_2$. Kronheimer and Nakajima provided a
systematic mathematical description of framed instantons on ALE spaces \cite{PeterBKronheimer:1990zmj,Bianchi:1995xd}. In
the present work, we instead follow a more explicit approach: we employ a
cohomogeneity-one (radial) ansatz constructed from the left-invariant one-forms on $S^3$ and
solve the self-duality equations directly for $SU(N)$ connections in the
minimal liftable sector $Q_{(N)}=1/2$. The resulting solutions determine
both their bulk topological charge and their asymptotic holonomy class.

The EH path integral prepares a state on the Hilbert space ${\cal H}_{Y_\infty}$ that lives on the asymptotic boundary $\mathbb{RP}^3$. In a semiclassical treatment, we may label the asymptotic states with the holonomy sectors \cite{Anber:2026pfg}:
\begin{equation}
 |\Psi_\ell\rangle
 \in{\cal H}_{Y_\infty},
 \qquad
 |\Psi_\ell\rangle
 =
 \sum_{\mathfrak{k}} Z^{(\ell;\mathfrak{k})}\,|\ell;\mathfrak{k}\rangle \,,\quad 
 \ell=0,1,\ldots,\left\lfloor\frac{N}{2}\right\rfloor.
\end{equation}
Here, $\ell$ labels the boundary-holonomy sector, whereas $\mathfrak{k}$ denotes any
additional quantum numbers needed to distinguish states within that sector.

To extract a scalar condensate from the state prepared by the EH path
integral, one must choose a bra state $\langle B|$ in the dual asymptotic
Hilbert space. Suppressing additional quantum numbers, we write
$\langle B|=\sum_\ell b_\ell^*\langle\ell|$ and define
\begin{equation}
 \langle{\cal O}\rangle_B
 \equiv
 \frac{\langle B|\Psi[{\cal O}]\rangle}
 {\langle B|\Psi[1]\rangle}.
\end{equation}
For ${\cal O}=\operatorname{tr}_{\mathbf N}(\lambda\lambda)$, the leading
semiclassical contribution to the numerator is supplied by the
$Q_{(N)}=1/2$ instanton. Its two adjoint gaugino zero modes are saturated by
the bilinear insertion, while its source-free amplitude vanishes because
the corresponding Grassmann integrations are unsaturated. The denominator
is therefore provided, at leading order, by the source-free path integral
in the trivial $Q_{(N)}=0$ sector. If $\ell_s$ denotes the boundary
holonomy selected by the fractional instanton, the resulting contribution
takes the schematic form
\begin{equation}
 \left.
 \left\langle
  \operatorname{tr}_{\mathbf N}(\lambda\lambda)
 \right\rangle_B
 \right|_{Q=1/2}
 \simeq
 \frac{
  b_{\ell_s}^*\,
  Z_{Q_{(N)}=1/2}^{(\ell_s)}
  [\operatorname{tr}_{\mathbf N}(\lambda\lambda)]
 }{
  b_0^*\,Z_{Q_{(N)}=0}^{(0)}[1]
 }.
\end{equation}
The instanton amplitude is obtained by integrating the classical weight $\exp\left[-\frac{8\pi^2Q_{(N)}}{g^2}\right]$,
the bosonic and fermionic collective-coordinate measures, and the regulated
nonzero-mode determinant over the instanton moduli space. The ratio
$b_{\ell_s}/b_0$ is fixed by the chosen asymptotic state or by a gluing
prescription; it is not determined by the local instanton calculation.
Thus, semiclassical instanton calculus computes the contribution associated
with a definite bulk saddle and boundary-holonomy sector, while its
interpretation as a vacuum condensate requires specifying how the
corresponding boundary state is projected.

In a nonsupersymmetric theory, the correlators
\begin{equation}
 \left\langle
 \prod_{i=1}^{N}
 \operatorname{tr}_{\mathbf N}(\lambda\lambda)(x_i)
 \right\rangle_B
\end{equation}
would generally depend on the insertion points $x_i$ and on the bolt size (resolution parameter)
$a$.  In a supersymmetric theory on EH space, however, the two preserved supercharges impose much
stronger constraints through supersymmetric Ward identities.  These
identities imply that separated correlators of the chiral gaugino
bilinears are independent of the insertion points $x_i$, depend holomorphically
on the complexified strong scale $\Lambda$, and are independent of the EH
resolution parameter $a$. While the first two statements are expected to hold in supersymmetric theories (as in the case of theories on $\mathbb R^4$), the last statement follows because the
variation of the EH metric with respect to $a$ is a normalizable
transverse-traceless deformation whose stress-tensor insertion is
supersymmetry-exact up to total derivatives.  These conclusions hold at
fixed boundary sector and assume a supersymmetric regulator and boundary
state, together with the absence of spacetime-boundary contributions.

Our first result concerns $SU(2)$ super-Yang--Mills theory. The minimal
EH instanton has topological charge $Q_{(2)}=1/2$, approaches the
nontrivial central holonomy \begin{equation}h_c=-I_2\end{equation} at infinity, and supports precisely
two adjoint gaugino zero modes, as follows from the APS index. The
insertion $\operatorname{tr}_{\mathbf2}(\lambda\lambda)$ saturates these
zero modes, and the remaining semiclassical calculation reduces to an
integral over four bosonic collective coordinates: one radial-size
modulus and three global color orientations. Assuming that the regulated
nonzero-mode determinant ratio is unity, as expected from supersymmetric
pairing in this central-holonomy background, the integral can be
performed explicitly and gives
\begin{equation}
 \left.
 \left\langle
  \operatorname{tr}_{\mathbf2}(\lambda\lambda)
 \right\rangle_B
 \right|_{Q=1/2}
 =
 \frac{b_1^*}{b_0^*}\,
 16\pi^2\Lambda^3,
 \label{eq:intro-SU2-condensate}
\end{equation}
where $b_1/b_0$ specifies the relative projection of the chosen boundary
bra onto the nontrivial- and trivial-holonomy sectors. For $SU(2)$, this ratio is fixed in the companion Letter~\cite{Anber:2026bsx}: on $\mathbb{RP}^3$ the center operator anticommutes with the chiral generator, the clustering vacua are the center-operator eigenstates, and $b_1/b_0=(-1)^k$, giving $\langle\operatorname{tr}_{\mathbf 2}(\lambda\lambda)\rangle_k=16\pi^2\Lambda^3e^{i(\theta+2\pi k)/2}$ for $k=0,1$.

The collective-coordinate integral is independent of the EH bolt radius $a$, a direct verification of the supersymmetric Ward identity. What is remarkable, however, is that modulo the relative projection,  $b_1/b_0$, the numerical coefficient of the condensate
agrees with the standard result obtained by holomorphic methods on
$\mathbb R^4$ and by controlled semiclassical calculations on other
compactifications.

The corresponding construction for $SU(N>2)$ is considerably more
subtle. A self-dual configuration with topological charge
$Q_{(N)}=1/2$ is obtained by embedding the EH $SU(2)$ instanton into a
single $SU(2)$ subgroup of $SU(N)$. The APS index of the adjoint Weyl
operator is ${\cal I}_{\rm adj}=2$, so this configuration has the correct
number of zero modes to contribute to the gaugino bilinear. Unlike the
$SU(2)$ solution, however, its asymptotic holonomy,
\begin{equation}
 h_c=\operatorname{diag}(-I_2,I_{N-2}),
\end{equation}
is noncentral. In this background, the adjoint action of the
noncentral holonomy is not uniform. Under parallel transport around the
closed contour near the bolt, the gaugino components lying within the
embedded $\mathfrak{su}(2)$ block or its centralizer return with gauge
holonomy $+1$, whereas the off-diagonal components connecting the
$\mathfrak{su}(2)$ block to the remaining $N-2$ color directions acquire
gauge holonomy $-1$. By contrast, a central holonomy acts trivially in the
adjoint representation, so every color component of the gaugino has
gauge holonomy $+1$.

A further difficulty arises from the continuum fluctuations in the
one-loop measure. For ${\cal N}=1$ $SU(N)$ super-Yang--Mills theory, the $\beta$-function coefficient is
$b_0=3N$, and a charge-$1/2$ saddle requires the Pauli-Villars regulator dependence to scale as
$\mu^{3N/2}$ for some regularization scale $\mu$. The four bosonic and two fermionic collective coordinates
inherited from the embedded $SU(2)$ solution supply only $\mu^3$. For
$N>2$, the remaining factor
\begin{equation}
 \mu^{\frac32(N-2)}
\end{equation}
must therefore arise from the regulated continuum fluctuations connecting
the embedded $SU(2)$ block to the remaining $N-2$ color directions.

Let ${\cal R}_{\rm nz}^{(N)}$ denote this fluctuation contribution. Within
the one-saddle semiclassical expansion, renormalization-group (RG) invariance 
organizes it in the canonical form
\begin{equation}
 {\cal R}_{\rm nz}^{(N)}(t,a;\mu,g)
 =
 \left[g^2(\mu)\right]^{-\frac{N-2}{2}}
 (\mu t)^{\frac32(N-2)}
 \widehat{\cal R}_N\left(\frac{a}{t}\right),
 \label{eq:intro-SUN-nonzero-mode-ratio}
\end{equation}
where $t$ is a parameter that has the dimension of the inverse mass (which is precisely the radial-size modulus of the $SU(N)$ instanton on EH space) and $\widehat{\cal R}_N$ is a finite dimensionless function determined
by the continuum spectral data. When this expression is combined with
the classical weight and the zero-mode Jacobians, the explicit
$\mu$- and $g(\mu)$-dependence assembles into the holomorphic factor
$(\Lambda^3)^{N/2}$. Dimensional analysis then gives the formal
single-root contribution
\begin{equation}
 \left.
 \left\langle
  \operatorname{tr}_{\boldsymbol N}(\lambda\lambda)(r)
 \right\rangle_B
 \right|_{Q=1/2}^{\rm sc}
 =
 \frac{b_1^*}{b_0^*}\,
 16\pi^2\Lambda^3
 (a\Lambda)^{\frac32(N-2)}
 \Phi_N\left(\frac{r}{a}\right),
 \label{eq:intro-SUN-single-root-amplitude}
\end{equation}
for some dimensionless function $\Phi_N$.

Equation~\eqref{eq:intro-SUN-single-root-amplitude} is a formal
single-saddle expression, not the complete boundary-sector-projected
condensate. The supersymmetric Ward identities constrain the latter:
provided that the regulator and boundary state preserve the two 
supercharges and that no spacetime-boundary
terms survive, the complete amplitude is independent of both the insertion
point and the EH resolution parameter $a$ and depends holomorphically on the
strong scale. In a chosen boundary sector, dimensional
analysis then allows
\begin{equation}
 \left\langle
  \operatorname{tr}_{\boldsymbol N}(\lambda\lambda)
 \right\rangle_{B}
 =
 C_{B}\Lambda^3,
 \label{eq:intro-SUN-complete-condensate-form}
\end{equation}
where the Ward identities do not determine the dimensionless coefficient
$C_{B}$.

If the Ward identities were assumed to apply separately to the isolated
single-root contribution in
\eqref{eq:intro-SUN-single-root-amplitude}, position independence would
require $\Phi_N$ to be constant, while $a$-independence would formally
force that constant to vanish for $N>2$. We do not interpret this as a
prediction that the complete condensate vanishes. Rather, it shows that
the isolated single-root semiclassical factorization cannot simultaneously
be complete and satisfy all the Ward identities. The unknown continuum
determinant, the noncompact region $t\gtrsim\Lambda^{-1}$ (where $t$ parametrizes the radial-size modulus) which extends well into the strongly-coupled regime, additional
configurations in the same boundary sector, or a moduli-space surface term
must be understood before the single-root sector can be completed. Thus,
the zero-mode calculation  does not determine the
$SU(N>2)$ bilinear condensate.

To examine a higher-rank background in which the index and RG counting
are instead saturated entirely by zero modes, we study a charge-one
instanton in $SU(4)$ super-Yang--Mills theory. It is constructed by
embedding two EH $SU(2)$ instantons along mutually orthogonal roots,
\begin{equation}
 A^{\rm B}
 =
 \operatorname{diag}
 \left(
  A_{\rm EH}(t_1),A_{\rm EH}(t_2)
 \right),
 \qquad
 Q_{(4)}=1,
 \qquad
 h_c=-I_4.
\end{equation}
Because the boundary holonomy is central, all global $SU(4)$ color
rotations preserve the asymptotic framing, and every color component of the gaugino has
gauge holonomy $+1$. At a generic point, the
configuration possesses two radial-size moduli and fourteen color
orientations, giving a sixteen-dimensional bosonic moduli space. The APS
index is ${\cal I}_{\rm adj}=8$: four gaugino zero modes are inherited
from the two diagonal $SU(2)$ blocks, while four occupy the off-diagonal
color blocks. Their combined Jacobians produce $\mu^{12}$, precisely
matching the $\beta$-function coefficient $b_0Q_{(4)}=12$. In contrast with the single-block background,
RG invariance therefore requires no additional scale-dependent
determinant.

Assuming cancellation of the regulated nonzero-mode determinants, we
saturate the eight Grassmann collective coordinates with four insertions
of
${\cal O}=\operatorname{tr}_{\mathbf4}(\lambda\lambda)$
and integrate over the moduli space of the two-block family. The
coincident-point calculation gives
\begin{equation}
 \left.
 \left\langle
 \prod_{i=1}^4 {\cal O}(x_i)
 \right\rangle_B
 \right|_{Q=1}
 =
 \frac{b_-^*}{b_+^*}\,
 \frac{1}{6}
 \left(16\pi^2\Lambda^3\right)^4,
 \label{eq:intro-SU4-four-bilinear-result}
\end{equation}
where $b_-/b_+$ specifies the relative projection onto the
$h_c=-I_4$ and $h_c=+I_4$ boundary-holonomy sectors.

The collective-coordinate integral over the two-block family is independent
of the insertion points and the bolt size, in agreement with the
supersymmetric Ward identities. The two-block family, however, does not give
the complete charge-one four-point function. The trivial-holonomy sector
$h_c=+I_4$ also supports charge-one instantons with eight adjoint gaugino
zero modes: an EH $SU(2)$ instanton with trivial holonomy, embedded in
$SU(4)$. Its moduli are the position on EH, the size, and eleven color
orientations, again giving a sixteen-dimensional moduli space. Denoting its
contribution by $c_0\left(16\pi^2\Lambda^3\right)^4$, the complete charge-one
four-point function is
\begin{equation}
 \left.
 \left\langle
 \prod_{i=1}^4 {\cal O}(x_i)
 \right\rangle
 \right|_{Q=1}
 =
 \left(c_0+\frac{b_-^*}{b_+^*}\,\frac{1}{6}\right)
 \left(16\pi^2\Lambda^3\right)^4 .
 \label{eq:intro-SU4-full-four-point}
\end{equation}
If the coincident products are related to the separated four-point function
without contact terms and the boundary bra selects a definite
cluster-decomposing vacuum with
$\langle{\cal O}\rangle^4=\left(16\pi^2\Lambda^3\right)^4$,
cluster decomposition requires
\begin{equation}
 c_0=1-\frac{b_-^*}{b_+^*}\,\frac16 ,
\end{equation}
that is, $c_0=5/6$ or $7/6$ for $b_-^*/b_+^*=\pm1$. The factor $1/6$ is
therefore not by itself a prediction for the bilinear condensate, nor a
clustering puzzle; the test of clustering in this sector is the value of
$c_0$, whose computation requires the zero modes of the embedded EH
instanton in the off-diagonal color blocks and is left for future work.
The two-block result also cannot be compared with the single-root saddle:
the latter belongs to a distinct noncentral-holonomy sector and, for
$SU(N>2)$, does not determine a bilinear condensate because its continuum
determinant and strong-coupling completion are unknown.

\paragraph{Chiral charges, center symmetry, and the mixed anomaly.}
A second set of results concerns the symmetries of the EH-prepared state, which we treat in Sec.~\ref{sec:EH-anomalies-APS}. Because the path integral prepares a state rather than a number, the Fujikawa phase of a boundary sector, such as $e^{-2\pi i/N}$ for the single-root saddle, is not by itself an 't~Hooft anomaly. We show that it counts the gaugino quanta created by the saddle above the Dirac sea of its holonomy sector. The conserved chiral operator, $\mathsf X=\exp[\frac{2\pi i}{2N}(Q_5-2N\operatorname{CS})]$, receives three contributions on the boundary state: the quanta, supplied by the APS index; the Chern--Simons (CS) term of the asymptotic connection, the boundary counterpart of the $\theta$-shift that accompanies a chiral rotation; and the charge of the Dirac sea, fixed by the spectral asymmetry of the Weyl operator on $\mathbb{RP}^3$. The bulk gauge term of the index is canceled by the Chern--Simons term and its $\eta$-invariant by the sea charge, so only the gravitational term survives. Consequently, the EH-prepared state is an eigenstate of $\mathsf X$ with the same eigenvalue in every boundary sector, as it must be, since the sectors are mixed by tunneling. 

For even $N$, the $\mathbb Z_2$ subgroup of the $\mathbb Z_N^{(1)}$ center symmetry acts on the untwisted $\mathbb{RP}^3$ Hilbert space through an operator $\mathsf U$ that maps the holonomy sector $\ell$ to $N/2-\ell$ and shifts the Chern--Simons invariant by $N/4$. The torsion of the boundary thus plays the role of the 't~Hooft flux, and one finds the operator algebra $\mathsf X\mathsf U=e^{-i\pi N/2}\,\mathsf U\mathsf X$, the mixed chiral--center anomaly, nontrivial for $N\equiv2\bmod4$~\cite{Anber:2026bsx}. In the bulk, $\mathsf U$ is a topological surface operator that can be pushed onto the bolt but not removed. It introduces no 't~Hooft twist: its flux through the bolt vanishes modulo $N$, the gauge bundle remains an honest $SU(N)$ bundle, and the adjoint index is unchanged, while the topological charge weighted by $\theta$ shifts by $N/4$. The anomaly can therefore be read off directly from the bulk: $\mathsf U$ maps the EH-prepared state to a state with the same gaugino content but opposite chiral charge, so the two are orthogonal. Gluing EH to its orientation reversal with $\mathsf U$ inserted on the common boundary, this orthogonality becomes the vanishing of the partition function on $S^2\times S^2$ in a genuine 't~Hooft-flux background: the twist that is invisible on EH appears only on the cycle that crosses the gluing slice. For $SU(2)$, the trivial and $Q=\frac12$ sectors form a $\mathsf U$-pair, and the anomaly fixes the relative sign of the condensates in the two chiral vacua. For $N>2$, the single-root sector is not paired with the trivial sector, and its phase remains a chiral charge rather than an anomaly.

\paragraph{In lieu of a conclusion.} The main lesson of these examples is that instanton calculus on EH naturally
computes boundary-conditioned amplitudes rather than expectation values in
an automatically selected vacuum. Subject to the absence of
spacetime-boundary  terms, supersymmetric Ward
identities require separated chiral correlators to be position
independent, holomorphic in the strong-coupling scale $\Lambda$, and independent of the EH resolution
parameter $a$. 

For $SU(2)$, the central holonomy, the absence of spectator fluctuations,
and an exact radial-profile identity lead, for a particular boundary
projection, to the striking value $16\pi^2\Lambda^3$. This agreement does
not by itself prove that EH prepares a standard vacuum~\cite{Anber:2026bsx}. For $SU(N>2)$, the single-root saddle has
the correct zero-mode counting, but its formal semiclassical scaling is
incompatible with simultaneous position and $a$ independence unless the
isolated contribution vanishes. We interpret this tension as evidence that
the unknown continuum determinant, the large-modulus region, additional
saddles, or a moduli-space surface term is essential, rather than as a
prediction that the complete condensate vanishes.

The $SU(4)$ two-root family provides a complementary result: integrating
over all collective coordinates of the constructed family yields the
nontrivial factor $1/6$ in the four-bilinear amplitude. This is not the
complete charge-one correlator, however. The trivial-holonomy sector
$h_c=+I_4$ supports charge-one instantons, namely an EH $SU(2)$ instanton
embedded in $SU(4)$, with the same number of gaugino zero modes. Their
contribution $c_0$ adds to the two-block one, and clustering requires
$c_0+\frac{b_-^*}{b_+^*}\frac16=1$. The factor $1/6$ is therefore a property
of the two-root boundary-conditioned amplitude; whether clustering holds
depends on $c_0$, which we leave for future work.
 For $SU(2)$, the Hamiltonian description of Ref.~\cite{Anber:2026bsx} relates these
amplitudes to definite chiral vacua; for $SU(N>2)$, a more complete Hamiltonian or gluing
description of the EH-prepared state is still needed to relate the
semiclassical amplitudes to definite chiral vacua and to clarify how the chiral
charges of the EH sectors are realized in the infrared.

\paragraph{Organization of the paper.}
The remainder of the paper is organized as follows. In
Sec.~\ref{EHSPACEALLTHAT}, we review the geometry and topology of
Eguchi--Hanson space and construct the self-dual $SU(2)$ and $SU(N)$
gauge backgrounds used throughout the paper. We also discuss their
topological charges, asymptotic holonomies, liftability, and their relation
to the corresponding Cartan-valued configurations. In
Sec.~\ref{sec:EH-boundary-anomalies-supersymmetry}, we formulate the EH
path integral as a state-preparation problem on the asymptotic
$\mathbb{RP}^3$ boundary. We describe the boundary-holonomy sectors,
the boundary conditions for adjoint fermions, the APS index and the
associated chiral response, the center symmetry on $\mathbb{RP}^3$ and the mixed
chiral--center anomaly realized on the EH-prepared state, and the supersymmetry Ward identities
relevant to the condensate calculations. Section~\ref{sec:SYM-condensate}
is devoted to the minimal $Q_{(2)}=1/2$ instanton in $SU(2)$
super-Yang--Mills theory. There, we construct its adjoint gaugino zero
modes, determine the bosonic and fermionic collective-coordinate
measures, and evaluate its contribution to the gaugino bilinear.
In Sec.~\ref{sec:SUN-condensates}, we turn to higher-rank theories. We
first examine the bilinear induced by a single-root $SU(2)$ embedding in
$SU(N)$ and explain the unresolved role of the continuum determinants.
We then study the charge-one, two-root background in $SU(4)$ and compute
the corresponding four-bilinear, eight-fermion amplitude.

The appendices contain the technical details underlying these results.
Appendix~\ref{Mathematical preliminaries} collects our differential-geometric,
spinorial, and self-duality conventions. Appendix~\ref{app:EH-gauge-multiplet-Ward-identities} constructs the gauge
kinetic and supercurrent multiplets on EH space and derives the Ward
identities governing position independence, holomorphy, and independence
of the resolution parameter $a$, including the possible roles of contact
terms and spacetime- or moduli-space-boundary contributions.
Appendix~\ref{sec:gauge-zero-modes} reviews the fluctuation equations and
the construction of gauge-field zero modes in a background gauge in curved space.
Appendix~\ref{app:SU2-condensate-details} gives the detailed derivation and
normalization of the $SU(2)$ fermionic and bosonic zero modes, together
with their collective-coordinate Jacobians. Finally,
Appendix~\ref{app:SU4-eight-fermion-details} develops the full
sixteen-dimensional moduli-space analysis of the two-block $SU(4)$
background, including the off-diagonal harmonic dressings, the
moduli-space metric and Haar measure, the eight normalized gaugino zero
modes, and the final integration over the collective-coordinate space.

\section{Eguchi--Hanson space and  background gauge fields}
\label{EHSPACEALLTHAT}

In this section, we review the geometric properties of the
Eguchi--Hanson space that will be needed throughout our analysis.  We
then discuss the construction and global properties of abelian
$U(1)$ backgrounds, together with Cartan-valued and genuinely
nonabelian $SU(2)$, $PSU(2)$, $SU(N)$ bundles, as well as the minimal Cartan-valued $PSU(N)$ bundles.
Particular attention is given to their self-duality, flux
quantization, topological charges, and global aspects of the bundles. The reader is also referred to Appendix \ref{Mathematical preliminaries} for a review of preliminary differential geometry used in this work. 

\subsection{Eguchi--Hanson space}

The Eguchi--Hanson (EH) gravitational instanton is a complete, noncompact,
Ricci-flat four-manifold with self-dual Riemann curvature
\cite{Eguchi:1978xp,Eguchi:1978gw}.  It is hyperk\"ahler and provides the
smooth resolution of the $\mathbb C^2/\mathbb Z_2$ orbifold singularity.  In
terms of the left-invariant one-forms $\{\sigma_x,\sigma_y,\sigma_z\}$ on $S^3$, the EH
metric is
\begin{equation}
 \label{EH_metric}
 ds^2_{\rm EH}
 =f^2(r)\,dr^2
 +r^2\left[\sigma_x^2+\sigma_y^2+f^{-2}(r)\sigma_z^2\right],
\end{equation}
where
\begin{equation}
 \label{f_factor}
 f^{-2}(r)=1-\frac{a^4}{r^4}.
\end{equation}
Our normalization of the one-forms is
\begin{align}
 \label{sigmas}
 \sigma_x&=\frac12\left(\sin\psi\,d\theta
             -\sin\theta\cos\psi\,d\varphi\right),\quad
 \sigma_y=\frac12\left(-\cos\psi\,d\theta
             -\sin\theta\sin\psi\,d\varphi\right),\\
 \sigma_z&=\frac12\left(d\psi+\cos\theta\,d\varphi\right),\nonumber
\end{align}
so that
\begin{equation}
 d\sigma_x=2\sigma_y\wedge\sigma_z,
 \qquad
 d\sigma_y=2\sigma_z\wedge\sigma_x,
 \qquad
 d\sigma_z=2\sigma_x\wedge\sigma_y.
 \label{eq:EH-Maurer-Cartan}
\end{equation}
The coordinate ranges are
\begin{equation}
 a\leq r<\infty,
 \qquad 0\leq\theta\leq\pi,
 \qquad 0\leq\varphi<2\pi,
 \qquad 0\leq\psi<2\pi.
 \label{eq:EH-coordinate-ranges}
\end{equation}

The parameter $a>0$ has dimensions of length and determines the size of the
exceptional two-sphere, the bolt. To see this,   near $r=a$, introduce
$\rho^2=a(r-a)$.  At fixed $(\theta,\varphi)$, the $(r,\psi)$ part of the
metric becomes
\begin{equation}
 ds^2_{(r,\psi)}=d\rho^2+\rho^2d\psi^2+O(\rho^4/a^2).
\end{equation}
Thus, the period $\psi\sim\psi+2\pi$ removes a conical singularity at the bolt.
The remaining metric at $r=a$ is that of $S^2$:
\begin{equation}
 ds^2_{S^2_{\rm bolt}}
 =\frac{a^2}{4}\left(d\theta^2+\sin^2\theta\,d\varphi^2\right),
 \label{eq:EH-bolt-metric}
\end{equation}
and thus $a$ is the bolt size.

Because the period of $\psi$ is half its usual value on the covering
three-sphere, the space at infinity is
\begin{equation}
 S^3_\infty/\mathbb Z_2\simeq\mathbb{RP}^3,
\end{equation}
and the metric approaches that of $\mathbb R^4/\mathbb Z_2$.  Consequently,
EH is asymptotically locally Euclidean (ALE), but not globally.  Notice that, with the coordinate
identification in \eqref{eq:EH-coordinate-ranges}, the formal $a\to0$ limit is
the singular orbifold $\mathbb R^4/\mathbb Z_2$, rather than its double cover
$\mathbb R^4$. Thus, $a$ is the resolution parameter of the orbifold singularity.

It will be useful to use  the orthonormal frame and  introduce the vielbeins
\begin{equation}
 e^0=f\,dr,
 \qquad e^1=r\sigma_x,
 \qquad e^2=r\sigma_y,
 \qquad e^3=rf^{-1}\sigma_z,
 \label{eq:EH-orthonormal-frame}
\end{equation}
and the orientation used throughout this work is
\begin{equation}
 \int_{S^3/\mathbb Z_2}
 \sigma_x\wedge\sigma_y\wedge\sigma_z=-\pi^2.
 \label{eq:EH-angular-volumebulk}
\end{equation}

An important feature of the EH space is that it supports abelian and
nonabelian self-dual gauge fields without modifying the
gravitational background, since their stress-energy tensor vanishes
identically. A regular abelian background is locally represented by
\cite{Eguchi:1978gw}
\begin{equation}
 A^{\rm B}
 ={\cal C}_{(1)}\frac{a^2}{r^2}\sigma_z,
 \label{form of ABUL}
\end{equation}
where ${\cal C}_{(1)}$ labels the $U(1)$ flux sector. Its globally defined
field strength is
\begin{align}
 F^{\rm B}=dA^{\rm B}
 ={}&-\frac{{\cal C}_{(1)}a^2}{r^3}dr\wedge
 \left(d\psi+\cos\theta\,d\varphi\right)
 +\frac{{\cal C}_{(1)}a^2}{2r^2}\sin\theta\,
 d\varphi\wedge d\theta
 \nonumber\\
 ={}&\frac{2{\cal C}_{(1)}a^2}{r^4}
 \left(e^3\wedge e^0+e^1\wedge e^2\right).
 \label{feildstrngth}
\end{align}
With the EH orientation adopted in \eqref{eq:EH-angular-volumebulk}, the Hodge
star acts as
\begin{equation}
 \star\left(e^3\wedge e^0\right)=e^1\wedge e^2,
 \qquad
 \star\left(e^1\wedge e^2\right)=e^3\wedge e^0.
 \label{eq:EH-Hodge-star-conventionbulk}
\end{equation}
It follows immediately that $F_{(1)}^{\rm B}$ is self-dual
\begin{equation}
 \star  F^{\rm B}= F^{\rm B},
\end{equation}
and hence,  the stress-energy tensor of $F^{\rm B}$ vanishes identically. Therefore, the background does not backreact on the EH metric.

 The curvature $F^{\rm B}$ is smooth, normalizable, and regular at the bolt, while for nonzero
${\cal C}$ the potential must be understood patchwise as a connection
on a nontrivial line bundle. Its bolt flux is
\begin{equation}
 \int_{S^2_{\rm bolt}}F^{\rm B}=-2\pi{\cal C}_{(1)},
\end{equation}
so ${\cal C}_{(1)}\in\mathbb Z$ for matter of minimal $U(1)$ charge one, while its topological charge is
\begin{eqnarray}
Q_{(1)}=\frac{1}{8\pi^2}\int_{\rm EH} F^{\rm B}\wedge F^{\rm B}=\frac{{\cal C}_{(1)}^2}{4}\,.
\label{U1TOPCHARGE}
\end{eqnarray}

The fractional topological charges admitted by EH space can be traced
to its nonunimodular intersection lattice and its nontrivial asymptotic
boundary.  The second homology is generated by the bolt
$\Sigma\simeq S^2$,
\begin{equation}
 H_2({\rm EH},\mathbb Z)
 \simeq
 \mathbb Z[\Sigma],
 \qquad
 [\Sigma]\cdot[\Sigma]=\mp2,
 \label{eq:EH-intersection-lattice}
\end{equation}
where the sign corresponds to the EH orientation.  As an
intersection lattice, $H_2({\rm EH},\mathbb Z)$ is isomorphic, up to
sign, to the $A_1\equiv \mathfrak{su}(2)$ root lattice.  Since its intersection matrix has
determinant two, this lattice is not unimodular.  Denoting the lattice by $L$ and the dual lattice (the weight lattice) by $L^\vee$,
one has
\begin{equation}
 L^\vee/L
 \simeq
 \mathbb Z_2,
 \label{eq:EH-discriminant-group}
\end{equation}
in agreement with the torsion group of the asymptotic boundary,
\begin{equation}
 H_1(\partial{\rm EH},\mathbb Z)
 =
 H_1(\mathbb{RP}^3,\mathbb Z)
 \simeq
 \mathbb Z_2.
 \label{eq:EH-boundary-torsion}
\end{equation}
Gauge fluxes carrying nontrivial asymptotic data are naturally measured
in the dual lattice $L^\vee$.  Their quadratic form is determined by the
inverse intersection matrix that yields a factor of
$1/2$. The factor of $1/4$ of the topological charge in (\ref{U1TOPCHARGE}) is attributed to the $1/2$ coming from the inverse intersection matrix, while another factor of $1/2$ comes from the definition of the topological charge.  The reader is referred to \cite{Anber:2026pfg} for a more thorough discussion of the abelian field theory on ALE spaces.

\subsection{$SU(2)$ and $PSU(2)$ background  gauge fields}

We now turn to gauge backgrounds with gauge groups $SU(2)$ and
$PSU(2)$. Such configurations can be constructed in two ways. The
first is to embed the $U(1)$ field (\ref{form of ABUL}) along the Cartan generator
of $\mathfrak{su}(2)$, with the global form of the bundle determined by
the quantized flux. The second is to seek a genuinely nonabelian
$\mathfrak{su}(2)$-valued connection. Substituting an
$SU(2)$-invariant ansatz into the self-duality condition of the gauge fields reduces the
problem to a coupled system of ordinary differential equations for
the radial gauge-field profiles.

\subsubsection*{Cartan-valued background}

We first consider a background along the Cartan generator of $SU(2)$,
\begin{equation}
 \label{abelianA}
 A^{\rm B}
 =m_{(2)}\frac{\tau_3}{2}\frac{a^2}{r^2}\sigma_z.
 \qquad m_{(2)}\in\mathbb Z,
\end{equation}
Here, we use $\tau_a$ for the Pauli matrices  of the $SU(2)$ color space; they obey
$\operatorname{tr}(\tau_a\tau_b)=2\delta_{ab}$.  Using
\eqref{eq:EH-Maurer-Cartan}, the field strength is
\begin{eqnarray}
 F^{\rm B}
 &=&\frac{2m_{(2)}a^2}{r^4}\frac{\tau_3}{2}
 \left(e^3\wedge e^0+e^1\wedge e^2\right)\,,
 \label{eq:EH-Cartan-field-strength-coordinate}
\end{eqnarray}
 its flux through the bolt is
\begin{equation}
 \int_{S^2_{\rm bolt}}F^{\rm B}
 =-2\pi m_{(2)}\frac{\tau_3}{2},
 \label{eq:EH-bolt-flux}
\end{equation}
while the exponentiation of the Lie-algebra-valued flux is
\begin{equation}
 \exp\left(i\int_{S^2_{\rm bolt}}F^{\rm B}\right)
 =(-1)^{m_{(2)}}I_2.
 \label{eq:EH-center-flux}
\end{equation}
Thus, $m_{(2)}\bmod2$ measures the $\mathbb Z_2^{(1)}$ center flux.  For odd
$m_{(2)}$ the configuration belongs naturally to a $PSU(2)=SU(2)/\mathbb Z_2$
bundle and does not lift to an $SU(2)$ bundle in the presence of fundamental
matter.  For even $m_{(2)}$ the center flux is trivial and such a lift is
possible. More on the difference between $SU(2)$ and $PSU(2)$ bundles is discussed below. 

The corresponding topological charge is
\begin{equation}
 \label{su2chargeQ2}
 Q_{(2)}
 =\frac{1}{8\pi^2}\int_{\rm EH}
   \operatorname{tr}\!\left(F^{\rm B}\wedge F^{\rm B}\right)
 =\frac{m_{(2)}^2}{8}.
\end{equation}
 The integer $m_{(2)}$ will also determine the
index of the adjoint Dirac operator and the allowed angular quantum numbers of fermions.

\subsubsection*{Nonabelian $SU(2)$ background}

To obtain a fully nonabelian $SU(2)$  solution, we may propose an ansatz for the background gauge field $A^{{\rm B}}$ and impose the self-duality condition
\begin{equation}
 \bar{\hat\sigma}^{ab}E_a^{\mu}E_b^{\nu}F_{\mu\nu}^{\rm B}=0,
 \label{eq:EH-gauge-self-duality}
\end{equation}
which amounts to projecting out the anti-self-dual field\footnote{Under the Euclidean rotation group
$\operatorname{Spin}(4)\simeq SU(2)_L\times SU(2)_R$, a two-form decomposes as
$\Lambda^2=\Lambda^2_+\oplus\Lambda^2_-\simeq(1,0)\oplus(0,1)$. Introducing the orthonormal-frame components
$F_{ab}^{\rm B}=E_a{}^\mu E_b{}^\nu F_{\mu\nu}^{\rm B}$, the two irreducible components are represented by the  spinors
$F_{\alpha\beta}^{\rm B}=(\hat\sigma^{ab})_{\alpha\beta}F_{ab}^{\rm B}$ and
$F_{\dot\alpha\dot\beta}^{\rm B}=(\bar{\hat\sigma}^{ab})_{\dot\alpha\dot\beta}F_{ab}^{\rm B}$.
With the orientation and sigma-matrix conventions used here,
$\hat\sigma^{ab}$ is self-dual and $\bar{\hat\sigma}^{ab}$ is anti-self-dual. Consequently, the self-duality condition
$F_{ab}^{\rm B}=\frac12\epsilon_{ab}{}^{cd}F_{cd}^{\rm B}$
is equivalent to the vanishing of the $(0,1)$ component,
$F_{\dot\alpha\dot\beta}^{\rm B}=0$, or, equivalently,
$\bar{\hat\sigma}^{ab}E_a{}^\mu E_b{}^\nu F_{\mu\nu}^{\rm B}=0$.}. 
Here
\begin{equation}
 \hat\sigma^a=(I_2,i\vec\sigma),
 \qquad
 \bar{\hat\sigma}^a=(I_2,-i\vec\sigma),
\end{equation}
where $\vec \sigma$ are the spin Pauli matrices and
\begin{eqnarray}
 \hat\sigma^{ab}
 \equiv\frac12\left(\hat\sigma^a\bar{\hat\sigma}^b
                   -\hat\sigma^b\bar{\hat\sigma}^a\right)\,,\quad
 \bar{\hat\sigma}^{ab}
 \equiv\frac12\left(\bar{\hat\sigma}^a\hat\sigma^b
                   -\bar{\hat\sigma}^b\hat\sigma^a\right).
 \label{eq:EH-sigma-ab}
\end{eqnarray}
The Latin letters $a,b=0,1,2,3$ are the orthonormal-frame indices, while the Greek letters $\mu,\nu$ denote $r,\theta,\varphi,\psi$, which are the curved-space indices. The inverse vielbein $E_a^{\mu}$ (defined via $e^a_\mu E_b^\mu=\delta^a_b$)  converts curved-space indices into orthonormal-frame
indices.  A self-dual Yang--Mills field automatically satisfies the
Yang--Mills equations by the Bianchi identity.  It also has a vanishing
stress-energy tensor
and therefore does not backreact on the EH metric.

Consider the cohomogeneity-one ansatz
\begin{equation}
 \label{generalansatzB}
 A^{\rm B}
 =G(r)\sigma_z\frac{\tau_3}{2}
 +F(r)\left(\sigma_x\frac{\tau_1}{2}
            +\sigma_y\frac{\tau_2}{2}\right),
\end{equation}
where both $G(r)$ and $F(r)$ depend only on one coordinate,  the radial direction. 
For Hermitian generators, the field strength is
$F^{\rm B}=dA^{\rm B}+iA^{\rm B}\wedge A^{\rm B}
=\sum_{a=1}^3F_a^{\rm B}\tau_a/2$.  Its color components are
\begin{align}
 F_1^{\rm B}
 &=F'\,dr\wedge\sigma_x+F(2-G)\sigma_y\wedge\sigma_z,
 \nonumber\\
 F_2^{\rm B}
 &=F'\,dr\wedge\sigma_y+F(2-G)\sigma_z\wedge\sigma_x,
 \\
 F_3^{\rm B}
 &=G'\,dr\wedge\sigma_z+(2G-F^2)\sigma_x\wedge\sigma_y.
 \nonumber
\end{align}
In the orthonormal frame, condition \eqref{eq:EH-gauge-self-duality} becomes
\begin{equation}
 {\cal A}(r)\left(\hat\sigma^1\frac{\tau_1}{2}
                  +\hat\sigma^2\frac{\tau_2}{2}\right)
 +{\cal B}(r)\hat\sigma^3\frac{\tau_3}{2}
=0,
\end{equation}
where
\begin{equation}
 {\cal A}(r)=\frac{F'}{fr}+\frac{fF}{r^2}(2-G),
 \qquad
 {\cal B}(r)=\frac{G'}{r}+\frac{2G-F^2}{r^2}.
\end{equation}
Consequently, the self-duality equations reduce to
\begin{equation}
 \label{basicDE}
 rF'+f^2F(2-G)=0,
 \qquad
 rG'+2G-F^2=0.
\end{equation}

The topological charge of the ansatz is a boundary term.  Defining
\begin{equation}
 {\cal K}(r)\equiv2F^2(r)+G^2(r)-F^2(r)G(r),
\end{equation}
one obtains
\begin{align}
 \label{general su2tc}
 Q_{(2)}
 &=\frac{1}{8\pi^2}\int_{\rm EH}
   \operatorname{tr}\!\left(F^{\rm B}\wedge F^{\rm B}\right)
 =\frac{1}{16\pi^2}\int_{\rm EH}F_a^{\rm B}\wedge F_a^{\rm B}
 \nonumber\\
 &=\frac18\left[\mathcal K(a)-\mathcal K(\infty)\right].
\end{align}
For the solutions considered below, $F(\infty)=G(\infty)=0$, and hence
\begin{equation}Q_{(2)}=\frac{{\cal K}(a)}{8}\,.\end{equation}

\subsubsection*{The minimal nonabelian solution}

The minimal nontrivial nonabelian solution is a one-parameter family that has
$Q_{(2)}=1/2$ and is given by
\begin{equation}
 \label{su2solsd}
 G(r)=\frac{2(t^2r^2+a^4)}{r^2(t^2+r^2)},
 \qquad
 F(r)=\frac{2\sqrt{t^4-a^4}}{r^2+t^2},
 \qquad
 t\in[a,\infty).
\end{equation}
Direct substitution verifies \eqref{basicDE}. At the bolt,
\begin{equation}
 G(a)=2,
 \qquad
 F^2(a)=4\frac{t^2-a^2}{t^2+a^2},
\end{equation}
so that ${\cal K}(a)=4$ independently of $t$ and therefore
\begin{equation}Q_{(2)}=\frac12.\end{equation} The parameter $t$ is the radial-size modulus of the gauge
configuration.

At the endpoint $t=a$, the solution becomes
\begin{equation}
 F(r)=0,
 \qquad
 G(r)=\frac{2a^2}{r^2},
\end{equation}
and \eqref{generalansatzB} reduces to the Cartan-valued background
\eqref{abelianA} with $m_{(2)}=2$. Since $m_{(2)}$ is even, the
corresponding obstruction class  vanishes,
\begin{equation}
 w_2=m_{(2)}\ {\rm mod}\ 2=0,
\end{equation}
and the gauge bundle admits an $SU(2)$ lift. Here, $w_2\in H^2(\mbox{EH},\mathbb Z_2)$ is the second
Stiefel--Whitney class of the $PSU(2)\simeq SO(3)$ gauge bundle. It
measures the obstruction to lifting $PSU(2)$ to an $SU(2)$ bundle:
$w_2=0$ if such a lift exists, whereas $w_2\neq0$ characterizes
a genuinely nonliftable $PSU(2)$ bundle. Because the entire
one-parameter family is obtained by a continuous deformation of this
endpoint without changing its asymptotic bundle data, every finite
$t\geq a$ belongs to the same liftable sector\footnote{The formal limit $t\rightarrow\infty$, however,  changes the
asymptotic behavior of the connection and it must therefore be treated
separately from the finite-$t$ family; see \cite{Boutaleb-Joutei:1979hdt}.}. Thus, in a theory with
global gauge group $SU(2)$, the solution is an admissible $SU(2)$
bundle. In a theory with global gauge group $PSU(2)$, the same
connection describes a liftable $PSU(2)$ bundle in the $w_2=0$
sector; it is not a genuinely nonliftable $PSU(2)$ configuration.

With the Yang--Mills normalization used in this work, the classical
action of the $Q_{(2)}=1/2$ instanton is
\begin{equation}
 S_{\rm cl}
 =\frac{8\pi^2}{g^2}Q_{(2)}
 =\frac{4\pi^2}{g^2},
 \label{eq:EH-minimal-classical-action}
\end{equation}
and $g$ is the gauge coupling constant. 
\subsection*{Liftable and nonliftable solutions}

It is important to distinguish the local self-duality equations from
the global properties of the gauge bundle. The equations
${\cal A}(r)={\cal B}(r)=0$ constrain only the local radial profiles
$F(r)$ and $G(r)$; by themselves, they do not determine whether the
connection is defined on an $SU(2)$ bundle or on a nonliftable
$PSU(2)$ bundle. The simplest branch is obtained by setting
\begin{equation}
 F(r)=0.
\end{equation}
The remaining equation then reduces to
\begin{equation}
 rG'(r)+2G(r)=0,
\end{equation}
whose solution is
\begin{equation}
 G(r)=m_{(2)}\frac{a^2}{r^2}.
 \label{eq:reducible-general-m-solution}
\end{equation}
This is precisely the Cartan-valued background discussed above. Its
global character is determined by the parity of $m_{(2)}$:
\begin{equation}
 w_2=\bigl(m_{(2)}\ {\rm mod}\ 2\bigr)\eta,
 \qquad
 \eta\in H^2({\rm EH},\mathbb Z_2).
\end{equation}
For even $m_{(2)}$, the obstruction vanishes and the corresponding
$PSU(2)$ bundle admits an $SU(2)$ lift. For odd $m_{(2)}$, one has
$w_2\neq0$, and the configuration is defined on a genuinely
nonliftable $PSU(2)$ bundle. Thus, the self-duality equations do
possess genuine $PSU(2)$ solutions. These solutions are nevertheless
reducible, since their connections take values only in a local
$\mathfrak u(1)\subset\mathfrak{su}(2)$ subalgebra. For example,
\begin{equation}
 F(r)=0,
 \qquad
 G(r)=\frac{a^2}{r^2}
\end{equation}
has $m_{(2)}=1$, $w_2\neq0$, and
$Q_{(2)}=1/8$. It is therefore a genuine $PSU(2)$ configuration
globally, even though it is Cartan-valued.

The situation is more restrictive for irreducible solutions with
$F(r)\neq0$. Introducing the radial coordinate $\rho$ and the
integration constant $\beta$ through \cite{Boutaleb-Joutei:1979hdt}
\begin{equation}
 \coth\rho=\frac{r^2}{a^2},
 \qquad \cosh\beta\equiv\frac{t^2}{\sqrt{t^4-a^4}}\,,
 \beta\geq0,
\end{equation}
the nonabelian solutions of the present cohomogeneity-one equations (\ref{basicDE})
take the form
\begin{equation}
 G(\rho)
 =
 2p\tanh\rho\,
 \coth\bigl[p(\rho+\beta)\bigr],
 \qquad
 F(\rho)
 =
 \frac{2p\sinh\rho}
 {\sinh\bigl[p(\rho+\beta)\bigr]}.
 \label{eq:general-p-nonabelian-solution}
\end{equation}
The bolt corresponds to $\rho\rightarrow\infty$, where
\begin{equation}
 G(a)=2p,
 \qquad
 F(\rho)\sim
 2p\,e^{-p\beta}e^{-(p-1)\rho}.
 \label{eq:general-p-bolt-behavior}
\end{equation}
Smooth extension through the bolt requires
$p\in\mathbb Z_{>0}$. Consequently, the flux parameter associated
with every smooth nonabelian solution in this family is
\begin{equation}
 m_{(2)}=G(a)=2p\in2\mathbb Z.
\end{equation}
These solutions therefore satisfy $w_2=0$ and admit an $SU(2)$
lift. The minimal member is $p=1$, for which
$m_{(2)}=2$ and $Q_{(2)}=1/2$.

One might attempt to obtain an odd value of $m_{(2)}$ by taking $p$
to be half-integral. This does not produce a smooth irreducible
solution. If $u$ denotes the proper radial distance from the bolt,
then $e^{-\rho}$ is proportional to $u$ near $u=0$, and hence
\begin{equation}
 F(u)\sim u^{p-1}.
\end{equation}
For $p=1/2$, the off-diagonal profile diverges at the bolt, while for
larger half-integral values of $p$ it contains a noninteger power of
$u$ and fails to extend smoothly through the bolt\footnote{ A smooth field configuration must be infinitely differentiable near the bolt.}. The only smooth
odd-$m_{(2)}$ configuration obtained from this branch arises in the
reducible limit $\beta\rightarrow\infty$, in which
$F(r)\rightarrow0$ and the solution returns to
\eqref{eq:reducible-general-m-solution}.

We therefore conclude that the present ansatz contains genuine
nonliftable $PSU(2)$ solutions for odd $m_{(2)}$, but only on its
reducible Cartan-valued branch. All smooth irreducible solutions with
$F(r)\neq0$ obtained from this ansatz lie in the even-$m_{(2)}$,
$w_2=0$ sector and are liftable to $SU(2)$. This conclusion applies
to the particular cohomogeneity-one ansatz used here; it does not
constitute a general nonexistence theorem for irreducible
$PSU(2)$ instantons on EH space. A more general nonliftable solution,
if it exists, would have to be formulated patchwise on the
nontrivial $SO(3)$ bundle and would obey global transition and bolt
regularity conditions not encoded by the two globally defined radial
functions $F(r)$ and $G(r)$ alone.

\subsection{$SU(N)$ backgrounds}
\label{subsec:SU-N-self-dual-instantons}
In this section, we embed the $SU(2)$ background in the $SU(N)$ group to obtain self-dual solutions carrying a topological charge of $1/2$.
We normalize the Hermitian generators in the fundamental
representation according to
\begin{equation}
 \operatorname{tr}_{\mathbf N}(T^aT^b)
 =
 \frac12\delta^{ab}.
\end{equation}
It is nevertheless convenient to formulate the construction in a
root-normalized Cartan--Weyl basis for $\mathfrak{su}(N)$,
\begin{eqnarray}
 [H_I,E_{\boldsymbol\beta}]
 =
 \beta_I E_{\boldsymbol\beta},\quad
 [E_{\boldsymbol\beta},E_{-\boldsymbol\beta}]
 =
 \boldsymbol\beta\cdot\boldsymbol H,\quad
 [E_{\boldsymbol\beta},E_{\boldsymbol\gamma}]
 =
 {\cal N}_{\boldsymbol\beta,\boldsymbol\gamma}
 E_{\boldsymbol\beta+\boldsymbol\gamma},
 \label{eq:SU-N-Cartan-Weyl-algebra}
\end{eqnarray}
where the last relation applies whenever
$\boldsymbol\beta+\boldsymbol\gamma$ is a root.  The Cartan--Weyl
elements are normalized as
\begin{eqnarray}
 \operatorname{tr}_{\mathbf N}(H_IH_J)=\delta_{IJ},\quad
 \operatorname{tr}_{\mathbf N}
 \left(E_{\boldsymbol\beta}E_{-\boldsymbol\gamma}\right)
 =
 \delta_{\boldsymbol\beta\boldsymbol\gamma}.
 \label{weylbasesnorm}
\end{eqnarray}
 For every root
$\boldsymbol\beta$, the associated Hermitian $SU(2)$ generators are
\begin{equation}
 T_{\boldsymbol\beta}^1
 =
 \frac12
 \left(
  E_{\boldsymbol\beta}+E_{-\boldsymbol\beta}
 \right),
 \qquad
 T_{\boldsymbol\beta}^2
 =
 \frac{1}{2i}
 \left(
  E_{\boldsymbol\beta}-E_{-\boldsymbol\beta}
 \right),
 \qquad
 T_{\boldsymbol\beta}^3
 =
 \frac12\boldsymbol\beta\cdot\boldsymbol H.
 \label{eq:SU-N-root-SU2-Hermitian-generators}
\end{equation}
For $\boldsymbol\beta^2=2$, these obey
$
 \operatorname{tr}_{\mathbf N}
 \left(
  T_{\boldsymbol\beta}^aT_{\boldsymbol\beta}^b
 \right)
 =
 \frac12\delta^{ab}.
 \label{eq:SU-N-root-SU2-generator-normalization}
$
The roots $\bm \beta$ and the simple roots $\bm\alpha$ of $\mathfrak{su}(N)$  may be represented in the hyperplane
$\sum_{i=1}^{N}x_i=0$ of $\mathbb R^N$ as
\begin{equation}
 \boldsymbol\beta_{ij}
 =
 \boldsymbol e_i-\boldsymbol e_j,
 \qquad
 1\leq i<j\leq N,
 \qquad
 \boldsymbol\alpha_i
 =
 \boldsymbol e_i-\boldsymbol e_{i+1},
 \qquad
 i=1,\ldots,N-1,
 \label{allroots}
\end{equation}
where $\{ \boldsymbol e_i\}$ are unit vectors on $\mathbb R^N$,
so that every root has length\footnote{With the roots normalized to have squared length
$\bm\beta^2=2$, the simply laced algebra $su(N)$ admits the standard
identification of the Cartan subalgebra with its dual, under which roots
coincide with coroots and weights coincide with coweights.}
$\boldsymbol\beta^2=2$.  The set of positive roots, denoted by
$\Delta_+$, contains $N(N-1)/2$ elements.

\paragraph{Cohomogeneity-one ansatz.}
Consider the gauge potential
\begin{equation}
 A^{\rm B}
 =
 \boldsymbol G(r)\cdot\boldsymbol H\,\sigma_z
 +
 P(r)\sigma_-
 +
 P^\dagger(r)\sigma_+,
 \qquad
 P(r)
 =
 \sum_{\boldsymbol\beta\in\Delta_+}
 F_{\boldsymbol\beta}(r)E_{\boldsymbol\beta}.
 \label{generalASUN}
\end{equation}
Here the functions $F_{\boldsymbol\beta}$ are taken to be real and
$\sigma_\pm=\sigma_x\pm i\sigma_y$.  With the convention
\begin{equation}
 F^{\rm B}=dA^{\rm B}+iA^{\rm B}\wedge A^{\rm B},
\end{equation}
the field strength is
\begin{align}
 F^{\rm B}
 ={}&
 \boldsymbol G'\cdot\boldsymbol H\,
 dr\wedge\sigma_z
 +
 P'\,dr\wedge\sigma_-
 +
 (P^\dagger)'\,dr\wedge\sigma_+
 \nonumber\\
 &+
 i\left(
  -\boldsymbol G\cdot\boldsymbol H
  +[P,P^\dagger]
 \right)
 \sigma_-\wedge\sigma_+
 \nonumber\\
 &+
 i\left(
  2P-[\boldsymbol G\cdot\boldsymbol H,P]
 \right)
 \sigma_-\wedge\sigma_z
 \nonumber\\
 &+
 i\left(
  -2P^\dagger-[\boldsymbol G\cdot\boldsymbol H,P^\dagger]
 \right)
 \sigma_+\wedge\sigma_z.
 \label{FBORI}
\end{align}

An important point is that the commutator in the second line is not
generically Cartan-valued.  Rather,
\begin{eqnarray}
 [P,P^\dagger]
 ={}
 \sum_{\boldsymbol\beta\in\Delta_+}
 F_{\boldsymbol\beta}^2\,
 \boldsymbol\beta\cdot\boldsymbol H
 +
 {\cal C},\quad
 {\cal C}
 \equiv{}
 \sum_{\substack{
  \boldsymbol\beta,\boldsymbol\gamma\in\Delta_+\\
  \boldsymbol\beta\neq\boldsymbol\gamma
 }}
 F_{\boldsymbol\beta}F_{\boldsymbol\gamma}
 [E_{\boldsymbol\beta},E_{-\boldsymbol\gamma}].
 \label{eq:SU-N-cross-root-commutator}
\end{eqnarray}
The term ${\cal C}$ is absent only for a restricted set of root
profiles.  In particular, it vanishes if a single root is excited, or
if the nonzero profiles are associated with pairwise strongly
orthogonal roots.

Using the orthonormal frame (\ref{eq:EH-orthonormal-frame}), 
the Cartan components of the self-duality condition  $\bar{\hat\sigma}^{ab}E_a^{\mu}E_b^{\nu}F_{\mu\nu}^{\rm B}=0$ give
\begin{equation}
 \frac{\boldsymbol G'}{r}
 +
 \frac{2\boldsymbol G}{r^2}
 -
 \frac{2}{r^2}
 \sum_{\boldsymbol\beta\in\Delta_+}
 F_{\boldsymbol\beta}^2\boldsymbol\beta
 =
 0,
 \label{eq:SU-N-Cartan-self-duality-equation}
\end{equation}
and
\begin{equation}
 \frac{F_{\boldsymbol\beta}'}{fr}
 +
 \frac{fF_{\boldsymbol\beta}}{r^2}
 \left(
  2-\boldsymbol G\cdot\boldsymbol\beta
 \right)
 =
 0,
 \qquad
 \boldsymbol\beta\in\Delta_+.
 \label{eq:SU-N-root-self-duality-equation}
\end{equation}
Finally, the non-Cartan part of $[P,P^\dagger]$ occurs in the
$e^1\wedge e^2$ component without a corresponding
$e^0\wedge e^3$ component.  Self-duality therefore also requires
\begin{equation}
 {\cal C}
 =
 \sum_{\substack{
  \boldsymbol\beta,\boldsymbol\gamma\in\Delta_+\\
  \boldsymbol\beta\neq\boldsymbol\gamma
 }}
 F_{\boldsymbol\beta}F_{\boldsymbol\gamma}
 [E_{\boldsymbol\beta},E_{-\boldsymbol\gamma}]
 =
 0.
 \label{eq:SU-N-cross-root-constraint}
\end{equation}
Equations
\eqref{eq:SU-N-Cartan-self-duality-equation},
\eqref{eq:SU-N-root-self-duality-equation}, and
\eqref{eq:SU-N-cross-root-constraint} are the complete conditions
within the ansatz \eqref{generalASUN}.  The first two equations alone
are sufficient only after the additional constraint
\eqref{eq:SU-N-cross-root-constraint} has been imposed.
\begin{figure}[t]
 \centering
 \includegraphics[width=\textwidth]
 {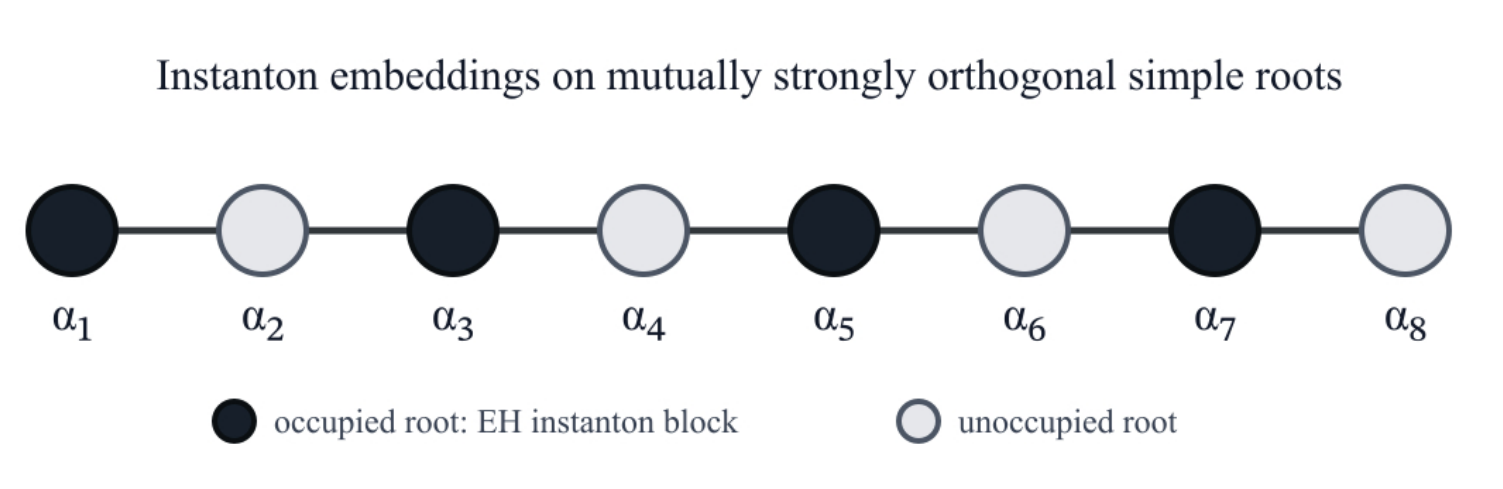}
 \caption{Dynkin-diagram representation of the instanton embedding, illustrated
 for $A_8\simeq\mathfrak{su}(9)$. The dark nodes denote the mutually strongly
 orthogonal simple roots $\alpha_1,\alpha_3,\alpha_5,\alpha_7$, each supporting
 an independent EH instanton block, while the light nodes denote
 the unoccupied simple roots.}
 \label{fig:strongly-orthogonal-root-instantons}
\end{figure}

A useful class of solutions is obtained by choosing a set
${\cal S}\subset\Delta_+$ of pairwise strongly orthogonal roots,
\begin{equation}
 [E_{\boldsymbol\beta},E_{-\boldsymbol\gamma}]=0,
 \qquad
 \boldsymbol\beta,\boldsymbol\gamma\in{\cal S},
 \qquad
 \boldsymbol\beta\neq\boldsymbol\gamma.
 \label{eq:SU-N-strongly-orthogonal-root-condition}
\end{equation}
For $SU(N)$, the roots
\begin{equation}
 \boldsymbol\alpha_1,\boldsymbol\alpha_3,
 \boldsymbol\alpha_5,\ldots
\end{equation}
provide a standard example and correspond to mutually disjoint
$SU(2)$ blocks; see Figure \ref{fig:strongly-orthogonal-root-instantons}.

\paragraph{Single-root embedding.}
The simplest solution is obtained by exciting only a simple root, e.g., 
$\boldsymbol\alpha_1=\boldsymbol e_1-\boldsymbol e_2$:
\begin{equation}
 \boldsymbol G
 =
 \frac{\boldsymbol\alpha_1}{2}G,
 \qquad
 F_{\boldsymbol\alpha_1}
 =
 \frac{F}{2},
 \qquad
 F_{\boldsymbol\beta}=0
 \quad
 \text{for }
 \boldsymbol\beta\neq\boldsymbol\alpha_1\,.
 \label{eq:SU-N-single-root-reduction}
\end{equation}
Because $\boldsymbol\alpha_1^2=2$, equations
\eqref{eq:SU-N-Cartan-self-duality-equation} and
\eqref{eq:SU-N-root-self-duality-equation} reduce to
\begin{equation}
 \frac{G'}{r}
 +
 \frac{2G-F^2}{r^2}
 =
 0,
 \qquad
 \frac{F'}{fr}
 +
 \frac{fF}{r^2}(2-G)
 =
 0.
 \label{generalsunsdb}
\end{equation}
These are precisely the $SU(2)$ self-duality equations \eqref{basicDE} discussed in the previous section.  Thus, one finds
\begin{equation}
 \boldsymbol G(r)
 =
 \boldsymbol\alpha_1
 \frac{t^2r^2+a^4}{r^2(r^2+t^2)},
 \qquad
 F_{\boldsymbol\alpha_1}(r)
 =
 \frac{\sqrt{t^4-a^4}}{r^2+t^2},
 \qquad
 t\in[a,\infty),
 \label{specialsunsun}
\end{equation}
with all other root profiles equal to zero.

For a solution satisfying
\eqref{eq:SU-N-cross-root-constraint}, define
\begin{equation}
 {\cal K}(r)
 =
 -\frac14\boldsymbol G(r)^2
 -
 \frac12
 \sum_{\boldsymbol\beta\in\Delta_+}
 F_{\boldsymbol\beta}(r)^2
 \left(
  2-\boldsymbol G(r)\cdot\boldsymbol\beta
 \right).
 \label{eq:SU-N-Chern-Simons-boundary-function}
\end{equation}
Direct calculation gives
\begin{equation}
 \operatorname{tr}_{\mathbf N}
 \left(F^{\rm B}\wedge F^{\rm B}\right)
 =
 -8\frac{d{\cal K}}{dr}\,
 dr\wedge\sigma_x\wedge\sigma_y\wedge\sigma_z.
 \label{eq:SU-N-topological-density-total-derivative}
\end{equation}
Using the orientation convention
$
 \int_{S^3/\mathbb Z_2}
 \sigma_x\wedge\sigma_y\wedge\sigma_z
 =
 -\pi^2,
$
the topological charge becomes
\begin{eqnarray}
 Q_{(N)}
 =
 \frac{1}{8\pi^2}
 \int_{\rm EH}
 \operatorname{tr}_{\mathbf N}
 \left(F^{\rm B}\wedge F^{\rm B}\right)
 =
 \left[
 -\frac14\boldsymbol G^2
 -
 \frac12
 \sum_{\boldsymbol\beta\in\Delta_+}
 F_{\boldsymbol\beta}^2
 \left(
  2-\boldsymbol G\cdot\boldsymbol\beta
 \right)
 \right]_{r=a}^{r=\infty}.
 \label{QSUN}
\end{eqnarray}

For the single-root solution \eqref{specialsunsun},
$
 \boldsymbol G(\infty)=0,
 F_{\boldsymbol\alpha_1}(\infty)=0,
$
whereas at the bolt
$
 \boldsymbol G(a)=\boldsymbol\alpha_1,
 \boldsymbol G(a)^2=2,
 \boldsymbol G(a)\cdot\boldsymbol\alpha_1=2.
$
The root contribution to \eqref{QSUN} therefore vanishes at the bolt,
independently of the value of $F_{\boldsymbol\alpha_1}(a)$, and
$
 {\cal K}(\infty)=0,
 {\cal K}(a)=-\frac12.
$
Consequently,
\begin{equation}
Q_{(N)}=\frac12.
 \label{eq:SU-N-single-root-topological-charge}
\end{equation}
This agrees with the direct embedding. Writing: 
\begin{equation}
 A_{(N)}^{\rm B}(t)
 =
 \begin{pmatrix}
  A_{(2)}^{\rm B}(t)&0\\
  0&0_{N-2}
 \end{pmatrix},
 \label{eq:SUN-single-block-embedding1}
\end{equation}
where $  A_{(2)}^{\rm B}(t)$ is given by (\ref{generalansatzB}, \ref{su2solsd}),
the trace over the active $2\times2$ block is
identical to the fundamental trace in the $SU(2)$ theory.

\paragraph{Embeddings along mutually orthogonal roots.}
The single-root construction can be extended by choosing a set of mutually
orthogonal roots
\begin{equation}
 {\cal S}
 =
 \left\{
  \boldsymbol\beta_1,\ldots,\boldsymbol\beta_s
 \right\},
 \qquad
 \boldsymbol\beta_I\cdot\boldsymbol\beta_J
 =
 2\delta_{IJ},
 \qquad
 s\leq\left\lfloor\frac{N}{2}\right\rfloor.
 \label{eq:SUN-mutually-orthogonal-roots}
\end{equation}
For $SU(N)$, such roots may be chosen to be mutually orthogonal simple roots (see Figure~\ref{fig:strongly-orthogonal-root-instantons})
\begin{equation}\boldsymbol\beta_I=\boldsymbol\alpha_{2I-1}=\boldsymbol e_{2I-1}-\boldsymbol e_{2I}.\end{equation}
Their associated $\mathfrak{su}(2)$ subalgebras commute, and hence one may
embed an independent EH solution in each two-dimensional block:
\begin{equation}
 A^{\rm B}
 =
 \sum_{I=1}^{s}
 \left[
  \frac{G_I(r)}{2}
  \boldsymbol\beta_I\cdot\boldsymbol H\,\sigma_z
  +
  \frac{F_I(r)}{2}
  \left(
   E_{\boldsymbol\beta_I}\sigma_-
   +E_{-\boldsymbol\beta_I}\sigma_+
  \right)
 \right].
 \label{eq:SUN-multiple-root-embedding}
\end{equation}
Because the different $\mathfrak{su}(2)$ factors commute, the field strength
decomposes as $F^{\rm B}=\sum_I F_I^{\rm B}$. Moreover, orthogonality of the
roots implies
\begin{equation}
 \operatorname{tr}_{\boldsymbol N}
 \left(
  F_I^{\rm B}\wedge F_J^{\rm B}
 \right)=0,
 \qquad I\neq J.
 \label{eq:SUN-multiple-root-cross-terms}
\end{equation}
The topological charge is therefore additive:
\begin{align}
 Q_{(N)}
 =
 \frac{1}{8\pi^2}
 \int_{\rm EH}
 \operatorname{tr}_{\boldsymbol N}
 \left(
  F^{\rm B}\wedge F^{\rm B}
 \right)
 =
 \sum_{I=1}^{s}
 \frac{1}{8\pi^2}
 \int_{\rm EH}
 \operatorname{tr}_{\boldsymbol N}
 \left(
  F_I^{\rm B}\wedge F_I^{\rm B}
 \right)
 =
 \frac12\sum_{I=1}^{s}p_I^2,
 \label{eq:SUN-multiple-root-topological-charge}
\end{align}
where $p_I$ labels the $SU(2)$ solution embedded along
$\boldsymbol\beta_I$. In particular, the profile (\ref{specialsunsun}) is replaced by $s$ identical profiles with distinct radial-size moduli $t_{I}$, $I=1,..,s$:
\begin{equation}
 \boldsymbol G_I(r)
 =
 \boldsymbol\beta_I
 \frac{t_I^2r^2+a^4}{r^2(r^2+t_I^2)},
 \qquad
 F_{\boldsymbol\beta_I}(r)
 =
 \frac{\sqrt{t_I^4-a^4}}{r^2+t_I^2},
 \qquad
 t_I \in[a,\infty),
 \label{specialsunsungen}
\end{equation}
This family of solutions is continuously connected to the Cartan-valued solutions at the endpoint $t_{I}=a$:
\begin{equation}
 A^{\rm B}=H_{\boldsymbol p}\frac{a^2}{r^2}\sigma_z
 =
 \sum_{I=1}^{s}
 p_I\boldsymbol\beta_I\cdot\boldsymbol H \frac{a^2}{r^2}\sigma_z,
 \qquad
 Q_{(N)}
 =
 \frac14\operatorname{tr}_{\boldsymbol N}
 H_{\boldsymbol p}^{\,2}
 =
 \frac12\sum_{I=1}^{s}p_I^2.
\end{equation}
Recalling that these are the minimal $SU(2)$ solutions with $p_I=1$, we find
\begin{equation}Q_{(N)}=\frac s2,\end{equation} an integer multiple of $1/2$.  
An important example we shall consider in this work is the two-root embedding (block-diagonal) background in $SU(4)$:
\begin{equation}
 A_{(4)}^{\rm B}(t_1,t_2)
 =
 \begin{pmatrix}
  A_{(2)}^{\rm B}(t_1)&0\\
  0&  A_{(2)}^{\rm B}(t_2)
 \end{pmatrix},
 \label{eq:SU4-two-block-embedding1}
\end{equation}
with topological charge $Q_{(4)}=1$.

\paragraph{Liftability of the orthogonal-root configurations.}

The liftability of the configurations associated with $s$ mutually
orthogonal roots can be seen most directly at their Cartan endpoint, where
the field strength is Cartan-valued. Its flux through the bolt is
\begin{equation}
 \int_{S^2_{\rm bolt}}F^{\rm B}
 =
 -2\pi
 \sum_{I=1}^{s}\bm \beta_{I}\cdot \bm H,
\end{equation}
and hence
\begin{equation}
 \exp\left(
  i\int_{S^2_{\rm bolt}}F^{\rm B}
 \right)
 =
 \exp\left(
  -2\pi i
  \sum_{I=1}^{s}\bm\beta_{I}\cdot \bm H
 \right)
 =
 I_N.
 \label{eq:SUN-orthogonal-root-bolt-flux-exponential}
\end{equation}
The last equality follows because the charge is a sum of
$SU(N)$ roots, each of which exponentiates to the identity. Equivalently,
the transition function between two patches covering the bolt is
single-valued as an $SU(N)$-valued function. The full nonabelian
orthogonal-root solutions are continuously connected to these Cartan
configurations without changing their bundle topology. Therefore,
\eqref{eq:SUN-orthogonal-root-bolt-flux-exponential} establishes that the
entire family defines globally well-defined, liftable $SU(N)$ bundles.

\subsection{Comments on the minimal nonliftable $PSU(N)$ backgrounds}
\label{sec:minimal-PSUN-background}

For completeness, we briefly describe the minimal Cartan-valued
background associated with a genuine $PSU(N)=SU(N)/\mathbb Z_N$
bundle. Let $H$ belong to the weight lattice of $SU(N)$, and
consider
\begin{equation}
A^{\rm B}
 =
 H\,\frac{a^2}{r^2}\sigma_z .
 \label{eq:PSUN-Cartan-background}
\end{equation}
Since the connection is Cartan valued, its field strength is
\begin{equation}
F^{\rm B}
 =
 H\,d\left(\frac{a^2}{r^2}\sigma_z\right).
 \label{eq:PSUN-Cartan-field-strength}
\end{equation}

For an $SU(N)$ bundle, $H$ must belong to the root lattice. A
$PSU(N)$ bundle, by contrast, only requires
\begin{equation}
 \exp(2\pi iH)\in\mathbb Z_N.
 \label{eq:PSUN-coweight-condition}
\end{equation}
If this center element is nontrivial, the corresponding $PSU(N)$
bundle cannot be lifted to an $SU(N)$ bundle. The minimum-norm
representative of the obstruction class $k\in\mathbb Z_N$ is the
fundamental weight
\begin{equation}
 \omega_k
 =
 \operatorname{diag}\left(
 \underbrace{\frac{N-k}{N},\ldots,\frac{N-k}{N}}_{k},
 \underbrace{-\frac{k}{N},\ldots,-\frac{k}{N}}_{N-k}
 \right),
 \qquad
 k=1,\ldots,N-1.
 \label{eq:PSUN-fundamental-coweight}
\end{equation}
It obeys
\begin{equation}
 \exp\left(2\pi i\omega_k\right)
 =
 e^{-2\pi ik/N}I_N.
 \label{eq:PSUN-coweight-center-element}
\end{equation}
Correspondingly, the exponentiated flux through the bolt is
\begin{equation}
 \exp\left(
 i\int_{S^2_{\rm bolt}}{F}^{\rm B}
 \right)
 =
 e^{2\pi ik/N}I_N.
 \label{eq:PSUN-bolt-center-flux}
\end{equation}
For $k\neq0$ modulo $N$, this flux defines a genuine nonliftable
$PSU(N)$ bundle. This is consistent with what one expects based on the second cohomology $H^2(\mathrm{EH}, \mathbb Z_N)\cong \mbox{Hom}(\mathbb Z,\mathbb Z_{N})\cong\mathbb Z_{N}$, which can be calculated via the universal coefficient theorem\footnote{According to the universal coefficient theorem,
\begin{equation}
0\longrightarrow
\mathrm{Ext}\!\left(H_1(\rm {EH},\mathbb Z),\mathbb Z_N\right)
\longrightarrow
H^2(\rm {EH},\mathbb Z_N)
\longrightarrow
\mathrm{Hom}\!\left(H_2(\rm {EH},\mathbb Z),\mathbb Z_N\right)
\longrightarrow 0.
\end{equation}
Since $H_1(\rm{ EH},\mathbb Z)=0$ and $H_2(\rm{ EH},\mathbb Z)\cong \mathbb Z$, we find
\begin{equation}
H^2(\mathrm{EH},\mathbb Z_N)
\cong
\mathrm{Hom}(\mathbb Z,\mathbb Z_N)
\cong
\mathbb Z_N.
\end{equation}}.

With the normalization
$
 \operatorname{tr}_{\mathbf N}(T_aT_b)
 =
 \frac12\delta_{ab},
$
the topological charge of a Cartan-valued background is
\begin{equation}
 Q^{PSU(N)}
 =
 \frac{1}{8\pi^2}
 \int_{\rm EH}
 \operatorname{tr}_{\mathbf N}
 \left(
 { F}^{\rm B}\wedge{ F}^{\rm B}
 \right)
 =
 \frac14\operatorname{tr}_{\mathbf N}(H^2)= \frac{k(N-k)}{4N}.
 \label{eq:PSUN-Cartan-topological-charge}
\end{equation}
The smallest charge among all nontrivial obstruction sectors is
therefore obtained for $k=1$ or $k=N-1$:
\begin{equation}
 Q_{\rm min}^{PSU(N)}
 =
 \frac{N-1}{4N}.
 \label{eq:PSUN-absolute-minimal-charge}
\end{equation}

The Atiyah--Patodi--Singer (APS) index of an adjoint Weyl fermion vanishes in a background
constructed from a fundamental weight with minimum charge $ Q_{\rm min}^{PSU(N)}$ (see Section \ref{sec:EH-anomalies-APS} for the computations of the index):
\begin{equation}
 {\cal I}_{\rm adj}
 \left[\omega_k\right]
 =
 0.
 \label{eq:PSUN-minimal-adjoint-index}
\end{equation}
 The cancellation
occurs between the fractional bulk contribution and the combined
gravitational and boundary $\eta$-invariant contribution. Together
with the standard vanishing theorem for the opposite chirality (see Footnote~\ref{vanishingtheorem}) and
the requirement of square-integrability, the vanishing APS index
implies
\begin{equation}
 \ker_{L^2} D_{\rm adj}
 =
 0.
 \label{eq:PSUN-no-adjoint-zero-modes}
\end{equation}
In particular, these minimal Cartan-valued $PSU(N)$ backgrounds do not
possess normalizable adjoint gaugino zero modes and therefore cannot
generate gaugino condensates.
We shall consequently restrict the remainder of the analysis to
liftable $SU(N)$ bundles and will not discuss the nonliftable $PSU(N)$
sectors further.

\section{Path integral, boundary states,  chiral anomalies, and supersymmetry on EH space}
\label{sec:EH-boundary-anomalies-supersymmetry}

Since EH is noncompact, its Euclidean path integral is not naturally a
scalar partition function.  To formulate the path integral more precisely,
we regulate the geometry at a large radius $r=R$ and denote the resulting
compact manifold by $X_R$.  Its boundary is
\begin{equation}
 Y_\infty\equiv\partial X_R
 \simeq S^3/\mathbb Z_2
 \simeq\mathbb{RP}^3.
\end{equation}
The limit $R\rightarrow\infty$ is understood at the end.  According to the
usual cutting-and-gluing interpretation of the Euclidean path integral \cite{Atiyah:1989vu},
integration over the fields on $X_R$, with fixed asymptotic boundary data,
prepares a state
\begin{equation}
 |\Psi\rangle\in{\cal H}_{Y_\infty}
\end{equation}
in the Hilbert space obtained by quantizing the theory on
$Y_\infty$~\cite{Anber:2026pfg}.  A scalar partition function is obtained
only after pairing this state with a state in the dual Hilbert space or
 gluing
$X_R$ to another four-manifold.

The relevant topology of the boundary is
\begin{equation}
 H_p(Y_\infty,\mathbb Z)
 \simeq
 \begin{cases}
  \mathbb Z,   & p=0,3,\\
  \mathbb Z_2, & p=1,\\
  0,           & p=2,
 \end{cases}\,,\quad  H^p(Y_\infty,\mathbb Z)
 \simeq
 \begin{cases}
  \mathbb Z,   & p=0,3,\\
  0,           & p=1,\\
  \mathbb Z_2, & p=2.
 \end{cases}
 \label{eq:EH-boundary-homology-cohomology}
\end{equation}
Thus, $Y_\infty$ has no ordinary two-cycle, but it possesses a torsion
one-cycle $c$ satisfying $2c=0$.  This cycle generates
\begin{equation}
 \pi_1(Y_\infty)=H_1(Y_\infty,\mathbb Z)\simeq\mathbb Z_2.
\end{equation}
The corresponding torsion class appears in
$H^2(Y_\infty,\mathbb Z)\simeq\mathbb Z_2$ through the universal
coefficient theorem.  
Although this torsion is invisible in de Rham cohomology, it is detected by
the holonomy of a flat connection.  In particular,
\begin{equation}
 H^1(Y_\infty,U(1))
 \simeq
 \operatorname{Hom}\!\left(
 H_1(Y_\infty,\mathbb Z),U(1)\right)
 \simeq\mathbb Z_2.
 \label{eq:EH-flat-boundary-connections}
\end{equation}
Therefore, an ordinary flat $U(1)$ connection may be locally pure gauge
while having holonomy $+1$ or $-1$ around $c$.  

Boundary holonomies of $SU(N)$ are more involved, which we now discuss, along with their roles in determining the boundary states in the path-integral computations.

\subsection{Boundary sectors of the $SU(N)$ backgrounds}
\label{subsec:SUN-boundary-holonomies}

A flat $SU(N)$ connection on the asymptotic boundary is characterized,
up to conjugation, by its holonomy $h_c$ around $c$. Since $c^2=1$,
the holonomy must satisfy
\begin{equation}
 h_c^2=I_N.
 \label{eq:SUN-boundary-holonomy-order-two}
\end{equation}
It can therefore be diagonalized with eigenvalues equal to $+1$ or
$-1$. The condition $\det h_c=1$ requires the multiplicity of the
eigenvalue $-1$ to be even. Consequently, the inequivalent conjugacy
classes may be represented by
\begin{equation}
 h_c^{(\ell)}
 =
 \operatorname{diag}
 \left(
  -I_{2\ell},I_{N-2\ell}
 \right),
 \qquad
 \ell=0,1,\ldots,\left\lfloor\frac{N}{2}\right\rfloor.
 \label{eq:SUN-boundary-holonomy-representatives}
\end{equation}
We denote the corresponding components of the boundary Hilbert space
by
\begin{equation}
|\Psi\rangle= |\ell;\mathfrak{k}\rangle \in
 {\cal H}_{Y_\infty}\,,\quad 
 \ell=0,1,\ldots,\left\lfloor\frac{N}{2}\right\rfloor.
 \label{eq:SUN-boundary-sector-labels}
\end{equation}
Here
$|\ell;\mathfrak{k}\rangle$ should be understood as a reference state in the
component with the corresponding asymptotic holonomy $ h_c^{(\ell)}$, with $\mathfrak{k}$ denoting other quantum numbers. 

For even $N$, the last component,
\begin{equation}
 h_c^{(N/2)}=-I_N\,,\quad  \left|\ell=\frac{N}{2};\mathfrak{k}\right\rangle \in
 {\cal H}_{Y_\infty},
 \label{eq:SUN-central-minus-boundary-sector}
\end{equation}
has central holonomy, meaning the holonomy belongs to the center of $SU(N)$. For example, the $SU(2)$ case  has precisely
two boundary holonomies represented by $+I_2$ and $-I_2$. While $SU(4)$ case has three boundary holonomies represented by $+I_4$, $\mbox{diag}(-I_2,I_2)$, and $-I_4$. In both cases, only $I_N$ and $-I_N$ are central. 

For odd $N$, the only central order-two
holonomy is $h_c=I_N$. 

Next, we discuss how the background bulk field selects the boundary holonomy. 

\paragraph{Cartan-valued backgrounds.}

A general liftable Cartan-valued $SU(N)$ background in the family of solutions considered in this work
 may be written as
\begin{equation}
 A^{\rm B}
 =
 H_{\boldsymbol n}
 \frac{a^2}{r^2}\sigma_z,
 \qquad
 H_{\boldsymbol n}
 =
 \operatorname{diag}(n_1,\ldots,n_N),
 \label{eq:SUN-Cartan-background}
\end{equation}
where
\begin{equation}
 n_i\in\mathbb Z,
 \qquad
 \sum_{i=1}^{N}n_i=0.
 \label{eq:SUN-Cartan-cocharacter-condition}
\end{equation}
 Upon restriction to the asymptotic boundary, only the parities of the
$n_i$ remain. The holonomy around the torsion cycle is
\begin{equation}
 h_c(\boldsymbol n)
 =
 \exp\left(i\pi H_{\boldsymbol n}\right)
 =
 \operatorname{diag}
 \left(
  (-1)^{n_1},\ldots,(-1)^{n_N}
 \right).
 \label{eq:SUN-Cartan-boundary-holonomy}
\end{equation}
Its determinant is unity because $\sum_i n_i=0$. Defining (remember that the total number of negative eigenvalues is even)
\begin{equation}
 2\ell(\boldsymbol n)
 =
 \#\left\{
  i\ \big|\ n_i\ {\rm is\ odd}
 \right\},
 \label{eq:SUN-Cartan-boundary-label}
\end{equation}
one obtains
\begin{equation}
 h_c(\boldsymbol n)
 \simeq
 h_c^{(\ell(\boldsymbol n))},
 \qquad
 |\ell;\mathfrak{k}\rangle
 \in
 {\cal H}_{Y_\infty}\,,\quad 
 \ell=0,1,\ldots,\left\lfloor\frac{N}{2}\right\rfloor.
 \label{eq:SUN-Cartan-selected-boundary-sector}
\end{equation}
Thus, different bulk flux vectors can select the same boundary
component: the boundary remembers only the reduction of
$\boldsymbol n$ modulo two, up to Weyl permutations.

\paragraph{Embedded nonabelian backgrounds.}

The simplest nonabelian $SU(N)$ background is obtained by embedding
the $SU(2)$ instanton into a two-dimensional subspace of the
fundamental representation:
\begin{equation}
 A_{(N)}^{\rm B}(t)
 =
 \begin{pmatrix}
  A_{(2)}^{\rm B}(t)&0\\
  0&0_{N-2}
 \end{pmatrix},
 \label{eq:SUN-single-block-embedding}
\end{equation}
where
$
 A_{(2)}^{\rm B}(t)
$ is given by (\ref{generalansatzB}, \ref{su2solsd}).
At the  endpoint $t=a$, the nonabelian solution reduces to:
\begin{equation}
 F(r;a)=0,
 \qquad
 G(r;a)=\frac{2a^2}{r^2},
\end{equation}
and hence
\begin{equation}
 A_{(N)}^{\rm B}(a)
 =
 \operatorname{diag}(1,-1,0,\ldots,0)
 \frac{a^2}{r^2}\sigma_z.
 \label{eq:SUN-single-block-abelian-endpoint}
\end{equation}
This is the Cartan-valued background with
\begin{equation}
 \boldsymbol n=(1,-1,0,\ldots,0).
\end{equation}
It follows that the entire finite-$t$ family has
\begin{equation}
 Q_{(N)}=\frac12,
 \qquad
 h_c^{(1)}
 \simeq
 \operatorname{diag}(-I_2,I_{N-2}),
 \qquad
 |\ell=1;\mathfrak{k}\rangle
 \in
 {\cal H}_{Y_\infty}.
 \label{eq:SUN-single-block-boundary-sector}
\end{equation}
Indeed, the conjugacy class of the boundary holonomy is discrete and
cannot change under a continuous deformation, e.g., scale transformation or rotations in the moduli space, that preserves
regularity at the bolt and the asymptotic boundary condition.

For $N>2$, the holonomy in
\eqref{eq:SUN-single-block-boundary-sector} is noncentral. Its
centralizer is
$
S\left(U(2)\times U(N-2)\right)
 \label{eq:SUN-single-block-boundary-centralizer}
$,
 and global color conjugations leave the conjugacy class $\ell=1$
unchanged.

\paragraph{Several nonabelian blocks.}

More general solutions can be constructed by placing mutually
commuting $SU(2)$ instantons in orthogonal two-dimensional blocks:
\begin{equation}
 A_{\boldsymbol p}^{\rm B}
 =
 \operatorname{diag}
 \left(
  A^{(p_1)}(t_1),\ldots,
  A^{(p_s)}(t_s),0_{N-2s}
 \right),
 \qquad
 s\leq\left\lfloor\frac{N}{2}\right\rfloor,
 \label{eq:SUN-multiple-block-background}
\end{equation}
where $p_j$ labels the corresponding $SU(2)$ family. At the reducible
endpoint, $t_{I}=a$, this solution is connected to
\begin{equation}
 H_{\boldsymbol p}
 =
 \operatorname{diag}
 \left(
  p_1,-p_1,\ldots,p_s,-p_s,
  0,\ldots,0
 \right).
 \label{eq:SUN-multiple-block-Cartan-endpoint}
\end{equation}
Its topological charge and boundary holonomy are
\begin{align}
 Q_{(N)}
 =
 \frac12\sum_{j=1}^{s}p_j^2,\quad
 h_c
 =
 \operatorname{diag}
 \left(
  (-1)^{p_1}I_2,\ldots,
  (-1)^{p_s}I_2,I_{N-2s}
 \right).
 \label{eq:SUN-multiple-block-charge-holonomy}
\end{align}
Consequently,
\begin{equation}
 \ell(\boldsymbol p)
 =
 \#\left\{
  j\ \big|\ p_j\ {\rm is\ odd}
 \right\}.
 \label{eq:SUN-multiple-block-boundary-label}
\end{equation}
In particular, if all the blocks are minimal, $p_j=1$, then
\begin{equation}
 Q_{(N)}=\frac{s}{2},
 \qquad
 h_c^{(s)}\simeq
 \operatorname{diag}(-I_{2s},I_{N-2s}),
 \qquad
| \ell=s;\mathfrak{k}\rangle
 \in
 {\cal H}_{Y_\infty}.
 \label{eq:SUN-minimal-multiple-block-sector}
\end{equation}
When $N$ is even and $s=N/2$, the holonomy becomes central:
\begin{equation}
 h_c^{(N/2)}=-I_N\,, \qquad
\left| \ell=\frac{N}{2};\mathfrak{k}\right\rangle
 \in
 {\cal H}_{Y_\infty}\,.
 \label{eq:SUN-full-block-central-holonomy}
\end{equation}
In this case, every global $SU(N)$ color rotation commutes with the
boundary holonomy. This is the mechanism that allows the
off-diagonal color orientations to become genuine collective
coordinates for the full block-diagonal configuration. This observation is crucial to constructing the bosonic moduli space, as we shall see in a detailed $SU(4)$ example.

\subsection{EH-prepared states and boundary-conditioned observables}
\label{sec:EH-boundary-states-chiral-vacua}

Let $X$ be a Euclidean cobordism whose oriented boundary is
\begin{equation}
 \partial X=\overline{Y}_1\sqcup Y_2.
\end{equation}
After specifying field configurations $\Phi_1$ and $\Phi_2$ on $Y_1$
and $Y_2$, respectively, the path integral defines the transition kernel \cite{Atiyah:1989vu}
\begin{equation}
 {\cal Z}_X[\Phi_2,\Phi_1]
 =
 \langle\Phi_2|U_X|\Phi_1\rangle.
\end{equation}
Equivalently, it defines a linear map
\begin{equation}
 U_X:{\cal H}_{Y_1}\longrightarrow{\cal H}_{Y_2}.
\end{equation}
If $X$ is closed, its path integral instead gives the scalar partition
function $Z[X]$.

When $X$ has only one boundary, $\partial X=Y$, it may be regarded as a
cobordism from the empty set to $Y$. Its path integral then prepares a
state 
\begin{equation}
 |\Psi_X\rangle\in{\cal H}_Y,
 \qquad
 \Psi_X[\Phi]
 =
 \int_{\left.\phi\right|_Y=\Phi}
 [{\cal D}\phi]\,e^{-S[\phi]}.
 \label{eq:path-integral-prepared-state}
\end{equation}
Because there is no initial boundary, independent initial data are replaced
by regularity conditions in the interior. The resulting state will be
referred to as a no-boundary state\footnote{The terminology follows the no-boundary proposal of Hartle and Hawking \cite{Hartle:1983ai}, in which smoothness of a Euclidean cap replaces initial data. Here, however, the geometry is a fixed background rather than being summed over, and the state is a quantum-field-theory state on the asymptotic $\mathbb{RP}^3$.}.

This is precisely the situation on EH space. Introducing a large-radius
cutoff, the truncated manifold
$
 X_R=\{a\leq r\leq R\}
$
has a single boundary
$
 \partial X_R=Y_R\simeq\mathbb{RP}^3.
$
The bolt at $r=a$ is a smooth interior submanifold, not an additional
boundary. No independent boundary condition is therefore imposed at
$r=a$; instead, smoothness of the fields near the bolt determines the
no-boundary state.

Formally, one may cut the geometry at an intermediate radius $r_0>a$.
The interior cap then prepares a state
\begin{equation}
 |0_{\rm NB}\rangle,
 \label{eq:EH-no-boundary-state}
\end{equation}
in the Hilbert space of the $\mathbb{RP}^3$ slice at $r_0$. Radial
evolution gives
\begin{equation}
 |\Psi_{Q,\ell};R\rangle
 =
 \Big[U_{\rm EH}(R,r_0)
 |0_{\rm NB}\rangle\Big]_{Q,\ell},
 \label{eq:EH-radially-evolved-state}
\end{equation}
with
\begin{equation}
U_{\rm EH}(R,r_0)
 =
 {\cal P}\exp\left[
  -\int_{\tau(r_0)}^{\tau(R)}
  d\tau\,H(\tau)
 \right],
 \label{eq:EH-radial-evolution-operator}
\end{equation}
where $[\cdots]_{Q,\ell}$ denotes the contribution of configurations with total topological charge $Q$ and asymptotic holonomy class $\ell$ at $r=R$. 
Here $\tau$ is proper radial time and $H(\tau)$ is the corresponding
radial Hamiltonian. We use this expression only as a formal
interpretation of the path integral; an explicit Hamiltonian construction
will not be attempted here. In particular, since $H(\tau)$ depends on
radial time, we do not assume that the prepared state is an instantaneous
Hamiltonian eigenstate or that radial evolution provides an ordinary
vacuum projection; see the discussion in Section~\ref{sec:EH-anomalies-APS} and Figure~\ref{fig:EH-histories} for an illustration.

Taking $R\rightarrow\infty$, the EH path integral with a possible
gauge-invariant bulk insertion ${\cal O}$ prepares a state
\begin{equation}
 |\Psi_{Q,\ell}[{\cal O}]\rangle.
 \label{eq:EH-prepared-state}
\end{equation}
A local bulk insertion does not alter the prescribed asymptotic holonomy.
The labels $Q$ and $\ell$ characterize the bulk saddle and its asymptotic
boundary data; neither label should be identified with a four-dimensional
chiral vacuum.

In a semiclassical description, we decompose the full asymptotic state into components associated with the allowed holonomy classes $|\ell\rangle$, $\ell=0,1,\ldots,\left\lfloor\frac{N}{2}\right\rfloor$.
Then,  the state prepared by the full path integral (obtained by summing over all the allowed topological charges) may be written as
\begin{equation}
 |\Psi[{\cal O}]\rangle
 =
 \sum_{\ell=0}^{\lfloor N/2\rfloor}
 |\Psi_\ell[{\cal O}]\rangle,
 \qquad
 |\Psi_\ell[{\cal O}]\rangle
 =
 \sum_Q
 |\Psi_{Q,\ell}[{\cal O}]\rangle,
 \label{eq:EH-state-holonomy-decomposition}
\end{equation}
where the sum over $Q$ is restricted to bulk sectors compatible with the
boundary class $\ell$. For example, the trivial configuration has
\begin{equation}
 Q_{(N)}=0,
 \qquad
 h_c^{(0)}=I_N,
 \qquad
 \ell=0,
 \label{eq:EH-trivial-saddle-data}
\end{equation}
whereas the single-root instanton has
\begin{equation}
 Q_{(N)}=\frac12,
 \qquad
 h_c^{(1)}\simeq\operatorname{diag}(-I_2,I_{N-2}),
 \qquad
 \ell=1.
 \label{eq:EH-single-root-saddle-data}
\end{equation}

To obtain a scalar amplitude from the EH-prepared state, one must choose a
bra in the dual asymptotic Hilbert space. Let
\begin{equation}
 \langle B|
 =
 \sum_{\ell=0}^{\lfloor N/2\rfloor}
 b_\ell^*\langle B_\ell|,
 \qquad
 \langle B_\ell|
 \in{\cal H}_{Y_\infty}^*,
 \label{eq:SUN-general-boundary-bra}
\end{equation}
where $\langle B_\ell|$ is a fixed reference bra in the $\ell$-th
component and $b_\ell$ specifies its relative weight. We define the
projected sector amplitudes by
\begin{equation}
 Z_Q^{(\ell)}[{\cal O}]
 \equiv
 \langle B_\ell|
 \Psi_{Q,\ell}[{\cal O}]
 \rangle,
 \qquad
 Z^{(\ell)}[{\cal O}]
 \equiv
 \sum_Q Z_Q^{(\ell)}[{\cal O}].
 \label{eq:EH-projected-sector-amplitudes}
\end{equation}
The normalized boundary-conditioned expectation value is then defined as
\begin{equation}
 \langle{\cal O}\rangle_B
\equiv
 \frac{
  \langle B|\Psi[{\cal O}]\rangle
 }{
  \langle B|\Psi[1]\rangle
 }
 =
 \frac{
  \displaystyle
  \sum_{\ell=0}^{\lfloor N/2\rfloor}
  b_\ell^*Z^{(\ell)}[{\cal O}]
 }{
  \displaystyle
  \sum_{\ell=0}^{\lfloor N/2\rfloor}
  b_\ell^*Z^{(\ell)}[1]
 }.
 \label{eq:SUN-boundary-bra-expectation-value}
\end{equation}
If supersymmetric Ward identities are to be applied, the chosen bra
components must in addition satisfy (see Section~\ref{sec:EH-preserved-supersymmetry})
\begin{equation}
 \langle B_\ell|\bar{\cal Q}_{\dot \alpha}=0,\quad \dot\alpha=\dot 1,\dot 2\,,
 \label{eq:EH-supersymmetric-boundary-bra}
\end{equation}
where $\bar{\cal Q}_{\dot \alpha}$ are the conserved supercharges,
and the asymptotic boundary conditions must eliminate any supercurrent
flux at infinity.

Suppose that, at leading semiclassical order, the numerator is supplied
by a saddle $\mathfrak s$ with charge $Q_{\mathfrak s}$ and boundary label
$\ell_{\mathfrak s}$, whereas the denominator is supplied by the trivial
saddle. The corresponding contribution is
\begin{equation}
 \left.\langle{\cal O}\rangle_B\right|_{\mathfrak s}
 =
 \frac{
  b_{\ell_{\mathfrak s}}^*
  Z_{Q_{\mathfrak s}}^{(\ell_{\mathfrak s})}[{\cal O}]
  +\cdots
 }{
  b_0^*Z_{Q=0}^{(0)}[1]
  +\cdots
 }.
 \label{eq:SUN-leading-boundary-bra-expectation-value}
\end{equation}
In sectors containing unsaturated fermion zero modes,
$Z_Q^{(\ell)}[1]=0$, so such sectors do not contribute to the leading
source-free denominator. We use the conventional normalization
\begin{equation}
 Z_{Q=0}^{(0)}[1]=1.
 \label{eq:EH-trivial-sector-normalization}
\end{equation}
The relative coefficient
$b_{\ell_{\mathfrak s}}/b_0$ is part of the boundary data and must be
fixed by a choice of boundary state or by a gluing prescription. In the
explicit condensate calculations below, we adopt the convention
\begin{equation}
 \frac{b_{\ell_{\mathfrak s}}}{b_0}=1.
 \label{eq:SUN-boundary-weight-convention}
\end{equation}
This fixes the relative normalization of the relevant holonomy
components, but it does not identify the chosen bra with a particular
four-dimensional chiral vacuum.

Equation
\eqref{eq:SUN-leading-boundary-bra-expectation-value} therefore defines a
boundary-conditioned instanton amplitude on EH space. Pure
${\cal N}=1$ $SU(N)$ super-Yang--Mills theory is expected to possess $N$
chiral vacua, but the present radial construction neither derives these
states nor determines their wavefunctional components in the different
holonomy sectors (for $SU(2)$, see Ref.~\cite{Anber:2026bsx}). To interpret the EH amplitude as a condensate in a
specific chiral vacuum, one would need additional data, such as a bra
explicitly prepared in that vacuum or an infinitesimal
chiral-symmetry-breaking source.

\subsection{Boundary conditions for adjoint fermions}
\label{sec:SUN-adjoint-boundary-conditions}

We now determine the global boundary conditions obeyed by an adjoint
fermion in the $SU(N)$ backgrounds studied above. These boundary
conditions determine the flat bundle twisting the Dirac operator on the
asymptotic boundary and, consequently, the boundary contribution to the
APS index.

Since the index is invariant under smooth deformations that preserve the
asymptotic holonomy, a nonabelian family may be continuously deformed to
its Cartan-valued endpoint, meaning that the deformation selects the point $t_{I}=a$ on the space of radial-size moduli. We denote the corresponding Cartan generator
at the bolt by 
\begin{equation}
 H_{\rm B}
 =
 \boldsymbol G(a)\cdot\boldsymbol H.
 \label{eq:SUN-Cartan-generator-at-bolt}
\end{equation}

Consider first the contractible loop $\ell_a$ near the bolt generated by
$\psi\rightarrow\psi+2\pi$ at fixed $(r,\theta,\varphi)$, with $r\simeq a$.
Since the $\psi$-circle collapses smoothly at the bolt, its spin holonomy must approach the identity in the shrinking-loop limit. Direct evaluation of the EH spin connection indeed gives
$\operatorname{Hol}_{\rm spin}(\ell_{a+\varepsilon})\to I_4$ as $\varepsilon\to0$; see \cite{Anber:2025gvb}.
However, the gauge holonomy is instead (use (\ref{eq:SUN-multiple-root-embedding}, \ref{specialsunsungen}))
\begin{equation}
 h_a=\exp\left(i\oint_{\ell_a}A^{\rm B}\right)
 =
 \exp(i\pi H_{\rm B}).
 \label{eq:SUN-bolt-gauge-holonomy}
\end{equation}
Therefore, an adjoint fermion transforms according to
\begin{equation}
 \lambda(\psi+2\pi)
 =
 h_a^{-1}\lambda(\psi)h_a.
 \label{eq:SUN-adjoint-bolt-transport}
\end{equation}

Expanding the adjoint fermion in the Cartan--Weyl basis,
\begin{equation}
 \lambda
 =
 \boldsymbol\lambda\cdot\boldsymbol H
 +
 \sum_{\boldsymbol\beta\in\Delta_+}
 \left(
  \lambda_{\boldsymbol\beta}^{+}E_{\boldsymbol\beta}
  +
  \lambda_{\boldsymbol\beta}^{-}E_{-\boldsymbol\beta}
 \right),
 \label{lambdasun}
\end{equation}
and using
$
 e^{-i\pi H_{\rm B}}
 E_{\boldsymbol\beta}
 e^{i\pi H_{\rm B}}
 =
 e^{-i\pi\boldsymbol\beta\cdot\boldsymbol G(a)}
 E_{\boldsymbol\beta},
$
we obtain
\begin{align}
 \boldsymbol\lambda(\psi+2\pi)
 =
 \boldsymbol\lambda(\psi),\quad
 \lambda_{\boldsymbol\beta}^{\pm}(\psi+2\pi)
 =
 \exp\left[
  \mp i\pi
  \boldsymbol\beta\cdot\boldsymbol G(a)
 \right]
 \lambda_{\boldsymbol\beta}^{\pm}(\psi).
 \label{eq:SUN-adjoint-bolt-boundary-conditions}
\end{align}
Thus, the Cartan components of $\lambda$ are periodic, while the root components are
periodic or antiperiodic according to whether
$\boldsymbol\beta\cdot\boldsymbol G(a)$ is even or odd. Indeed, global consistency
requires
\begin{equation}
 \boldsymbol\beta\cdot\boldsymbol G(a)\in\mathbb Z
 \qquad
 \text{for every root }\boldsymbol\beta,
 \label{eq:SUN-adjoint-global-consistency}
\end{equation}
otherwise, the fermions cannot have integer or half-integer total quantum numbers.

At infinity, the constant-$r$ slices approach
$
 Y_\infty
 \simeq
 S^3/\mathbb Z_2
 \simeq
 \mathbb{RP}^3.
$
Let $c$ denote the noncontractible torsion cycle generated along
$\psi\rightarrow\psi+2\pi$. Although the local gauge potential vanishes
as $r^{-2}$, its holonomy is not necessarily trivial. Applying Stokes'
theorem to the cigar-shaped surface extending from $c$ to the
bolt gives
\begin{equation}
 h_c
 =
 \exp\left(
  i\oint_{c}A^{\rm B}
 \right)
 =
 \exp(i\pi H_{\rm B}).
 \label{eq:SUN-asymptotic-gauge-holonomy}
\end{equation}
The gauge holonomy acting on the adjoint fermion is consequently
\begin{equation}
 \operatorname{Hol}_{\rm adj}
 =
 \operatorname{Ad}_{h_c}.
 \label{eq:SUN-adjoint-boundary-holonomy}
\end{equation}

The gravitational holonomy of a Dirac fermion around $c$ is
\begin{equation}
 \exp\left[
  -\frac12\oint_{c}
  \omega_{ab}\Sigma^{ab}
 \right]
 =
 -\gamma_5\,,
 \label{eq:EH-asymptotic-spin-holonomy}
\end{equation}
where $\omega_{ab}$ is the spin-connection one-form and $\Sigma^{ab}=\mbox{diag}\left( \hat\sigma^{ab}, \bar{\hat\sigma}^{ab}\right)/2$.
Thus, if the adjoint Weyl fermion is temporarily embedded into a
four-component spinor $\Psi_{\rm adj}$, its complete asymptotic boundary
condition is
\begin{equation}
 \Psi_{\rm adj}(\psi+2\pi)
 =
 -\gamma_5\,
 \operatorname{Ad}_{h_c}
 \Psi_{\rm adj}(\psi).
 \label{generalfermionhol}
\end{equation}
 Further details 
can be found in
Ref.~\cite{Anber:2025dxy}.

\subsection{Chiral and center symmetries on EH space: APS index, boundary sectors, and the mixed anomaly}
\label{sec:EH-anomalies-APS}
\label{sec:center-on-boundary-sectors}

Chiral anomalous response on EH space originates from the action of a global chiral transformation $\lambda\mapsto e^{2\pi i\beta_\chi}\lambda$ on the functional measure of the four-dimensional bulk adjoint fermions. Using the Fujikawa method, the response appears as a phase ${\cal A}$ of the path integral, given by the parameter $\beta_\chi$ multiplied by the adjoint Dirac index ${\cal I}_{\rm adj}$ on EH space \cite{Anber:2025dxy}:
\begin{eqnarray}
\mbox{P.I.}\mapsto {\cal A}~\mbox{P.I.}\,,\quad {\cal A}= \exp\left[
  -2\pi i\beta_\chi
  {\cal I}_{\rm adj}({\rm EH})
 \right]\,.
\end{eqnarray}
On a closed manifold, such a phase is an 't~Hooft anomaly, provided that it cannot be removed by a local counterterm. On EH space, however, the path integral prepares a state on the asymptotic $\mathbb{RP}^3$ rather than computing a number, and the meaning of ${\cal A}$ must be reconsidered. The aim of this subsection is to settle this question. We proceed in four steps. First, we compute the APS index in every boundary holonomy sector. Second, we construct the operator $\mathsf X$ that implements the discrete chiral symmetry on the boundary Hilbert space and compute its eigenvalue on the EH-prepared state. This eigenvalue receives three contributions: the gaugino quanta created by the bulk, which are counted by ${\cal A}$, the Chern--Simons term of the boundary gauge field, and the charge of the gaugino Dirac sea on $\mathbb{RP}^3$. Together they give a single eigenvalue, the same in every boundary sector. Third, we construct the operator $\mathsf U$ that implements the part of the center symmetry acting on $\mathbb{RP}^3$ and derive the algebra of $\mathsf U$ and $\mathsf X$, which realizes the mixed anomaly. Finally, we show how this algebra is read off from the EH-prepared state, with the chiral charges supplied by the bulk index.

\paragraph{The APS index on EH space.}
According to the Atiyah--Patodi--Singer (APS) index theorem \cite{Atiyah:1976jg,Atiyah:1976qjr,Atiyah:1975jf}, the Dirac index receives contributions from the bulk characteristic classes and from the $\eta$-invariant of the $\mathbb{RP}^3$ boundary (a detailed analysis is carried out in Ref.~\cite{Anber:2025dxy}):
\begin{eqnarray}
 {\cal I}_{\rm adj}({\rm EH})
 ={}
 \frac{1}{8\pi^2}
 \int_{\rm EH}
 \operatorname{tr}_{\rm adj}
 \left(
  F_{}\wedge F_{}
 \right)
 +
 \frac{N^2-1}{24\cdot8\pi^2}
  \int_{\rm EH}
  \operatorname{tr}({\cal R}\wedge{\cal R})
 -
 \eta_{\rm adj}(\mathbb{RP}^3).
 \label{eq:EH-adjoint-APS-general}
\end{eqnarray}
Using
$
 \operatorname{tr}_{\rm adj}(T^aT^b)
 =
 N\delta^{ab}
 =
 2N\operatorname{tr}_{\boldsymbol N}(T^aT^b),
$
and
$
 Q_{(N)}
 =
 \frac{1}{8\pi^2}
 \int_{\rm EH}
 \operatorname{tr}_{\boldsymbol N}
 \left(
  F_{}\wedge F_{}
 \right),
$
the gauge contribution becomes
$
 \frac{1}{8\pi^2}
 \int_{\rm EH}
 \operatorname{tr}_{\rm adj}
 \left(
  F_{}\wedge F_{}
 \right)
 =
 2NQ_{(N)}.
$
With the EH orientation adopted in this work, the gravitational topological charge is
$
 \frac{1}{24\cdot8\pi^2}
 \int_{\rm EH}
 \operatorname{tr}({\cal R}\wedge{\cal R})
 =
 -\frac18,
$
so the gravitational contribution is
$
 -\frac{N^2-1}{8}.
$

Since $\operatorname{Hol}_{\rm adj}$ has eigenvalues $\pm1$, we define the signature of the holonomy:
\begin{equation}
 \operatorname{sgn}
 \left(
  \operatorname{Hol}_{\rm adj}
 \right)
 =
 n_+-n_-
 =
 \operatorname{tr}_{\rm adj}
 \left(
  \operatorname{Ad}_{h_c}
 \right),
 \label{eq:EH-adjoint-holonomy-signature}
\end{equation}
where $n_+$ and $n_-$ are the numbers of adjoint components with holonomy $+1$ and $-1$, respectively.  Each adjoint component contributes $\pm\frac18$ to the $\eta$-invariant, the sign being fixed by the holonomy, and there are no boundary zero modes  \cite{Anber:2025gvb}.  A component with holonomy $+1$ contributes $-1/8$ to the $\eta$-invariant, whereas a component with holonomy $-1$ contributes $+1/8$~\cite{Anber:2025dxy}:
\begin{equation}
 \eta_{\rm adj}(\mathbb{RP}^3)
 =
 -\frac18
 \operatorname{sgn}
 \left(
  \operatorname{Hol}_{\rm adj}
 \right).
 \label{eq:EH-adjoint-eta-signature}
\end{equation}
Combining all contributions gives
\begin{equation}
 {\cal I}_{\rm adj}
 =
 2NQ_{(N)}
 +
 \frac{
  -(N^2-1)
  +
  \operatorname{sgn}
  (\operatorname{Hol}_{\rm adj})
 }{8}.
 \label{eq:EH-adjoint-APS-practical}
\end{equation}
The boundary term is essential: the bulk charge can be fractional, whereas the complete APS index is an integer. For the self-dual backgrounds considered here, the opposite-chirality Weyl operator has no normalizable kernel, so the index equals the number of normalizable adjoint zero modes.

\paragraph{The index in every boundary sector.}
Recall that the boundary holonomy is conjugate to one of the representatives $h_c^{(\ell)}=\operatorname{diag}(-I_{2\ell},I_{N-2\ell})$, $\ell=0,1,\ldots,\lfloor N/2\rfloor$, and that $Q_{(N)}\equiv\ell/2$ in the sector $\ell$. The adjoint generators with holonomy $-1$ are the off-diagonal ones connecting the $-I_{2\ell}$ and $I_{N-2\ell}$ blocks, of which there are $4\ell(N-2\ell)$. Hence
\begin{equation}\label{importantetaadj}
 \operatorname{sgn}
 \left(
  \operatorname{Hol}_{\rm adj}
 \right)
 =N^2-1-8\ell(N-2\ell)=(N-4\ell)^2-1,
\end{equation}
and \eqref{eq:EH-adjoint-APS-practical} gives
\begin{equation}
 {\cal I}_{\rm adj,\ell}=2NQ_{(N)}-\ell(N-2\ell)\ \in\ 2\ell^2+2N\mathbb Z .
 \label{eq:center-sector-index-general}
\end{equation}
The gravitational and $\eta$-invariant contributions combine to $-\ell(N-2\ell)$, which depends only on the boundary sector, while shifting $Q_{(N)}$ by an integer shifts the index by a multiple of $2N$. We now discuss the cases relevant for the condensates.

\emph{The single-root embedding.} The nonabelian instanton embedded along $\boldsymbol\alpha_1=\boldsymbol e_1-\boldsymbol e_2$ is continuously connected to the Cartan configuration
\begin{equation}
 H_{\boldsymbol n}
 =
 \operatorname{diag}(1,-1,0,\ldots,0).
\end{equation}
It has
\begin{equation}
 Q_{(N)}=\frac12,
 \qquad
 h_c^{(1)}
 =
 \operatorname{diag}(-I_2,I_{N-2}),
 \label{eq:EH-SUN-single-root-charge-holonomy}
\end{equation}
i.e., $\ell=1$. In this case,
\begin{equation}
 \mathfrak{su}(N)
 =
 \mathfrak g_+\oplus\mathfrak g_-,
 \qquad
 \operatorname{Ad}_{h_c}\big|_{\mathfrak g_\pm}
 =
 \pm I,
 \label{eq:EH-single-root-adjoint-decomposition}
\end{equation}
with block-diagonal and off-diagonal sectors
\begin{align}
 \mathfrak g_+
 &=
 \mathfrak{s}\bigl(\mathfrak u(2)\oplus\mathfrak u(N-2)\bigr),
 &
 \dim\mathfrak g_+
 &=N^2-4N+7,
 \nonumber\\
 \mathfrak g_-
 &=
 (\boldsymbol 2,\overline{\boldsymbol{N-2}})
 \oplus
 (\overline{\boldsymbol 2},\boldsymbol{N-2}),
 &
 \dim\mathfrak g_-
 &=4(N-2).
 \label{eq:EH-single-root-adjoint-sectors}
\end{align}
Hence $\operatorname{sgn}(\operatorname{Hol}_{\rm adj})=N^2-8N+15$, and the APS index is
\begin{align}
 {\cal I}_{\rm adj}
 =N+(2-N)=2.
 \label{indexmastermix}
\end{align}
Thus, the single-root embedding supports exactly two normalizable adjoint zero modes for every $N$; the gravitational and boundary contributions change the naive bulk result $N$ to the physical index $2$. These zero modes are localized near the bolt and do not escape to the boundary, as is shown by explicitly solving the Dirac equation in Appendix~\ref{app:EH-SU2-gaugino-zero-modes}.

\emph{Embeddings along mutually orthogonal roots.} Consider $s$ mutually commuting minimal $SU(2)$ instantons embedded along pairwise orthogonal roots. At their Cartan-valued endpoint,
\begin{equation}
 H_s
 =
 \operatorname{diag}
 \left(
  1,-1,\ldots,1,-1,0,\ldots,0
 \right),
 \qquad
 s\leq\left\lfloor\frac{N}{2}\right\rfloor,
\end{equation}
with
\begin{equation}
 Q_{(N)}=\frac{s}{2},
 \qquad
 h_c^{(s)}
 =
 \operatorname{diag}(-I_{2s},I_{N-2s}),
\end{equation}
i.e., $\ell=s$. Then $\operatorname{sgn}(\operatorname{Hol}_{\rm adj})=(N-4s)^2-1$ and
\begin{align}
 {\cal I}_{\rm adj}
 =2s^2.
 \label{eq:EH-SUN-multiple-root-adjoint-index}
\end{align}
In particular, the two-block $SU(4)$ solution has
\begin{equation}
 Q_{(4)}=1,
 \qquad
 h_c^{(2)}=-I_4,
 \qquad
 {\cal I}_{\rm adj}=8,
\end{equation}
which is crucial to constructing the multi-fermion condensate in the $SU(4)$ gauge group.

\emph{Central holonomy.} An important simplification occurs when the asymptotic holonomy $h_c$ is central, $\ell=0$ or $\ell=N/2$. Since $\operatorname{Ad}_{h_c}=\mathbf 1_{\rm adj}$, the holonomy acts trivially on every adjoint component of the boundary Dirac operator, and the gravitational contribution cancels the $\eta$-invariant contribution. The APS index then reduces to ${\cal I}_{\rm adj}=2NQ_{(N)}$ for every topological sector compatible with the central holonomy. For noncentral holonomy, by contrast, $\operatorname{Ad}_{h_c}$ acts nontrivially on part of the adjoint representation, and the two contributions no longer cancel.

\paragraph{Boundary sectors and the EH-prepared state.}
The boundary Hilbert space decomposes into the holonomy sectors \eqref{eq:SUN-boundary-sector-labels},
\begin{equation}
 {\cal H}_{Y_\infty}=\bigoplus_{\ell=0}^{\lfloor N/2\rfloor}{\cal H}_\ell ,
 \label{eq:center-sector-decomposition}
\end{equation}
where ${\cal H}_\ell$ is spanned by the states $|\ell;\mathfrak k\rangle$ supported on the holonomy class $h_c^{(\ell)}$, including all their excitations. The EH path integral prepares a state in ${\cal H}_{Y_\infty}$, which we decompose accordingly,
\begin{equation}
 |\Psi_{\rm EH}\rangle=\sum_{\ell=0}^{\lfloor N/2\rfloor}|\Psi_\ell\rangle,
 \qquad |\Psi_\ell\rangle\in{\cal H}_\ell .
 \label{eq:center-EH-state-decomposition}
\end{equation}
Fixing the holonomy in the sector $\ell$ means, in the context of the path integral on EH space, integrating over all bulk gauge fields approaching $h_c^{(\ell)}$ and summing over all compatible topological charges. Since the bolt carries no Chern--Simons contribution, a smooth configuration on EH satisfies
\begin{equation}
 Q_{(N)}=\frac{1}{8\pi^2}\int_{\rm EH}\operatorname{tr}F\wedge F\equiv\operatorname{CS}[A_\infty],
 \label{eq:center-Q-equals-CS}
\end{equation}
where $\operatorname{CS}$ is the Chern--Simons functional of the asymptotic connection,
\begin{equation}
\operatorname{CS}[A]=\frac{1}{8\pi^2}\int_{\mathbb{RP}^3}\operatorname{tr}\left(A\wedge dA-\frac{2i}{3}A\wedge A\wedge A\right),
 \label{eq:center-CS-def}
\end{equation}
normalized so that $d\operatorname{CS}=\frac{1}{8\pi^2}\operatorname{tr}F\wedge F$.  Each component $|\Psi_\ell\rangle$ is therefore built from configurations with $Q_{(N)}\in\frac{\ell}{2}+\mathbb Z$. For the single-root sector, for example, these charges form the lattice\footnote{Fixing the single-root holonomy
$h^{(1)}_c=\operatorname{diag}(-I_2,I_{N-2})$ fixes only the fractional part of the topological charge, namely $Q=\frac12\ {\rm mod}\,\mathbb Z$. Within the general Cartan family
$H=\operatorname{diag}(n_1,\ldots,n_N)$, with $n_i\in\mathbb Z$ and $\sum_i n_i=0$, this holonomy requires exactly two of the $n_i$ to be odd, while $Q=\frac14\sum_i n_i^2$. For $N\geq4$, the Cartan family realizes every positive charge $Q=k+\frac12$, $k\geq0$. For $SU(2)$, however, the Cartan charges are restricted to $Q=n^2/2$ with $n$ odd, while for $SU(3)$ they obey $2Q=a^2+ab+b^2$ with $a$ and $b$ odd. These lower-rank restrictions are properties of the Cartan ansatz. General nonabelian configurations can shift $Q$ by an arbitrary integer without changing the asymptotic holonomy.}
$Q_{(N)}=k+\frac12$, $k\in\mathbb Z$, so that schematically the projected-sector amplitude is (see the discussion around Eq.~(\ref{eq:EH-projected-sector-amplitudes}))
\begin{equation}
 Z_{\rm EH}^{(\ell=1)}
 =
 \sum_{k\in\mathbb Z}
 Z_{Q=k+1/2}^{(\ell=1)}.
\end{equation}
Components with unsaturated fermion zero modes are understood in the presence of fermionic sources.

\begin{figure}[t]
\centering
\includegraphics[width=\textwidth]{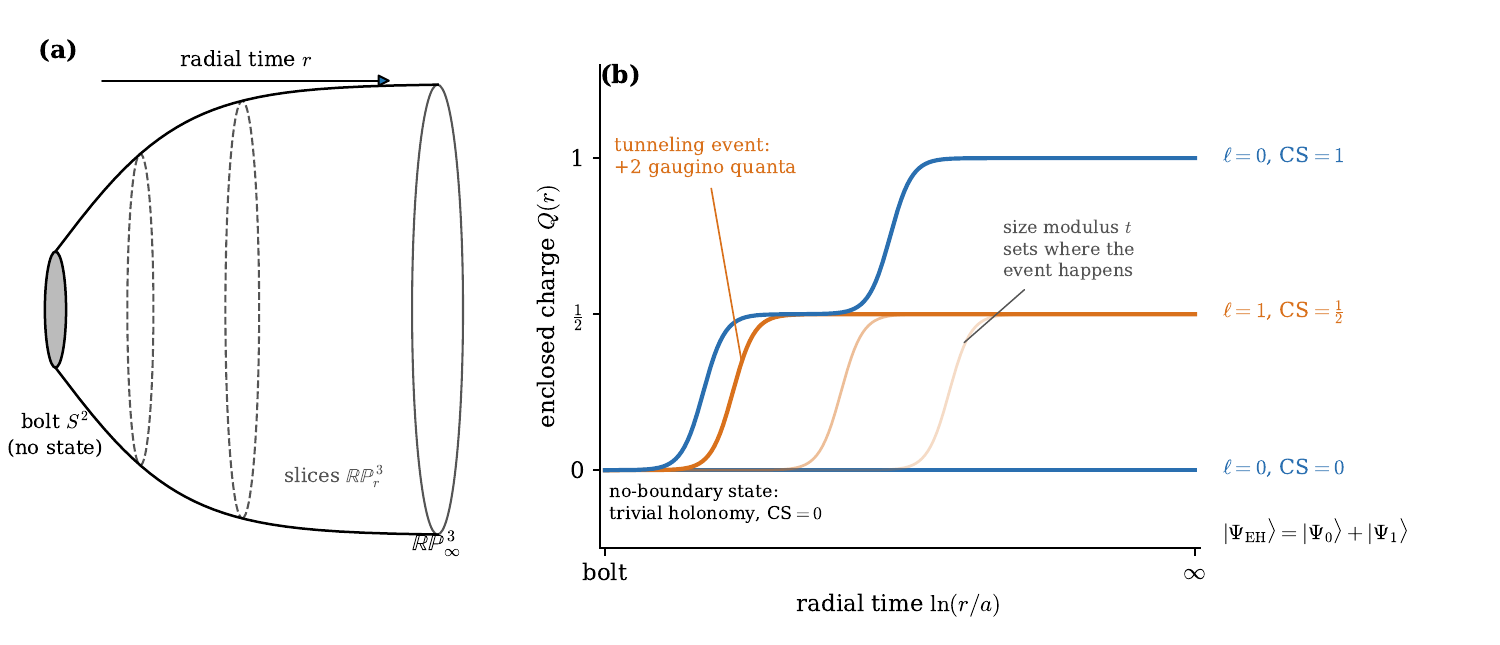}
\caption{(a) EH space as a cap: radial slices $\mathbb{RP}^3_r$ degenerate to the bolt, where there is no state, and approach $\mathbb{RP}^3_\infty$ at infinity. (b) Radial histories contributing to the $SU(2)$ EH-prepared state, shown through the enclosed charge $Q(r)$. Every history starts from the no-boundary state near the bolt, with trivial holonomy and $\operatorname{CS}=0$, and reaches its asymptotic sector through fractional tunneling events in radial time; each event with $\Delta Q=\frac12$ creates two gaugino quanta, and its position is set by the size modulus $t$. Histories ending at $Q\in\mathbb Z$ build $|\Psi_0\rangle$, those ending at $Q\in\frac12+\mathbb Z$ build $|\Psi_1\rangle$.}
\label{fig:EH-histories}
\end{figure}

 The sectors $\ell$ label classical vacua, not superselection sectors. All $SU(N)$ bundles on $\mathbb{RP}^3$ are trivial \cite{Freed:1992vw}, and the space of connections is connected, so different flat connections are separated only by potential barriers, which fractional tunneling events cross. On the cylinder $\mathbb R\times\mathbb{RP}^3$, such events connect any two sectors, with $Q\equiv(\ell'-\ell)/2\pmod 1$; the minimal one, with $Q=\frac12$, is known in closed form~\cite{Anber:2026bsx}. Consequently, $\ell$ is not conserved, and any conserved charge must be a single operator on the whole of ${\cal H}_{Y_\infty}$ rather than one defined sector by sector.
On EH space, the same events appear in a different guise. Since EH has a single boundary, it does not compute amplitudes between sectors; it prepares a boundary state. Each term in the path integral can be read as a radial history that starts from the no-boundary state near the bolt, where smoothness forces trivial holonomy and vanishing Chern--Simons number, and reaches its asymptotic sector through fractional tunneling events in radial time. The minimal EH instanton is such an event: in the gauge regular at the bolt, its holonomy around $c$ runs from $I$ to $-I$ and its enclosed charge from $0$ to $\frac12$ as $r$ increases. The path integral sums over all these histories, so the boundary state is a superposition with components in every $\ell$, each reached by tunneling from the same starting point. Nothing mixes at the boundary itself; see Figure~\ref{fig:EH-histories}.

\paragraph{The chiral operator $\mathsf X$ and the chiral charge of the EH-prepared state.}
In pure ${\cal N}=1$ $SU(N)$ super-Yang--Mills theory, the ABJ anomaly reduces the classical $U(1)_\chi$ chiral symmetry to the discrete chiral symmetry $\mathbb Z_{2N}^{\chi}$, generated by
\begin{equation}
 \lambda\longmapsto e^{\frac{2\pi i}{2N}}\lambda,\qquad A\longmapsto A,
 \qquad
 \beta_\chi=\frac{1}{2N}.
 \label{eq:center-chiral-on-fields}
\end{equation}
The identification of $\mathbb Z_{2N}^{\chi}$ as the microscopic chiral symmetry follows from the theory on a closed spin manifold: for an ordinary $SU(N)$ bundle, $Q\in\mathbb Z$ and ${\cal I}_{\rm adj}=2NQ$, so the fermionic measure is invariant under \eqref{eq:center-chiral-on-fields}.

On ${\cal H}_{Y_\infty}$, the naive generator $e^{\frac{2\pi i}{2N}Q_5}$, where $Q_5$ is the gauge-invariant axial charge (the gaugino number), is not conserved, because 
\begin{equation}
d\star J_5=\frac{2N}{8\pi^2}\operatorname{tr}F\wedge F+\frac{N^2-1}{24\cdot8\pi^2}\operatorname{tr}{\cal R}\wedge{\cal R},
\end{equation}
 with $2N=2T(\text{adj})$. Since the metric is a fixed background, the conserved charge is $Q_5-2N\operatorname{CS}$, and the operator that implements \eqref{eq:center-chiral-on-fields} on the Hilbert space is
\begin{equation}
 \mathsf X=\exp\left[\frac{2\pi i}{2N}\left(Q_5-2N\operatorname{CS}\right)\right].
 \label{eq:center-X-generator}
\end{equation}
It is invariant under single-valued large gauge transformations, which shift $\operatorname{CS}$ by an integer and change $\mathsf X$ by $e^{-2\pi i}=1$, so it is well defined on the physical Hilbert space and commutes with the Hamiltonian. Moreover, $\mathsf X$ maps each sector to itself,
\begin{equation}
 \mathsf X\,{\cal H}_\ell={\cal H}_\ell .
 \label{eq:center-X-on-sectors}
\end{equation}
Indeed, $Q_5$ acts only on the gaugino, while $\operatorname{CS}[A]$ acts on wavefunctionals by multiplication, $\Psi[A,\lambda]\to e^{-2\pi i\operatorname{CS}[A]}\Psi[A,\lambda]$. Neither changes the gauge field. Because the sectors are connected by tunneling, $Q_5$ in \eqref{eq:center-X-generator} must be the complete axial charge, including the charge of the gaugino Dirac sea, which changes by spectral flow from one sector to another; only then does $\mathsf X$ commute with the full Hamiltonian.

We now compute the eigenvalue of $\mathsf X$ on the EH-prepared state. On a state in the sector $\ell$, the axial charge is the charge of the gaugino Dirac sea of that sector plus the number of gaugino quanta above it, so that\footnote{The anomaly equation $d\star J_5=\frac{2N}{8\pi^2}\operatorname{tr}F\wedge F+\frac{N^2-1}{24\cdot8\pi^2}\operatorname{tr}{\cal R}\wedge{\cal R}$ is local and contains no boundary term when $J_5$ is the complete current, including the contribution of the Dirac sea. If instead the current is normal-ordered with respect to the Dirac sea of a given boundary sector, its divergence acquires a boundary term $-\eta_{\rm adj}\,\delta(r-R)$, and its integrated charge is the APS index ${\cal I}_{\rm adj}$, the number of quanta. The two descriptions differ by the sea charge, $Q_5^{\rm tot}=Q_5^{\rm n.o.}+\eta_{\rm adj}$. Only the complete charge defines a single operator on ${\cal H}_{Y_\infty}$ that is conserved across sectors connected by tunneling, and it is this charge that enters $\mathsf X$.}
\begin{equation}
 Q_5-2N\operatorname{CS}
 =\underbrace{Q_5^{\rm sea}(\ell)}_{\text{Dirac sea}}
 +\underbrace{N_{\rm quanta}}_{\text{path integral}}
 -\underbrace{2N\operatorname{CS}}_{\theta\text{-shift}} .
 \label{eq:center-X-bookkeeping}
\end{equation}
Each term is supplied by a single computation.

\emph{The quanta: the path integral.} A chiral rotation \eqref{eq:center-chiral-on-fields} of the gaugino in the EH path integral restricted to the sector $\ell$ multiplies it by the Fujikawa phase. By \eqref{eq:center-sector-index-general}, every term in the sum over $Q_{(N)}\in\frac\ell2+\mathbb Z$ acquires the same phase, since integer shifts of $Q_{(N)}$ change the index by multiples of $2N$:
\begin{equation}
 {\cal A}_\ell=\exp\left(-\frac{2\pi i}{2N}\,{\cal I}_{{\rm adj},\ell}\right)=e^{-2\pi i\,\ell^2/N}.
 \label{eq:center-sector-charges}
\end{equation}
Rotating the gaugino field acts on the gaugino quanta but not on the filled Dirac sea, which is part of the definition of the boundary condition. The phase \eqref{eq:center-sector-charges} therefore records that $|\Psi_\ell\rangle$ carries $N_{\rm quanta}={\cal I}_{{\rm adj},\ell}$ gaugino quanta above the Dirac sea of its sector\footnote{The quanta and the zero modes are related (the same number) but not identical. The zero modes are normalizable solutions of the bulk Dirac equation, localized near the bolt, and enter the path integral as Grassmann collective coordinates; the quanta are Fock excitations of the boundary Hilbert space above the Dirac sea of the sector. The index fixes the net number of the latter, quanta minus antiquanta, as the index counts zero modes of one chirality minus the other. The mechanism is the same as for the 't~Hooft vertex on $\mathbb R^4$, where the zero modes of an instanton encode the fermions created by spectral flow in the transition between vacua. Without insertions, the Grassmann integration over the zero modes is saturated only by the boundary gaugino data, through the fermion propagators connecting the zero modes to the asymptotic slice, so that $|\Psi_\ell\rangle$ has no overlap with the bare Fock vacuum of its sector but does overlap with states containing ${\cal I}_{{\rm adj},\ell}$ additional quanta. If instead an operator such as $\operatorname{tr}_{\mathbf N}(\lambda\lambda)$ is inserted in the bulk, it absorbs the zero modes locally; the corresponding state then has no extra quanta, and the chiral charge is carried by the insertion. This is the amplitude $Z^{(1)}[\operatorname{tr}\lambda\lambda]$ that enters the condensate.}; rotating the argument of the wavefunctional produces the inverse of the phase $e^{2\pi i\beta_\chi N_{\rm quanta}}$. We refer to ${\cal A}_\ell$ as the zero-mode charge of the sector. For the single-root sector, the minimal saddle has ${\cal I}_{\rm adj}^{(\ell=1)}=2$, and throughout the fixed-holonomy sector
\begin{equation}
 {\cal I}_{\rm adj}\left(k+\frac12,h_c^{(\ell=1)}\right)
 =
 2+2Nk,
 \label{eq:EH-SUN-fixed-holonomy-index}
\end{equation}
so that
\begin{align}
 {\cal A}_{h_c}
 &=
 \exp\left[
 -2\pi i\frac{1}{2N}
 {\cal I}_{\rm adj}\left(k+\frac12,h_c^{(\ell=1)}\right)
 \right]
 =
 \exp\left(-\frac{2\pi i}{N}\right).
 \label{eq:EH-SUN-chiral-anomaly}
\end{align}
The phase is well defined for the complete dynamical sum at fixed boundary holonomy; it is not an artifact of retaining only the minimal charge-$1/2$ saddle. Because every source-free term in this sector possesses unsaturated fermion zero modes, the phase \eqref{eq:EH-SUN-chiral-anomaly} should be understood as the transformation of the regulated fermionic measure, or of the generating functional before the fermionic sources are set to zero.

\emph{The Chern--Simons term: the $\theta$-shift.} The fermion rotation alone is not a symmetry. On a closed manifold it must be accompanied by the shift $\theta\to\theta+2\pi\cdot2N\beta_\chi$, which compensates the bulk contribution $2NQ$ to the Fujikawa phase. On EH, the $\theta$-weight is $e^{i\theta Q_{(N)}}=e^{i\theta\operatorname{CS}[A_\infty]}$ by \eqref{eq:center-Q-equals-CS}, so the compensating shift acts on the boundary state as multiplication by a function of $\operatorname{CS}[A_\infty]$. This is the term $-2N\operatorname{CS}$ in \eqref{eq:center-X-generator}: the path-integral counterpart of $\mathsf X$ is the fermion rotation combined with the $\theta$-shift.

\emph{The Dirac sea: the boundary spectrum.} The charge of the filled Dirac sea is invisible to the field rotation and must be supplied by the spectrum of the Weyl operator on $\mathbb{RP}^3$ in the holonomy background $h_c^{(\ell)}$. On the round $S^3$ of radius $R$, its eigenvalues are $\pm(n+\frac32)/R$, with degeneracy $(n+1)(n+2)$, symmetric under $\lambda\to-\lambda$. The antipodal map, combined with the holonomy of a given adjoint component, keeps on $\mathbb{RP}^3$ the positive modes of one parity of $n$ and the negative modes of the other. The spectrum becomes asymmetric, and its $\eta$-invariant is $\mp\frac18$ in the normalization of \eqref{eq:EH-adjoint-eta-signature}. The sign and the normalization of the sea charge follow from the anomaly equation itself. Integrating $d\star J_5$ over EH, whose only boundary is $\mathbb{RP}^3$, the total axial charge of the prepared state is $Q_5=2NQ_{(N)}-\frac{N^2-1}{8}$, while the number of quanta is ${\cal I}_{{\rm adj},\ell}$, given by \eqref{eq:EH-adjoint-APS-general}. Their difference is
\begin{equation}
 Q_5^{\rm sea}(\ell)=\eta_{\rm adj}=-\frac{N^2-1}{8}+\ell(N-2\ell).
 \label{eq:center-sea-charge}
\end{equation}

\emph{The eigenvalue.} Inserting the three contributions into \eqref{eq:center-X-bookkeeping} and using the APS theorem ${\cal I}_{{\rm adj},\ell}=2NQ_{(N)}-\frac{N^2-1}{8}-\eta_{\rm adj}$,
\begin{equation}
 Q_5-2N\operatorname{CS}
 =\eta_{\rm adj}+\Big(2NQ_{(N)}-\frac{N^2-1}{8}-\eta_{\rm adj}\Big)-2NQ_{(N)}
 =-\frac{N^2-1}{8}.
 \label{eq:center-X-cancellation}
\end{equation}
Each term of the APS index is canceled by one entry of \eqref{eq:center-X-bookkeeping}: the bulk gauge contribution $2NQ_{(N)}$ by the Chern--Simons term, and the boundary $\eta$-invariant by the charge of the Dirac sea. Only the gravitational contribution survives, and it is the same in every sector. Hence
\begin{equation}
 \mathsf X|\Psi_\ell\rangle=\chi_0\,|\Psi_\ell\rangle\quad\text{for all }\ell,
 \qquad
 \mathsf X|\Psi_{\rm EH}\rangle=\chi_0\,|\Psi_{\rm EH}\rangle ,
 \qquad
 \chi_0=e^{-2\pi i(N^2-1)/16N}.
 \label{eq:center-X-on-EH}
\end{equation}
The EH-prepared state is an eigenstate of $\mathsf X$ with a single eigenvalue because $\mathsf X$ is conserved along the entire radial evolution, including the cap\footnote{This can also be seen directly from charge conservation. For every configuration in the EH path integral, integrating the anomaly equation over the region between the bolt and the asymptotic slice relates $Q_5-2N\operatorname{CS}$ on $\mathbb{RP}^3$ to the charge entering through an inner boundary and to the bulk sources. EH has no inner boundary, since the bolt is an interior submanifold, and the gauge source is exactly cancelled by the Chern--Simons term, leaving only the gravitational density, which is the same for all configurations. A component of the EH-prepared state carrying a different value of $\mathsf X$ would therefore require charge to flow in from an inner boundary or to be created by a sector-dependent source, and neither is available.}: the components with $\ell\neq0$ are generated from the no-boundary state by fractional tunneling events, each of which creates exactly the gaugino quanta that compensate its change of Chern--Simons number and sea charge, while the cap, having no inner boundary, can contribute only the sector-independent gravitational term\footnote{The constant $\chi_0$ can be absorbed by completing $\mathsf X$ with the gravitational Chern--Simons term, $Q_5-2N\operatorname{CS}-(N^2-1)\operatorname{CS}_g$, with $d\operatorname{CS}_g=\frac{1}{24\cdot8\pi^2}\operatorname{tr}{\cal R}\wedge{\cal R}$. With the framing induced from EH, $\operatorname{CS}_g(\mathbb{RP}^3)=-\frac18$, and the EH-prepared state becomes exactly neutral. Since the metric is a background, $\operatorname{CS}_g$ is a c-number on ${\cal H}_{Y_\infty}$; this choice changes $\mathsf X$ by a constant phase and affects none of the statements below.}. Two examples illustrate the bookkeeping:
\begin{table}[h]
\centering
\begin{tabular}{c|c|c|c|c}
 & $Q_5^{\rm sea}=\eta_{\rm adj}$ & $N_{\rm quanta}={\cal I}_{\rm adj}$ & $2N\operatorname{CS}$ & $Q_5-2N\operatorname{CS}$\\
\hline
$SU(2)$, $\ell=0$ & $-\frac38$ & $0$ & $0$ & $-\frac38$\\
$SU(2)$, $\ell=1$ & $-\frac38$ & $2$ & $2$ & $-\frac38$\\
$SU(4)$, $\ell=0$ & $-\frac{15}8$ & $0$ & $0$ & $-\frac{15}8$\\
$SU(4)$, $\ell=1$ (single root) & $+\frac18$ & $2$ & $4$ & $-\frac{15}8$
\end{tabular}
\caption{The three contributions to the $\mathsf X$-charge of the lowest EH saddles.}
\label{tab:X-bookkeeping}
\end{table}
For $SU(2)$ the sea charge is the same in both sectors, since $\operatorname{Ad}(-I_2)=1$, and the two zero modes of the $Q=\frac12$ saddle are compensated by $\operatorname{CS}=\frac12$. For $SU(4)$ the sea charge itself shifts by $2$ between the sectors.

The perturbative ground state of the sector $\ell$, the gaugino Fock vacuum $|\Omega_\ell\rangle$, carries only the sea charge \eqref{eq:center-sea-charge} and $\operatorname{CS}\equiv\ell/2$. Its $\mathsf X$-charge is therefore sector dependent,
\begin{equation}
 \mathsf X|\Omega_\ell\rangle=\chi_0\,e^{-2\pi i\ell^2/N}|\Omega_\ell\rangle=\chi_0\,{\cal A}_\ell\,|\Omega_\ell\rangle ,
 \label{eq:center-X-on-Fock}
\end{equation}
and the ${\cal I}_{{\rm adj},\ell}$ quanta of $|\Psi_\ell\rangle$ contribute exactly the compensating factor ${\cal A}_\ell^{-1}$. This shows how the zero-mode charge ${\cal A}_\ell$ enters: it is the charge of $|\Psi_\ell\rangle$ measured relative to $|\Omega_\ell\rangle$, not its $\mathsf X$-eigenvalue.

\emph{Comparison with $\mathbb R^4$.} The same bookkeeping applies on $\mathbb R^4$ with $S^3$ slices, but there two of its three terms vanish and the third is invisible. The Weyl spectrum on the round $S^3$ around a flat connection is symmetric, so the sea charge vanishes; flat space has no gravitational term; and the vacua $|n\rangle$ have integer $\operatorname{CS}=n$. Because $\operatorname{CS}$ is an integer on the vacua $|n\rangle$, the factor $e^{-2\pi i\operatorname{CS}}$ in $\mathsf X$ is trivial on them, and the naive rotation $e^{\frac{2\pi i}{2N}Q_5}$ and $\mathsf X$ assign the same eigenvalues to these states, even though only $\mathsf X$ is conserved as an operator. The distinction between the Fujikawa phase and the $\mathsf X$-charge therefore never arises in practice. On twisted $\mathbb T^3$~\cite{Cox:2021vsa}, $\operatorname{CS}$ becomes fractional while the sea charge still vanishes. $\mathbb{RP}^3$ is the case in which all three contributions are nonzero: $\operatorname{CS}$ is fractional because of the torsion, the spectrum is asymmetric because of the antipodal projection, and the curvature of EH supplies the gravitational term.

\paragraph{Charges versus anomalies.}
The single-root configuration defines a liftable $SU(N)$ bundle: its center-obstruction class vanishes, and no background two-form gauge field for $\mathbb Z_N^{(1)}$ has been activated. The fractional topological charge originates from the Chern--Simons invariant of the flat connection $h_c$ on $\mathbb{RP}^3$, which is possible because of the torsion of the boundary. Correspondingly, the APS index entering \eqref{eq:EH-SUN-chiral-anomaly} contains the bulk gauge contribution, the gravitational contribution, and the $\eta$-invariant of the Dirac operator on the asymptotic boundary, and only their complete combination is integer-valued. The phase \eqref{eq:EH-SUN-chiral-anomaly} is therefore not an 't~Hooft anomaly. It counts the gaugino quanta of the state prepared in that sector, and, once the Dirac sea and the Chern--Simons term are included, the EH-prepared state carries the same $\mathsf X$-charge in every sector. An anomaly arises only when a symmetry acting on the $\mathbb{RP}^3$ Hilbert space fails to commute with $\mathsf X$, so that it maps states of one chiral charge to states of another by a state-independent projective phase. The only candidate is the center symmetry, to which we now turn.

\paragraph{Which center elements act on $\mathbb{RP}^3$.}
The $\mathbb Z_N^{(1)}$ center symmetry acts on ${\cal H}_{Y_\infty}$ by multiplying the holonomy around the torsion cycle $c$ of $\mathbb{RP}^3$ by a center element,
\begin{equation}
 h_c\longrightarrow\omega^k h_c,\qquad \omega=e^{2\pi i/N},\quad k=0,\ldots,N-1.
 \label{eq:SUN-center-on-holonomy}
\end{equation}
Since $c^2=1$, an admissible boundary holonomy must satisfy \eqref{eq:SUN-boundary-holonomy-order-two}. The image $\omega^kh_c$ satisfies this condition only if $(\omega^kh_c)^2=\omega^{2k}I_N=I_N$, i.e., $\omega^{2k}=1$. Hence only the subgroup $\mathbb Z_{\gcd(2,N)}\subset\mathbb Z_N$ acts on the $\mathbb{RP}^3$ Hilbert space.

For odd $N$, this subgroup is trivial. The $\lfloor N/2\rfloor+1$ boundary sectors are then not related by any symmetry. For instance, for $SU(3)$ the sectors $h_c=I_3$ and $h_c=\operatorname{diag}(-I_2,1)$ are distinct boundary conditions, and no operator on ${\cal H}_{Y_\infty}$ maps one into the other. For even $N$ the subgroup is $\mathbb Z_2$, generated by $\omega^{N/2}=-I_N$. Unless stated otherwise, we take $N$ even in what follows.

\paragraph{The center operator $\mathsf U$.}
We first recall the $SU(2)$ construction~\cite{Anber:2026bsx}. Write a point of the covering $S^3$ as a unit quaternion and parametrize it by Euler angles,
\begin{equation}
 g(\theta,\varphi,\psi)
 =e^{i\varphi\tau_3/2}\,e^{i\theta\tau_2/2}\,e^{i(\psi+\pi)\tau_3/2}\in SU(2),
 \qquad 0\le\theta\le\pi,\quad 0\le\varphi<2\pi,\quad 0\le\psi<4\pi,
 \label{eq:center-odd-map-g}
\end{equation}
where $\tau_a$ are the Pauli matrices. The map $g$ is the identity map $S^3\to SU(2)\cong S^3$, so it has degree one. Its Maurer--Cartan form is
\begin{equation}
 g^{-1}dg=i\,\tau_a\sigma_a ,
 \label{eq:center-odd-map-MC}
\end{equation}
with the left-invariant one-forms $\sigma_a$ of \eqref{sigmas}. The antipodal map $x\to-x$ of $S^3$ is $\psi\to\psi+2\pi$, and $\mathbb{RP}^3$ is obtained by restricting to $0\le\psi<2\pi$, as in \eqref{eq:EH-coordinate-ranges}. Under this shift,
\begin{equation}
 g(\theta,\varphi,\psi+2\pi)=-\,g(\theta,\varphi,\psi),
 \label{eq:center-odd-map-odd}
\end{equation}
so $g$ is odd under the antipodal map. Thus $g$ is not a single-valued map $\mathbb{RP}^3\to SU(2)$, while $g\,dg^{-1}$ and $\operatorname{Ad}g$ are single-valued, since the sign cancels.

For $SU(N)$ with even $N$ we take $N/2$ copies of $g$ along the diagonal,
\begin{equation}
 G=\operatorname{diag}\left(g,g,\ldots,g\right)=g\otimes I_{N/2}\in SU(N),
 \qquad
 G(-x)=-G(x)=\omega^{N/2}G(x).
 \label{eq:center-odd-map-G}
\end{equation}
We define $\mathsf U$ as the operator on ${\cal H}_{Y_\infty}$ that implements the transformation of the boundary fields
\begin{equation}
 \mathsf U:\qquad
 A\longmapsto A^G=G\,A\,G^{-1}-i\,G\,dG^{-1},
 \qquad
 \lambda\longmapsto G\,\lambda\,G^{-1},
 \label{eq:center-U-on-fields}
\end{equation}
i.e., $(\mathsf U\Psi)[A,\lambda]=\Psi[A^{G^{-1}},G^{-1}\lambda G]$ on wavefunctionals. Because $-I_N$ is central, $\operatorname{Ad}G$ and $G\,dG^{-1}$ are single-valued on $\mathbb{RP}^3$. Therefore \eqref{eq:center-U-on-fields} maps $SU(N)$ configurations on $\mathbb{RP}^3$, adjoint fermions included, to configurations on $\mathbb{RP}^3$. Since $G$ itself is not single-valued on $\mathbb{RP}^3$, $\mathsf U$ is not a gauge transformation; it is a genuine global symmetry, the generator of the $\mathbb Z_2\subset\mathbb Z_N^{(1)}$ that survives on $\mathbb{RP}^3$. Changing $G$ by a single-valued $SU(N)$ gauge transformation changes $\mathsf U$ only by a gauge transformation, so $\mathsf U$ is well defined on the physical Hilbert space and commutes with the Hamiltonian. Its square is implemented by $G^2=g^2\otimes I_{N/2}$, which is single-valued on $\mathbb{RP}^3$: each block $g^2$ has degree two on $S^3$, hence degree one on $\mathbb{RP}^3$. Thus $\mathsf U^2$ is a large gauge transformation that shifts the Chern--Simons number by $N/2$ (recall that $N$ is even). Physical states are invariant under it once the $\theta$ angle has been rotated into the fermions, so that $\mathsf U^2=1$ on ${\cal H}_{Y_\infty}$.

\paragraph{Action of $\mathsf U$ on the boundary sectors.}
Parallel transport of $A^G$ along a path from $x$ to $y$ is $G(y)\,W_A(y,x)\,G(x)^{-1}$, where $W_A$ is the parallel transport of $A$. Along $c$, which runs from $\psi$ to $\psi+2\pi$, \eqref{eq:center-odd-map-odd} gives
\begin{equation}
 h_c[A^G]=G(\psi+2\pi)\,h_c[A]\,G(\psi)^{-1}
 =-\,G(\psi)\,h_c[A]\,G(\psi)^{-1}\simeq -\,h_c[A].
 \label{eq:center-U-on-holonomy}
\end{equation}
Up to conjugation, $\mathsf U$ multiplies the holonomy by $-I_N$, as it should. On the representatives \eqref{eq:SUN-boundary-holonomy-representatives},
\begin{equation}
 -h_c^{(\ell)}=\operatorname{diag}\left(I_{2\ell},-I_{N-2\ell}\right)\simeq h_c^{(N/2-\ell)},
 \qquad\text{so}\qquad
 \mathsf U\,|\ell;\mathfrak k\rangle=\left|\tfrac N2-\ell;\mathfrak k'\right\rangle,
 \quad
 \mathsf U\,{\cal H}_\ell={\cal H}_{N/2-\ell} .
 \label{eq:SUN-center-on-sectors}
\end{equation}
Here $\mathfrak k'$ denotes the image of the remaining quantum numbers. Thus $\mathsf U$ pairs the sectors
\begin{equation}
 \ell\longleftrightarrow\frac N2-\ell,\qquad \ell=0,1,\ldots,\frac N2 .
 \label{eq:center-sector-pairing}
\end{equation}
In particular, it exchanges the trivial sector $\ell=0$ with the central sector $\ell=N/2$, $h_c=-I_N$, of \eqref{eq:SUN-full-block-central-holonomy}. When $N\equiv0\bmod4$, the sector $\ell=N/4$ is mapped to itself. The simplest example is the trivial sector itself: acting on $A=0$ gives the flat connection
\begin{equation}
 \mathsf U:\quad 0\longmapsto -i\,G\,dG^{-1}=\operatorname{diag}\left(-i\,g\,dg^{-1},\ldots,-i\,g\,dg^{-1}\right),
 \label{eq:center-U-on-trivial}
\end{equation}
which is $N/2$ copies of the nontrivial $SU(2)$ flat connection and has holonomy $-I_N$. This is precisely the boundary value of the configuration with all $s=N/2$ blocks minimal in \eqref{eq:SUN-minimal-multiple-block-sector}.

\paragraph{Shift of the Chern--Simons invariant.}
The Chern--Simons functional enters in two places: the chiral operator \eqref{eq:center-X-generator} contains it explicitly, and, through \eqref{eq:center-Q-equals-CS}, its value on the boundary connection, which is defined only modulo $1$ since large gauge transformations on $\mathbb{RP}^3$ shift it by integers, fixes the fractional part of the bulk topological charge, while the integer part depends on how the bulk is filled. Since $\mathsf U$ leaves $Q_5$ invariant, its commutation with $\mathsf X$, as well as its effect on the bulk topological sectors, is controlled by the change of $\operatorname{CS}$. Under \eqref{eq:center-U-on-fields}, on a closed three-manifold, the standard gauge variation of the Chern--Simons functional leaves only the winding term,
\begin{equation}
 \Delta\operatorname{CS}\equiv\operatorname{CS}[A^G]-\operatorname{CS}[A]
 =\pm\frac{1}{24\pi^2}\int_{\mathbb{RP}^3}\operatorname{tr}\left(G^{-1}dG\right)^3 ,
 \label{eq:center-CS-shift-def}
\end{equation}
where the overall sign depends on orientation conventions. The integrand is built from $G^{-1}dG$. This combination is invariant under $G\to-G$, so it is a well-defined three-form on $\mathbb{RP}^3$ even though $G$ is not. To evaluate it, we use \eqref{eq:center-odd-map-MC}. With $\operatorname{tr}(\tau_a\tau_b\tau_c)=2i\epsilon_{abc}$, one finds $\operatorname{tr}\left(g^{-1}dg\right)^3=12\,\sigma_x\wedge\sigma_y\wedge\sigma_z$, so each block contributes $\frac{1}{24\pi^2}\int\operatorname{tr}(g^{-1}dg)^3=\pm1$ on $S^3$ ($0\le\psi<4\pi$), i.e., $\deg g=1$. On $\mathbb{RP}^3$ ($0\le\psi<2\pi$) the integral covers half of $S^3$, so each block contributes $\pm\frac12$. Summing over the $N/2$ blocks,
\begin{equation}
 \Delta\operatorname{CS}=\pm\frac12\deg_{S^3}G=\pm\frac N4
 \equiv\frac N4\pmod 1 .
 \label{eq:SUN-center-CS-shift}
\end{equation}
The result is defined mod $1$, because composing $G$ with a single-valued large gauge transformation shifts $\Delta\operatorname{CS}$ by an integer\footnote{Since $\pi_1(SU(N))=\pi_2(SU(N))=0$ and $\pi_3(SU(N))=\mathbb Z$, homotopy classes of single-valued maps $h:\mathbb{RP}^3\to SU(N)$ are labelled by the degree $\deg h=\frac{1}{24\pi^2}\int_{\mathbb{RP}^3}\operatorname{tr}(h^{-1}dh)^3\in\mathbb Z$, and a large gauge transformation shifts $\operatorname{CS}$ by $\deg h$. Because $g$ is odd under the antipodal map, the generator on $\mathbb{RP}^3$ is not $g$, as on $S^3$, but $h_1=\operatorname{diag}(g^2,I_{N-2})$, which is even and has degree $2$ on $S^3$, hence degree $1$ on $\mathbb{RP}^3$. Single-valued maps conjugate the holonomy, so they preserve the sector $\ell$; within each sector, the classical vacua with $\operatorname{CS}=\frac\ell2+n$ are related by large gauge transformations, and $\mathsf X$ is invariant under them. The torsion of $\mathbb{RP}^3$ also allows ``half-degree'' maps, which are odd under the antipodal map, $M(-x)=\Omega M(x)$. When $\Omega$ is central, as for $G=g\otimes I_{N/2}$ with $\Omega=-I_N$, the adjoint action is single-valued and $M$ implements the center symmetry $\mathsf U$. When $\Omega$ is not central, as for $K=\operatorname{diag}(g,I_{N-2})$ with $\Omega=\operatorname{diag}(-I_2,I_{N-2})$ and $N>2$, the off-diagonal components of $A$ and $\lambda$ change sign, and $K$ does not act on the configuration space of $\mathbb{RP}^3$ at all, even though it maps $A=0$ to the single-root flat connection. The obstruction is entirely due to the $x$-dependent sign of $g$; constant color rotations, such as the orientation moduli in $K_{h_c}$ we shall use in the next section, are trivially single-valued and act on the full configuration space. In all cases $M^2$ is single-valued and is an ordinary large gauge transformation; for instance $G^2$ has degree $N/2$, consistent with $\mathsf U^2$.}. Moreover, $\pm N/4$ coincide mod $1$ for even $N$, so the sign convention drops out. As a check, the minimal block-diagonal representatives \eqref{eq:SUN-minimal-multiple-block-sector} show that each block with holonomy $-I_2$ contributes $\frac12$. Hence the sector $\ell$ has $\operatorname{CS}\equiv\ell/2\pmod 1$, and its image $N/2-\ell$ has $\operatorname{CS}\equiv N/4-\ell/2$. The difference is $N/4-\ell\equiv N/4\pmod 1$, the same for every sector, in agreement with \eqref{eq:SUN-center-CS-shift}.

\paragraph{The algebra of $\mathsf U$ and $\mathsf X$.}
The gaugino transforms under $\mathsf U$ by $\operatorname{Ad}G$, which commutes with the chiral rotation \eqref{eq:center-chiral-on-fields}, so $Q_5$ is $\mathsf U$-invariant; this includes the sea charge, since $\operatorname{Ad}(-h_c)=\operatorname{Ad}h_c$ leaves the boundary spectrum unchanged. The Chern--Simons term, however, is shifted by \eqref{eq:SUN-center-CS-shift}. Hence $\mathsf U\mathsf X\mathsf U^{-1}=e^{-2\pi i\Delta\operatorname{CS}}\mathsf X$, i.e.,
\begin{equation}
 \mathsf X\,\mathsf U\,\mathsf X^{-1}\mathsf U^{-1}=e^{-i\pi N/2}
 =\begin{cases}
 -1,& N\equiv2\bmod4,\\
 +1,& N\equiv0\bmod4.
 \end{cases}
 \label{eq:SUN-center-chiral-algebra}
\end{equation}
For $N\equiv2\bmod4$, $\mathsf X$ and $\mathsf U$ anticommute. Hence, if $|\psi\rangle\in{\cal H}_\ell$ has chiral charge $x$, $\mathsf X|\psi\rangle=x|\psi\rangle$, then $\mathsf U|\psi\rangle\in{\cal H}_{N/2-\ell}$ has chiral charge $-x$. Moreover, no single nonzero state is a simultaneous eigenstate of $\mathsf X$ and $\mathsf U$: if $\mathsf X|\psi\rangle=x|\psi\rangle$ and $\mathsf U|\psi\rangle=u|\psi\rangle$, then \eqref{eq:SUN-center-chiral-algebra} gives $xu=-xu$. Thus, the algebra of $\mathsf U$ and $\mathsf X$ cannot be implemented on a one-dimensional Hilbert space; the minimum dimension is $2$. This is the restriction of the familiar mixed anomaly between $\mathbb Z_{2N}^{\chi}$ and $\mathbb Z_N^{(1)}$ \cite{Gaiotto:2017yup,Cox:2021vsa} to the $\mathbb Z_2$ subgroup of the center visible on $\mathbb{RP}^3$, realized on the untwisted Hilbert space; the torsion of the boundary replaces the 't~Hooft flux that would otherwise be required. For $N\equiv0\bmod4$, $\mathsf U$ and $\mathsf X$ commute, and no projective phase is visible on $\mathbb{RP}^3$ with a purely temporal background.

\paragraph{Bulk realization of $\mathsf U$.}
Before computing the action of $\mathsf U$ and $\mathsf X$ on the EH-prepared state, we explain how $\mathsf U$ is realized on the EH path integral. Acting with $\mathsf U$ on the boundary data, $A_\infty\to A_\infty^G$, any bulk configuration whose boundary value is $A_\infty^G$ must have, by \eqref{eq:center-Q-equals-CS} and \eqref{eq:SUN-center-CS-shift},
\begin{equation}
 Q'\equiv\operatorname{CS}[A^G_\infty]=Q+\frac N4\pmod 1 .
 \label{eq:center-Q-shift}
\end{equation}
For $SU(2)$, this relates the sectors $Q\in\mathbb Z$ and $Q\in\frac12+\mathbb Z$. Since no bulk $SU(N)$ gauge transformation can shift $Q$ by a fraction, $\mathsf U$ cannot be realized as a change of variables among $SU(N)$ configurations on EH.

Indeed, the transformation \eqref{eq:center-U-on-fields} depends on $G$ only up to sign, so what must be extended into the bulk is the adjoint action of $G$. Away from the bolt this is immediate: ${\rm EH}\setminus{\rm bolt}\simeq(a,\infty)\times\mathbb{RP}^3$, and the same transformation can be used on every radial slice. Across the bolt it is impossible. Along the loop $c$, $G$ returns to itself only up to the center element $-I_N$, and this sign cannot be removed by deforming the loop within $\mathbb{RP}^3$. In EH, however, the $\psi$-circle collapses at the bolt, so $c$ bounds a disk, and a smooth extension over this disk would force the sign to be trivial. The transformation therefore becomes singular on the bolt, where it carries the center element $-I_N=\omega^{N/2}$; this singularity is the surface operator discussed next.

The center symmetry is nevertheless a symmetry of the full four-dimensional theory, so this obstruction has a bulk meaning. A one-form symmetry acts through topological surface operators. On ${\cal H}_{Y_\infty}$, $\mathsf U$ is the surface operator wrapping the generator of $H_2(\mathbb{RP}^3;\mathbb Z_N)=\mathbb Z_2$ at fixed radial time. Insert it on a slice at large radius and push it inward. Since it is topological, it can be deformed freely, but it cannot be removed, because in EH this cycle is equivalent to $\frac N2$ times the bolt, which is a nontrivial two-cycle.

\begin{figure}[t]
\centering
\includegraphics[width=0.85\textwidth]{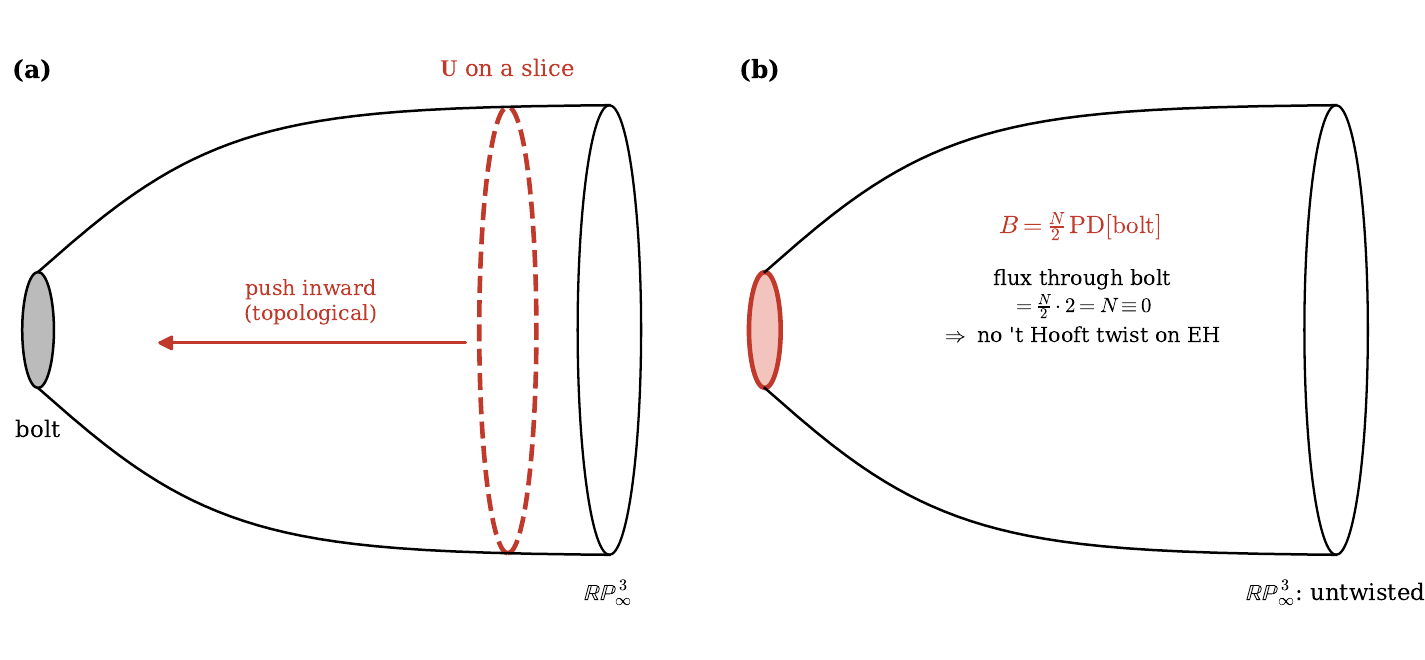}
\caption{(a) The center operator $\mathsf U$ is a topological surface operator on a radial slice of EH and can be pushed inward. (b) On the bolt it becomes the $\mathbb Z_N$ background $B=\frac N2\,{\rm PD}[{\rm bolt}]$, whose flux through the bolt is $\frac N2\cdot2=N\equiv0\pmod N$; EH therefore carries no 't~Hooft twist, and the asymptotic $\mathbb{RP}^3$ remains untwisted.}
\label{fig:EH-U-push}
\end{figure}

It is important to stress what this operator does \emph{not} do: it does not introduce an 't~Hooft twist. Written as a $\mathbb Z_N$ two-form field, the surface operator pushed onto the bolt is $B=\frac N2\,{\rm PD}[{\rm bolt}]$ (where $\rm PD$ is the Poincar\'e dual), a background for the $\mathbb Z_2$ subgroup of $\mathbb Z_N^{(1)}$ generated by $\omega^{N/2}$; see Figure~\ref{fig:EH-U-push}. Its flux through the bolt, the only two-cycle of EH, is $\frac N2\cdot[{\rm bolt}]\cdot[{\rm bolt}]=N\equiv0\pmod N$, and we used the fact that the intersection number of the bolt in the EH space, in the orientation used in this work, is $+2$. Hence the gauge bundle carries no 't~Hooft flux, it is a genuine $SU(N)$ bundle, and the boundary remains in the untwisted Hilbert space ${\cal H}_{Y_\infty}$. The bulk field is the same genuine $SU(N)$ configuration as before; $B$ merely records the location of the surface operator, across which the boundary data of the $SU(N)$ field are multiplied by the center element $-I_N$ along $c$. Writing $|\Psi_{\rm EH}[B]\rangle$ for the state prepared with the surface operator inserted, we thus have
\begin{equation}
 \mathsf U\,|\Psi_{\rm EH}[0]\rangle=|\Psi_{\rm EH}[B]\rangle .
 \label{eq:center-U-on-EH-state}
\end{equation}

This is most transparent in the Cartan backgrounds \eqref{eq:SUN-Cartan-background}. There, the action of $\mathsf U$ can be encoded smoothly by shifting every bolt flux, $n_i\to n_i+1$. This flips the sign of every eigenvalue of the boundary holonomy, $h_c\to-h_c$, while leaving all differences $n_i-n_j$ unchanged. The shift adds to the curvature a piece proportional to the identity matrix. This piece is pure bookkeeping for the boundary flip: it commutes with everything, no adjoint field sees it, and the traceless part of the field, and hence all bulk physics, including the adjoint APS index, is unchanged\footnote{The shift $n_i\to n_i+1$ gives $\sum_i(n_i+1)=N\neq0$, so this smooth description of the $\mathsf U$-image is not an $SU(N)$ connection but a $U(N)$ representative of it: the $SU(N)$ connection is tensored with a line bundle carrying unit flux through the bolt on every eigenvalue. This auxiliary $U(1)$ factor is not dynamical; it only encodes the boundary flip, equivalently to the surface-operator description in the main text. Its determinant has flux $N\equiv0\pmod N$, so no 't~Hooft flux is introduced, and the traceless part of the connection, which is all that the adjoint fields see, is the original $SU(N)$ one. It is then useful to distinguish two instanton numbers. Splitting the curvature of the $U(N)$ representative as $F=\hat F+\frac1N(\operatorname{tr}F)\,\bm1$ with $\operatorname{tr}\hat F=0$, we define
\begin{equation*}
Q_{(N)}=\frac1{8\pi^2}\int\operatorname{tr}F\wedge F,\qquad
Q_{\rm adj}=\frac1{8\pi^2}\int\operatorname{tr}\hat F\wedge\hat F
=Q_{(N)}-\frac1N\,\frac1{8\pi^2}\int\operatorname{tr}F\wedge\operatorname{tr}F .
\end{equation*}
For a genuine $SU(N)$ configuration, $\operatorname{tr}F=0$ and the two coincide; this is the charge $Q_{(N)}$ appearing in \eqref{eq:center-sector-index-general}. The charge $Q_{(N)}$ is the one weighted by $\theta$ and equal to $\operatorname{CS}[A_\infty]$ modulo $1$. The adjoint fermions couple to $F$ only through commutators $[F,\lambda]$, so the identity piece drops out and the APS index depends only on $Q_{\rm adj}$: $\mathcal I=2NQ_{\rm adj}-\ell(N-2\ell)$. In the Cartan backgrounds,
\begin{equation*}
Q_{(N)}=\frac14\sum_i n_i^2,\qquad 2N\,Q_{\rm adj}=\frac12\sum_{i<j}(n_i-n_j)^2 ,
\end{equation*}
so under $n_i\to n_i+1$ one finds $\Delta Q_{(N)}=\frac14\bigl(2\sum_i n_i+N\bigr)=\frac N4$, carried entirely by the auxiliary $U(1)$ factor, while $Q_{\rm adj}$ is unchanged. The boundary term is unchanged as well, since $\operatorname{Ad}(-h_c)=\operatorname{Ad}h_c$, and indeed $(\frac N2-\ell)\,2\ell=\ell(N-2\ell)$.}. What does change is the charge $Q_{(N)}$ weighted by $\theta$, which shifts by $\frac N4$, in agreement with \eqref{eq:center-Q-shift}. Equivalently, in the presence of $B$ the fractional part of $Q_{(N)}$ is $\frac{N-1}{2N}\int\mathcal P(B)$ (with $\mathcal P(B)$ is the Pontryagin square and the periods of $B$ normalized to be integers mod $N$), and $\int_{\rm EH}\mathcal P(B)=\left(\frac N2\right)^2[{\rm bolt}]\cdot[{\rm bolt}]=\frac{N^2}{2}$, giving $\frac{N(N-1)}{4}\equiv\frac N4\pmod 1$.

The case of $SU(2)$ makes the consequences explicit. Here $\ell(N-2\ell)=0$ in both sectors, so $\mathcal I=4Q_{\rm adj}$. The lowest honest configurations are listed in the upper part of Table~\ref{tab:SU2-U-action}, and their $\mathsf U$-images in the lower part. The images are written in the smooth Cartan form $n_i+1$; in the bulk they are the same $SU(2)$ fields as in the upper part.
\begin{table}[h]
\centering
\begin{tabular}{c|c|c|c|c}
configuration & sector $\ell$ & $Q_{(N)}$ & $Q_{\rm adj}$ & index $\mathcal I$\\
\hline
$(0,0)$ & $0$ & $0$ & $0$ & $0$\\
$(1,-1)$ & $1$ & $\frac12$ & $\frac12$ & $2$\\
\hline\hline
$\mathsf U:(0,0)\to(1,1)$ & $1$ & $\frac12$ & $0$ & $0$\\
$\mathsf U:(1,-1)\to(2,0)$ & $0$ & $1$ & $\frac12$ & $2$
\end{tabular}
\caption{Action of $\mathsf U$ on the lowest $SU(2)$ Cartan configurations on EH. Upper part: honest configurations. Lower part: their $\mathsf U$-images.}
\label{tab:SU2-U-action}
\end{table}
Three features stand out. First, the sector flips, since every holonomy eigenvalue $(-1)^{n_i}$ changes sign. Second, $Q_{(N)}$, and hence $\operatorname{CS}$, shifts by $\frac12=\frac N4$, so the $\theta$-weight acquires the phase $e^{i\theta/2}$. Third, the index is unchanged, since $n_1-n_2$ is unchanged. Consequently, the $\mathsf U$-image in the sector $\ell=1$ carries no zero modes, whereas honest configurations there carry $2\pmod 4$; conversely, the image in $\ell=0$ carries two zero modes, whereas honest configurations there carry $0\pmod 4$. Compare now the two states that land in the same sector, say the honest $(1,-1)$ and the image $(1,1)$ in $\ell=1$. They have the same sea charge and the same $\operatorname{CS}$ modulo $1$, but their numbers of quanta differ by $2$. The operator $\mathsf X$ of \eqref{eq:center-X-generator} therefore assigns them phases differing by $e^{2\pi i\cdot 2/4}=-1$, which is precisely the statement $\mathsf X\mathsf U=-\mathsf U\mathsf X$. In the language of Table~\ref{tab:X-bookkeeping}, the image has $Q_5-2N\operatorname{CS}=-\frac38-2$ instead of $-\frac38$. For general $N$ the same mechanism holds: $\mathsf U$ leaves $Q_5$ unchanged and shifts $\operatorname{CS}$ by $\frac N4$, producing the phase $e^{-i\pi N/2}$ of \eqref{eq:SUN-center-chiral-algebra}.

Consequently, the components of the EH-prepared state in paired sectors are not $\mathsf U$-images of each other. The vector $\mathsf U|\Psi_\ell\rangle\in{\cal H}_{N/2-\ell}$ is the corresponding component of $|\Psi_{\rm EH}[B]\rangle$, whereas $|\Psi_{N/2-\ell}\rangle$ is built from honest $SU(N)$ configurations without the surface operator. For $SU(2)$, for example, $\mathsf U|\Psi_0\rangle$ is unsuppressed, whereas $|\Psi_1\rangle$ is weighted by at least $e^{-8\pi^2/(2g^2)}$ from the $Q=\frac12$ saddle. The EH-prepared state is thus not $\mathsf U$-invariant, even though $\mathsf U$ is an exact symmetry, because the bolt can absorb the center-symmetry surface operator. As we now show, the anomaly constrains this state completely at the level of its chiral charges, and in fact forces $\mathsf U|\Psi_{\rm EH}\rangle$ to be orthogonal to $|\Psi_{\rm EH}\rangle$ for $N\equiv2\bmod4$.

\paragraph{Reading the anomaly from the EH-prepared state.}
We can now compute $\mathsf X\mathsf U$ and $\mathsf U\mathsf X$ on the full EH-prepared state \eqref{eq:center-EH-state-decomposition}. No normalization of the components or choice of basis is needed. By \eqref{eq:center-X-on-EH}, $\mathsf U\mathsf X|\Psi_{\rm EH}\rangle=\chi_0\,\mathsf U|\Psi_{\rm EH}\rangle$. The operator $\mathsf U$ maps $|\Psi_\ell\rangle$ to a vector $\mathsf U|\Psi_\ell\rangle\in{\cal H}_{N/2-\ell}$, which by \eqref{eq:center-U-on-EH-state} is the corresponding component of the EH state prepared with the surface operator inserted. Its chiral charge follows from \eqref{eq:center-X-bookkeeping}: the number of quanta is unchanged, since the index is unchanged (Table~\ref{tab:SU2-U-action}); the sea charge is unchanged, since $\operatorname{Ad}(-h_c)=\operatorname{Ad}h_c$; and $\operatorname{CS}$ is shifted by $\frac N4$. Hence
\begin{equation}
 \mathsf X\,\mathsf U|\Psi_\ell\rangle=e^{-i\pi N/2}\,\chi_0\;\mathsf U|\Psi_\ell\rangle ,
 \label{eq:center-X-on-U-components}
\end{equation}
and therefore
\begin{equation}
 \mathsf X\,\mathsf U\,|\Psi_{\rm EH}\rangle=e^{-i\pi N/2}\,\mathsf U\,\mathsf X\,|\Psi_{\rm EH}\rangle ,
 \label{eq:center-algebra-on-EH}
\end{equation}
which is \eqref{eq:SUN-center-chiral-algebra} realized on the EH-prepared state. For $SU(2)$, $\mathsf X|\Psi_{\rm EH}\rangle=\chi_0|\Psi_{\rm EH}\rangle$ while $\mathsf X\,\mathsf U|\Psi_{\rm EH}\rangle=-\chi_0\,\mathsf U|\Psi_{\rm EH}\rangle$.

The bulk supplies more than a realization of an operator identity. For $N\equiv2\bmod4$, $|\Psi_{\rm EH}\rangle$ and $\mathsf U|\Psi_{\rm EH}\rangle=|\Psi_{\rm EH}[B]\rangle$ are eigenstates of the unitary operator $\mathsf X$ with the distinct eigenvalues $\chi_0$ and $-\chi_0$, hence
\begin{equation}
 \langle\Psi_{\rm EH}|\Psi_{\rm EH}[B]\rangle=0 .
 \label{eq:center-EH-orthogonal}
\end{equation}
The same holds sector by sector: ${\cal H}_{N/2-\ell}$ contains the honest component $|\Psi_{N/2-\ell}\rangle$, of charge $\chi_0$, and the $\mathsf U$-image $\mathsf U|\Psi_\ell\rangle$, of charge $-\chi_0$. For $SU(2)$ these are the honest configuration $(1,-1)$ and the image $(1,1)$ of Table~\ref{tab:SU2-U-action}. Thus the EH-prepared state is not $\mathsf U$-invariant as a consequence of the anomaly, and not merely because of the different suppressions of its components.

The zero-mode charges \eqref{eq:center-sector-charges} show how this comes about. For every $\ell$,
\begin{equation}
 \frac{{\cal A}_{N/2-\ell}}{{\cal A}_\ell}
 =\exp\left[-\frac{2\pi i}{N}\left(\left(\tfrac N2-\ell\right)^2-\ell^2\right)\right]
 =e^{-i\pi N/2}.
 \label{eq:center-charge-ratio}
\end{equation}
By \eqref{eq:center-X-on-Fock}, the Fock vacua $|\Omega_\ell\rangle$ and $|\Omega_{N/2-\ell}\rangle$, which are $\mathsf U$-images of each other, therefore carry $\mathsf X$-charges in the ratio $e^{-i\pi N/2}$, exactly as the algebra \eqref{eq:SUN-center-chiral-algebra} requires. The quanta created by the EH saddles carry the compensating charge ${\cal A}_\ell^{-1}$, so that the honest components all share the eigenvalue $\chi_0$, whereas the $\mathsf U$-image of $|\Psi_\ell\rangle$ carries the quanta of the sector $\ell$ on top of the Fock vacuum of the sector $N/2-\ell$, and its charge is off by $e^{-i\pi N/2}$. This does not depend on the details of the $\eta$-invariant: since $\operatorname{Ad}(-h_c)=\operatorname{Ad}h_c$, the gravitational and $\eta$-invariant contributions are identical in paired sectors, and the ratio is fixed by $\Delta\operatorname{CS}=\frac N4$. This is how the anomaly is read from the bulk with the help of the boundary: the index of the EH path integral in each sector fixes the chiral charges of the boundary components, and the center symmetry relates the sectors in pairs.

For example, for $SU(6)$ the zero-mode charges are ${\cal A}_\ell=e^{-i\pi\ell^2/3}=(1,e^{-i\pi/3},e^{-4i\pi/3},-1)$ for $\ell=0,1,2,3$, and both pairs $0\leftrightarrow3$ and $1\leftrightarrow2$ have ratio $-1$. For $SU(4)$ they are $(1,-i,1)$: the pair $0\leftrightarrow2$ has ratio $+1$, and the self-paired sector $\ell=1$ is consistent with $\mathsf U$ commuting with $\mathsf X$.

\paragraph{Zero-mode saturation and relevance for the condensate.}
The gaugino bilinear carries chiral charge two and transforms as
\begin{equation}
 \operatorname{tr}_{\mathbf N}(\lambda\lambda)
 \longrightarrow
 \exp\left(\frac{2\pi i}{N}\right)
 \operatorname{tr}_{\mathbf N}(\lambda\lambda).
\end{equation}
Its phase cancels that of the fermionic measure in the single-root sector. In the minimal $Q=1/2$ sector, the bilinear also absorbs the two adjoint zero modes, so the corresponding bilinear amplitude is invariant under $\mathbb Z_{2N}^{\chi}$. This zero-mode saturation establishes that the single-root sector is allowed to contribute, but it does not by itself prove that its completed contribution is nonzero or determine its coefficient. Moreover, a nonvanishing projected amplitude is a matrix element between the EH-prepared state and the chosen boundary bra; it represents a vacuum condensate only once that bra is identified with a chiral vacuum (see the precise definition of what we mean by condensate in Sec.~\ref{sec:EH-boundary-states-chiral-vacua}). A nonvanishing condensate in this sense would realize the familiar spontaneous breaking
\begin{equation}
 \mathbb Z_{2N}^{\chi}
 \longrightarrow
 \mathbb Z_2,
\end{equation}
with $N$ chiral vacua.

More than one boundary sector enters the condensate. A chiral vacuum on $\mathbb{RP}^3$ is a superposition over boundary sectors. In the ratio that defines the condensate, the bilinear is supplied by a sector with fractional $Q$, while the normalization is supplied by the trivial sector. For $SU(2)$ these two sectors form a $\mathsf U$-pair. The relative phase of the two components of a chiral vacuum is therefore fixed by \eqref{eq:SUN-center-chiral-algebra}, and this is why the two chiral vacua carry condensates of opposite sign. For $N>2$ the single-root sector that supplies the bilinear is not $\mathsf U$-paired with the trivial sector. The relative phase is then fixed by the $\mathsf X$-charge of the Fock vacuum $|\Omega_1\rangle$ relative to $|\Omega_0\rangle$, which by \eqref{eq:center-X-on-Fock} is the zero-mode charge ${\cal A}_1=e^{-2\pi i/N}$ of that sector, rather than by \eqref{eq:SUN-center-chiral-algebra}. We summarize the action of $\mathsf U$ for the lowest even ranks:
\begin{itemize}
 \item $SU(2)$: $\ell=0\leftrightarrow\ell=1$, with $\Delta\operatorname{CS}\equiv\frac12$ and phase $-1$. The pair consists of the trivial sector and the sector of the $Q=\frac12$ instanton.
 \item $SU(4)$: $\ell=0\leftrightarrow\ell=2$, while $\ell=1$ is mapped to itself. Here $\Delta\operatorname{CS}\equiv0$ and there is no phase. The trivial sector and the central sector of the two-block saddle are paired with equal zero-mode charges. The single-root sector $\ell=1$, with zero-mode charge $-i$, is not related to the trivial sector by any symmetry.
 \item $SU(6)$: $\ell=0\leftrightarrow\ell=3$ and $\ell=1\leftrightarrow\ell=2$, with $\Delta\operatorname{CS}\equiv\frac12$ and phase $-1$. In particular, the single-root sector $\ell=1$ is paired with $\ell=2$, not with the trivial sector.
\end{itemize}
In general, the single-root sector $\ell=1$ is paired with $\ell=N/2-1$, which coincides with the trivial sector only for $SU(2)$. In all cases, the phase \eqref{eq:EH-SUN-chiral-anomaly} of the single-root sector is a chiral charge, namely the zero-mode charge of that sector; the anomaly resides in the operator algebra \eqref{eq:SUN-center-chiral-algebra}. For $SU(2)$, the single-root sector is the $\mathsf U$-partner of the trivial sector, so the algebra fixes the ratio of their charges to the anomalous phase $e^{-i\pi N/2}=-1$. For $N>2$, no symmetry relates the single-root sector to the trivial sector, and its phase is not constrained by the anomaly. For $N\equiv2\bmod4$, the anomaly is visible instead in every $\mathsf U$-pair, through \eqref{eq:center-X-on-U-components} and \eqref{eq:center-EH-orthogonal}. For $SU(2)$, our calculation therefore realizes the chiral symmetry breaking directly: the EH saddle fixes the magnitude of the condensate, the anomaly fixes the relative sign in the two vacua, and the vacuum itself is selected by an infinitesimal source, as discussed in Sec.~\ref{sec:SYM-condensate}. For $SU(N>2)$, the APS index and the chiral charges remain robust, whereas the magnitude, and even the nonvanishing, of the single-root contribution require the complete continuum determinant, control of possible boundary and moduli-space surface terms, and a specification of the boundary state. 

\paragraph{Remark: the closed-manifold form of the orthogonality.}
Equation \eqref{eq:center-EH-orthogonal} has a direct four-dimensional interpretation. The orientation-reversed copy $\overline{\rm EH}$ prepares the bra $\langle\Psi_{\rm EH}|$, so the overlap is the partition function on the double of EH with $\mathsf U$ inserted on the gluing slice,
\begin{equation}
 \langle\Psi_{\rm EH}|\mathsf U|\Psi_{\rm EH}\rangle
 =Z\big[{\rm EH}\cup_{\mathbb{RP}^3}\overline{\rm EH}\,;\ \mathsf U\big],
 \qquad
 {\rm EH}\cup_{\mathbb{RP}^3}\overline{\rm EH}\cong S^2\times S^2,
 \label{eq:center-double-EH}
\end{equation}
where the identification of the double with $S^2\times S^2$ is standard; see, e.g., \cite{Anderson:1990sgi}.
To describe fluxes on the glued manifold, let $S^2_A$ and $S^2_B$ be the two factors of $S^2\times S^2$, with intersection numbers $S^2_A\cdot S^2_A=S^2_B\cdot S^2_B=0$ and $S^2_A\cdot S^2_B=1$. The two bolts are the combinations
\begin{equation}
 S^2({\rm EH})=S^2_A+S^2_B,\qquad S^2(\overline{\rm EH})=S^2_A-S^2_B,
 \label{eq:center-bolts-AB}
\end{equation}
with self-intersections $+2$ and $-2$, respectively, in the orientation used throughout this work. 

Normalize the $\mathbb Z_N$ two-form field $B$ of the surface operator to have integer periods mod $N$, as before. Pushing the surface operator into either half turns it into $\frac N2$ times the corresponding bolt, so $B=\frac N2\,{\rm PD}\big(S^2({\rm EH})\big)$, equivalently $\frac N2\,{\rm PD}\big(S^2(\overline{\rm EH})\big)$, since the two differ by $N\,{\rm PD}(S^2_B)\equiv0\pmod N$. Since $B$ is concentrated on the EH bolt, its flux through any surface $S$ is $\frac N2$ times the number of times $S$ crosses that bolt, $\int_SB=\frac N2\,S^2({\rm EH})\cdot S$. Through the EH bolt itself this is its self-intersection, $(S^2_A+S^2_B)^2=2S^2_A\cdot S^2_B=2$, and through the mirror bolt, which lies in the other half and never meets it, it is zero:
\begin{equation*}
 \int_{S^2({\rm EH})}B=\tfrac N2\cdot 2=N\equiv0,\qquad
 \int_{S^2(\overline{\rm EH})}B=0 ,
\end{equation*}
so neither half carries an 't~Hooft flux. The sphere $S^2_B$, however, meets the EH bolt once, so
\begin{equation}
 \int_{S^2_B}B=\frac N2\,(S^2_A+S^2_B)\cdot S^2_B=\frac N2\not\equiv0\pmod N .
\end{equation}
On the closed manifold, $B$ is therefore a genuine 't~Hooft flux, $w_2\neq0$, that neither half can see on its own.

The consequences follow as usual. Since $\int\mathcal P(B)=\left(\frac N2\right)^2\big(S^2({\rm EH})\big)^2=\frac{N^2}{2}$, the topological charge in this background obeys $Q\equiv\frac{N(N-1)}{4}\equiv\frac N4\pmod 1$. For $N\equiv2\bmod4$, $Q\in\frac12+\mathbb Z$, and the discrete chiral rotation multiplies the partition function by $e^{2\pi i\cdot 2NQ/2N}=-1$; equivalently, the gaugino zero modes are unsaturated. Hence 
\begin{equation}
Z_{S^2\times S^2}[B]=0,
\end{equation}
 which is \eqref{eq:center-EH-orthogonal} cut open along $\mathbb{RP}^3$. The overlap becomes nonzero once an operator of compensating chiral charge is inserted; for $SU(2)$, $\langle\Psi_{\rm EH}|\mathsf U|\Psi_{\rm EH}[\operatorname{tr}\lambda\lambda]\rangle$ need not vanish. For $N\equiv0\bmod4$, $Q\in\mathbb Z$ and there is no constraint, in agreement with \eqref{eq:SUN-center-chiral-algebra}.

\subsection{Supersymmetry and Ward identities on Eguchi--Hanson space}
\label{sec:EH-preserved-supersymmetry}

EH space preserves two supercharges in
${\cal N}=1$ supersymmetry. With the orientation and sigma-matrix
conventions used in this paper, its spin connection admits two linearly
independent covariantly constant dotted spinors, whereas no nonzero
covariantly constant undotted spinor exists:
\begin{equation}
 \nabla_\mu\bar\epsilon_{\dot\alpha}=0,
 \qquad
 \epsilon_\alpha=0\,, \quad \alpha,\dot\alpha=1,2\,.
 \label{eq:EH-covariantly-constant-spinors}
\end{equation}
Reversing the orientation interchanges the two chiralities. Although the
parallel spinors are not square-integrable on the noncompact EH space,
this is immaterial because supersymmetry parameters are nondynamical and
need not be normalizable.

A detailed derivation of the off-shell supersymmetry transformations, the
invariance and closure of the Euclidean SYM action, the component
descent identities, and Ward identities is given in
Appendix~\ref{app:EH-gauge-multiplet-Ward-identities}. Here we record only
the main results. We assume that the
regulator, functional measure, asymptotic boundary conditions, and the
boundary bra and ket preserve both dotted supercharges:
\begin{equation}
 \langle B|\bar{\cal Q}_{\dot\alpha}=0,
 \qquad
 \bar{\cal Q}_{\dot\alpha}|\Psi\rangle=0.
 \label{eq:EH-supersymmetric-Ward-identity}
\end{equation}
We also assume that no supercurrent, spacetime-boundary, or moduli-space
surface term survives in the corresponding Ward identities. The explicit
normalizable fermion zero modes satisfy the required asymptotic falloff, as shown
in Appendix~\ref{app:EH-SU2-gaugino-zero-modes}, while higher fermion (scattering) modes do not contribute to the surface term.

\paragraph{Position independence.}

Consider the chiral gaugino bilinear
\begin{equation}
 {\cal O}(x)
 =
 \operatorname{tr}_{\mathbf N}
 \bigl(\lambda\lambda\bigr)(x).
 \label{eq:EH-chiral-bilinear}
\end{equation}
The component calculation in
Appendix~\ref{app:EH-gauge-multiplet-Ward-identities} establishes the
off-shell descent identities
\begin{equation}
 [\bar{\cal Q}_{\dot\alpha},{\cal O}]=0,
 \qquad
 \nabla_\mu{\cal O}
 =
 \frac{i}{4}
 e^a{}_\mu\bar{\hat\sigma}_a^{\dot\alpha\alpha}
 \left\{
  \bar{\cal Q}_{\dot\alpha},
  \psi_{{\cal S}\alpha}
 \right\},
 \label{eq:EH-bilinear-descent-summary}
\end{equation}
where $\psi_{{\cal S}\alpha}$ is the fermionic component of the chiral
gauge kinetic multiplet. Thus, the derivative of ${\cal O}$ is exact with
respect to the preserved dotted supercharges. This result does not require
a globally defined undotted supercharge or a translation symmetry of the
EH metric.

For mutually separated insertions, the Ward identity therefore gives
\begin{equation}
 \nabla_\mu^{(i)}
 \left\langle
  \prod_{j=1}^{p}{\cal O}(x_j)
 \right\rangle_B
 =0.
 \label{eq:EH-position-independent-chiral-correlator}
\end{equation}
In particular, $\partial_r\langle{\cal O}(r)\rangle_B=0$ for $r>a$.
This statement by itself does not imply independence of the EH parameter $a$; the independence of $a$ will be discussed below. 

\paragraph{Holomorphy.}
Let
\begin{equation}
 {\cal G}_B
 =
 \left\langle
  \prod_{i=1}^{p}{\cal O}_i
 \right\rangle_B
\end{equation}
be a normalized correlator of chiral operators. In a holomorphic field
normalization, and with the boundary-state coefficients held fixed, the
anti-holomorphic part of the SYM action obeys
\begin{equation}
 \frac{\partial S}{\partial\bar\tau}
 =
 -\frac{1}{16\pi i}
 \left\{
  \bar{\cal Q}_{\dot1},
  [\bar{\cal Q}_{\dot2},\overline{\cal V}]
 \right\}
 +
 \frac{\partial S_{\rm surf}}{\partial\bar\tau},
 \qquad
 \overline{\cal V}
 =
 \int_{\rm EH}d^4x\sqrt g\,
 \operatorname{tr}_{\mathbf N}(\bar\lambda\bar\lambda)\,,\quad  \tau=\frac\theta{2\pi}+\frac{4\pi i}{g^2}.
 \label{eq:EH-antiholomorphic-action-Q-exact}
\end{equation}
If the surface term vanishes, the annihilation of the states by the dotted supercharges gives
\begin{equation}
 \frac{\partial{\cal G}_B}{\partial\bar\tau}=0.
 \label{eq:EH-chiral-correlator-holomorphy}
\end{equation}
Thus the boundary-conditioned chiral correlators are holomorphic in
$\tau$. Introducing the holomorphic strong scale
\begin{equation}
 \Lambda_{\rm h}^{3N}
 =
 \mu^{3N}e^{2\pi i\tau},
 \label{holomorphicLambda}
\end{equation}
this condition becomes
\begin{equation}
 \frac{\partial{\cal G}_B}
 {\partial\bar\Lambda_{\rm h}}
 =
 0.
 \label{eq:EH-correlator-holomorphic-Lambda}
\end{equation}
Fractional powers of $\Lambda_{\rm h}^{3N}$ are locally holomorphic after
choosing the branch associated with the relevant fractional
topological-charge sector.

\paragraph{Dependence on the Eguchi--Hanson resolution parameter.} The dependence on the EH resolution parameter $a$ is constrained by a
supersymmetric Ward identity.  This question is distinct from position
independence, since differentiating with respect to $a$ inserts the
stress-energy tensor.  A smooth comparison of EH metrics with different
values of $a$ is obtained by introducing a radial coordinate $\varrho$ of
fixed range,
\begin{equation}
 r^4=\varrho^4+a^4,
 \qquad
 0\leq\varrho<\infty,
\end{equation}
and defining
\begin{equation}
 h_{\mu\nu}
 =
 \left.
 \frac{\partial g_{\mu\nu}}{\partial a}
 \right|_{\varrho},
\end{equation}
where $g_{\mu\nu}$ is the EH metric.
As shown in Appendix~\ref{app:EH-a-independence}, this deformation is
smooth at the bolt, normalizable, and transverse traceless:
\begin{equation}
 \nabla_\nu h^{\mu\nu}=0,
 \qquad
 g^{\mu\nu}h_{\mu\nu}=0,
 \qquad
 h_{ab}=O\left(\frac{a^3}{r^4}\right)
 \quad
 (r\rightarrow\infty).
 \label{eq:bulk-EH-a-deformation-properties}
\end{equation}

The  Ferrara--Zumino, or more generally ${\cal S}$-, multiplet
relates the stress tensor to the variation of a local supercurrent
operator under the two dotted supercharges preserved on EH.  After
contraction with the deformation
\eqref{eq:bulk-EH-a-deformation-properties}, the relevant component
identity takes the form
\begin{equation}
 \frac12 h^{\mu\nu}T_{\mu\nu}
 =
 \left\{
  \bar{\cal Q}_{\dot\alpha},
  {\cal V}_a^{\dot\alpha}
 \right\} -{i\over4}\nabla_\nu
  \left(h^{\mu\nu}j^{(R)}_\mu\right),
 \label{eq:bulk-EH-a-exactness}
\end{equation}
where $T_{\mu\nu}$ is the stress tensor, $j^{(R)}_\mu$ is the $U(1)_R$ current, ${\cal V}_a^{\dot\alpha}
 ={1\over8i}
 h^{\mu\rho}\bar{\hat\sigma}_\rho^{\dot\alpha\alpha}
 S_{\mu\alpha}$, $S_{\mu\alpha}$ is the supercurrent, and a sum over the preserved dotted supercharges is understood.

For a normalized correlator of chiral operators,
\begin{equation}
 {\cal G}_B(a)
 =
 \left\langle
  \prod_i{\cal O}_i(x_i)
 \right\rangle_{B},
 \qquad
 \left[
  \bar{\cal Q}_{\dot\alpha},{\cal O}_i
 \right]=0,
\end{equation}
metric variation gives
\begin{equation}
 \frac{\partial{\cal G}_B}{\partial a}
 =
 -\frac12
 \int_{\rm EH} d^4y\,\sqrt{g}\,
 h_{\mu\nu}(y)
 \left[
  \left\langle
   T^{\mu\nu}(y)
   \prod_i{\cal O}_i(x_i)
  \right\rangle_B
  -
  \left\langle T^{\mu\nu}(y)\right\rangle_B
  \left\langle\prod_i{\cal O}_i(x_i)\right\rangle_B
 \right].
 \label{eq:bulk-EH-a-variation}
\end{equation}
The supersymmetry-exact term in
\eqref{eq:bulk-EH-a-exactness} vanishes by the dotted Ward identity.
There is no contribution from the bolt, which is a smooth interior
two-sphere rather than a boundary, while the falloff in
\eqref{eq:bulk-EH-a-deformation-properties} makes the asymptotic surface
term coming from the divergence of the $U(1)_R$ current vanish.  Therefore,
provided that the regulator and boundary state preserve the dotted
supercharges and that no contact or noncompact moduli-space surface term
survives,
\begin{equation}
 \frac{\partial}{\partial a}
 \left\langle
  \prod_i{\cal O}_i(x_i)
 \right\rangle_B
 =
 0.
 \label{eq:bulk-EH-chiral-a-independence}
\end{equation}

The anomaly of the classical continuous $U(1)_R$ symmetry does not by
itself invalidate this conclusion.  Quantum mechanically one must use
the renormalized Ferrara--Zumino or ${\cal S}$-multiplet rather than the
classical $R$-multiplet; the anomaly is then encoded in its chiral anomaly
operator.  The transverse-traceless projection above, along with the fast fall-off of $h_{\mu\nu}$, ensures the validity of (\ref{eq:bulk-EH-chiral-a-independence}); note that this argument does not use the conservation of the $U(1)_R$ current.  Equation
\eqref{eq:bulk-EH-chiral-a-independence} should nevertheless be
understood as a statement about the complete regulated amplitude in a
supersymmetric boundary sector, not necessarily about an isolated saddle
or an incomplete zero-mode approximation.  Any residual $a$-dependence
in such a partial contribution must be canceled by omitted determinants
or additional saddles, or else signal a boundary, contact, or
moduli-space surface contribution to the Ward identity.

\paragraph{Supersymmetric pairing.} Finally, the preserved supersymmetry formally pairs the nonzero bosonic and fermionic fluctuations. If the paired operators have compatible
domains and identical regulated spectral densities, a supersymmetric
regulator gives a unit determinant ratio. In the $SU(2)$ calculation
below, we shall assume that this cancellation continues to hold on EH space,
while treating the bosonic and fermionic zero modes separately
\footnote{\label{footnotedet}On a noncompact manifold, formal
supersymmetric pairing does not by itself establish the cancellation of
the regulated determinants. The fluctuation operators have continuous
spectra, and differences in their continuum spectral densities or
asymptotic scattering data can leave finite or logarithmic contributions~\cite{Weinberg:1979ma}. For example, explicit supersymmetric calculations around
 monopole-instantons on $\mathbb R^3\times S^1$ show
that nonzero-mode determinants need not cancel despite the preservation
of supersymmetry~\cite{Poppitz:2012sw,Anber:2014lba}. Establishing the
cancellation on EH would therefore require an explicit calculation of
the regulated relative determinants, or equivalently of the corresponding
heat kernels or scattering phase shifts. We do not attempt such calculations here.}.
The zero modes must instead be treated through their
collective-coordinate Jacobians. In particular, the APS index gives two
normalizable undotted gaugino zero modes in the minimal
$Q_{(2)}=1/2$ background, whereas the preserved supersymmetry parameters
have the opposite chirality. Supersymmetry of the complete quantum
calculation further requires a regulator and boundary conditions that
preserve the same supercharges; these choices are assumed throughout.

Likewise, for the two-block embedding in the $SU(4)$ theory, we shall assume that the nonzero-mode determinants cancel supersymmetrically. By contrast, for the single-root embedding in $SU(N>2)$, renormalization-group consistency with the exact supersymmetric $\beta$-function requires a nontrivial contribution from the determinants. Consequently, the bilinear condensate cannot be determined from the zero-mode calculation alone.

\section{The bilinear condensate in $SU(2)$ super-Yang--Mills theory}
\label{sec:SYM-condensate}

Having built all the machinery needed to perform our calculations, we now evaluate the contribution of the minimal EH instanton to
the gaugino bilinear ${\cal O}(x)=\operatorname{tr}_{\mathbf2}
  (\lambda\lambda)(x)$ in pure ${\cal N}=1$ $SU(2)$ super-Yang--Mills (SYM)
theory. After eliminating the auxiliary field, the Euclidean action is
\begin{align}
 S_{\rm SYM}
 ={}&
 \frac{1}{g^2}
 \int_{\rm EH}d^4x\sqrt{g}\,
 \operatorname{tr}_{\mathbf 2}
 \left[
  \frac12F_{\mu\nu}F^{\mu\nu}
  +2i\bar\lambda_{\dot\alpha}
  (\bar{\hat\sigma}^{a})^{\dot\alpha\alpha}
  E_a^{\mu}D_\mu^{\rm B}\lambda_\alpha
 \right]
-\frac{i\theta}{8\pi^2}
 \int_{\rm EH}
 \operatorname{tr}_{\mathbf 2}(F\wedge F),
 \label{symaction2}
\end{align}
where
\begin{equation}
 D^{\rm B}_\mu\lambda
 =
 \partial_\mu\lambda+i[A_\mu^{\rm B},\lambda]
 +\frac12\Xi_\mu\lambda,
 \qquad
 \Xi_\mu
 =
 \frac12\omega_{\mu ab}\hat\sigma^{ab},
 \label{diraceqlambsusy}
\end{equation}
and $\omega_{\mu ab}$ is the spin-connection one-form. 
The  topological charge and asymptotic holonomy of the minimal $SU(2)$ background (\ref{generalansatzB}, \ref{su2solsd}) are
\begin{equation}
 Q_{(2)}=\frac12\,,\quad  h_c^{(1)}=-I_2.
 \label{eq:SYM-minimal-Cartan-endpoint}
\end{equation}
Its classical action is therefore
\begin{equation}
 S_{\rm cl}
 =
 \frac{4\pi^2}{g^2}
 -\frac{i\theta}{2}.
 \label{eq:SYM-minimal-background-action}
\end{equation}
The action of the $SU(N)$ gauge group is identical to (\ref{symaction2}) after replacing $ \operatorname{tr}_{\mathbf 2}$ with $ \operatorname{tr}_{\mathbf N}$.

\subsection{Adjoint fermion zero modes}

The adjoint holonomy of the nonabelian $Q_{(2)}=1/2$ solution is trivial, and the APS
index is
$
{\cal I}_{\rm adj}
 =2T_{\rm adj}Q_{(2)}=4\times\frac12=2.
$
The opposite-chirality Weyl operator has no normalizable kernel\footnote{\label{vanishingtheorem}The absence of normalizable dotted zero modes follows from the  Lichnerowicz  identity
\begin{equation}
 {\cal D}_{+}^{\dagger}{\cal D}_{+}\bar\lambda
 =
 -D_\mu D^\mu\bar\lambda
 +\frac14R\,\bar\lambda
 -\frac{i}{2}\bar{\hat\sigma}^{\mu\nu}
 [F_{\mu\nu}^{\rm B},\bar\lambda],
\end{equation}
where ${\cal D}_{+}=\hat\sigma^\mu D_\mu$ and
$D_\mu\bar\lambda=\nabla_\mu\bar\lambda
+i[A_\mu^{\rm B},\bar\lambda]$. Since EH space is Ricci-flat, we have
$R=0$. Also, self-duality of $F^{\rm B}_{\mu\nu}$  implies
$\bar{\hat\sigma}^{\mu\nu}F_{\mu\nu}^{\rm B}= \bar{\hat\sigma}^{ab}E_a^{\mu}E_b^{\nu}F_{\mu\nu}^{\rm B}=0$. For an $L^2$-integrable zero
mode, integration by parts therefore gives
\begin{equation}
 0
 =
 \|{\cal D}_{+}\bar\lambda\|^2
 =
 \int_{\rm EH}d^4x\sqrt g\,
 \operatorname{tr}
 \left[
  (D_\mu\bar\lambda)^\dagger D^\mu\bar\lambda
 \right],
\end{equation}
where the boundary term vanishes under the $L^2$ normalizability condition.
Hence $D_\mu\bar\lambda=0$, so $\bar\lambda$ is covariantly constant
and has constant pointwise norm. Since EH has infinite volume, a
nonzero covariantly constant spinor cannot be square-integrable.
Therefore, $\bar\lambda=0$.}.  Hence the
background supports precisely two adjoint gaugino zero modes.  Their direct
derivation, including the bolt boundary condition, separated Weyl equations,
and normalizability analysis, is given in
Appendix~\ref{app:EH-SU2-gaugino-zero-modes}.

For a reference gauge orientation, the two modes can be collected into
\begin{align}
 \lambda_1
 =\mathcal F_3(r,t)\frac{\tau_3}{2}\eta_1
 +\mathcal F_{\pm}(r,t)\frac{\tau_-}{2}\eta_2,\quad
 \lambda_2
=-\mathcal F_3(r,t)\frac{\tau_3}{2}\eta_2
 +\mathcal F_{\pm}(r,t)\frac{\tau_+}{2}\eta_1,
 \label{fzeromodest}
\end{align}
where $\tau_\pm=(\tau_1\pm i\tau_2)/2$, and $\eta_1,\eta_2$ are independent
Grassmann collective coordinates. These fermion zero modes satisfy periodic boundary conditions (as $\psi$ varies between $0$ and $2\pi$) at the bolt because the background is central.  The radial profiles are
\begin{align}
 \mathcal F_3(r,t)
 =\frac{\sqrt{(t^4-a^4)(r^4-a^4)}}
 {r(r^2+t^2)^2},\quad
 \mathcal F_{\pm}(r,t)
 =\frac{2(t^2r^2+a^4)}{r(r^2+t^2)^2}.
 \label{eq:SYM-zero-mode-profiles}
\end{align}
Under a global orientation $U$, the modes transform as
$\lambda\mapsto U\lambda U^{-1}$; the gauge-invariant bilinear is therefore
independent of the orientation.

With $\epsilon^{12}=+1$, one finds
\begin{align}
 \operatorname{tr}_{\mathbf 2}(\lambda\lambda)
\equiv
 \operatorname{tr}_{\mathbf 2}
 \left(\epsilon^{\alpha\beta}\lambda_\beta\lambda_\alpha\right)
 =\frac12\left(2\mathcal F_3^2+\mathcal F_{\pm}^2\right)
 \eta_1\eta_2.
 \label{eq:SYM-zero-mode-bilinear}
\end{align}

\subsection{Collective-coordinates measures}

The minimal $Q_{(2)}=1/2$ instanton supports four normalizable bosonic
zero modes. The number of bosonic moduli is double the number of the fermion zero modes, as we show in general in Appendix \ref{sec:gauge-zero-modes}. One bosonic mode corresponds to variations of the radial-size modulus 
$t\in[a,\infty)$, while the other three are generated by global
$SU(2)$ color rotations. In contrast to a BPST instanton on
$\mathbb R^4$, there are no independent translational collective
coordinates, since translations are incompatible with the fixed EH
geometry and its $\mathbb Z_2$ asymptotic identification. Tangent vectors
to the physical moduli space are defined modulo infinitesimal gauge
transformations. We choose their representatives to be orthogonal to
the small gauge orbits by imposing the background gauge condition
\begin{equation}
 D_\mu^{\rm B}\delta A^\mu=0\,, \quad  D_\mu^{\rm B}=\nabla_\mu+i[A_\mu^{\rm B},\,\,\cdot ]\,.
 \label{eq:background-gauge-conditionbulk}
\end{equation}

To construct the moduli-space metric, we differentiate the instanton
connection $A_\mu^{\rm B}$ with respect to $t$ and to the three color-orientation
coordinates. The scale variation $\delta_{(0)}A$ automatically satisfies the background
gauge condition. Under a $SU(2)$ global color rotation 
\begin{eqnarray}
A^{\rm B}\mapsto A_U^{\rm B}=U^\dagger A^{\rm B} U\,.
\end{eqnarray}
Parametrizing $U$ using the three Pauli matrices $\tau_a$, $a=1,2,3$, then under infinitesimal rotations
 we have  $\delta_{(a)}^{}A_{\mu}^{}= i[A^{\rm B}_\mu,\tau_a]=D_\mu^{\rm B}\tau_a$. 
 The variations generated by global color rotations must
 be supplemented by compensating gauge transformations so that they
are transverse to the small gauge orbit. Thus, we need to solve for $\Phi_a(r)$
\begin{equation}
 \delta_{(a)}A_\mu
 =
 D_\mu^{\rm B}\Phi_a,
 \qquad
 D_{\rm B}^\mu D_\mu^{\rm B}\Phi_a=0,
 \label{eq:global-orientation-zero-modesbulk}
\end{equation}
with the boundary conditions $\Phi_a(r\rightarrow \infty)=\tau_a$.

The inner products of the resulting
four transverse zero modes (Gram matrix) define the metric on the bosonic moduli space (see \cite{Dorey:2002ik,Vandoren:2008xg} for reviews of instanton calculus):
\begin{equation}
 {\cal G}_{IJ}
 =
 \frac{2}{g^2}
 \int_{\rm EH}d^4x\sqrt{g_{\rm EH}}\,
 g^{\mu\nu}
 \operatorname{tr}_{\mathbf 2}
 \left(
 \delta_{(I)}A_\mu\delta_{(J)}A_\nu
 \right),
 \qquad
 I,J=0,1,2,3.
 \label{eq:bosonic-moduli-space-metric}
\end{equation}
The detailed construction and normalization of these modes are given in
Appendices~\ref{sec:gauge-zero-modes} and
\ref{sec:bosonic-zero-modes-Q-half}.

For a generic value $t>a$, the instanton is irreducible. A global
transformation $U\in SU(2)$ and the transformation $-U$ act identically on
the gauge field because they differ by the center. The space of generic
color orientations is therefore
\begin{equation}
 \frac{SU(2)}{\mathbb Z_2}
 \simeq SO(3)
 \simeq\mathbb{RP}^3.
\end{equation}
Denoting the left-invariant one-forms on this orientation space by
$\Sigma_i$, the zero-mode inner products give the metric on the bosonic moduli space $\cal M_{\rm b}$
\begin{equation}
 ds_{\cal M_{\rm b}}^2
 =
 \frac{8\pi^2}{g^2}
 \left[
  \frac{dt^2}{1-a^4/t^4}
  +t^2\left(\Sigma_1^2+\Sigma_2^2\right)
  +t^2\left(1-\frac{a^4}{t^4}\right)\Sigma_3^2
 \right].
 \label{eq:SYM-bosonic-moduli-space-metric}
\end{equation}
Thus, up to the overall factor $8\pi^2/g^2$, the bosonic moduli-space
metric is precisely the EH metric, with the instanton scale $t$ playing
the role of its radial coordinate.  At $t=a$, the
coefficient of $\Sigma_3^2$ vanishes, and one angular direction collapses,
leaving an $S^2$ bolt. This is the reducible Cartan endpoint of the
instanton family. The apparent degeneration of the orientation coordinates at this
point is therefore a smooth bolt of the moduli space.

The determinant of
\eqref{eq:SYM-bosonic-moduli-space-metric} is particularly simple:
the factors involving $1-a^4/t^4$ cancel between the radial direction and
the collapsing angular direction, giving
\begin{equation}
 \sqrt{\det{\cal G}_{\rm b}}
 =
 \left(\frac{8\pi^2}{g^2}\right)^2t^3.
\end{equation}
Including the standard Gaussian normalization $(2\pi)^{-2}$ associated with the four
bosonic zero modes, the collective-coordinate measure becomes
\begin{equation}
 d{\cal M}_{\rm b}
 =
 \frac{\sqrt{\det{\cal G}_{\rm b}}}{(2\pi)^2}
 \,dt\,dV_{SO(3)}
 =
 \frac{16\pi^2t^3}{g^4}\,
 dt\,dV_{SO(3)},
 \qquad
 \int_{SO(3)}dV_{SO(3)}=\pi^2.
 \label{eq:SYM-bosonic-measure}
\end{equation}

The adjoint Weyl operator has two normalizable zero modes, in agreement
with the APS index ${\cal I}_{\rm adj}=2$. Expanding the gaugino field
along these modes introduces two Grassmann collective coordinates
$\eta_1$ and $\eta_2$, as in (\ref{fzeromodest}). Their overlap matrix ${\cal U}_F$ is obtained by
integrating the pointwise inner products of the unnormalized fermion zero
modes over EH space:
\begin{equation}
 {\cal U}_F^{ij}
 =\frac{2}{g^2}\int_{\rm EH}d^4x\sqrt g\,
 \operatorname{tr}_{\mathbf 2}\left[
 \epsilon^{\alpha\beta}
 \frac{\partial\lambda_\alpha}{\partial\eta_i}
 \frac{\partial\lambda_\beta}{\partial\eta_j}\right].
\end{equation}
 Converting from the mode coefficients to the
Grassmann collective coordinates gives the measure over the fermionic moduli space ${\cal M}_{\rm f}$
\begin{equation}
 d{\cal M}_{\rm f}
 =
 \frac{d\eta_2d\eta_1}{\sqrt{\det{\cal U}_F}}
 =
 \frac{2g^2}{\pi^2t^2}\,
 d\eta_2d\eta_1.
 \label{eq:SYM-fermionic-measure}
\end{equation}
The explicit 
evaluation of ${\cal U}_F$ is presented in
Appendix~\ref{Fermionic measure}. Equations
\eqref{eq:SYM-bosonic-measure} and
\eqref{eq:SYM-fermionic-measure} contain all the collective-coordinate
Jacobians required for the semiclassical condensate calculation.

\subsection{The condensate}

Combining the classical weight, the Pauli--Villars factor $\mu^3$, the two
collective-coordinate measures, and the bilinear insertion gives, according to the definition (\ref{eq:SUN-leading-boundary-bra-expectation-value}), and adopting the normalization $Z^{(0)}_{Q=0}[1]=1$ and, for the moment, the boundary-state convention $b_1/b_0=1$, whose meaning is discussed below, we obtain
\begin{align}
 \left\langle\operatorname{tr}_{\mathbf 2}(\lambda\lambda)(r)\right\rangle_{B}
 ={}&\mu^3e^{-4\pi^2/g^2(\mu)+i\theta/2}
 \int d{\cal M}_{\rm f}\,d{\cal M}_{\rm b}\,
 \operatorname{tr}_{\mathbf 2}(\lambda\lambda)(r)
 \nonumber\\
 ={}&16\pi^2\frac{\mu^3}{g^2(\mu)}
 e^{-4\pi^2/g^2(\mu)+i\theta/2}
 \int_a^\infty dt\,t
 \left[2\mathcal F_3^2(r,t)+\mathcal F_{\pm}^2(r,t)\right].
 \label{eq:SYM-condensate-before-t-integral}
\end{align}
The remaining radial-size modulus integral obeys the exact identity
\begin{equation}
 {\cal J}(r,a)\equiv \int_a^\infty dt\,t
 \left[2\mathcal F_3^2(r,t)+\mathcal F_{\pm}^2(r,t)\right]=1.
 \label{eq:SYM-scale-integral-identity}
\end{equation}
In particular, the result is independent of the insertion point $r$ and the
EH resolution parameter $a$, as expected from the supersymmetric Ward identities.

In the Pauli--Villars, or equivalently canonical NSVZ scheme \cite{Terning:2006bq}, we define the
two branches of the strong scale by\footnote{Note that this differs from the holomorphic strong coupling introduced in (\ref{holomorphicLambda}) by the prefactor $1/g^2$.}
\begin{equation}
 \Lambda_k^3
 \equiv
 \frac{\mu^3}{g^2(\mu)}
 \exp\left[-\frac{4\pi^2}{g^2(\mu)}
 +\frac{i}{2}(\theta+2\pi k)\right],
 \qquad k=0,1,
 \label{eq:SYM-PV-strong-scale}
\end{equation}
where $\mu$ is the Pauli-Villars regularization scale. 
The one-instanton result is therefore\footnote{Ref.~\cite{Bianchi:1995ad} (see also~\cite{Bianchi:1995xd}) reports the coefficient $8\pi^2$
for the EH gaugino condensate. The apparent factor-of-two difference
may be attributed to the normalization of the Weyl bilinear. In the
present work, we define
\begin{equation}
 \operatorname{tr}_{\mathbf2}(\lambda\lambda)
 \equiv
 \operatorname{tr}_{\mathbf2}
 \left(
 \epsilon^{\alpha\beta}\lambda_\beta\lambda_\alpha
 \right)
 =
 -2\operatorname{tr}_{\mathbf2}(\lambda_1^a\lambda_2^b T^aT^b),\quad \operatorname{tr}_{\mathbf2}(T^aT^b)=\delta_{ab}/2\,,
\end{equation}
whereas 
Ref.~\cite{Bianchi:1995ad} does not define this spinor
contraction explicitly enough. In addition, their normalization of the strong scale
$\Lambda_{{\cal N}=1}$ is not stated explicitly.}
\begin{equation}
 \left\langle\operatorname{tr}_{\mathbf 2}(\lambda\lambda)\right\rangle_{B,k}
 =16\pi^2\Lambda_k^3,
 \qquad k=0,1.
 \label{eq:SYM-final-condensate}
\end{equation}

It is important to state precisely what \eqref{eq:SYM-final-condensate} represents. The EH path integral without insertions does not prepare a chiral vacuum. Its components in the two sectors carry the same $\mathsf X$-charge, cf.\ \eqref{eq:center-X-on-EH}, and since $Z^{(1)}[1]=0$, its projection onto the two lowest states of $\mathbb{RP}^3$ is proportional to the perturbative state $|0\rangle$, i.e., to the equal superposition $|k{=}0\rangle+|k{=}1\rangle$ of the two chiral vacua $|k\rangle=\big(|0\rangle+(-1)^k|1\rangle\big)/\sqrt2$, with $|1\rangle=\mathsf U|0\rangle$~\cite{Anber:2026bsx}. The quantity computed above is therefore a boundary-conditioned amplitude: the bilinear is supplied by the $Q=\frac12$ saddle in the sector $\ell=1$, the normalization by the trivial saddle in $\ell=0$, and their relative weight by the bra. Choosing the bra to be a chiral vacuum, $\langle B|=\langle k|$, gives $b_1/b_0=(-1)^k$, and the factor $(-1)^k=e^{i\pi k}$ turns $\Lambda_0^3$ into $\Lambda_k^3$; this is the origin of the two branches in \eqref{eq:SYM-final-condensate}. The convention $b_1/b_0=1$ used above corresponds to $k=0$. The two condensates differ by a sign, as required by the mixed anomaly, under which $\mathsf X$ maps $|k\rangle$ to $|k+1\rangle$. Which vacuum is realized is not decided by the EH geometry, but, as in any instance of spontaneous symmetry breaking, by an infinitesimal source: a small gaugino mass $m$ splits the two vacua, $E_k-E_0\propto-\operatorname{Re}\big(m\,\langle\operatorname{tr}_{\mathbf2}(\lambda\lambda)\rangle_k\big)$, and a long cylinder attached to the EH boundary then projects the EH-prepared state onto the selected vacuum, with which it always has nonzero overlap. In this sense, \eqref{eq:SYM-final-condensate} is the gaugino condensate in the chiral vacuum $|k\rangle$, with the magnitude $16\pi^2|\Lambda^3|$ supplied by the EH saddle and the phase by the choice of vacuum.

The numerical coefficient $16\pi^2$ agrees with the exact result
obtained by holomorphic methods on $\mathbb R^4$, as well as with
semiclassical calculations on $\mathbb R^3\times S^1$ and
$\mathbb T^4$. Although the radial-size modulus extends over the
noncompact range $t\in[a,\infty)$, this does not obstruct an attempt to
the determination of the coefficient semiclassically. For $a\Lambda\ll1$,
one may place the insertion in the weak-coupling region
$a\leq r\ll\Lambda^{-1}$. As shown below, the integral is then
parametrically saturated by configurations with
$t\ll\Lambda^{-1}$, while the contribution from
$t\gtrsim\Lambda^{-1}$ is suppressed by $O((r\Lambda)^2)$. Thus, subject
to the assumed supersymmetric cancellation of the regulated nonzero-mode
determinants, most of the coefficient $16\pi^2$ (i.e. $16\pi^2-\epsilon$, with $0<\epsilon\ll1$) can be extracted within a
controlled semiclassical regime. The supersymmetric Ward identity then
extends this value to arbitrary insertion points through radial position
independence, including points for which a direct evaluation of the
collective-coordinate integral would probe strongly coupled
configurations. The semiclassical approximation is therefore not
uniformly reliable over the entire radial-size-modulus integral, but the
position-independent condensate can nevertheless be determined by
evaluating it in the parametrically weakly coupled region.

\paragraph{Weak-coupling saturation of the $SU(2)$ result.}

To test our assertion and to determine which range of $t$ saturates the result (\ref{eq:SYM-scale-integral-identity}) (to a high accuracy), introduce an
intermediate cutoff $T>a$ and write
\begin{equation}
 {\cal J}(r,a)
 =
 {\cal J}^{<}(r,a;T)
 +
 {\cal J}^{>}(r,a;T),
\end{equation}
where ${\cal J}^{>}$ denotes the contribution from $t\geq T$. Direct
integration gives
\begin{align}
 {\cal J}^{>}(r,a;T)
 ={}&
 \frac{3r^4-a^4}
 {r^2(r^2+T^2)}
 -
 \frac{3(r^4-a^4)}
 {(r^2+T^2)^2}
+
 \frac{(r^4-a^4)^2}
 {r^2(r^2+T^2)^3},
 \label{eq:SU2-generic-r-large-t-tail}
\end{align}
and hence
\begin{equation}
 {\cal J}^{<}(r,a;T)
 =
 1-{\cal J}^{>}(r,a;T).
\end{equation}
For $T\gg r$, the large-modulus tail behaves as
\begin{equation}
 {\cal J}^{>}(r,a;T)
 =
 \frac{3r^4-a^4}{r^2T^2}
 +O\left(\frac{r^4}{T^4}\right).
 \label{eq:SU2-generic-r-tail-asymptotics}
\end{equation}

Suppose now that the insertion lies in the weak-curvature-scale region
$r\Lambda\ll1$. One may then choose an intermediate scale satisfying
\begin{equation}
 r\ll t_{\rm w}\ll\Lambda^{-1}.
\end{equation}
Throughout $a\leq t\leq t_{\rm w}$, the characteristic scale $1/t$ is
larger than $\Lambda$ and the running gauge coupling is parametrically
small. Equations
\eqref{eq:SU2-generic-r-large-t-tail} and
\eqref{eq:SU2-generic-r-tail-asymptotics} then show that
\begin{equation}
 {\cal J}^{<}(r,a;t_{\rm w})
 =
 1-
 O\left(\frac{r^2}{t_{\rm w}^2}\right).
\end{equation}
For example, choosing
\begin{equation}
 t_{\rm w}=\sqrt{\frac{r}{\Lambda}}
\end{equation}
gives
\begin{equation}
 {\cal J}^{>}(r,a;t_{\rm w})
 =
 \left[
 3-\left(\frac{a}{r}\right)^4
 \right]r\Lambda
 +O\left((r\Lambda)^2\right).
\end{equation}
The part of the formal integral lying beyond the strong-coupling scale is
even more suppressed:
\begin{align}
 {\cal J}^{>}(r,a;\Lambda^{-1})
 &=
 \left(
 3r^2-\frac{a^4}{r^2}
 \right)\Lambda^2
 +O\left((r\Lambda)^4\right)
&=
 O\left((r\Lambda)^2\right).
 \label{eq:SU2-generic-r-strong-tail}
\end{align}

Therefore, for $r\Lambda\ll1$, the integral producing the coefficient
$16\pi^2$ is saturated by radial-size moduli satisfying
$t\ll\Lambda^{-1}$, where the gauge coupling is weak.

\section{Gaugino condensates in $SU(N)$ theories}
\label{sec:SUN-condensates}

The appearance of the coefficient $16\pi^2$ in the $SU(2)$ gaugino
bilinear is striking, since it agrees with the coefficient obtained on
$\mathbb R^4$. As shown above, this coefficient is parametrically
saturated by weakly coupled radial-size moduli when the insertion lies in
the weak-coupling region, while supersymmetric position independence
extends the result to arbitrary insertion points. To determine whether
this agreement persists for higher-rank groups, we examine two additional
examples: the bilinear amplitude induced by a single-root embedding in
$SU(N)$ with $N>2$, and the eight-fermion amplitude induced by a
two-root embedding in $SU(4)$. These examples show that the $SU(2)$ result
does not provide a universal prescription for condensates on EH space.
In the single-root background, the noncentral holonomy requires a
nontrivial continuum determinant deterring us from computing the bilinear condensate in the single-root embedding background, whereas the central two-block background contributes a factor $1/6$, which
must be combined with the charge-one trivial-holonomy sector before
clustering can be tested.

\subsection{The bilinear condensate in the single-root embedding of $SU(N)$}
\label{subsec:SUN-single-root-condensate}

We now study the bilinear condensate induced by the $SU(N)$ background obtained by
embedding the EH $SU(2)$ instanton along the simple root
$
 \boldsymbol\alpha_1
 =
 \boldsymbol e_1-\boldsymbol e_2
$ that was discussed in Section \ref{subsec:SU-N-self-dual-instantons}. As a reminder, this solution carries topological charge $Q_{(N)}=1/2$ and has asymptotic holonomy $ h_c
 =
 \operatorname{diag}(-I_2,I_{N-2})$. The steps required to obtain the condensate are closely related to those in the $SU(2)$ case. 

\paragraph{Explicit zero modes.}
The solution of the Weyl equation in the specified background gives two normalizable zero modes, exactly as expected from the APS index and the vanishing theorem, Footnote~\ref{vanishingtheorem}. The two zero modes may be assembled using two independent Grassmann
collective coordinates $\eta_1$ and $\eta_2$:
\begin{align}
 \lambda_1
 &=
 \frac12{\cal F}_3(r,t)\eta_1\,
 \boldsymbol\alpha_1\cdot\boldsymbol H
 +
 \frac12{\cal F}_{\pm}(r,t)\eta_2\,
 E_{-\boldsymbol\alpha_1},
 \nonumber\\
 \lambda_2
 &=
 -\frac12{\cal F}_3(r,t)\eta_2\,
 \boldsymbol\alpha_1\cdot\boldsymbol H
 +
 \frac12{\cal F}_{\pm}(r,t)\eta_1\,
 E_{\boldsymbol\alpha_1}.
 \label{eq:SUN-single-root-zero-modes}
\end{align}
The radial profiles of ${\cal F}_3$ and ${\cal F}_{\pm}$ are given in (\ref{eq:SYM-zero-mode-profiles}). Notice that these zero modes obey periodic boundary conditions at the bolt, even though the asymptotic holonomy is noncentral in $SU(N)$. Indeed, the holonomy takes the form
$
h_c=\operatorname{diag}(-I_2,I_{N-2}),
$
and the zero modes lie entirely within the $\mathfrak{su}(2)$ subalgebra associated with the root $\boldsymbol{\alpha}_1$. Since the restriction of $h_c$ to this active $2\times2$ block is the central element $-I_2$, it acts trivially on the embedded adjoint fermions. With the fundamental trace normalization,
$
 \operatorname{tr}_{\boldsymbol N}
 \left[
  \left(
   \boldsymbol\alpha_1\cdot\boldsymbol H
  \right)^2
 \right]
 =
 2,
 \operatorname{tr}_{\boldsymbol N}
 \left(
  E_{\boldsymbol\alpha_1}
  E_{-\boldsymbol\alpha_1}
 \right)
 =
 1,
$
the gauge-invariant bilinear becomes
\begin{align}
 \operatorname{tr}_{\boldsymbol N}(\lambda\lambda)
 &\equiv
 \operatorname{tr}_{\boldsymbol N}
 \left(
  \epsilon^{\alpha\beta}
  \lambda_\beta\lambda_\alpha
 \right)
=
 \frac12
 \left(
  2{\cal F}_3^2
  +{\cal F}_{\pm}^2
 \right)
 \eta_1\eta_2.
 \label{eq:SUN-single-root-bilinear}
\end{align}

\paragraph{The bosonic moduli space.}

The single-root embedding has one radial-size modulus  and three color
orientations when the asymptotic holonomy $ h_c
 =
 \operatorname{diag}(-I_2,I_{N-2})$ is held fixed. To see this, we first note that the subgroup preserving this representative of the boundary holonomy is
\begin{equation}
 K_{h_c}
 =
 S\left(U(2)\times U(N-2)\right).
\end{equation}
The stabilizer of the full embedded connection inside $K_{h_c}$ is
\begin{equation}
 H_A
 =
 S\left(U(1)\times U(N-2)\right).
\end{equation}
Consequently, the allowed color-orientation orbit at fixed boundary
framing is
\begin{equation}
 \frac{K_{h_c}}{H_A}
 \simeq
 \frac{SU(2)}{\mathbb Z_2}
 \simeq
 SO(3).
 \label{eq:SUN-single-root-fixed-boundary-orbit}
\end{equation}
The framed bosonic moduli space is therefore
\begin{equation}
 {\cal M}_{{\rm b},h_c}
 \simeq
\mbox{EH},
 \qquad
 \dim_{\mathbb R}{\cal M}_{{\rm b},h_c}=4.
 \label{eq:SUN-single-root-framed-moduli-space}
\end{equation}

By contrast, the larger orbit
\begin{equation}
 \frac{
  SU(N)
 }{
  SU(N-2)\times U(1)
 }
\end{equation}
has dimension $4N-5$. Together with the radial-size modulus, it gives
$4(N-1)$ parameters. The additional $4(N-2)$ rotations mix the
embedded two-dimensional subspace with its complement and transform
the fixed representative of the boundary holonomy:
\begin{equation}
 h_c\longrightarrow U^{-1}h_cU.
\end{equation}
They therefore move between differently framed boundary data and must
not be integrated as collective coordinates in a path integral with
fixed $h_c$. The correct fixed-boundary moduli space is
\eqref{eq:SUN-single-root-framed-moduli-space}.

Since the allowed zero modes are entirely contained in the embedded
$\mathfrak{su}(2)$ subalgebra, their norms are the same as in the
$SU(2)$ calculation. The bosonic and fermionic measures are
\begin{equation}
 d{\cal M}_{\rm b, h_c}
 =
 \frac{16\pi^2t^3}{g^4}\,
 dt\,dV_{SO(3)},
 \qquad
 \int_{SO(3)}dV_{SO(3)}=\pi^2,
 \label{eq:SUN-single-root-bosonic-measure}
\end{equation}
and
\begin{equation}
 d{\cal M}_{\rm f}
 =
 \frac{2g^2}{\pi^2t^2}\,
 d\eta_2d\eta_1.
 \label{eq:SUN-single-root-fermionic-measure}
\end{equation}


\paragraph{Semiclassical bilinear amplitude.}

The single-root saddle belongs to the boundary component $\ell=1$,
whereas the trivial saddle belongs to $\ell=0$. Its contribution therefore
depends on the choice of boundary bra. At leading semiclassical order, according to (\ref{eq:EH-trivial-sector-normalization}), we
have
\begin{equation}
 \left.
 \left\langle
  \operatorname{tr}_{\boldsymbol N}(\lambda\lambda)(r)
 \right\rangle_B
 \right|_{Q=1/2}
 =
 \frac{b_1^*}{b_0^*}
 \frac{
  Z_{Q=1/2}^{(1)}
  [\operatorname{tr}_{\boldsymbol N}(\lambda\lambda)(r)]
 }{
  Z_{Q=0}^{(0)}[1]
 }.
 \label{eq:SUN-single-root-boundary-normalization}
\end{equation}
We adopt the boundary-state convention $b_1/b_0=1$ and the normalization
$Z_{Q=0}^{(0)}[1]=1$.

Let $\Delta_N(t,a;\mu,g)$ denote the regulated contribution of the
nonzero modes outside the embedded $\mathfrak{su}(2)$ sector. Combining the classical weight, the
zero-mode measures, and the bilinear insertion gives the formal
single-saddle expression
\begin{align}
 \left.
 \left\langle
  \operatorname{tr}_{\boldsymbol N}(\lambda\lambda)(r)
 \right\rangle_B
 \right|_{Q=1/2}
 ={}&
 16\pi^2
 \frac{\mu^3}{g^2(\mu)}
 \exp\left[
  -\frac{4\pi^2}{g^2(\mu)}
  +\frac{i\theta}{2}
 \right]
 \nonumber\\
 &\times
 \int_a^\infty dt\,t
 \left[
  2{\cal F}_3^2(r,t)
  +{\cal F}_{\pm}^2(r,t)
 \right]
 \Delta_N(t,a;\mu,g),
 \label{eq:SUN-single-root-formal-amplitude}
\end{align}
where $\mu$ is the Pauli-Villars regularization scale.

For $N=2$,  the supersymmetric
nonzero-mode cancellation gives
\begin{equation}
 \Delta_2(t,a;\mu,g)=1.
\end{equation}
The identity
\begin{equation}
 \int_a^\infty dt\,t
 \left[
  2{\cal F}_3^2(r,t)
  +{\cal F}_{\pm}^2(r,t)
 \right]
 =
 1
\end{equation}
then reproduces the $SU(2)$ result (\ref{eq:SYM-final-condensate}).

For $N>2$, the nonzero-mode contribution cannot be set equal to unity. The
off-diagonal adjoint multiplets have no normalizable zero modes, but their
continuum fluctuations must supply the regulator-scale and coupling
dependence missing from the embedded $\mathfrak{su}(2)$ collective-coordinate
measure. At the level of the one-saddle semiclassical expansion, RG
bookkeeping requires
\begin{equation}
 \Delta_N(t,a;\mu,g)
 =
 \left[g^2(\mu)\right]^{-\frac{N-2}{2}}
 (\mu t)^{\frac32(N-2)}
 \widehat\Delta_N\left(\frac{a}{t}\right),
 \label{eq:SUN-single-root-spectator-determinant}
\end{equation}
where $\widehat\Delta_N$ is a finite dimensionless function determined by
the continuum spectral data. The allocation of the explicit power of
$g^2(\mu)$ among the determinant, anomalous field normalizations, and the
canonical collective-coordinate measure is scheme-dependent; only their
completed product in
\eqref{eq:SUN-single-root-formal-amplitude} is meaningful.

Introduce the RG-invariant supersymmetric scale
\begin{equation}
 \Lambda_k^3
 =
 \frac{\mu^3}{g^2(\mu)}
 \exp\left[
  -\frac{8\pi^2}{Ng^2(\mu)}
  +\frac{i}{N}(\theta+2\pi k)
 \right],
 \qquad
 k=0,\ldots,N-1.
 \label{eq:SUN-canonical-strong-scale}
\end{equation}
On the branch appropriate to the $Q_{(N)}=1/2$ saddle,
\begin{equation}
 \frac{\mu^3}{g^2(\mu)}
 e^{-4\pi^2/g^2(\mu)+i\theta/2}
 \,
 \mu^{\frac32(N-2)}
 \left[g^2(\mu)\right]^{-\frac{N-2}{2}}
 =
 \left(\Lambda_0^3\right)^{N/2}.
\end{equation}
The formal single-root contribution may therefore be written as
\begin{align}
 \left.
 \left\langle
  \operatorname{tr}_{\boldsymbol N}(\lambda\lambda)(r)
 \right\rangle_B
 \right|_{Q=1/2}
 ={}&
 16\pi^2
 \left(\Lambda_0^3\right)^{N/2}
 {\cal I}_N(r,a),
 \label{eq:SUN-single-root-RG-improved-amplitude}
\end{align}
where
\begin{align}
 {\cal I}_N(r,a)
 ={}&
 \int_a^\infty dt\,
 t^{\frac{3N}{2}-2}
 \left[
  2{\cal F}_3^2(r,t)
  +{\cal F}_{\pm}^2(r,t)
 \right]
 \widehat\Delta_N\left(\frac{a}{t}\right).
 \label{eq:SUN-single-root-RG-integral}
\end{align}
No separate factor of the running canonical coupling $g(1/t)$ appears in
this holomorphic organization of the one-saddle expression, as required from the supersymmetric Ward identities.

It is important to distinguish
\eqref{eq:SUN-single-root-RG-improved-amplitude} from the complete
boundary-sector-projected condensate. The Ward identities derived in Sec.~\ref{sec:EH-preserved-supersymmetry}
constrain the complete normalized amplitude, rather than an arbitrarily
truncated saddle contribution. Provided that the regulator and boundary
state preserve the two dotted supersymmetries and that no contact,
spacetime-boundary, or moduli-space-boundary terms survive, they imply for the complete amplitude defined by (\ref{eq:SUN-boundary-bra-expectation-value})
\begin{equation}
 \partial_r
 \left\langle
  \operatorname{tr}_{\boldsymbol N}(\lambda\lambda)(r)
 \right\rangle_B
 =
 \partial_a
 \left\langle
  \operatorname{tr}_{\boldsymbol N}(\lambda\lambda)(r)
 \right\rangle_B
 =
 0,
\end{equation}
together with holomorphic dependence on the strong scale. If no other
dimensionful parameter is available, the complete condensate in a chosen
boundary sector and vacuum branch must consequently have the form
\begin{equation}
 \left\langle
  \operatorname{tr}_{\boldsymbol N}(\lambda\lambda)
 \right\rangle_{B}
 =
 C_{B}\Lambda_k^3,
 \label{eq:SUN-single-root-complete-condensate-form}
\end{equation}
where the dimensionless coefficient $C_{B}$ is not fixed by the Ward
identities. 

The formal single-root expression exhibits a tension for
$N>2$. Within the semiclassical factorization
\eqref{eq:SUN-single-root-RG-improved-amplitude}, and in the absence of
any scale other than $a$ and the insertion point, dimensional analysis
gives
\begin{equation}
 {\cal I}_N(r,a)
 =
 a^{\frac32(N-2)}
 \Phi_N\left(\frac{r}{a}\right).
 \label{eq:SUN-single-root-dimensional-form}
\end{equation}
Equivalently,
\begin{equation}
 \left.
 \left\langle
  \operatorname{tr}_{\boldsymbol N}(\lambda\lambda)(r)
 \right\rangle_B
 \right|_{Q=1/2}
 =
 16\pi^2\Lambda_0^3
 (a\Lambda_0)^{\frac32(N-2)}
 \Phi_N\left(\frac{r}{a}\right).
 \label{eq:SUN-single-root-holomorphic-scaling}
\end{equation}
If the Ward identities applied separately to this isolated
$Q=1/2$ contribution, position independence would first require
\begin{equation}
 \Phi_N\left(\frac{r}{a}\right)=C_N.
\end{equation}
For $N>2$, imposing $a$-independence on the same isolated
single-root expression would formally require
\begin{equation}
 0
 =
 \frac{\partial}{\partial a}
 \left[
  C_Na^{\frac32(N-2)}
 \right]
 =
 \frac32(N-2)C_N
 a^{\frac32(N-2)-1}.
 \label{eq:SUN-single-root-a-independence-tension}
\end{equation}
Although this equation would force $C_N=0$ if all the preceding assumptions
held separately for the isolated saddle, it should not be interpreted as a
prediction that the complete condensate vanishes. Rather, it demonstrates
that the semiclassical single-root factorization cannot simultaneously be
complete and obey the full $a$-independence Ward identity for $N>2$. The
most likely sources of its incompleteness are the noncompact large-$t$
region that probes the strong-coupling region $t\gtrsim\Lambda^{-1}$, the regulated continuum determinant, additional saddles in
the same boundary sector, or a moduli-space surface term. After these
effects are included, the complete boundary-sector-projected amplitude may
consistently take the form $C_{B}\Lambda_k^3$. The single-root
calculation therefore neither determines the coefficient $C_{B}$ nor
establishes that it vanishes.

\subsection{The eight-fermion amplitude in the two-root $SU(4)$ background}
\label{subsec:SU4-eight-fermion-main-text}

We next consider the charge-one $SU(4)$ configuration obtained by embedding
two commuting EH instantons along the mutually orthogonal roots
$\boldsymbol\alpha_1$ and $\boldsymbol\alpha_3$.  A representative of this
family is
\begin{equation}
 A^{\rm B}(t_1,t_2,U)
 =
 U^\dagger
 \begin{pmatrix}
  A_{\rm EH}^{\rm B}(t_1)&0\\
  0&A_{\rm EH}^{\rm B}(t_2)
 \end{pmatrix}
 U,
 \qquad
 U\in SU(4),
 \qquad
 t_1,t_2\in[a,\infty),
 \label{eq:main-SU4-two-block-background}
\end{equation}
where $t_1$ and $t_2$ are the two radial-size moduli. 
Each diagonal block carries charge $1/2$, so that
\begin{equation}
 Q_{(4)}=1,
 \qquad
 S_{\rm cl}=\frac{8\pi^2}{g^2}-i\theta.
 \label{eq:main-SU4-two-block-classical-data}
\end{equation}
The asymptotic holonomy is the central element
\begin{equation}
 h_c^{(\ell=2)}=-I_4, \quad \mbox{corresponding to the boundary state}\,,\quad |\ell=2;k\rangle \in {\cal H}_{\infty}\,.
 \label{eq:main-SU4-two-block-holonomy}
\end{equation}
Consequently, every global $SU(4)$ rotation preserves the boundary
holonomy sector.  This is the crucial difference from the single-root
embedding, whose noncentral holonomy forbids rotations that mix the active
two-dimensional color subspace with its complement.

The APS index of the adjoint Weyl operator in
\eqref{eq:main-SU4-two-block-background} is
$
 {\cal I}_{\rm adj}=8.
$
The vanishing theorem for the opposite-chirality operator, Footnote~\ref{vanishingtheorem}, then implies that
the background supports precisely eight normalizable gaugino zero modes.
The minimal local operator that can absorb them is therefore a product of
four bilinears,
\begin{equation}
 {\cal G}_4(x_1,x_2,x_3,x_4)
 =
 \left\langle
  \prod_{\ell=1}^{4}{\cal O}(x_\ell)
 \right\rangle,
 \qquad
 {\cal O}(x)
 =
 \operatorname{tr}_{\mathbf4}(\lambda\lambda)(x).
 \label{eq:main-SU4-four-bilinear-correlator}
\end{equation}
The construction of the bosonic and fermionic collective
coordinates is a formidable task that was possible with the help of OpenAI ChatGPT, Sol 5.6. We independently verified all the calculations, including all numerical coefficients.  The detailed calculations are given in
Appendices~\ref{sec:SU4-moduli-orthogonality}--\ref{subsec:SU4-four-bilinear-integration}.
Here we summarize the ingredients needed for the final integration.

\begin{figure}[t]
 \centering
 \includegraphics[width=\textwidth]
 {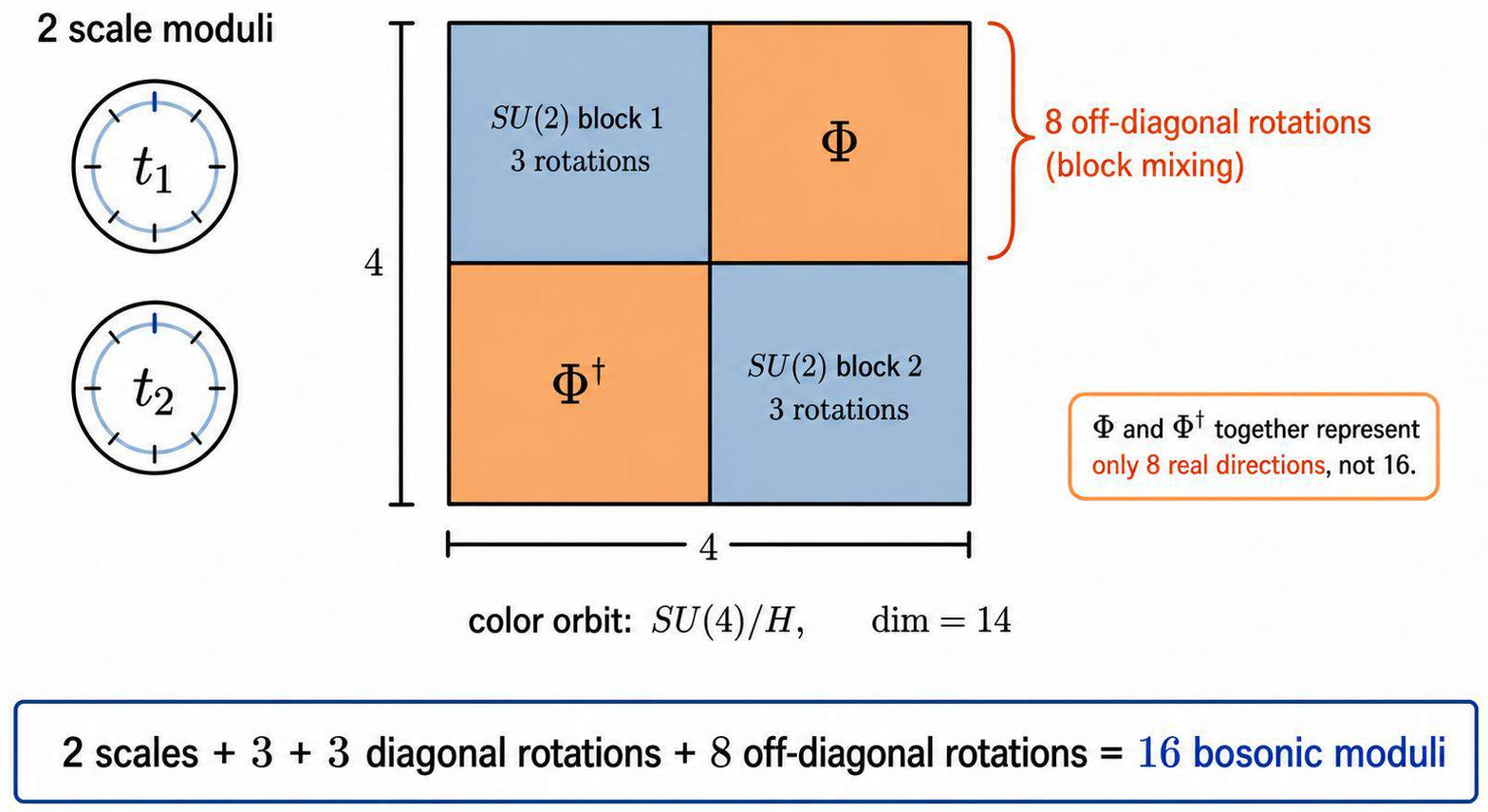}
 \caption{Decomposition of the bosonic collective coordinates of the
 two-block $SU(4)$ instanton. The two diagonal $SU(2)$ blocks possess
 independent radial-size (scale) moduli $t_1$ and $t_2$, together with three color
 orientations for each block. The off-diagonal complex matrix $\Phi$
 contains eight real parameters and generates the rotations that mix
 the two blocks; its Hermitian conjugate $\Phi^\dagger$ introduces no
 additional independent directions. The complete bosonic moduli space
 therefore has dimension $2+3+3+8=16$, while the color orbit
 $SU(4)/H$ is fourteen-dimensional.}
 \label{fig:SU4-two-block-moduli}
\end{figure}

\paragraph{Bosonic collective coordinates.}

At a generic point $t_1\neq t_2$, the stabilizer of the background is
\begin{equation}
 H
 =
 \left\{
  \operatorname{diag}(z_1I_2,z_2I_2):
  z_1,z_2\in U(1),\ (z_1z_2)^2=1
 \right\}
 \simeq U(1)\times\mathbb Z_2.
 \label{eq:main-SU4-stabilizer}
\end{equation}
Since the $SU(4)$ center acts trivially on the background by conjugation, the stabilizer of the
background necessarily contains
$Z(SU(4))\simeq\mathbb Z_4$.  The elements $\pm I_4$ belong to $U(1)$, whereas the remaining central
elements $\pm iI_4$ lie in 
$\mathbb Z_2$.  
The color orbit $SU(4)/H$ is fourteen-dimensional.  Together with the two
radial-size moduli, this gives a sixteen-dimensional bosonic moduli space; see Figure~\ref{fig:SU4-two-block-moduli}.  Six
angular directions rotate the two diagonal $SU(2)$ blocks independently (those are dealt with in the $SU(2)$ case discussed above),
while the remaining eight are cross-block rotations.

We next explain how the cross-block rotations are represented by genuine
zero modes satisfying the background gauge condition. Let $X$ be a constant off-diagonal
generator of the  $\mathfrak{su}(4)$ algebra. The infinitesimal global transformation
$U=I_4+i\varepsilon X+O(\varepsilon^2)$ produces, via (\ref{eq:main-SU4-two-block-background}),
\begin{equation}
 \delta_X^{(0)}A_\mu
 =
 i[A_\mu^{\rm B},X]
 =
 D_\mu^{\rm B} X.
 \label{eq:main-SU4-undressed-global-rotation}
\end{equation}
Because this deformation is obtained by differentiating a family of
gauge-conjugate self-dual connections, it satisfies the linearized
self-duality condition. In general, however, it does not satisfy the
background-gauge condition $D^\mu\delta A_\mu=0$. We must therefore add a compensator $\Theta_X(r)$ to enforce the background gauge condition, such that
\begin{equation}
 \delta_XA_\mu=D_\mu^{\rm B}\Theta_X,
 \qquad
 D^2_{\rm B}\Theta_X=0,
 \qquad
 \Theta_X(r)\longrightarrow X
 \quad\text{as}\quad r\longrightarrow\infty.
 \label{eq:main-SU4-harmonic-dressing}
\end{equation}
 Since the asymptotic holonomy of the two-block
background is the central element $-I_4$, every
$X\in\mathfrak{su}(4)$ preserves the selected boundary sector. For the
eight cross-block rotations, we can choose the Pauli matrices along with the identity matrix as a basis of the generators $X$. Thus, we  take $X=X_s^A$, where $s=0,1,2,3$
labels the matrices $M_s=(I_2,\tau_1,\tau_2,\tau_3)$ and
$A={\rm R},{\rm I}$ distinguishes their real and imaginary Hermitian
embeddings in $\mathfrak{su}(4)$; see Eq.~(\ref{eq:SU4-offdiag-real-generators}) for their explicit construction. 

It remains to verify that imposing background gauge conditions does not mix these
eight modes with the two radial-size modes or the six rotations internal to
the diagonal $SU(2)$ blocks. Introduce the block-parity involution
\begin{equation}
 \Gamma=
 \begin{pmatrix}
  I_2&0\\
  0&-I_2
 \end{pmatrix},
 \qquad
 {\cal P}(Y)=\Gamma Y\Gamma.
 \label{eq:main-SU4-block-parity}
\end{equation}
The background is even under ${\cal P}$, while block-diagonal and
off-diagonal matrices have parity $+1$ and $-1$, respectively. Since
\begin{equation}
 [D_\mu^{\rm B},{\cal P}]=0,
 \qquad
 [D^2_{\rm B},{\cal P}]=0,
\end{equation}
the harmonic dressing preserves the parity of its boundary generator $X_s^A$.
Consequently, the solutions $\Theta_s^A$ with off-diagonal boundary
values remain off-diagonal. Similarly, the gauge-fixing equations for the radial-size modes
and internal block rotations are even, so their background-gauge
representatives remain block diagonal. This establishes the separation
of the two sectors after the compensating gauge transformations have
been included.

Finally, the Yang--Mills inner product is invariant under ${\cal P}$.
If $a_+$ is block diagonal and $a_-$ is off diagonal, then
\begin{align}
 \langle a_+,a_-\rangle_{\rm YM}
 &=
 \langle{\cal P}a_+,{\cal P}a_-\rangle_{\rm YM}
 =
 -\langle a_+,a_-\rangle_{\rm YM}.
\end{align}
It follows that
\begin{equation}
 \langle a_+,a_-\rangle_{\rm YM}=0.
 \label{eq:main-SU4-block-parity-orthogonality}
\end{equation}
Thus all mixed Gram-matrix elements (of the bosonic moduli space) between the block-diagonal and
cross-block sectors vanish, and the bosonic Gram determinant factorizes
into the product of the two corresponding determinants.


Writing
\begin{equation}
 p_i=\frac{t_i^2}{a^2},
 \qquad
 u_i=p_i+\sqrt{p_i^2-1},\quad i=1,2,
 \label{eq:main-SU4-p-u-variables}
\end{equation}
the cross-block norms are encoded in
\begin{align}
 d_0
 &=
 a^2
 \frac{(u_1-u_2)^2(u_1u_2+1)}
 {4u_1u_2(u_1+u_2)},
\quad
 d_\perp
 =
 a^2\frac{p_1+p_2}{2},
\quad
 d_3
 =
 a^2
 \frac{(u_1+u_2)(u_1u_2-1)^2}
 {4u_1u_2(u_1u_2+1)}.
 \label{eq:main-SU4-d-functions}
\end{align}
The complete bosonic measure is
\begin{equation}
 d{\cal M}_{\rm b}
 =
 \frac{1}{2!}
 \frac{C^8}{(2\pi)^8}
 t_1^3t_2^3d_0d_\perp^2d_3\,
 dt_1dt_2\,dV_{SU(4)/H}\,, \quad  C=\frac{8\pi^2}{g^2}.
 \label{eq:main-SU4-bosonic-measure}
\end{equation}
The factor $1/2!$ removes the redundancy under exchange of the two
identical blocks. The quotient group volume is
\begin{equation}
 \operatorname{Vol}\left(\frac{SU(4)}{H}\right)
 =\frac{\pi^8}{12}.
 \label{eq:main-SU4-orbit-volume}
\end{equation}

\paragraph{Fermionic collective coordinates.}

Four zero modes are inherited directly from the two diagonal $SU(2)$
blocks.  The other four lie in the upper-right and lower-left $SU(4)$ color blocks.
They are generated from the bifundamental harmonic scalar
\begin{equation}
 \Phi_0(r)=h_0(r)I_2,
 \qquad
 (D^{12})^2\Phi_0=0,
 \qquad
 h_0(r\rightarrow\infty)=1,
 \label{eq:main-SU4-bifundamental-scalar}
\end{equation}
through the factorization map
\begin{equation}
 \Psi_\alpha{}^{\dot\beta}
 =
 \hat\sigma^a_{\alpha}{}^{\dot\beta}D_a^{12}\Phi_0,
 \label{eq:main-SU4-scalar-fermion-map}
\end{equation}
where $ D_\mu^{12}\Phi
 =
 \partial_\mu\Phi
 +iA_{1\mu}\Phi
 -i\Phi A_{2\mu}$. 
$\Phi_0$ is the compensator $\Theta_X(r)$ associated to the identity element $X=I_2$ that parametrizes the off-diagonal $\mathfrak{su}(4)$, and $h_0$ is given by \begin{eqnarray}
 h_0(x)
 =
 \frac{(x+c_0)}{
 \sqrt{(x+p_1)(x+p_2)}}\,,\quad  c_0
=
 \frac{u_1u_2+1}{u_1+u_2}\,.
\end{eqnarray}
The self-duality of the two-block field strengths guarantees that
$\bar{\hat\sigma}^{\,a}D_a^{12}\Psi=0$.  The two values of $\dot\beta$
give two upper-right modes, and the conjugate bifundamental equation gives
two independent lower-left modes.  Together with the four diagonal modes,
they exhaust the adjoint index ${\cal I}_{adj}=8$.

Notice all eight fermion zero modes in the two-block $SU(4)$ embedding obey periodic boundary conditions at the bolt. In this case, the asymptotic holonomy is
$
h_c=\operatorname{diag}(-I_2,-I_2)=-I_4,
$
which is central in $SU(4)$ and therefore acts trivially on the entire adjoint representation. Consequently, both the four zero modes inherited from the two diagonal $SU(2)$ blocks and the four bifundamental zero modes connecting them are periodic.

Combining the four diagonal modes with the four cross-block modes gives the measure over the fermionic moduli space
\begin{equation}
 d{\cal M}_{\rm f}
 =
 \mu^{12}
 \frac{g^8}{16\pi^8t_1^2t_2^2d_0^2}\,
 d^4\eta\,d^4\xi,
 \label{eq:main-SU4-fermionic-measure}
\end{equation}
where the Grassmann variables $\eta_{i1},\eta_{i2}$ are associated with the two diagonal blocks and
$\xi_1,\ldots,\xi_4$ multiply the cross-block modes.

\paragraph{Saturation of the zero modes.}

In terms of the unnormalized zero-mode basis, the bilinear decomposes as
\begin{equation}
 {\cal O}
 =
 q_1\eta_{11}\eta_{12}
 +q_2\eta_{21}\eta_{22}
 +{\cal O}_\times,
 \qquad
 q_i
 =
 {\cal F}_{3i}^2+\frac12{\cal F}_{\pm i}^2\,, \quad i=1,2.
 \label{eq:main-SU4-bilinear-decomposition}
\end{equation}
Defining
\begin{equation}
 A_0=\frac{h_0'}{f},
 \qquad
 B_0=\frac{h_0(F_1-F_2)}{2r},
 \qquad
 C_0=\frac{h_0f(G_1-G_2)}{2r},
 \qquad
 {\cal P}_0=A_0^2+2B_0^2+C_0^2,
 \label{eq:main-SU4-P0-definition}
\end{equation}
where $F_{1,2}$ and $G_{1,2}$ are given by Eq.~(\ref{su2solsd}), 
the cross-block bilinear satisfies
\begin{equation}
 {\cal O}_\times^2
 =
 -32{\cal P}_0^2\,
 \xi_1\xi_2\xi_3\xi_4.
 \label{eq:main-SU4-cross-bilinear-square}
\end{equation}
The unique term in ${\cal O}^4$ that saturates all eight Grassmann
coordinates contains one bilinear from each diagonal block and two
cross-block bilinears.  Its multinomial coefficient is
$4!/(1!1!2!)=12$, and hence
\begin{align}
 \left.{\cal O}^4\right|_{\eta^4\xi^4}
 =
 -384q_1q_2{\cal P}_0^2\,
 \eta_{11}\eta_{12}\eta_{21}\eta_{22}
 \xi_1\xi_2\xi_3\xi_4.
 \label{eq:main-SU4-Grassmann-saturation}
\end{align}
The overall sign is fixed by the chosen orientation of the fermionic
measure and has no invariant meaning by itself.

\paragraph{Radial-size moduli integral and result.}
The final step is to integrate over the radial size moduli. It can be shown numerically that the integral is independent of $r$. To obtain the result analytically, we may set $r=a$ in the integrand.  At $r=a$,
\begin{equation}
 q_i(a)=\frac{2a^2}{(t_i^2+a^2)^2},
 \qquad
 {\cal P}_0(a)
 =
 \frac{32u_1u_2(u_1-u_2)^2}
 {a^2(u_1+u_2)^2(u_1+1)^2(u_2+1)^2}.
 \label{eq:main-SU4-bolt-insertion-data}
\end{equation}
After carrying out the Grassmann and color-orientation integrals, the
remaining scale integral is
\begin{equation}
 {\cal J}_4
 =
 \int_a^\infty dt_1
 \int_a^\infty dt_2\,
 t_1t_2
 \frac{d_3d_\perp^2}{d_0}\,
 q_1(a)q_2(a){\cal P}_0(a)^2.
 \label{eq:main-SU4-scale-integral}
\end{equation}
Introducing
\begin{equation}
 y_i=\frac{u_i-1}{u_i+1},
 \qquad
 0\leq y_i<1,
 \end{equation}
eliminates all dependence on the EH resolution parameter and gives
\begin{align}
 {\cal J}_4
 &=
 4\int_0^1dy_1\int_0^1dy_2\,
 y_1y_2(y_1-y_2)^2(y_1+y_2)^2=\frac16.
 \label{eq:main-SU4-scale-integral-result}
\end{align}
The apparent singularity at $t_1=t_2$ is canceled by the vanishing of
${\cal P}_0^2/d_0$, and the coincident-scale locus is integrable. Even though we have performed the above calculations at the specific point $r=a$, we have checked that the factor of $1/6$ is independent of $r$.

Assuming a supersymmetric Pauli--Villars regulator and cancellation of the
nonzero-mode determinants, the two-block saddle therefore contributes
\begin{equation}
{\cal G}_4^{\rm B}
 =
 \frac{2^{15}\pi^8}{3}
 \frac{\mu^{12}}{g^8}
 \exp\left(-\frac{8\pi^2}{g^2}+i\theta\right).
 \label{eq:main-SU4-PV-amplitude}
\end{equation}
Using the definition of the strong scale $\Lambda$
\begin{equation}
 \Lambda^{12}
 =
 \frac{\mu^{12}}{g^8}
 \exp\left(-\frac{8\pi^2}{g^2}+i\theta\right),
 \label{eq:main-SU4-PV-scale}
\end{equation}
this becomes
\begin{equation}
  {\cal G}_4^{\rm B}(x_1,x_2,x_3,x_4)
  =
  \frac16
  \left(16\pi^2\Lambda_{}^3\right)^4.
 \label{eq:main-SU4-final-four-bilinear-result}
\end{equation}
The displayed normalization adopts the boundary-state convention
$b_2/b_0=1$ and $Z_0^{(0)}[1]=1$.  For a general boundary bra, the right-hand
side of \eqref{eq:main-SU4-final-four-bilinear-result} is multiplied by
$b_2^*/b_0^*$.

Although the collective-coordinate integral in
\eqref{eq:main-SU4-final-four-bilinear-result} extends into regions where
one or both radial-size moduli are strongly coupled, the factor $1/6$
can be extracted within a parametrically weakly coupled regime. Indeed,
for $a\Lambda\ll1$, introducing a scale
$a\ll T\ll\Lambda^{-1}$ and restricting both moduli to $t_i\leq T$
gives
\begin{equation}
 {\cal J}_4^{<}(T)
 =
 \frac16
 \left(\frac{T^2-a^2}{T^2+a^2}\right)^4
 =
 \frac16
 \left[
  1-O\left(\frac{a^2}{T^2}\right)
 \right].
\end{equation}
The contribution from the region in which at least one modulus satisfies
$t_i\gtrsim\Lambda^{-1}$ is only of order $(a\Lambda)^2$. The factor
$1/6$ is therefore parametrically saturated by weakly coupled
configurations. Yet, this conclusion does not mean that the calculations are performed in a controlled-weak-coupling limit, since the moduli can still probe the strongly coupled region. 

The supersymmetric Ward identity derived above implies
position independence of separated chiral correlators, provided that no
boundary or moduli-space surface terms are present. Consequently, if the
coincident product used in the calculation is related to the separated
four-point function without an additional contact term, the result may
be continued to widely separated insertion points. The two-block family
is, however, not the only charge-one contribution to the four-point
function. The trivial-holonomy sector $h_c=+I_4$ supports charge-one
instantons with the same number, eight, of adjoint gaugino zero modes:
an EH $SU(2)$ instanton with trivial holonomy embedded in $SU(4)$, whose
sixteen moduli are its position on EH, its size, and eleven color
orientations. Writing its contribution as $c_0\left(16\pi^2\Lambda^3\right)^4$,
with the normalization $Z_0^{(0)}[1]=1$ used above, the complete charge-one
four-point function is
\begin{equation}
 {\cal G}_4^{\rm full}
 =
 \left(c_0+\frac{b_2^*}{b_0^*}\,\frac16\right)
 \left(16\pi^2\Lambda^3\right)^4 .
\end{equation}
In a definite cluster-decomposing vacuum with
$\langle{\cal O}\rangle^4=\left(16\pi^2\Lambda^3\right)^4$, clustering
therefore requires $c_0=1-\frac{b_2^*}{b_0^*}\frac16$, i.e.\ $c_0=5/6$
or $7/6$ for $b_2^*/b_0^*=\pm1$. The factor $1/6$ by itself neither
determines the bilinear condensate nor signals a clustering puzzle;
deciding whether clustering holds requires computing $c_0$, which we
leave for future work. Neither contribution needs to coincide with the
bilinear contribution obtained from the single-block embedding. The
single-block saddle has charge $Q=1/2$ and noncentral asymptotic holonomy
$\operatorname{diag}(-I_2,I_2)$, whereas the two-block saddle has charge
$Q=1$ and central asymptotic holonomy $-I_4$. They therefore define
amplitudes in distinct boundary-holonomy sectors and are weighted by
different components of the EH boundary state. Cluster decomposition
constrains the full correlator and one-point function evaluated in the
same normalized, cluster-decomposing state; it does not require separate
fixed-holonomy or fixed-topological-charge contributions to factorize.
Consequently, any mismatch between the single-block and two-block
calculations is not by itself an inconsistency.

{\bf \flushleft{Acknowledgments:}} We thank Stefano Cremonesi and Erich Poppitz for discussions. ChatGPT 5.6, developed by OpenAI, and Claude 5.5, developed by Anthropic,  were used as aids in drafting and in calculations. The author takes full responsibility for the correctness of the results. This work was supported by the STFC under grant ST/X000591/1.

\appendix

\section{Mathematical preliminaries}
\label{Mathematical preliminaries}

In this section, we collect the geometric and spinorial conventions
needed in our discussion of the Eguchi--Hanson (EH) space.  For a general
review, see Ref.~\cite{Eguchi:1980jx}.

Let
\begin{equation}
 e^a=e^a{}_\mu dx^\mu,
 \qquad
 E_a=E_a{}^\mu\partial_\mu
\end{equation}
denote the vielbein one-forms and their dual vector fields.  The
Euclidean metric and its inverse are
\begin{equation}
 g_{\mu\nu}
 =
 \delta_{ab}e^a{}_\mu e^b{}_\nu,
 \qquad
 g^{\mu\nu}
 =
 \delta^{ab}E_a{}^\mu E_b{}^\nu.
\end{equation}
Latin indices $a,b,\ldots=0,1,2,3$ denote tangent-space indices, while
Greek indices $\mu,\nu,\ldots$ denote curved-space indices.  The
vielbeins and inverse vielbeins satisfy
\begin{equation}
 E_a{}^\mu e^b{}_\mu=\delta_a^b,
 \qquad
 E_a{}^\mu e^a{}_\nu=\delta^\mu_\nu.
 \label{invrsedef}
\end{equation}

We define the spin-connection one-forms by
\begin{equation}
 \omega^a{}_b
 =
 \omega^a{}_{b\mu}dx^\mu.
\end{equation}
For a torsion-free geometry, Cartan's first structure equation is
\begin{equation}
 de^a+\omega^a{}_b\wedge e^b=0.
 \label{tortionfree}
\end{equation}
Metric compatibility implies
\begin{equation}
 \omega_{\mu ab}=-\omega_{\mu ba},
 \qquad
 \omega_{\mu ab}
 \equiv
 \delta_{ac}\omega^c{}_{b\mu}.
 \label{eq:spin-connection-antisymmetry}
\end{equation}
The curvature two-form is
\begin{equation}
 R^a{}_b
 =
 d\omega^a{}_b+\omega^a{}_c\wedge\omega^c{}_b
 =
 \frac12R^a{}_{b\mu\nu}dx^\mu\wedge dx^\nu
 =
 \frac12R^a{}_{bcd}e^c\wedge e^d.
 \label{eq:curvature-two-form}
\end{equation}

\paragraph{Orientation and duality conventions.}

The left-invariant one-forms used in the EH metric satisfy
\begin{equation}
 \sigma_x\wedge\sigma_y\wedge\sigma_z
 =
 -\frac18\sin\theta\,
 d\theta\wedge d\varphi\wedge d\psi,
\end{equation}
and hence
\begin{equation}
 \int_{S^3/\mathbb Z_2}
 \sigma_x\wedge\sigma_y\wedge\sigma_z
 =
 -\pi^2.
 \label{eq:EH-angular-volume}
\end{equation}
We orient the EH manifold by
\begin{align}
 \operatorname{vol}_{\rm EH}
 &=
 \frac{r^3}{8}\sin\theta\,
 dr\wedge d\theta\wedge d\varphi\wedge d\psi
 \nonumber\\
 &=
 -r^3dr\wedge
 \sigma_x\wedge\sigma_y\wedge\sigma_z
 =
 -e^0\wedge e^1\wedge e^2\wedge e^3.
 \label{eq:EH-volume-orientation}
\end{align}
Thus, the ordered orthonormal coframe
$(e^0,e^1,e^2,e^3)$ is negatively oriented with respect to
$\operatorname{vol}_{\rm EH}$.  The components of the geometric
Levi-Civita tensor in this frame therefore satisfy
\begin{equation}
 \varepsilon_{0123}=-1,
 \label{eq:EH-orientation-convention}
\end{equation}
with all other components fixed by total antisymmetry.

It is useful to distinguish two different duality operations.  The
internal dual acts on the pair of tangent-space indices carried by the
curvature:
\begin{equation}
 \left(\star_{\rm int}R\right)_{ab}
 \equiv
 \frac12\varepsilon_{ab}{}^{cd}R_{cd}.
 \label{eq:internal-dual-curvature}
\end{equation}
By contrast, the ordinary Hodge star acts on the differential-form
indices:
\begin{equation}
 \left(\star R_{ab}\right)_{cd}
 \equiv
 \frac12\varepsilon_{cd}{}^{ef}R_{abef}.
 \label{eq:form-dual-curvature}
\end{equation}
In particular, the Hodge star acts on the basis two-forms  as
\begin{equation}
 \star\left(e^3\wedge e^0\right)=e^1\wedge e^2,
 \qquad
 \star\left(e^1\wedge e^2\right)=e^3\wedge e^0.
 \label{eq:EH-Hodge-star-convention}
\end{equation}

Unless stated otherwise, self-duality of the gravitational curvature
will refer to the internal duality operation
\eqref{eq:internal-dual-curvature}.  Thus, the curvature is internally
self-dual or anti-self-dual when
\begin{equation}
 \star_{\rm int}R=\pm R,
 \label{eq:internal-curvature-duality}
\end{equation}
respectively.  

The internal dual may similarly be applied to the spin connection:
\begin{equation}
 \left(\star_{\rm int}\omega\right)_{ab}
 \equiv
 \frac12\varepsilon_{ab}{}^{cd}\omega_{cd}.
 \label{eq:internal-dual-spin-connection}
\end{equation}
In a frame in which
\begin{equation}
 \star_{\rm int}\omega=\pm\omega,
\end{equation}
the curvature has the same internal chirality.  

For the EH vielbein used in this work, the nonvanishing spin-connection
components obey
\begin{equation}
 \omega_{10}=\omega_{23},
 \qquad
 \omega_{20}=\omega_{31},
 \qquad
 \omega_{30}=\omega_{12}.
 \label{eq:EH-spin-connection-duality-relations}
\end{equation}
Together with \eqref{eq:EH-orientation-convention}, these relations
imply
\begin{equation}
 \star_{\rm int}\omega=\omega,
 \qquad
 \star_{\rm int}R=R.
 \label{eq:EH-internal-self-duality}
\end{equation}
Thus, the EH spin connection and curvature are internally self-dual in
our orientation convention.

\paragraph{Spinorial conventions.}

We use the Euclidean sigma matrices
\begin{equation}
 \hat\sigma^a=(I_2,i\vec{\sigma}),
 \qquad
 \bar{\hat\sigma}^a=(I_2,-i\vec{\sigma}),
\end{equation}
which obey
\begin{equation}
 \hat\sigma^a\bar{\hat\sigma}^b
 +\hat\sigma^b\bar{\hat\sigma}^a
 =
 2\delta^{ab}I_2,
 \qquad
 \bar{\hat\sigma}^a\hat\sigma^b
 +\bar{\hat\sigma}^b\hat\sigma^a
 =
 2\delta^{ab}I_2.
 \label{eq:Euclidean-sigma-Clifford-identities}
\end{equation}
Their antisymmetrized products are defined by
\begin{equation}
 \hat\sigma^{ab}
 \equiv
 \frac12
 \left(
 \hat\sigma^a\bar{\hat\sigma}^b
 -\hat\sigma^b\bar{\hat\sigma}^a
 \right),
 \qquad
 \bar{\hat\sigma}^{ab}
 \equiv
 \frac12
 \left(
 \bar{\hat\sigma}^a\hat\sigma^b
 -\bar{\hat\sigma}^b\hat\sigma^a
 \right).
 \label{eq:Euclidean-sigma-generators}
\end{equation}
With the geometric convention
$\varepsilon_{0123}=-1$, these generators satisfy
\begin{equation}
 \frac12\varepsilon^{ab}{}_{cd}\hat\sigma^{cd}
 =
 \hat\sigma^{ab},
 \qquad
 \frac12\varepsilon^{ab}{}_{cd}\bar{\hat\sigma}^{cd}
 =
 -\bar{\hat\sigma}^{ab}.
 \label{eq:sigma-generator-duality}
\end{equation}
Thus, $\hat\sigma^{ab}$ and $\bar{\hat\sigma}^{ab}$ furnish the
self-dual and anti-self-dual chiral representations, respectively.

The sigma-matrix identities needed below are
\begin{align}
 \hat\sigma^{ab}\hat\sigma^c
 &=
 \delta^{bc}\hat\sigma^a
 -\delta^{ac}\hat\sigma^b
 -\varepsilon^{abcd}\hat\sigma^d,
 \nonumber\\
 \hat\sigma^c\bar{\hat\sigma}^{ab}
 &=
 \delta^{ca}\hat\sigma^b
 -\delta^{cb}\hat\sigma^a
 -\varepsilon^{cabd}\hat\sigma^d.
 \label{eq:EH-sigma-matrix-identities}
\end{align}

We define the two chiral projections of the spin connection by
\begin{equation}
 \Xi_\mu
 \equiv
 \frac12\omega_{\mu ab}\hat\sigma^{ab},
 \qquad
 \bar\Xi_\mu
 \equiv
 \frac12\omega_{\mu ab}\bar{\hat\sigma}^{ab}.
 \label{eq:chiral-spin-connections}
\end{equation}
Because the EH spin connection is internally self-dual, its
anti-self-dual projection vanishes:
\begin{equation}
 \bar\Xi_\mu=0,
 \label{eq:EH-vanishing-dotted-spin-connection}
\end{equation}
whereas $\Xi_\mu$ is generally nonzero.  The identities above also
give
\begin{equation}
 \omega_{\mu bc}\hat\sigma^c
 =
 -\frac12\Xi_\mu\hat\sigma^b
 +\frac12\hat\sigma^b\bar\Xi_\mu.
 \label{eq:spin-connection-sigma-identity}
\end{equation}
On the EH background, this reduces to
\begin{equation}
 \omega_{\mu bc}\hat\sigma^c
 =
 -\frac12\Xi_\mu\hat\sigma^b.
 \label{eq:EH-spin-connection-sigma-identity}
\end{equation}

\paragraph{Curved-space and tangent-space covariant derivatives.}

For a curved-space covector $A_\nu$,
\begin{equation}
 \nabla_\mu A_\nu
 =
 \partial_\mu A_\nu
 -\Gamma^\lambda_{\mu\nu}A_\lambda,
 \label{covofA}
\end{equation}
where the Levi-Civita connection is
\begin{equation}
 \Gamma^\lambda_{\mu\nu}
 =
 \frac12g^{\lambda\rho}
 \left(
 \partial_\mu g_{\rho\nu}
 +\partial_\nu g_{\rho\mu}
 -\partial_\rho g_{\mu\nu}
 \right).
\end{equation}
The vielbein and inverse vielbein satisfy the  postulates
\begin{align}
 \partial_\mu e^a{}_\nu
 -\Gamma^\lambda_{\mu\nu}e^a{}_\lambda
 +\omega^a{}_{b\mu}e^b{}_\nu
 &=0,
 \nonumber\\
 \partial_\mu E_a{}^\nu
 +\Gamma^\nu_{\mu\lambda}E_a{}^\lambda
 -\omega^b{}_{a\mu}E_b{}^\nu
 &=0.
 \label{eq:tetrad-postulates}
\end{align}
It follows that
\begin{equation}
 \omega^a{}_{b\mu}
 =
 e^a{}_\nu\nabla_\mu E_b{}^\nu
 =
 e^a{}_\nu
 \left(
 \partial_\mu E_b{}^\nu
 +\Gamma^\nu_{\mu\lambda}E_b{}^\lambda
 \right).
 \label{omegaab}
\end{equation}

Finally, writing $A_\nu=e^a{}_\nu A_a$, the tetrad postulate gives
\begin{align}
 \nabla_\mu A_\nu
 &=
 e^a{}_\nu
 \left(
 \partial_\mu A_a
 -\omega^b{}_{a\mu}A_b
 \right)
 \nonumber\\
 &=
 e^a{}_\nu
 \left(
 \partial_\mu A_a
 +\omega_{\mu a}{}^bA_b
 \right).
 \label{alterA}
\end{align}
The equality between the two lines follows from
$\omega_{\mu ab}=-\omega_{\mu ba}$.  Equation \eqref{alterA} applies
componentwise to Lie-algebra-valued one-forms and will be used below
to convert the gauge-fluctuation equation from curved-space components
to tangent-space components.


\section{The gauge kinetic multiplet and chiral Ward identities on
Eguchi--Hanson space}
\label{app:EH-gauge-multiplet-Ward-identities}

In this appendix, we derive the Ward identities used in the main text
directly in components. Although two of these identities (the holomorphy and position-independence identities) are standard in supersymmetric field theory; see, for example, \cite{Amati:1988ft}; it is useful to derive them explicitly in the present setting. EH space is curved and preserves only two supercharges, so the derivation ensures that no flat-space arguments specific to rigid supersymmetry on $\mathbb R^4$ are being used implicitly.  We first
fix the Euclidean vector-multiplet conventions, verify the invariance of the
action term by term, and then construct the chiral and anti-chiral gauge
kinetic multiplets.  The Ward identities of the position-independence, holomorphy, and $a$-independence will follow
as descents of these multiplets.

The local component identities are off-shell statements.  Their use inside
expectation values additionally requires a supersymmetric regulator and
measure, supersymmetric boundary states, and, in some cases, the absence of boundary surface terms.  These additional assumptions will be stated
explicitly when the Ward identities are derived.

\subsection{Spinor conventions and preserved supersymmetries}

We use
\begin{equation}
 \epsilon^{12}= \epsilon^{\dot 1\dot 2}=+1,
 \qquad
 \epsilon_{21}= \epsilon_{\dot 2\dot 1}=+1,
 \qquad
 \epsilon_{\alpha\beta}\epsilon^{\beta\gamma}
 =\delta_\alpha{}^\gamma,
 \label{eq:app-Ward-spinor-epsilon}
\end{equation}
and the same conventions for dotted indices.  Indices are raised and lowered
according to
\begin{equation}
 \lambda^\alpha=\epsilon^{\alpha\beta}\lambda_\beta,
 \qquad
 \lambda_\alpha=\epsilon_{\alpha\beta}\lambda^\beta.
\end{equation}
One also has
\begin{equation}
 \bar\zeta\bar{\hat\sigma}^{a}\chi
 =-\chi\hat\sigma^a\bar\zeta,
 \label{eq:app-Ward-Grassmann-vector-identity}
\end{equation}
whereas the scalar contraction is symmetric,
\begin{equation}
 \chi\psi=\psi\chi.
 \label{eq:app-Ward-Grassmann-scalar-identity}
\end{equation}

Let $e^a{}_{\mu}$ and $E_a{}^\mu$, as adopted in Appendix \ref{Mathematical preliminaries},  be the vielbein and inverse vielbein,
\begin{equation}
 e^a{}_{\mu}E_b{}^\mu=\delta^a{}_b,
 \qquad
 e^a{}_{\mu}E_a{}^\nu=\delta_\mu{}^\nu.
 \label{eq:app-Ward-frame-inverses}
\end{equation}
We also recall the definition of the orthonormal frame sigma matrices
\begin{equation}
 \hat\sigma^a=(I_2,i\vec\sigma),
 \qquad
 \bar{\hat\sigma}^a=(I_2,-i\vec\sigma),
 \label{eq:app-Ward-frame-sigma-definition}
\end{equation}
We use the geometric orientation of the main text,
\begin{equation}
 \varepsilon_{0123}=-1,
 \qquad
 \varepsilon^{0123}=-1,
 \label{eq:app-Ward-frame-orientation}
\end{equation}
where the second equality follows because the frame metric is Euclidean.
We define
\begin{equation}
 \hat\sigma^{ab}
 =\frac12
 \left(
  \hat\sigma^a\bar{\hat\sigma}^b
  -\hat\sigma^b\bar{\hat\sigma}^a
 \right),
 \qquad
 \bar{\hat\sigma}^{ab}
 =\frac12
 \left(
  \bar{\hat\sigma}^a\hat\sigma^b
  -\bar{\hat\sigma}^b\hat\sigma^a
 \right).
 \label{eq:app-Ward-frame-sigma-generators}
\end{equation}
No curved-index sigma matrix will be used in the component calculation.
The relevant frame identities are
\begin{align}
 \bar{\hat\sigma}^{a\,\dot\alpha\alpha}
 \hat\sigma^b_{\alpha\dot\beta}
 +
 \bar{\hat\sigma}^{b\,\dot\alpha\alpha}
 \hat\sigma^a_{\alpha\dot\beta}
 &=2\delta^{ab}\delta^{\dot\alpha}{}_{\dot\beta},
 \label{eq:app-Ward-Clifford}\\
 \hat\sigma^{ab}{}_{\alpha}{}^{\gamma}
 \hat\sigma_{b\,\beta\dot\alpha}
 &=-\delta_\alpha{}^\gamma
 \hat\sigma^a_{\beta\dot\alpha}
 +2\delta_\beta{}^\gamma
 \hat\sigma^a_{\alpha\dot\alpha},
 \label{eq:app-Ward-three-sigma}\\
 \hat\sigma^a_{\alpha\dot\alpha}
 \hat\sigma_{a\,\beta\dot\beta}
 &=2\epsilon_{\alpha\beta}
 \epsilon_{\dot\alpha\dot\beta}.
 \label{eq:app-Ward-Fierz}
\end{align}
Equation~\eqref{eq:app-Ward-three-sigma} follows directly from the definition
\begin{equation}
 \hat\sigma^{ab}
 =\frac12
 \left(
  \hat\sigma^a\bar{\hat\sigma}^b
  -\hat\sigma^b\bar{\hat\sigma}^a
 \right)
\end{equation}
and the Clifford relation~\eqref{eq:app-Ward-Clifford}.  Indeed, Lorentz
covariance allows only
\begin{equation}
 \hat\sigma^{ab}{}_{\alpha}{}^{\gamma}
 \hat\sigma_{b\,\beta\dot\alpha}
 =A\delta_\alpha{}^\gamma
 \hat\sigma^a_{\beta\dot\alpha}
 +B\delta_\beta{}^\gamma
 \hat\sigma^a_{\alpha\dot\alpha}.
\end{equation}
Taking the trace over $\alpha$ and $\gamma$ gives $2A+B=0$, while setting
$\gamma=\beta$ and using
$\hat\sigma^{ab}\hat\sigma_b=3\hat\sigma^a$ gives $A+2B=3$.
Therefore $A=-1$ and $B=2$, proving
\eqref{eq:app-Ward-three-sigma}.

Let $\omega_{\mu ab}=-\omega_{\mu ba}$ denote the spin connection of the
chosen orthonormal frame.  Its two chiral projections are
\begin{equation}
 \Xi_\mu
 =\frac12\omega_{\mu cd}\hat\sigma^{cd},
 \qquad
 \bar\Xi_\mu
 =\frac12\omega_{\mu cd}\bar{\hat\sigma}^{cd}.
 \label{eq:app-Ward-frame-chiral-spin-connections}
\end{equation}
For an adjoint undotted spinor and an adjoint dotted spinor, respectively,
we retain curved-space indices on the covariant derivatives and define
\begin{align}
 D_\mu\lambda_\alpha
 &={}
 \partial_\mu\lambda_\alpha
 +\frac12(\Xi_\mu)_\alpha{}^\beta\lambda_\beta
 +i[A_\mu,\lambda_\alpha],
 \nonumber\\
 D_\mu\bar\lambda_{\dot\alpha}
 &={}
 \partial_\mu\bar\lambda_{\dot\alpha}
 +\frac12(\bar\Xi_\mu)_{\dot\alpha}{}^{\dot\beta}
  \bar\lambda_{\dot\beta}
 +i[A_\mu,\bar\lambda_{\dot\alpha}].
 \label{eq:app-Ward-coordinate-spinor-derivatives}
\end{align}
All sigma matrices nevertheless retain tangent-space indices.  Thus a
spinor derivative is contracted as
\begin{equation}
 \bar\lambda\bar{\hat\sigma}^aE_a{}^\mu D_\mu\lambda,
 \label{eq:app-Ward-hybrid-Dirac-contraction}
\end{equation}
and a two-form is contracted with a Lorentz generator as
\begin{equation}
 \hat\sigma^{ab}E_a{}^\mu E_b{}^\nu F_{\mu\nu}.
 \label{eq:app-Ward-hybrid-two-form-contraction}
\end{equation}
The spin-connection identity used in differentiating a bispinor follows
directly from~\eqref{eq:app-Ward-frame-chiral-spin-connections}.  Indeed,
\begin{align}
 &\frac12\Xi_\mu\hat\sigma_b
 -\frac12\hat\sigma_b\bar\Xi_\mu
=
 \frac14\omega_{\mu cd}
 \left(
  \hat\sigma^{cd}\hat\sigma_b
  -\hat\sigma_b\bar{\hat\sigma}^{cd}
 \right)
 =-\omega_{\mu b}{}^c\hat\sigma_c.
 \label{eq:app-Ward-frame-soldering-derivation}
\end{align}
In the last equality, one substitutes the two triple-sigma identities that
follow from~\eqref{eq:app-Ward-Clifford}; the terms proportional to
$\omega_{\mu cd}\delta^{cd}$ vanish by antisymmetry, and the remaining two
terms combine into $-\omega_{\mu b}{}^c\hat\sigma_c$.  Equivalently,
\begin{equation}
 \omega_{\mu b}{}^c\hat\sigma_c
 =-\frac12\Xi_\mu\hat\sigma_b
 +\frac12\hat\sigma_b\bar\Xi_\mu\,.
 \label{eq:app-Ward-frame-spin-soldering-identity}
\end{equation}
Together with the vielbein postulate (\ref{eq:tetrad-postulates}), we find
\begin{align}
 &\nabla^{}_\mu
 \left(e^a{}_{\nu}\hat\sigma_a\right)
 \nonumber\\
 &\quad\equiv
 \partial_\mu\left(e^a{}_{\nu}\hat\sigma_a\right)
 -\Gamma^\rho_{\mu\nu}e^a{}_{\rho}\hat\sigma_a
 +\frac12\Xi_\mu e^a{}_{\nu}\hat\sigma_a
 -\frac12e^a{}_{\nu}\hat\sigma_a\bar\Xi_\mu
 =0.
 \label{eq:app-Ward-coordinate-soldering-constancy}
\end{align}
The conjugate identity is
\begin{align}
 &\nabla^{}_\mu
 \left(e^a{}_{\nu}\bar{\hat\sigma}_a\right)
 \nonumber\\
 &\quad=
 \partial_\mu\left(e^a{}_{\nu}\bar{\hat\sigma}_a\right)
 -\Gamma^\rho_{\mu\nu}e^a{}_{\rho}\bar{\hat\sigma}_a
 +\frac12\bar\Xi_\mu e^a{}_{\nu}\bar{\hat\sigma}_a
 -\frac12e^a{}_{\nu}\bar{\hat\sigma}_a\Xi_\mu
 =0.
 \label{eq:app-Ward-coordinate-conjugate-soldering-constancy}
\end{align}
Thus the matrices in~\eqref{eq:app-Ward-frame-sigma-definition} are
ordinary constant matrices, while the spin connection rotates their frame
and spinor indices in precisely compensating ways.

With the orientation used in this paper, the EH spin connection is internally
self-dual.  Its dotted projection therefore vanishes:
\begin{equation}
 \bar\Xi_\mu
 =\frac12\omega_{\mu bc}\bar{\hat\sigma}^{bc}=0.
 \label{eq:app-Ward-vanishing-dotted-connection}
\end{equation}
Consequently, the dotted spin bundle admits two linearly independent
parallel spinors,
\begin{equation}
 \nabla_\mu\bar\epsilon_{\dot\alpha}
 \equiv
  \partial_\mu\bar\epsilon_{\dot\alpha}
  +\frac12(\bar\Xi_\mu)_{\dot\alpha}{}^{\dot\beta}
   \bar\epsilon_{\dot\beta}
 =\partial_\mu\bar\epsilon_{\dot\alpha}=0.
 \label{eq:app-Ward-parallel-dotted-spinors}
\end{equation}
However, there is no nonzero parallel undotted spinor:  
\begin{equation}
 \epsilon_\alpha=0.
 \label{eq:app-Ward-no-undotted-spinor}
\end{equation}
Thus, the globally preserved supersymmetries on EH are precisely the two
dotted ones associated with~\eqref{eq:app-Ward-parallel-dotted-spinors}.

\subsection{Euclidean vector multiplet and the auxiliary-field convention}

The off-shell Euclidean vector multiplet consists of
\begin{equation}
 \left(A_\mu,\lambda_\alpha,
 \bar\lambda_{\dot\alpha},D\right).
\end{equation}
All four fields are Lie-algebra valued.  In particular,
\begin{equation}
 D=D^AT^A,
 \qquad
 \operatorname{tr}_{\mathbf N}(T^AT^B)
 =\frac12\delta^{AB}.
 \label{eq:app-Ward-generator-normalization}
\end{equation}
The fields $\lambda$ and $\bar\lambda$ are independent Euclidean Grassmann
variables.  They are not related by a Lorentzian reality condition.

We use the action
\begin{align}
 S_{\rm SYM}^{\rm off}
 ={}&
 \frac1{g^2}
 \int_{\rm EH}d^4x\sqrt g\,
 \operatorname{tr}_{\mathbf N}
 \left[
  \frac12F_{\mu\nu}F^{\mu\nu}
  +2i\bar\lambda\bar{\hat\sigma}^a
   E_a{}^\mu D_\mu\lambda
  -D^2
 \right]
 \nonumber\\
 &-
 \frac{i\theta}{8\pi^2}
 \int_{\rm EH}
 \operatorname{tr}_{\mathbf N}(F\wedge F).
 \label{eq:app-Ward-off-shell-action}
\end{align}
The gauge potential and curvature retain their coordinate components
$A_\mu$ and $F_{\mu\nu}$; the vielbein occurs only where a coordinate index
must be contracted with a tangent-space sigma matrix.

Under general rigid supersymmetry transformations, one has \cite{Wess:1992cp}
\begin{align}
 \delta A_\mu
 &=-i\epsilon\hat\sigma_a\bar\lambda\,e^a{}_{\mu}
 -i\lambda\hat\sigma_a\bar\epsilon\,e^a{}_{\mu},
 \nonumber\\
 \delta\lambda_\alpha
 &=\frac12
 \left(\hat\sigma^{ab}\epsilon\right)_\alpha
 E_a{}^\mu E_b{}^\nu F_{\mu\nu}
 +i\epsilon_\alpha D,
 \nonumber\\
 \delta\bar\lambda_{\dot\alpha}
 &=-\frac12
 \left(\bar{\hat\sigma}^{ab}\bar\epsilon\right)_{\dot\alpha}
 E_a{}^\mu E_b{}^\nu F_{\mu\nu}
 +i\bar\epsilon_{\dot\alpha}D,
 \nonumber\\
 \delta D
 &=-\epsilon\hat\sigma^aE_a{}^\mu D_\mu\bar\lambda
 +(D_\mu\lambda)E_a{}^\mu\hat\sigma^a\bar\epsilon.
 \label{eq:app-Ward-vector-transformations}
\end{align}

On EH, one sets $\epsilon=0$ and takes $\bar\epsilon$ to satisfy
\eqref{eq:app-Ward-parallel-dotted-spinors}.  The preserved transformations
are then
\begin{align}
 \delta_{\bar\epsilon}A_\mu
 &=-i\lambda\hat\sigma_a\bar\epsilon\,e^a{}_{\mu},
 \label{eq:app-Ward-dotted-A}\\
 \delta_{\bar\epsilon}\lambda
 &=0,
 \label{eq:app-Ward-dotted-lambda}\\
 \delta_{\bar\epsilon}\bar\lambda
 &=-\frac12\bar{\hat\sigma}^{ab}\bar\epsilon
 E_a{}^\mu E_b{}^\nu F_{\mu\nu}
 +i\bar\epsilon D,
 \label{eq:app-Ward-dotted-lambdabar}\\
 \delta_{\bar\epsilon}D
 &=(D_\mu\lambda)E_a{}^\mu\hat\sigma^a\bar\epsilon.
 \label{eq:app-Ward-dotted-D}
\end{align}

The advantage of retaining coordinate indices is particularly clear in the
field-strength variation.  For an adjoint coordinate covector $V_\nu$, set
\begin{equation}
 {\cal D}_\mu V_\nu
 =\partial_\mu V_\nu
 -\Gamma^\rho_{\mu\nu}V_\rho
 +i[A_\mu,V_\nu].
 \label{eq:app-Ward-coordinate-vector-derivative}
\end{equation}
Since the Levi--Civita connection is symmetric in its lower indices,
\begin{equation}
 \delta F_{\mu\nu}
 ={\cal D}_\mu\delta A_\nu
 -{\cal D}_\nu\delta A_\mu.
 \label{eq:app-Ward-frame-field-strength-variation-general}
\end{equation}
 Using
\eqref{eq:app-Ward-coordinate-soldering-constancy},
\begin{align}
 {\cal D}_\mu
 \left(\lambda e^a{}_{\nu}\hat\sigma_a\bar\epsilon\right)
 ={}&(D_\mu\lambda)e^a{}_{\nu}\hat\sigma_a\bar\epsilon
 +\lambda\nabla^{}_\mu
  \left(e^a{}_{\nu}\hat\sigma_a\right)\bar\epsilon
 +\lambda e^a{}_{\nu}\hat\sigma_a
  \nabla_\mu\bar\epsilon
 \nonumber\\
 ={}&(D_\mu\lambda)e^a{}_{\nu}\hat\sigma_a\bar\epsilon.
 \label{eq:app-Ward-frame-product-derivative}
\end{align}
The second equality uses 
$\nabla^{}_\mu(e^a{}_{\nu}\hat\sigma_a)=0$ and
$\nabla_\mu\bar\epsilon=0$.
Substitution into
\eqref{eq:app-Ward-frame-field-strength-variation-general} gives
\begin{equation}
 \delta_{\bar\epsilon}F_{\mu\nu}
 =-i
 \left[
  (D_\mu\lambda)e^a{}_{\nu}\hat\sigma_a\bar\epsilon
  -(D_\nu\lambda)e^a{}_{\mu}\hat\sigma_a\bar\epsilon
 \right].
 \label{eq:app-Ward-dotted-F}
\end{equation}

\subsection{Term-by-term invariance of the action}

We now verify the invariance of 
\eqref{eq:app-Ward-off-shell-action} under the dotted supersymmetry generators.
Write
\begin{equation}
 X_\mu\equiv D_\mu\lambda.
 \label{eq:app-Ward-X-definition}
\end{equation}

\paragraph{Auxiliary-field terms.}

The part of the gaugino kinetic-term variation generated by the
$i\bar\epsilon D$ term in~\eqref{eq:app-Ward-dotted-lambdabar} is
\begin{align}
 \left.
 \delta_{\bar\epsilon}
 \operatorname{tr}_{\mathbf N}
 \left(2i\bar\lambda\bar{\hat\sigma}^a
 E_a{}^\mu X_\mu\right)
 \right|_D
 &=2i\operatorname{tr}_{\mathbf N}
 \left[(i\bar\epsilon D)\bar{\hat\sigma}^a
 E_a{}^\mu X_\mu\right]
 \nonumber\\
 &=-2\operatorname{tr}_{\mathbf N}
 \left(D\bar\epsilon\bar{\hat\sigma}^a
 E_a{}^\mu X_\mu\right)
 \nonumber\\
 &=+2\operatorname{tr}_{\mathbf N}
 \left(DX_\mu E_a{}^\mu\hat\sigma^a\bar\epsilon\right),
 \label{eq:app-Ward-kinetic-D-variation}
\end{align}
where the last equality follows from
\eqref{eq:app-Ward-Grassmann-vector-identity}.  On the other hand,
\begin{align}
 \delta_{\bar\epsilon}
 \operatorname{tr}_{\mathbf N}(-D^2)
 &=-\operatorname{tr}_{\mathbf N}
 \left[(\delta_{\bar\epsilon}D)D
 +D(\delta_{\bar\epsilon}D)\right]
 \nonumber\\
 &=-2\operatorname{tr}_{\mathbf N}
 \left(D\delta_{\bar\epsilon}D\right)
 \nonumber\\
 &=-2\operatorname{tr}_{\mathbf N}
 \left(DX_\mu E_a{}^\mu\hat\sigma^a\bar\epsilon\right).
 \label{eq:app-Ward-D-square-variation}
\end{align}
Equations~\eqref{eq:app-Ward-kinetic-D-variation} and
\eqref{eq:app-Ward-D-square-variation} cancel exactly:
\begin{equation}
 \left.
 \delta_{\bar\epsilon}
 \operatorname{tr}_{\mathbf N}
 \left(
  2i\bar\lambda\bar{\hat\sigma}^a
  E_a{}^\mu D_\mu\lambda-D^2
 \right)
 \right|_D=0.
 \label{eq:app-Ward-D-cancellation}
\end{equation}

\paragraph{Gauge kinetic and $F$-dependent gaugino terms.}

Using~\eqref{eq:app-Ward-dotted-F}, the Yang--Mills term varies as
\begin{align}
 \delta_{\bar\epsilon}
 \operatorname{tr}_{\mathbf N}
 \left(\frac12F_{\mu\nu}F^{\mu\nu}\right)
 &=\operatorname{tr}_{\mathbf N}
 \left(F^{\mu\nu}\delta_{\bar\epsilon}F_{\mu\nu}\right)
 \nonumber\\
 &=-2i\operatorname{tr}_{\mathbf N}
 \left(F^{\mu\nu}X_\mu
 e^a{}_{\nu}\hat\sigma_a\bar\epsilon\right).
 \label{eq:app-Ward-F-square-variation}
\end{align}

The $F$-dependent part of the variation of $\bar\lambda$ gives
\begin{align}
 \left.
 \delta_{\bar\epsilon}
 \operatorname{tr}_{\mathbf N}
 \left(2i\bar\lambda\bar{\hat\sigma}^a
 E_a{}^\mu X_\mu\right)
 \right|_F
 =-i\operatorname{tr}_{\mathbf N}
 \left[
  \left(\bar{\hat\sigma}^{cd}\bar\epsilon\right)
  E_c{}^\rho E_d{}^\sigma F_{\rho\sigma}
  \bar{\hat\sigma}^aE_a{}^\mu X_\mu
 \right].
 \label{eq:app-Ward-kinetic-F-start}
\end{align}
The relevant sigma identity is
\begin{equation}
 \hat\sigma^a\bar{\hat\sigma}^{cd}
 =\delta^{ac}\hat\sigma^d
 -\delta^{ad}\hat\sigma^c
 -\varepsilon^{acdb}\hat\sigma_b.
 \label{eq:app-Ward-sigma-F-identity}
\end{equation}
The index movement and Grassmann reordering needed before applying this
matrix identity are
\begin{align}
 &\left(\bar{\hat\sigma}^{cd}\bar\epsilon\right)_{\dot\alpha}
 \bar{\hat\sigma}^{a\,\dot\alpha\alpha}
 E_a{}^\mu X_{\mu\alpha}
 \nonumber\\
 &\hspace{12mm}=
 -E_a{}^\mu X_\mu{}^\alpha
 \left(
  \hat\sigma^a\bar{\hat\sigma}^{cd}
 \right)_{\alpha\dot\alpha}
 \bar\epsilon^{\dot\alpha}.
 \label{eq:app-Ward-F-term-index-reordering}
\end{align}
The minus sign on the right-hand side is essential.  It follows by first
interchanging the two Grassmann-odd objects $X_\mu$ and $\bar\epsilon$ and
then using the spinor metrics to put the remaining matrix product in the
order $\hat\sigma^a\bar{\hat\sigma}^{cd}$.  Applying
\eqref{eq:app-Ward-sigma-F-identity}, and using the antisymmetry of
$F_{\rho\sigma}$, equation~\eqref{eq:app-Ward-kinetic-F-start} becomes
\begin{align}
 \left.
 \delta_{\bar\epsilon}
 \operatorname{tr}_{\mathbf N}
 \left(2i\bar\lambda\bar{\hat\sigma}^a
 E_a{}^\mu X_\mu\right)
 \right|_F
 ={}&
 2i\operatorname{tr}_{\mathbf N}
 \left(F^{\mu\nu}X_\mu
 e^a{}_{\nu}\hat\sigma_a\bar\epsilon\right)
 \nonumber\\
 &-2i\operatorname{tr}_{\mathbf N}
 \left(\widetilde F^{\mu\nu}X_\mu
 e^a{}_{\nu}\hat\sigma_a\bar\epsilon\right),
 \label{eq:app-Ward-kinetic-F-reduced}
\end{align}
where
\begin{equation}
 \varepsilon^{\mu\nu\rho\sigma}
 \equiv
 E_a{}^\mu E_b{}^\nu E_c{}^\rho E_d{}^\sigma
 \varepsilon^{abcd},
 \qquad
 \widetilde F^{\mu\nu}
 \equiv\frac12\varepsilon^{\mu\nu\rho\sigma}F_{\rho\sigma}.
\end{equation}
The first term in~\eqref{eq:app-Ward-kinetic-F-reduced} cancels
\eqref{eq:app-Ward-F-square-variation}.  The remaining term is a total
derivative.  Indeed, the Bianchi identity gives
\begin{equation}
 {\cal D}_\mu\widetilde F^{\mu\nu}=0,
\end{equation}
where ${\cal D}_\mu$ now includes the Levi--Civita connections on the two
upper coordinate indices and the adjoint gauge connection.
Since $\nabla_\mu\bar\epsilon=0$, it follows that
\begin{align}
 \operatorname{tr}_{\mathbf N}
 \left(\widetilde F^{\mu\nu}X_\mu
 e^a{}_{\nu}\hat\sigma_a\bar\epsilon\right)
 &={\nabla}_\mu
 \operatorname{tr}_{\mathbf N}
 \left(\widetilde F^{\mu\nu}\lambda
 e^a{}_{\nu}\hat\sigma_a\bar\epsilon\right).
 \label{eq:app-Ward-dual-F-divergence}
\end{align}
For clarity, the coordinate divergence on the right-hand side means
\begin{equation}
 \nabla_\mu J^\mu
 =\frac1{\sqrt g}\partial_\mu
 \left(\sqrt g\,J^\mu\right).
 \label{eq:app-Ward-frame-divergence}
\end{equation}
Thus the last term integrates to the ordinary flux through the asymptotic
boundary.

\paragraph{Cubic gaugino term.}

There is one further contribution because $D_\mu\lambda$ contains the gauge
connection.  Since $\delta\lambda=0$,
\begin{equation}
 \delta_{\bar\epsilon}(D_\mu\lambda)
 =i[\delta_{\bar\epsilon}A_\mu,\lambda]
 =[\lambda e^a{}_{\mu}\hat\sigma_a\bar\epsilon,\lambda].
\end{equation}
The resulting cubic term is proportional to
\begin{equation}
 \operatorname{tr}_{\mathbf N}
 \left[
  \bar\lambda\bar{\hat\sigma}^aE_a{}^\mu
  [\lambda e^b{}_{\mu}\hat\sigma_b\bar\epsilon,\lambda]
 \right].
 \label{eq:app-Ward-cubic-term}
\end{equation}
Writing $\lambda=\lambda^AT^A$ and using
\eqref{eq:app-Ward-Fierz}, the spinor part of
\eqref{eq:app-Ward-cubic-term} is proportional to
\begin{equation}
 \lambda^{B\alpha}\lambda^C_{\alpha}.
\end{equation}
By~\eqref{eq:app-Ward-Grassmann-scalar-identity}, this bilinear is symmetric
under $B\leftrightarrow C$.  The color factor is proportional to the
antisymmetric structure constant $f^{ABC}$.  Hence
\begin{equation}
 f^{ABC}\lambda^{B\alpha}\lambda^C_{\alpha}=0,
\end{equation}
and the cubic term vanishes identically.

Combining the preceding results gives
\begin{align}
 \delta_{\bar\epsilon}
 \operatorname{tr}_{\mathbf N}
 \left[
  \frac12F_{\mu\nu}F^{\mu\nu}
  +2i\bar\lambda\bar{\hat\sigma}^a
   E_a{}^\mu D_\mu\lambda
  -D^2
 \right]
 =-2i\nabla_\mu
 \operatorname{tr}_{\mathbf N}
 \left(
  \widetilde F^{\mu\nu}\lambda
  e^a{}_{\nu}\hat\sigma_a\bar\epsilon
 \right).
 \label{eq:app-Ward-local-action-variation}
\end{align}

\paragraph{Topological term.}

Using $\delta F=D(\delta A)$ and the Bianchi identity $DF=0$,
\begin{align}
 \delta\operatorname{tr}_{\mathbf N}(F\wedge F)
 &=2\operatorname{tr}_{\mathbf N}(D\delta A\wedge F)
 \nonumber\\
 &=2d\operatorname{tr}_{\mathbf N}(\delta A\wedge F).
 \label{eq:app-Ward-theta-variation}
\end{align}
Thus, every uncancelled term in the action variation is a boundary term:
\begin{equation}
 \delta_{\bar\epsilon}S_{\rm SYM}^{\rm off}
 =\int_{\partial{\rm EH}}d\Sigma_\mu\,
 {\cal J}^{\mu}_{\bar\epsilon}.
 \label{eq:app-Ward-action-boundary-variation}
\end{equation}

The explicit normalizable zero modes satisfy the required asymptotic falloff. For the localized instanton background, the connection profiles behave as $r^{-2}$, but the orthonormal components of the field strength fall as $F^{\rm B}_{ab}=O(r^{-4})$. A continuum Dirac eigenmode behaves asymptotically as $r^{-1}H_\nu(kr)=O(r^{-3/2})$, where $H_\nu$ is the Hankel function and $k$ is the momentum of the scattering wave, so the background--fermion contribution to the integrated surface term vanishes as $R^{-5/2}$. Terms involving both bosonic and fermionic continuum fluctuations are not controlled by this pointwise estimate; their boundary pairing is assumed to vanish under the compatible supersymmetric boundary conditions used to define the regulated fluctuation operators.

\subsection{Components of the gauge kinetic multiplets}

We now construct the components of the chiral superfield needed in the Ward identities.  This is a
local component construction and does not require a globally defined
undotted supersymmetry on EH.  More explicitly, at an arbitrary point
$p$ we may choose a normal orthonormal frame and extend an arbitrary
undotted test spinor so that
$\omega_{\mu ab}(p)=0$ and $\nabla_\mu\epsilon(p)=0$.  The calculation
below then determines tensorial component identities at $p$.  Since
$p$ is arbitrary, those identities hold in every frame throughout EH.
This device defines the components of the chiral gauge kinetic multiplet;
it does not assert the existence of a globally conserved undotted charge. We shall show the well-known result that the chiral superfields have lower, middle, and upper components given by
\begin{eqnarray}
\left( {\cal O},  \psi_{{\cal S}\alpha},  F_{\cal S} \right)
\end{eqnarray}
where the various components are given by (\ref{eq:app-Ward-O-definition}, \ref{eq:app-Ward-psi-S}, \ref{eq:app-Ward-F-S}). In particular, the lowest component ${\cal O}$ is the fermion bilinear and the upper component $F_{\cal S} $ is the kinetic term. Identical results hold for the anti-chiral superfield. 

Define
\begin{equation}
 {\cal O}
 =\operatorname{tr}_{\mathbf N}(\lambda\lambda)
 =\operatorname{tr}_{\mathbf N}
 \left(\epsilon^{\alpha\beta}
 \lambda_\beta\lambda_\alpha\right).
 \label{eq:app-Ward-O-definition}
\end{equation}
Applying the local undotted part of
\eqref{eq:app-Ward-vector-transformations} gives
\begin{align}
 \delta_\epsilon{\cal O}
 &=2\operatorname{tr}_{\mathbf N}
 \left[(\delta_\epsilon\lambda)^\beta\lambda_\beta\right]
 \nonumber\\
 &=\operatorname{tr}_{\mathbf N}
  \left[
  (\hat\sigma^{ab}\epsilon)^\beta
  \lambda_\beta E_a{}^\mu E_b{}^\nu F_{\mu\nu}
  +2i\epsilon^\beta\lambda_\beta D
 \right].
 \label{eq:app-Ward-delta-O-first}
\end{align}
For a traceless two-by-two matrix
$M_\alpha{}^\beta$, the bispinor
$M^{\alpha\beta}=\epsilon^{\alpha\gamma}M_\gamma{}^\beta$ is symmetric.
Since $\hat\sigma^{ab}$ is traceless, this gives
\begin{equation}
 (\hat\sigma^{ab}\epsilon)^\beta\lambda_\beta
 =-\epsilon^\alpha
 (\hat\sigma^{ab}\lambda)_\alpha.
 \label{eq:app-Ward-spinor-interchange}
\end{equation}
Substituting~\eqref{eq:app-Ward-spinor-interchange} into
\eqref{eq:app-Ward-delta-O-first}, one obtains
\begin{equation}
 \delta_\epsilon{\cal O}
 =\epsilon^\alpha\psi_{{\cal S}\alpha},
 \end{equation}
where we defined
\begin{equation}
 \psi_{{\cal S}\alpha}
 \equiv 2\operatorname{tr}_{\mathbf N}
 \left[
  i\lambda_\alpha D
  -\frac12
  (\hat\sigma^{ab}\lambda)_\alpha
  E_a{}^\mu E_b{}^\nu F_{\mu\nu}
 \right].
 \label{eq:app-Ward-psi-S}
\end{equation}
Equivalently, if $\chi_{{\cal S}\alpha}$ denotes the conventionally
normalized middle component,
\begin{equation}
 \psi_{{\cal S}\alpha}=\sqrt2\chi_{{\cal S}\alpha},
\end{equation}
and hence
\begin{equation}
 \chi_{{\cal S}\alpha}
 =\sqrt2\operatorname{tr}_{\mathbf N}
 \left[
  i\lambda_\alpha D
  -\frac12
  (\hat\sigma^{ab}\lambda)_\alpha
  E_a{}^\mu E_b{}^\nu F_{\mu\nu}
 \right].
 \label{eq:app-Ward-chi-S}
\end{equation}

We next derive the highest component.  Set $\bar\epsilon=0$ in
\eqref{eq:app-Ward-vector-transformations}.  Then
\begin{align}
 \delta_\epsilon\lambda_\alpha
 &=\frac12
 (\hat\sigma^{cd}\epsilon)_\alpha
 E_c{}^\rho E_d{}^\sigma F_{\rho\sigma}
 +i\epsilon_\alpha D,
 \\
 \delta_\epsilon D
 &=-\epsilon\hat\sigma^cE_c{}^\rho D_\rho\bar\lambda,
 \\
 \delta_\epsilon F_{\mu\nu}
 &=-i\epsilon\hat\sigma_a e^a{}_{\nu}D_\mu\bar\lambda
 +i\epsilon\hat\sigma_a e^a{}_{\mu}D_\nu\bar\lambda.
 \label{eq:app-Ward-undotted-component-variations}
\end{align}
Varying~\eqref{eq:app-Ward-psi-S} gives four terms:
\begin{align}
 \delta_\epsilon\psi_{{\cal S}\alpha}
 ={}&2i\operatorname{tr}_{\mathbf N}
 \left[(\delta_\epsilon\lambda_\alpha)D\right]
 +2i\operatorname{tr}_{\mathbf N}
 \left[\lambda_\alpha\delta_\epsilon D\right]
 \nonumber\\
 &-\operatorname{tr}_{\mathbf N}
  \left[
  (\hat\sigma^{ab}\delta_\epsilon\lambda)_\alpha
  E_a{}^\mu E_b{}^\nu F_{\mu\nu}
 \right]
 -\operatorname{tr}_{\mathbf N}
  \left[
  (\hat\sigma^{ab}\lambda)_\alpha
  E_a{}^\mu E_b{}^\nu\delta_\epsilon F_{\mu\nu}
 \right].
 \label{eq:app-Ward-delta-psi-undotted-four-terms}
\end{align}
The terms proportional to $DF_{\mu\nu}$ are
\begin{equation}
 i(\hat\sigma^{ab}\epsilon)_\alpha
 E_a{}^\mu E_b{}^\nu
 \operatorname{tr}_{\mathbf N}(F_{\mu\nu}D)
 -i(\hat\sigma^{ab}\epsilon)_\alpha
 E_a{}^\mu E_b{}^\nu
 \operatorname{tr}_{\mathbf N}(DF_{\mu\nu})=0.
 \label{eq:app-Ward-FD-cancellation-chiral}
\end{equation}
The $D^2$ term is
\begin{equation}
 -2\epsilon_\alpha
 \operatorname{tr}_{\mathbf N}(D^2).
 \label{eq:app-Ward-D2-chiral-top}
\end{equation}
For the purely bosonic field-strength term, the product identity
\begin{align}
 &\frac12
 \left(
  \hat\sigma^{ab}\hat\sigma^{cd}\epsilon
 \right)_\alpha
 \operatorname{tr}_{\mathbf N}
 \left(
  E_a{}^\mu E_b{}^\nu E_c{}^\rho E_d{}^\sigma
  F_{\mu\nu}F_{\rho\sigma}
 \right)
 \nonumber\\
 &\hspace{12mm}=
 -\epsilon_\alpha\operatorname{tr}_{\mathbf N}
 \left[
  F_{\mu\nu}F^{\mu\nu}
  +\frac12\varepsilon^{\mu\nu\rho\sigma}
   F_{\mu\nu}F_{\rho\sigma}
 \right]
 \label{eq:app-Ward-sigma-square-chiral}
\end{align}
gives
\begin{equation}
 2\epsilon_\alpha\operatorname{tr}_{\mathbf N}
 \left[
  \frac12F_{\mu\nu}F^{\mu\nu}
  +\frac14\varepsilon^{\mu\nu\rho\sigma}
   F_{\mu\nu}F_{\rho\sigma}
 \right].
 \label{eq:app-Ward-F2-chiral-top}
\end{equation}
It remains to retain every fermion-derivative term.  Write
$Y_\mu\equiv D_\mu\bar\lambda$.  The second term in
\eqref{eq:app-Ward-delta-psi-undotted-four-terms} gives
\begin{align}
 2i\operatorname{tr}_{\mathbf N}
 \left(\lambda_\alpha\delta_\epsilon D\right)
 &=2i\epsilon^\beta
 \operatorname{tr}_{\mathbf N}
 \left(
  \lambda_\alpha\hat\sigma^a_{\beta\dot\beta}
  E_a{}^\mu Y_\mu{}^{\dot\beta}
 \right).
 \label{eq:app-Ward-chiral-top-D-derivative-piece}
\end{align}
Here the sign changes when the Grassmann-odd parameter is moved through
$\lambda_\alpha$.  The fourth term gives
\begin{align}
 &-\operatorname{tr}_{\mathbf N}
  \left[
  (\hat\sigma^{ab}\lambda)_\alpha
  E_a{}^\mu E_b{}^\nu\delta_\epsilon F_{\mu\nu}
 \right]
 \nonumber\\
 &\quad=
 -2i\epsilon^\beta
 \hat\sigma^{ab}{}_{\alpha}{}^\gamma
 \hat\sigma_{b\,\beta\dot\beta}
 \operatorname{tr}_{\mathbf N}
 \left(\lambda_\gamma E_a{}^\mu
 Y_\mu{}^{\dot\beta}\right)
 \nonumber\\
 &\quad=
 2i\epsilon^\beta
 \operatorname{tr}_{\mathbf N}
 \left(
  \lambda_\alpha\hat\sigma^a_{\beta\dot\beta}
  E_a{}^\mu Y_\mu{}^{\dot\beta}
 \right)
 -4i\epsilon^\beta
 \hat\sigma^a_{\alpha\dot\beta}
 \operatorname{tr}_{\mathbf N}
 \left(\lambda_\beta E_a{}^\mu
 Y_\mu{}^{\dot\beta}\right),
 \label{eq:app-Ward-chiral-top-F-derivative-piece}
\end{align}
where~\eqref{eq:app-Ward-three-sigma} was used in the last line.  The
two-component Schouten identity, with the Grassmann ordering displayed, is
\begin{equation}
 \epsilon^\beta\lambda_\alpha
 \hat\sigma^a_{\beta\dot\beta}
 -(\epsilon^\beta\lambda_\beta)
 \hat\sigma^a_{\alpha\dot\beta}
 =\epsilon_\alpha\lambda^\beta
 \hat\sigma^a_{\beta\dot\beta}.
 \label{eq:app-Ward-chiral-Schouten}
\end{equation}
Adding~\eqref{eq:app-Ward-chiral-top-D-derivative-piece} and
\eqref{eq:app-Ward-chiral-top-F-derivative-piece}, and then applying
\eqref{eq:app-Ward-chiral-Schouten}, gives
\begin{equation}
 4i\epsilon_\alpha
 \operatorname{tr}_{\mathbf N}
 \left(\lambda\hat\sigma^aE_a{}^\mu
 D_\mu\bar\lambda\right).
 \label{eq:app-Ward-fermion-chiral-top}
\end{equation}
Equations~\eqref{eq:app-Ward-D2-chiral-top},
\eqref{eq:app-Ward-F2-chiral-top}, and
\eqref{eq:app-Ward-fermion-chiral-top} combine into
\begin{equation}
 \delta_\epsilon\psi_{{\cal S}\alpha}
 =2\epsilon_\alpha F_{\cal S},
 \label{eq:app-Ward-chiral-second-variation}
\end{equation}
where
\begin{align}
 F_{\cal S}
 =\operatorname{tr}_{\mathbf N}
 \left[
  \frac12F_{\mu\nu}F^{\mu\nu}
  +\frac14\varepsilon^{\mu\nu\rho\sigma}
   F_{\mu\nu}F_{\rho\sigma}
  -D^2
  +2i\lambda\hat\sigma^aE_a{}^\mu
   D_\mu\bar\lambda
 \right].
 \label{eq:app-Ward-F-S}
\end{align}
In particular, equations~\eqref{eq:app-Ward-D2-chiral-top} and
\eqref{eq:app-Ward-chiral-second-variation} show directly that the signs of
$i\lambda_\alpha D$ and $-D^2$ belong to the same multiplet convention.

The anti-chiral lowest component is
\begin{equation}
 \bar{\cal O}
 =\operatorname{tr}_{\mathbf N}(\bar\lambda\bar\lambda)
 =\operatorname{tr}_{\mathbf N}
 \left(
  \epsilon^{\dot\alpha\dot\beta}
  \bar\lambda_{\dot\beta}\bar\lambda_{\dot\alpha}
 \right).
 \label{eq:app-Ward-O-bar}
\end{equation}
Using~\eqref{eq:app-Ward-dotted-lambdabar} gives
\begin{equation}
 \delta_{\bar\epsilon}\bar{\cal O}
 =\bar\epsilon^{\dot\alpha}
 \bar\psi_{{\cal S}\dot\alpha},
\end{equation}
where
\begin{equation}
 \bar\psi_{{\cal S}\dot\alpha}
 =2\operatorname{tr}_{\mathbf N}
 \left[
  i\bar\lambda_{\dot\alpha}D
  -\frac12
  (\bar\lambda\bar{\hat\sigma}^{ab})_{\dot\alpha}
  E_a{}^\mu E_b{}^\nu F_{\mu\nu}
 \right].
 \label{eq:app-Ward-psi-S-bar}
\end{equation}
Thus,
\begin{equation}
 \bar\chi_{{\cal S}\dot\alpha}
 =\sqrt2\operatorname{tr}_{\mathbf N}
 \left[
  i\bar\lambda_{\dot\alpha}D
  -\frac12
  (\bar\lambda\bar{\hat\sigma}^{ab})_{\dot\alpha}
  E_a{}^\mu E_b{}^\nu F_{\mu\nu}
 \right].
 \label{eq:app-Ward-chi-S-bar}
\end{equation}

Varying~\eqref{eq:app-Ward-psi-S-bar} with a dotted parameter produces four
types of terms: $D^2$, $F D$, $F^2$, and a gaugino derivative.  The two
mixed terms are, with their spinor indices kept in the same order,
\begin{align}
 \left.\delta_{\bar\epsilon}
 \bar\psi_{{\cal S}\dot\alpha}\right|_{FD}
 ={}&-i\operatorname{tr}_{\mathbf N}
  \left[
  (\bar{\hat\sigma}^{ab}\bar\epsilon)_{\dot\alpha}
  E_a{}^\mu E_b{}^\nu F_{\mu\nu}D
 \right]
+i\operatorname{tr}_{\mathbf N}
  \left[
  (\bar{\hat\sigma}^{ab}\bar\epsilon)_{\dot\alpha}
  E_a{}^\mu E_b{}^\nu DF_{\mu\nu}
 \right]=0.
 \label{eq:app-Ward-antichiral-FD-cancellation}
\end{align}
In the second term we used
$(\bar\epsilon\bar{\hat\sigma}^{ab})_{\dot\alpha}
=-(\bar{\hat\sigma}^{ab}\bar\epsilon)_{\dot\alpha}$;
cyclicity of the color trace then gives the cancellation.  The $D^2$ term
is
\begin{equation}
 2i\operatorname{tr}_{\mathbf N}
 \left[(i\bar\epsilon_{\dot\alpha}D)D\right]
 =-2\bar\epsilon_{\dot\alpha}
 \operatorname{tr}_{\mathbf N}(D^2).
\end{equation}
The $F^2$ term is
\begin{align}
 \left.\delta_{\bar\epsilon}
 \bar\psi_{{\cal S}\dot\alpha}\right|_{F^2}
 ={}&\frac12\operatorname{tr}_{\mathbf N}
 \left[
  \left[
   (\bar{\hat\sigma}^{cd}\bar\epsilon)
   \bar{\hat\sigma}^{ab}
  \right]_{\dot\alpha}
  E_c{}^\rho E_d{}^\sigma E_a{}^\mu E_b{}^\nu
  F_{\rho\sigma}F_{\mu\nu}
 \right].
 \label{eq:app-Ward-antichiral-F2-start}
\end{align}
The product of the two anti-self-dual sigma matrices gives
\begin{equation}
 \left[
  (\bar{\hat\sigma}^{cd}\bar\epsilon)
  \bar{\hat\sigma}^{ab}
 \right]_{\dot\alpha}
 =-\left(
  \bar{\hat\sigma}^{ab}\bar{\hat\sigma}^{cd}
  \bar\epsilon
 \right)_{\dot\alpha}.
 \label{eq:app-Ward-antichiral-row-column-conversion}
\end{equation}
Here the right action is defined explicitly by
\begin{equation}
 (\bar\zeta\bar{\hat\sigma}^{ab})_{\dot\alpha}
 \equiv
 \bar\zeta_{\dot\beta}
 (\bar{\hat\sigma}^{ab})^{\dot\beta}{}_{\dot\alpha}
 =-
 (\bar{\hat\sigma}^{ab})_{\dot\alpha}{}^{\dot\beta}
 \bar\zeta_{\dot\beta}.
 \label{eq:app-Ward-antichiral-right-action}
\end{equation}
The final equality follows by raising the first index of the right-acting
generator with $\epsilon^{\dot\alpha\dot\beta}$ and using the symmetry of
$(\bar{\hat\sigma}^{ab})_{\dot\alpha\dot\beta}$.  Applying this equality to
$\bar\zeta=\bar{\hat\sigma}^{cd}\bar\epsilon$ proves
\eqref{eq:app-Ward-antichiral-row-column-conversion}.  Thus the matrix identity
needed in~\eqref{eq:app-Ward-antichiral-F2-start} is
\begin{align}
 &-\frac12
 \left(
  \bar{\hat\sigma}^{ab}\bar{\hat\sigma}^{cd}
  \bar\epsilon
 \right)_{\dot\alpha}
 E_a{}^\mu E_b{}^\nu E_c{}^\rho E_d{}^\sigma
 \operatorname{tr}_{\mathbf N}(F_{\mu\nu}F_{\rho\sigma})
 \nonumber\\
 &\hspace{15mm}=
 \bar\epsilon_{\dot\alpha}\operatorname{tr}_{\mathbf N}
 \left[
  F_{\mu\nu}F^{\mu\nu}
  -\frac12\varepsilon^{\mu\nu\rho\sigma}
  F_{\mu\nu}F_{\rho\sigma}
 \right].
 \label{eq:app-Ward-barsigma-square}
\end{align}
This follows by inserting the definition of
$\bar{\hat\sigma}^{ab}$ twice, applying the frame Clifford algebra, and
using the anti-self-duality of $\bar{\hat\sigma}^{ab}$.  Notice that the
sign of the term containing $\varepsilon^{\mu\nu\rho\sigma}$ is opposite to the
corresponding chiral identity~\eqref{eq:app-Ward-sigma-square-chiral}.
Finally, the two terms containing $X_\mu=D_\mu\lambda$ are
\begin{align}
 \left.\delta_{\bar\epsilon}
 \bar\psi_{{\cal S}\dot\alpha}\right|_{X}
 ={}&2i\operatorname{tr}_{\mathbf N}
  \left[
  \bar\lambda_{\dot\alpha}X_\mu
  E_a{}^\mu\hat\sigma^a\bar\epsilon
 \right]
 \nonumber\\
 &+2i\operatorname{tr}_{\mathbf N}
  \left[
  (\bar\lambda\bar{\hat\sigma}^{ab})_{\dot\alpha}
  E_a{}^\mu X_\mu\hat\sigma_b\bar\epsilon
 \right].
 \label{eq:app-Ward-antichiral-X-two-pieces}
\end{align}
Substitution of the definitions of
$\bar{\hat\sigma}^{ab}$, followed by the Clifford relation and graded
cyclicity, gives the explicit spinor reduction
\begin{align}
 &\operatorname{tr}_{\mathbf N}
  \left[
  \bar\lambda_{\dot\alpha}X_\mu
  E_a{}^\mu\hat\sigma^a\bar\epsilon
 +(\bar\lambda\bar{\hat\sigma}^{ab})_{\dot\alpha}
  E_a{}^\mu X_\mu\hat\sigma_b\bar\epsilon
 \right]
 \nonumber\\
 &\hspace{18mm}=
 -2\bar\epsilon_{\dot\alpha}
 \operatorname{tr}_{\mathbf N}
 \left(X_\mu E_a{}^\mu\hat\sigma^a\bar\lambda\right).
 \label{eq:app-Ward-antichiral-X-reduction}
\end{align}
Therefore,
\begin{equation}
 -4i\bar\epsilon_{\dot\alpha}
 \operatorname{tr}_{\mathbf N}
 \left[(D_\mu\lambda)E_a{}^\mu
 \hat\sigma^a\bar\lambda\right].
\end{equation}
Consequently,
\begin{equation}
 \delta_{\bar\epsilon}\bar\psi_{{\cal S}\dot\alpha}
 =2\bar\epsilon_{\dot\alpha}\bar F_{\cal S},
 \label{eq:app-Ward-second-antichiral-variation}
\end{equation}
where
\begin{align}
 \bar F_{\cal S}
 =\operatorname{tr}_{\mathbf N}
 \left[
  \frac12F_{\mu\nu}F^{\mu\nu}
  -\frac14\varepsilon^{\mu\nu\rho\sigma}
   F_{\mu\nu}F_{\rho\sigma}
  -D^2
  -2i(D_\mu\lambda)E_a{}^\mu
   \hat\sigma^a\bar\lambda
 \right].
 \label{eq:app-Ward-F-S-bar}
\end{align}

After integrating the fermion term in~\eqref{eq:app-Ward-F-S} by parts,
\begin{equation}
 \int_{\rm EH}d^4x\sqrt g\,
 \lambda\hat\sigma^aE_a{}^\mu D_\mu\bar\lambda
 =\int_{\rm EH}d^4x\sqrt g\,
 \bar\lambda\bar{\hat\sigma}^aE_a{}^\mu D_\mu\lambda
 +\text{surface term},
 \label{eq:app-Ward-fermion-integration-by-parts}
\end{equation}
and using
\begin{equation}
 -(D_\mu\lambda)E_a{}^\mu\hat\sigma^a\bar\lambda
 =\bar\lambda\bar{\hat\sigma}^aE_a{}^\mu D_\mu\lambda,
\end{equation}
the action becomes
\begin{equation}
 S_{\rm SYM}^{\rm off}
 =\frac1{8\pi i}
 \left[
  \tau\int_{\rm EH}d^4x\sqrt g\,F_{\cal S}
  -\bar\tau\int_{\rm EH}d^4x\sqrt g\,\bar F_{\cal S}
 \right]+S_{\rm surf},
 \label{eq:app-Ward-action-as-F-terms}
\end{equation}
where
\begin{equation}
 \tau=\frac\theta{2\pi}+\frac{4\pi i}{g^2}.
\end{equation}
Indeed,
\begin{equation}
 \frac{\tau-\bar\tau}{8\pi i}=\frac1{g^2},
 \qquad
 \frac{\tau+\bar\tau}{8\pi i}
 =-\frac{i\theta}{8\pi^2},
\end{equation}
where the second coefficient multiplies
$\int\operatorname{tr}_{\mathbf N}(F\wedge F)$.

\subsection{First descent and position independence}

We define the charge normalization by moving the constant Grassmann
parameter to the left and writing
\begin{equation}
 \delta_{\bar\epsilon}{\cal X}
 =\bar\epsilon^{\dot\alpha}
 \left[
  \bar{\cal Q}_{\dot\alpha},{\cal X}
 \right\}_{\rm gr},
 \label{eq:app-Ward-charge-normalization}
\end{equation}
where $[\ ,\ \}_{\rm gr}$ is a commutator for bosonic ${\cal X}$ and an
anticommutator for fermionic ${\cal X}$.  This convention removes any
ambiguity associated with writing the parameters on different sides in the
component formulas.

Because~\eqref{eq:app-Ward-dotted-lambda} gives
$\delta_{\bar\epsilon}\lambda=0$, one has
\begin{equation}
 \delta_{\bar\epsilon}{\cal O}=0,
 \qquad
 [\bar{\cal Q}_{\dot\alpha},{\cal O}]=0.
 \label{eq:app-Ward-O-Q-closed}
\end{equation}

We next vary~\eqref{eq:app-Ward-psi-S} directly.  Since
$\delta_{\bar\epsilon}\lambda=0$, only $D$ and $F_{\mu\nu}$ vary.  The
auxiliary-field contribution is
\begin{align}
 \left.
 \delta_{\bar\epsilon}\psi_{{\cal S}\alpha}
 \right|_D
 &=2i\operatorname{tr}_{\mathbf N}
 \left[
  \lambda_\alpha X_\mu{}^\beta
 \right]
 E_a{}^\mu\hat\sigma^a_{\beta\dot\alpha}
 \bar\epsilon^{\dot\alpha}
 \nonumber\\
 &=-2iE_a{}^\mu\hat\sigma^a_{\beta\dot\alpha}
 \bar\epsilon^{\dot\alpha}
 \operatorname{tr}_{\mathbf N}
 \left[X_\mu{}^\beta\lambda_\alpha\right].
 \label{eq:app-Ward-descent-D-piece}
\end{align}
The second equality follows from graded cyclicity.  The field-strength
contribution is
\begin{align}
 \left.
 \delta_{\bar\epsilon}\psi_{{\cal S}\alpha}
 \right|_F
 =-2i
 \hat\sigma^{ab}{}_{\alpha}{}^\gamma
 \hat\sigma_{b\,\beta\dot\alpha}
 E_a{}^\mu
 \bar\epsilon^{\dot\alpha}
 \operatorname{tr}_{\mathbf N}
 \left[X_\mu{}^\beta\lambda_\gamma\right].
 \label{eq:app-Ward-descent-F-piece}
\end{align}
Using~\eqref{eq:app-Ward-three-sigma}, this becomes
\begin{align}
 \left.
 \delta_{\bar\epsilon}\psi_{{\cal S}\alpha}
 \right|_F
 ={}&2iE_a{}^\mu\hat\sigma^a_{\beta\dot\alpha}
 \bar\epsilon^{\dot\alpha}
 \operatorname{tr}_{\mathbf N}
 \left[X_\mu{}^\beta\lambda_\alpha\right]
 \nonumber\\
 &-4iE_a{}^\mu\hat\sigma^a_{\alpha\dot\alpha}
 \bar\epsilon^{\dot\alpha}
 \operatorname{tr}_{\mathbf N}
 \left[X_\mu{}^\beta\lambda_\beta\right].
 \label{eq:app-Ward-descent-F-reduced}
\end{align}
The first term in~\eqref{eq:app-Ward-descent-F-reduced} cancels
\eqref{eq:app-Ward-descent-D-piece}, leaving
\begin{equation}
 \delta_{\bar\epsilon}\psi_{{\cal S}\alpha}
 =-4iE_a{}^\mu\hat\sigma^a_{\alpha\dot\alpha}
 \bar\epsilon^{\dot\alpha}
 \operatorname{tr}_{\mathbf N}
 \left[(D_\mu\lambda)^\beta\lambda_\beta\right].
 \label{eq:app-Ward-descent-intermediate}
\end{equation}

The derivative of ${\cal O}$ is
\begin{align}
 \nabla_\mu{\cal O}
 ={}&\operatorname{tr}_{\mathbf N}
 \left[
  \epsilon^{\alpha\beta}(D_\mu\lambda_\beta)\lambda_\alpha
 \right]
 +\operatorname{tr}_{\mathbf N}
 \left[
  \epsilon^{\alpha\beta}\lambda_\beta(D_\mu\lambda_\alpha)
 \right]
 \nonumber\\
 ={}&2\operatorname{tr}_{\mathbf N}
 \left[(D_\mu\lambda)^\alpha\lambda_\alpha\right].
 \label{eq:app-Ward-derivative-O}
\end{align}
Here $\nabla_\mu{\cal O}=\partial_\mu{\cal O}$ because ${\cal O}$
is both a Lorentz and gauge scalar.  To obtain the second line, move
$D_\mu\lambda_\alpha$ through
$\lambda_\beta$ in the second term and then interchange the dummy indices
$\alpha$ and $\beta$.  The Grassmann minus sign and the antisymmetry of
$\epsilon^{\alpha\beta}$ cancel.

Combining~\eqref{eq:app-Ward-descent-intermediate} and
\eqref{eq:app-Ward-derivative-O} gives
\begin{equation}
 \delta_{\bar\epsilon}\psi_{{\cal S}\alpha}
 =-2i\hat\sigma^a_{\alpha\dot\alpha}E_a{}^\mu
 \bar\epsilon^{\dot\alpha}\nabla_\mu{\cal O}.
 \label{eq:app-Ward-component-chiral-descent}
\end{equation}
Equivalently,
\begin{equation}
 \left\{
  \bar{\cal Q}_{\dot\alpha},
  \psi_{{\cal S}\alpha}
 \right\}
 =-2i\hat\sigma^a_{\alpha\dot\alpha}E_a{}^\mu
 \nabla_\mu{\cal O}.
 \label{eq:app-Ward-charge-chiral-descent}
\end{equation}
This is an off-shell operator identity.

Contracting~\eqref{eq:app-Ward-charge-chiral-descent} with
$e^a{}_{\mu}\bar{\hat\sigma}_a^{\dot\alpha\alpha}$ gives
\begin{equation}
 \nabla_\mu{\cal O}
 =\frac{i}{4}
 e^a{}_{\mu}\bar{\hat\sigma}_a^{\dot\alpha\alpha}
 \left\{
  \bar{\cal Q}_{\dot\alpha},
  \psi_{{\cal S}\alpha}
 \right\}.
 \label{eq:app-Ward-derivative-Q-exact}
\end{equation}

Let the boundary bra and the EH-prepared ket preserve both dotted charges:
\begin{equation}
 \langle B|\bar{\cal Q}_{\dot\alpha}=0,
 \qquad
 \bar{\cal Q}_{\dot\alpha}|\Psi\rangle=0.
 \label{eq:app-Ward-supersymmetric-states}
\end{equation}
If the functional measure and regulator preserve these charges and no
boundary or moduli-space surface term is generated, then
\begin{equation}
 \left\langle
  \left\{\bar{\cal Q}_{\dot\alpha},{\cal V}\right\}
 \right\rangle_B=0
 \label{eq:app-Ward-exact-insertion-zero}
\end{equation}
for every operator ${\cal V}$ in the domain of the Ward identity.

For mutually separated insertions of ${\cal O}$, every factor is
$\bar{\cal Q}$-closed.  Therefore,
\begin{align}
 \nabla_\mu^{(i)}
 \left\langle\prod_{j=1}^p{\cal O}(x_j)\right\rangle_B
 &=\frac{i}{4}
 e^a{}_{\mu}\bar{\hat\sigma}_a^{\dot\alpha\alpha}
 \left\langle
  \left\{
   \bar{\cal Q}_{\dot\alpha},
   \psi_{{\cal S}\alpha}(x_i)
   \prod_{j\ne i}{\cal O}(x_j)
  \right\}
 \right\rangle_B
 \nonumber\\
 &=0.
 \label{eq:app-Ward-position-independence}
\end{align}
Thus, separated chiral correlators are position independent even though EH
has no translation symmetry.  The result follows from the local descent and
the two dotted Ward identities, not from a translation generator.

For a scalar one-point function, the radial component of this coordinate
identity gives directly
\begin{equation}
 \partial_r
 \langle{\cal O}(r)\rangle_B=0,
 \qquad r>a.
 \label{eq:app-Ward-radial-independence}
\end{equation}
This statement does not imply independence of the EH parameter $a$, since
changing $a$ changes the background geometry.  It also does not identify a
renormalized coincident product $[\mathcal O(x)]^p$ with a separated
correlator unless possible contact terms have been fixed.

\subsection{Second descent and holomorphy}

The first anti-chiral descent is
\begin{equation}
 [\bar{\cal Q}_{\dot\alpha},\bar{\cal O}]
 =\bar\psi_{{\cal S}\dot\alpha}
 =\sqrt2\bar\chi_{{\cal S}\dot\alpha}.
 \label{eq:app-Ward-antichiral-first-descent}
\end{equation}
Equation~\eqref{eq:app-Ward-second-antichiral-variation} supplies the second
descent.  With the spinor convention~\eqref{eq:app-Ward-spinor-epsilon}, its
independent component may be written as
\begin{equation}
 \left\{
  \bar{\cal Q}_{\dot1},
  \bar\psi_{{\cal S}\dot2}
 \right\}=2\bar F_{\cal S}.
\end{equation}
Consequently,
\begin{equation}
 \bar F_{\cal S}
 =\frac12
 \left\{
  \bar{\cal Q}_{\dot1},
  [\bar{\cal Q}_{\dot2},\bar{\cal O}]
 \right\}.
 \label{eq:app-Ward-Fbar-double-Q}
\end{equation}

Consider the normalized correlator
\begin{equation}
 {\cal G}_B
 =\langle X\rangle_B,
 \qquad
 X=\prod_{i=1}^p{\cal O}(x_i).
 \label{eq:app-Ward-G-definition}
\end{equation}
We hold the coefficients defining the boundary bra fixed while differentiating
with respect to the coupling.  We also use holomorphic field normalization,
so that the composite $X$ has no explicit $\bar\tau$ dependence; otherwise
an additional term $\langle\partial_{\bar\tau}X\rangle_B$ must be retained.
For a normalized path integral,
\begin{align}
 \frac{\partial{\cal G}_B}{\partial\bar\tau}
 ={}&-\left\langle
  X\frac{\partial S}{\partial\bar\tau}
 \right\rangle_B
 +\langle X\rangle_B
 \left\langle\frac{\partial S}{\partial\bar\tau}\right\rangle_B.
 \label{eq:app-Ward-normalized-derivative}
\end{align}
This follows by differentiating both the numerator and denominator of
\begin{equation}
 \langle X\rangle_B
 =\frac{\int_B{\cal D}\Phi\,X e^{-S}}
 {\int_B{\cal D}\Phi\,e^{-S}}.
\end{equation}

From~\eqref{eq:app-Ward-action-as-F-terms},
\begin{equation}
 \frac{\partial S}{\partial\bar\tau}
 =-\frac1{8\pi i}
 \int_{\rm EH}d^4x\sqrt g\,\bar F_{\cal S}
 +\frac{\partial S_{\rm surf}}{\partial\bar\tau}.
 \label{eq:app-Ward-S-bartau}
\end{equation}
Define
\begin{equation}
 \overline{\cal V}
 =\int_{\rm EH}d^4x\sqrt g\,\bar{\cal O}.
 \label{eq:app-Ward-Vbar-definition}
\end{equation}
Using~\eqref{eq:app-Ward-Fbar-double-Q}, one obtains
\begin{equation}
 \frac{\partial S}{\partial\bar\tau}
 =-\frac1{16\pi i}
 \left\{
  \bar{\cal Q}_{\dot1},
  [\bar{\cal Q}_{\dot2},\overline{\cal V}]
 \right\}
 +\frac{\partial S_{\rm surf}}{\partial\bar\tau}.
 \label{eq:app-Ward-S-bartau-exact}
\end{equation}

Assume that the last surface term vanishes.  Since every factor in $X$ is
$\bar{\cal Q}$-closed,
\begin{align}
 \left\langle
  X\frac{\partial S}{\partial\bar\tau}
 \right\rangle_B
 &=-\frac1{16\pi i}
 \left\langle
  \left\{
   \bar{\cal Q}_{\dot1},
   X[\bar{\cal Q}_{\dot2},\overline{\cal V}]
  \right\}
 \right\rangle_B
 \nonumber\\
 &=0.
 \label{eq:app-Ward-X-Sbar-zero}
\end{align}
The same Ward identity with $X=1$ gives
\begin{equation}
 \left\langle\frac{\partial S}{\partial\bar\tau}\right\rangle_B=0.
 \label{eq:app-Ward-Sbar-zero}
\end{equation}
Both terms in~\eqref{eq:app-Ward-normalized-derivative} therefore vanish:
\begin{equation}
 \frac{\partial{\cal G}_B}{\partial\bar\tau}=0.
 \label{eq:app-Ward-holomorphy}
\end{equation}

In terms of the holomorphic scale
\begin{equation}
 \Lambda_{\rm h}^{3N}
 =\mu^{3N}e^{2\pi i\tau},
\end{equation}
equation~\eqref{eq:app-Ward-holomorphy} is equivalently
\begin{equation}
 \frac{\partial{\cal G}_B}{\partial\bar\Lambda_{\rm h}}=0.
\end{equation}
Fractional powers are locally holomorphic after choosing the branch
associated with the corresponding fractional topological-charge sector.

\subsection{Dependence on the Eguchi--Hanson resolution parameter}
\label{app:EH-a-independence}

We now examine whether the supersymmetric Ward identities constrain the
dependence of a chiral correlator on the EH resolution parameter $a$. This is
logically distinct from position independence: differentiating with respect
to $a$ inserts the stress-energy tensor, rather than a derivative of a local
operator. We first identify the corresponding smooth metric deformation and
then determine its supersymmetry transformation using the supercurrent
multiplet.

The coordinate range $r\geq a$ in \eqref{EH_metric} depends on $a$, so
$\partial_a g_{\mu\nu}$ at fixed $r$ is not a smooth comparison of the
metrics at their bolts. To identify all metrics with $a>0$ on the same
manifold, introduce a radial coordinate $\varrho$ of fixed range by
\begin{equation}
 r^4=\varrho^4+a^4,
 \qquad
 0\leq\varrho<\infty,
 \qquad
 R(\varrho,a)\equiv(\varrho^4+a^4)^{1/4}.
 \label{eq:app-Ward-fixed-EH-radial-coordinate}
\end{equation}
Here $\varrho$ is the coordinate, while $R(\varrho,a)$ is only a metric
function. In terms of these quantities, the EH metric and an orthonormal
frame are
\begin{align}
 ds^2_{\mathrm{EH}}
 &={\varrho^2\over R^2}\,d\varrho^2
 +R^2(\sigma_x^2+\sigma_y^2)
 +{\varrho^4\over R^2}\sigma_z^2,
 \label{eq:app-Ward-fixed-EH-metric}\\
 e^0&={\varrho\over R}d\varrho,\quad
 e^1=R\sigma_x,\quad
 e^2=R\sigma_y,\quad
 e^3={\varrho^2\over R}\sigma_z.
 \label{eq:app-Ward-fixed-EH-frame}
\end{align}
The bolt is at $\varrho=0$ for every $a$. Since
\begin{equation}
 \left.\frac{\partial R}{\partial a}\right|_{\varrho}
 ={a^3\over R^3},
\end{equation}
the deformation
\begin{equation}
 h_{\mu\nu}
 \equiv
 \left.\frac{\partial g_{\mu\nu}}{\partial a}\right|_{\varrho}
 \label{eq:app-Ward-a-deformation-definition}
\end{equation}
has orthonormal-frame components
\begin{equation}
 h_{ab}
 ={2a^3\over R^4}
 \operatorname{diag}(-1,1,1,-1).
 \label{eq:app-Ward-a-deformation-frame}
\end{equation}
It is therefore traceless. It is also transverse. To verify the nontrivial
radial component explicitly, set
\begin{equation}
 A={\varrho\over R},
 \qquad B=R,
 \qquad C={\varrho^2\over R},
 \qquad q={2a^3\over R^4},
\end{equation}
so that $h^a{}_b=\operatorname{diag}(-q,q,q,-q)$. Direct evaluation gives
\begin{align}
 \nabla_\nu h^\nu{}_{\varrho}
 &={d(-q)\over d\varrho}
 +2{B'\over B}(-q-q)
 +{C'\over C}(-q+q)
 \nonumber\\
 &={8a^3\varrho^3\over R^8}
 -{8a^3\varrho^3\over R^8}=0.
 \label{eq:app-Ward-a-deformation-radial-divergence}
\end{align}
The angular components vanish by the diagonal $SU(2)$-invariant form of the
deformation. Hence
\begin{equation}
  \nabla_\nu h^{\mu\nu}=0,
  \qquad
  g^{\mu\nu}h_{\mu\nu}=0.
 \label{eq:app-Ward-a-deformation-TT}
\end{equation}
Moreover, $h_{ab}=O(a^3/r^4)$ at infinity and is normalizable. Indeed,
$h_{ab}h^{ab}=16a^6/R^8$, while the volume form is proportional to
$\varrho^3d\varrho\wedge\sigma_x\wedge\sigma_y\wedge\sigma_z$, so the
radial norm is finite. Thus varying $a$ produces a smooth normalizable
transverse-traceless deformation and does not change the leading boundary
metric on $\mathbb{RP}^3$.

We next use the local supercurrent multiplet. Every four-dimensional
${\cal N}=1$ theory possesses an ${\cal S}$-multiplet containing the
supersymmetry current $S_{\mu\alpha}$ and the symmetric conserved
stress-energy tensor $T_{\mu\nu}$. Classically, pure SYM may be improved to
the $R$-multiplet. This use of the $R$-multiplet is an operator statement and
does not require turning on a background $U(1)_R$ connection on EH. In the
Euclidean normalization used here, its relevant local transformation is \cite{Dumitrescu:2011iu}
\begin{equation}
 \left\{
  \bar{\cal Q}_{\dot\alpha},S_{\mu\alpha}
 \right\}
 =2i\hat\sigma^\nu_{\alpha\dot\alpha}
 \widetilde{\cal T}_{\mu\nu},
 \label{eq:app-Ward-supercurrent-stress-transformation}
\end{equation}
where\footnote{The absence of a globally defined undotted supercharge on EH does not imply that the local operator $S_{\mu\alpha}$ is absent.  Just as the stress tensor $T_{\mu\nu}$ is defined by varying the action with respect to the background metric even when the manifold admits no translations, the supercurrent is defined by coupling the theory to a background gravitino and setting the latter to zero after differentiation:
$
S^\mu{}_\alpha=(1/\sqrt g)\,\delta S_{\rm coupled}/\delta\psi_\mu^\alpha|_{\psi=\bar\psi=0}.
$
A global undotted charge would additionally require a parallel undotted spinor, since
$
Q[\epsilon]=\int_\Sigma d\Sigma_\mu\,\epsilon^\alpha S^\mu{}_\alpha
$
is conserved only when $\nabla_\mu\epsilon^\alpha=0$.  No such spinor exists on EH, so $S_{\mu\alpha}$ remains a globally defined local vector--spinor operator but does not generate a global charge $Q_\alpha$.  The relation $\{\bar{\cal Q}_A,S_{\mu\alpha}\}\sim\bar\epsilon_A^{\dot\alpha}\hat\sigma^\nu_{\alpha\dot\alpha}T_{\mu\nu}+\cdots$ therefore uses only the preserved dotted charge $\bar{\cal Q}_A$; it is a local descent relation, not the global flat-space algebra $\{\bar Q_{\dot\alpha},Q_\alpha\}\sim P_{\alpha\dot\alpha}$.}
\begin{align}
 \widetilde{\cal T}_{\mu\nu}
 ={}&T_{\mu\nu}
 +{i\over4}\varepsilon_{\mu\nu\rho\lambda}{\cal F}^{\rho\lambda}
 -{i\over4}\varepsilon_{\mu\nu\rho\lambda}
   \nabla^\rho j^{(R)\lambda}
 +{i\over2}\nabla_\nu j^{(R)}_\mu.
 \label{eq:app-Ward-improved-stress-tensor}
\end{align}
Here,  ${\cal F}_{\mu\nu}$ (not to be confused with the Yang--Mills field strength) is
the closed two-form in the same multiplet.   In pure super-Yang--Mills
theory, the ${\cal S}$-multiplet can be improved to the
Ferrara--Zumino multiplet, for which 
\begin{equation}
 {\cal F}_{\mu\nu}=0.
\end{equation}

The supercurrent is
\begin{equation}
 S_{\mu\alpha}
 =
 -\frac{i}{g^2}
 \operatorname{tr}_{\mathbf N}
 \left[
  F_{ab}
  \left(
   \hat\sigma^{ab}\hat\sigma_\mu\bar\lambda
  \right)_\alpha
 \right].
 \label{eq:explicit-SYM-supercurrent}
\end{equation}
We also defined the stress-energy tensor by
\begin{equation}
 \delta_g S_{\rm SYM}
 =
 -\frac12
 \int d^4x\,\sqrt g\,
 T^{\mu\nu}\delta g_{\mu\nu}.
 \label{eq:SYM-stress-tensor-definition}
\end{equation}
The corresponding symmetric stress tensor is
\begin{align}
 T_{\mu\nu}
 ={}&
 \frac{2}{g^2}
 \operatorname{tr}_{\mathbf N}
 \left[
  F_{\mu\rho}F_\nu{}^\rho
  -\frac14g_{\mu\nu}F_{\rho\sigma}F^{\rho\sigma}
 \right]
 \nonumber\\
 &+
 \frac{i}{g^2}
 \operatorname{tr}_{\mathbf N}
 \left[
  \bar\lambda\bar{\hat\sigma}_{(\mu}
  \overleftrightarrow D_{\nu)}\lambda
  -
  g_{\mu\nu}
  \bar\lambda\bar{\hat\sigma}^{\rho}
  \overleftrightarrow D_{\rho}\lambda
 \right],
 \label{eq:explicit-SYM-stress-tensor}
\end{align}
where
\begin{equation}
 \bar\lambda\bar{\hat\sigma}_\mu
 \overleftrightarrow D_\nu\lambda
 \equiv
 \bar\lambda\bar{\hat\sigma}_\mu D_\nu\lambda
 -
 (D_\nu\bar\lambda)\bar{\hat\sigma}_\mu\lambda,
\end{equation}
and symmetrization has unit weight:
\begin{equation}
 X_{(\mu}Y_{\nu)}
 \equiv
 \frac12
 \left(
  X_\mu Y_\nu+X_\nu Y_\mu
 \right).
\end{equation}
The term in the second line of
\eqref{eq:explicit-SYM-stress-tensor} proportional to $g_{\mu\nu}$
vanishes upon imposing the gaugino equations of motion and, independently,
drops out when the stress tensor is contracted with a traceless metric
deformation.

Finally, the classical $U(1)_R$ current is
\begin{equation}
 j_\mu^{(R)}
 =
 \frac{2}{g^2}
 \operatorname{tr}_{\mathbf N}
 \left(
  \lambda\hat\sigma_\mu\bar\lambda
 \right).
 \label{eq:explicit-SYM-R-current}
\end{equation}
Using the classical gaugino equations of motion,
\begin{equation}
 \bar{\hat\sigma}^{\mu}D_\mu\lambda=0,
 \qquad
 (D_\mu\bar\lambda)\bar{\hat\sigma}^{\mu}=0,
\end{equation}
one finds
\begin{equation}
 \nabla^\mu j_\mu^{(R)}=0.
\end{equation}
Quantum mechanically, $j_\mu^{(R)}$ is anomalous and these expressions
must be understood as components of the renormalized
Ferrara--Zumino multiplet.

Contracting \eqref{eq:app-Ward-supercurrent-stress-transformation} with
$\bar{\hat\sigma}_\rho^{\dot\alpha\alpha}$ and using
$\bar{\hat\sigma}_\rho^{\dot\alpha\alpha}
\hat\sigma^\nu_{\alpha\dot\alpha}=2\delta_\rho{}^\nu$ gives
\begin{equation}
 \widetilde{\cal T}_{\mu\rho}
 ={1\over4i}\bar{\hat\sigma}_\rho^{\dot\alpha\alpha}
 \left\{\bar{\cal Q}_{\dot\alpha},S_{\mu\alpha}\right\}.
 \label{eq:app-Ward-inverted-supercurrent-stress}
\end{equation}
Define
\begin{equation}
 {\cal V}_a^{\dot\alpha}
 ={1\over8i}
 h^{\mu\rho}\bar{\hat\sigma}_\rho^{\dot\alpha\alpha}
 S_{\mu\alpha}.
 \label{eq:app-Ward-Va-definition}
\end{equation}
The symmetry of $h^{\mu\nu}$ eliminates the two terms in
\eqref{eq:app-Ward-improved-stress-tensor} that contain
$\varepsilon_{\mu\nu\rho\lambda}$. Using in addition
\eqref{eq:app-Ward-a-deformation-TT}, we obtain
\begin{align}
 {1\over2}h^{\mu\nu}T_{\mu\nu}
 &=
 \left\{
  \bar{\cal Q}_{\dot\alpha},
  {\cal V}_a^{\dot\alpha}
 \right\}
 -{i\over4}\nabla_\nu
  \left(h^{\mu\nu}j^{(R)}_\mu\right).
 \label{eq:app-Ward-a-local-exactness}
\end{align}
Thus the contraction of the stress tensor with the EH resolution modulus is
supersymmetry-exact up to a total derivative. A sum over the two dotted
charges is understood.

We now apply this identity to the normalized condensate
\begin{equation}
 {\cal C}_B(a)
 =\left\langle{\cal O}(x)\right\rangle_{B},
 \qquad
 {\cal O}=\operatorname{tr}_{\mathbf N}(\lambda\lambda).
\end{equation}
The operator ${\cal O}$ contains no curved vector index and has no explicit
classical $a$-dependence under the spin-bundle identification defined by the
parallel frame. Differentiating the normalized path integral therefore
gives
\begin{equation}
 {\partial{\cal C}_B\over\partial a}
 =-{1\over2}\int_{\rm EH}d^4y\sqrt g\,
 h_{\mu\nu}(y)
 \left\langle T^{\mu\nu}(y){\cal O}(x)\right\rangle_{B,\mathrm{conn}}.
 \label{eq:app-Ward-condensate-a-derivative}
\end{equation}
Here the subscript ``$\mathrm{conn}$'' denotes the connected insertion
with respect to the stress tensor,
\begin{equation}
 \left\langle T^{\mu\nu}(y)X\right\rangle_{B,\mathrm{conn}}
 =
 \left\langle T^{\mu\nu}(y)X\right\rangle_B
 -
 \left\langle T^{\mu\nu}(y)\right\rangle_B
 \left\langle X\right\rangle_B,
 \qquad
 X=\prod_i{\cal O}_i(x_i).
\end{equation}
Substituting \eqref{eq:app-Ward-a-local-exactness}, and using
$[\bar{\cal Q}_{\dot\alpha},{\cal O}]=0$, the bulk term becomes the
expectation value of
$\{\bar{\cal Q}_{\dot\alpha},
  {\cal V}_a^{\dot\alpha}{\cal O}\}$. It vanishes by
\eqref{eq:app-Ward-exact-insertion-zero}. Consequently,
\begin{align}
 {\partial{\cal C}_B\over\partial a}
 ={}&{i\over4}
 \lim_{R_\infty\to\infty}
 \int_{S^3_{R_\infty}/\mathbb Z_2}
 d\Sigma_\nu\,
 h^{\mu\nu}
 \left\langle j^{(R)}_\mu(y){\cal O}(x)\right\rangle_{B,\mathrm{conn}}
 \nonumber\\
 &\quad+
 \text{contact, regulator, and moduli-space surface terms}.
 \label{eq:app-Ward-condensate-a-surface}
\end{align}
There is no bolt term, because the bolt is a smooth interior two-sphere and
the deformation \eqref{eq:app-Ward-a-deformation-frame} is regular there.
Since $h_{ab}=O(a^3/r^4)$ and the area of the asymptotic lens space grows as
$r^3$, the displayed surface integral vanishes whenever the current matrix
element is bounded or decays at infinity. Under the supersymmetric
asymptotic conditions assumed in this appendix, and in the absence of the
other terms displayed in \eqref{eq:app-Ward-condensate-a-surface}, one finds
\begin{equation}
 {\partial\over\partial a}
 \left\langle
  \operatorname{tr}_{\mathbf N}(\lambda\lambda)(x)
 \right\rangle_B=0.
 \label{eq:app-Ward-condensate-a-independence}
\end{equation}

The status of this conclusion should be kept explicit. The local bulk
identity \eqref{eq:app-Ward-a-local-exactness} is a classical operator
identity. In the quantum theory, the anomalous continuous $U(1)_R$ current
must be replaced by the Ferrara--Zumino or universal ${\cal S}$-multiplet.
The supersymmetry current remains conserved, and the anomaly modifies trace,
improvement, and contact operators; the transverse-traceless projection in
\eqref{eq:app-Ward-a-deformation-TT} removes the bulk trace and longitudinal
pieces. Nevertheless, a regulated Ward-identity analysis is required to
exclude anomalous contact or boundary terms. Likewise, in a semiclassical
collective-coordinate representation, a supersymmetry-exact term may reduce
to a surface contribution at the noncompact large-size boundary of moduli
space. Equation \eqref{eq:app-Ward-condensate-a-independence} therefore
constrains the complete supersymmetric boundary-sector amplitude, subject to
the stated Ward-identity hypotheses; it need not hold for an incomplete
single-saddle or zero-mode contribution. In particular, residual
$a$-dependence in such a contribution must be canceled by the missing
determinants, additional saddles, or a nonvanishing boundary or moduli-space
surface term.

\section{Gauge-field zero modes about self-dual background}
\label{sec:gauge-zero-modes}

In this appendix, we study infinitesimal gauge-field fluctuations about a self-dual
background $A_\mu^{\rm B}$.  The normalizable solutions of the
linearized self-duality equations determine the tangent space to the
instanton moduli space and therefore enter the semiclassical collective
coordinate measure.

We decompose the gauge potential as
\begin{equation}
 A_\mu=A_\mu^{\rm B}+\delta A_\mu.
\end{equation}
To linear order in $\delta A_\mu$, the fluctuation of the field
strength is
\begin{equation}
 \delta F_{\mu\nu}
 =
 D_\mu^{\rm B}\delta A_\nu
 -D_\nu^{\rm B}\delta A_\mu,
 \label{eq:linearized-field-strength}
\end{equation}
where
\begin{equation}
 D_\mu^{\rm B}\delta A_\nu
 \equiv
 \nabla_\mu\delta A_\nu
 +i[A_\mu^{\rm B},\delta A_\nu].
 \label{eq:background-covariant-vector-derivative}
\end{equation}
Since the background is self-dual, preservation of self-duality at
linear order requires the anti-self-dual part of $\delta F$ to vanish.
With the orientation and sigma-matrix conventions of
Appendix~\ref{Mathematical preliminaries}, this condition is
\begin{equation}
 \bar{\hat\sigma}^{ab}
 E_a^\mu E_b^\nu
 \delta F_{\mu\nu}
 =
 0.
 \label{eq:linearized-self-duality-spinor}
\end{equation}
Using the antisymmetry of $\bar{\hat\sigma}^{ab}$, this is equivalently
\begin{equation}
 \bar{\hat\sigma}^{ab}
 E_a^\mu E_b^\nu
 D_\mu^{\rm B}\delta A_\nu
 =
 0.
 \label{shortCSfirst}
\end{equation}

We supplement the linearized self-duality equation by the background
gauge condition
\begin{equation}
 D_\mu^{\rm B}\delta A^\mu=0.
 \label{eq:background-gauge-condition}
\end{equation}
Subject to the appropriate falloff conditions, this makes the
fluctuations orthogonal to infinitesimal gauge transformations.  It
fixes the gauge up to residual transformations whose parameters obey
the background covariant Laplace equation.

The identity
\begin{equation}
 \bar{\hat\sigma}^a\hat\sigma^b
 =
 \delta^{ab}I_2+\bar{\hat\sigma}^{ab}
\end{equation}
allows us to combine
\eqref{shortCSfirst} and
\eqref{eq:background-gauge-condition} into the single first-order
equation
\begin{equation}
 \bar{\hat\sigma}^a\hat\sigma^b
 E_a^\mu E_b^\nu
 D_\mu^{\rm B}\delta A_\nu
 =
 0.
 \label{shortCS}
\end{equation}

We next rewrite this equation in tangent-space components.  Writing
\begin{equation}
 \delta A_\nu=e^c{}_\nu\delta A_c,
\end{equation}
Eq.~\eqref{alterA} gives
\begin{equation}
 \nabla_\mu\delta A_\nu
 =
 e^c{}_\nu
 \left(
 \partial_\mu\delta A_c
 +\omega_{\mu c}{}^b\delta A_b
 \right).
 \label{eq:gauge-fluctuation-tangent-derivative}
\end{equation}
Substituting this relation into \eqref{shortCS} and using
$E_b^\nu e^c{}_\nu=\delta_b^c$, we obtain
\begin{equation}
 \bar{\hat\sigma}^a\hat\sigma^c E_a^\mu
 \left[
 \left(
 \partial_\mu+i[A_\mu^{\rm B},\,]
 \right)\delta A_c
 -\omega_{\mu bc}\delta A_b
 \right]
 =
 0.
 \label{almost there}
\end{equation}

Introduce the bispinor representation of the tangent-space
fluctuation,
\begin{equation}
 \delta A_{\beta\dot\alpha}
 \equiv
 \left[\hat\sigma^c\right]_{\beta\dot\alpha}\delta A_c,
 \qquad
 \delta A_c
 =
 \frac12
 \left[\bar{\hat\sigma}_c\right]^{\dot\alpha\beta}
 \delta A_{\beta\dot\alpha}.
 \label{eq:gauge-vector-bispinor-map}
\end{equation}
The identity \eqref{eq:spin-connection-sigma-identity} then gives the
general spinorial form of \eqref{almost there}:
\begin{equation}
 \bar{\hat\sigma}^aE_a^\mu
 \left[
 \left(
 \partial_\mu+i[A_\mu^{\rm B},\,]
 \right)\delta A
 +\frac12\Xi_\mu\delta A
 -\frac12\delta A\,\bar\Xi_\mu
 \right]
 =
 0.
 \label{eq:general-bosonic-deformation-spinor}
\end{equation}
Here $\Xi_\mu$ acts on the undotted index from the left, whereas
$\bar\Xi_\mu$ acts on the dotted index from the right.

For the EH spin connection, internal self-duality implies
\begin{equation}
 \bar\Xi_\mu=0.
\end{equation}
Consequently, the fluctuation equation reduces to
\begin{equation}
 \left[\bar{\hat\sigma}^{a}\right]^{\dot\beta\beta}
 E_a^\mu
 \left[
 \left(
 \partial_\mu+i[A_\mu^{\rm B},\,]
 \right)
 \delta A_{\beta\dot\alpha}
 +
 \frac12
 \Xi_{\mu\beta}{}^\gamma
 \delta A_{\gamma\dot\alpha}
 \right]
 =
 0.
 \label{bosonicmodeseq}
\end{equation}
For every fixed value of the dotted index $\dot\alpha$, this is
precisely the same Weyl equation obeyed by an adjoint fermion zero-mode
wavefunction:
\begin{equation}
 \left[\bar{\hat\sigma}^{a}\right]^{\dot\beta\beta}
 E_a^\mu
 \left[
 \left(
 \partial_\mu+i[A_\mu^{\rm B},\,]
 \right)\lambda_\beta
 +
 \frac12\Xi_{\mu\beta}{}^\gamma\lambda_\gamma
 \right]
 =
 0.
 \label{eq:adjoint-Weyl-equation-comparison}
\end{equation}

Let $\lambda_\beta^{(A)}$, with
$A=1,\ldots,{\cal I}_{\rm adj}$, denote a basis of normalizable
commuting wavefunctions underlying the fermion zero modes.  The
Grassmann collective coordinates multiplying these wavefunctions are
not included in $\lambda_\beta^{(A)}$.  Since $\bar\Xi_\mu=0$, the
dotted spin bundle admits two constant basis spinors
$\eta_{\dot\alpha}^{(s)}$, $s=1,2$.  The complexified bosonic
deformation equation therefore has solutions
\begin{equation}
 \delta A_{\beta\dot\alpha}^{(A,s)}
 =
 \lambda_\beta^{(A)}
 \eta_{\dot\alpha}^{(s)}.
 \label{eq:complex-bosonic-modes-from-fermions}
\end{equation}

The free dotted index should not be interpreted as producing two
independent complex physical gauge fields for every fermion mode.  The
physical gauge fluctuation must satisfy the Hermiticity condition
\begin{equation}
 \delta A_\mu^\dagger=\delta A_\mu.
 \label{eq:gauge-fluctuation-Hermiticity}
\end{equation}
The Hermiticity condition relates the two dotted columns and reduces
the complexified solutions to the correct real tangent space of the
instanton moduli space.  Provided the opposite-chirality kernel and
the obstruction space vanish, as they do for the minimal EH
background considered here,
\begin{equation}
 \dim_{\mathbb R}{\cal M}_{\rm bos}
 =
 2\,\dim_{\mathbb C}\ker\slashed D_{\rm adj}
 =
 2{\cal I}_{\rm adj}.
 \label{eq:bosonic-fermionic-zero-mode-counting}
\end{equation}
Thus, for ${\cal I}_{\rm adj}=2$, there are four real bosonic zero
modes.

\section{The bilinear condensate in $SU(2)$}
\label{app:SU2-condensate-details}

In this appendix, we work out the details needed to compute the bilinear gaugino condensate in the $SU(2)$ case. The instanton calculus methods used here are reviewed in \cite{Dorey:2002ik,Vandoren:2008xg}. 
\subsection{Adjoint fermion zero modes}
\label{app:EH-SU2-gaugino-zero-modes}

This section derives the adjoint gaugino zero modes used in
Sec.~\ref{sec:SYM-condensate}.  We work in the $Q_{(2)}=1/2$ background
\eqref{su2solsd}.

\subsection*{Bolt boundary condition and APS index}

At the bolt, the nonabelian background obeys $G(a)=2$.  Along the circle
$\psi\to\psi+2\pi$ at fixed $(r,\theta,\varphi)$, its gauge holonomy is
\begin{equation}
 U_a
 =\exp\left(i\int_0^{2\pi}d\psi\,A_\psi^{\rm B}\right)
 =\exp(i\pi\tau_3)=-I_2.
 \label{eq:app-SU2-bolt-holonomy}
\end{equation}
The holonomy is central and therefore acts trivially on the adjoint gaugino:
\begin{equation}
 \lambda(\psi+2\pi)=U_a^\dagger\lambda(\psi)U_a=\lambda(\psi).
 \label{eq:app-SU2-adjoint-periodicity}
\end{equation}

For the adjoint representation, the APS index in the present sector is
\begin{align}
 \operatorname{ind}{\cal D}_{\rm adj}
 &=2T_{\rm adj}Q_{(2)}
 +\frac{-\dim(\operatorname{adj})+\mbox{sgn}(\operatorname{Hol})}{8}
 \nonumber\\
 &=4\left(\frac12\right)+\frac{-3+3}{8}=2.
 \label{eq:app-SU2-APS-index}
\end{align}
Here $T_{\rm adj}=2$ in the convention
$\operatorname{tr}_{\mathbf 2}(T_aT_b)=\delta_{ab}/2$.  The boundary contribution
vanishes because the adjoint holonomy is trivial.

\subsection*{Separated Weyl equations}

Introduce
\begin{equation}
 \sigma_\pm=\sigma_x\pm i\sigma_y,
 \qquad
 \tau_\pm=\frac{\tau_1\pm i\tau_2}{2},
\end{equation}
and expand the adjoint Weyl spinor as
\begin{equation}
 \lambda
 =\lambda_H\frac{\tau_3}{2}
 +\lambda_+\frac{\tau_+}{2}
 +\lambda_-\frac{\tau_-}{2}.
 \label{eq:app-SU2-Cartan-Weyl-expansion}
\end{equation}
For the background \eqref{generalansatzB}, direct evaluation of the
commutator gives
\begin{align}
 [A^{\rm B},\lambda]
 ={}&\left(G\sigma_z\lambda_+-F\sigma_-\lambda_H\right)
       \frac{\tau_+}{2}
 +\left(-G\sigma_z\lambda_-+F\sigma_+\lambda_H\right)
       \frac{\tau_-}{2}
 \nonumber\\
 &+\frac{F}{2}\left(\sigma_-\lambda_--\sigma_+\lambda_+\right)
       \frac{\tau_3}{2}.
 \label{eq:app-SU2-background-commutator}
\end{align}

Let $\lambda_{1A}$ and $\lambda_{2A}$, with
$A\in\{H,+,-\}$, denote the two Weyl components, and define
\begin{equation}
 {\cal D}_r
 \equiv\frac1f\frac{d}{dr}+\frac{2f}{r}+\frac{1}{fr}.
 \label{eq:app-SU2-radial-operator}
\end{equation}
The Weyl equation following from \eqref{diraceqlambsusy} then becomes
\begin{align}
 0={}&\left({\cal D}_r+\frac{2f}{r}\hat{\cal J}_z
 +\frac{fG}{r}\right)\lambda_{1+}
 -\frac{2}{r}\hat{\cal J}_-\lambda_{2+},
 \nonumber\\
 0={}&-\frac{2}{r}\hat{\cal J}_+\lambda_{1+}
 -\frac{2F}{r}\lambda_{1H}
 +\left({\cal D}_r-\frac{2f}{r}\hat{\cal J}_z
 -\frac{fG}{r}\right)\lambda_{2+},
 \label{eq:app-SU2-Weyl-plus}
\end{align}
\begin{align}
 0={}&\left({\cal D}_r+\frac{2f}{r}\hat{\cal J}_z
 -\frac{fG}{r}\right)\lambda_{1-}
 -\frac{2}{r}\hat{\cal J}_-\lambda_{2-}
 +\frac{2F}{r}\lambda_{2H},
 \nonumber\\
 0={}&-\frac{2}{r}\hat{\cal J}_+\lambda_{1-}
 +\left({\cal D}_r-\frac{2f}{r}\hat{\cal J}_z
 +\frac{fG}{r}\right)\lambda_{2-},
 \label{eq:app-SU2-Weyl-minus}
\end{align}
and
\begin{align}
 0={}&\left({\cal D}_r+\frac{2f}{r}\hat{\cal J}_z\right)\lambda_{1H}
 -\frac{2}{r}\hat{\cal J}_-\lambda_{2H}
 -\frac{F}{r}\lambda_{2+},
 \nonumber\\
 0={}&-\frac{2}{r}\hat{\cal J}_+\lambda_{1H}
 +\left({\cal D}_r-\frac{2f}{r}\hat{\cal J}_z\right)\lambda_{2H}
 +\frac{F}{r}\lambda_{1-}.
 \label{eq:app-SU2-Weyl-Cartan}
\end{align}
Because the holonomy of the background is central, $h_c=-I_2$, each component of the fermions $\lambda_H$ and $\lambda_{\pm}$ satisfies the periodic boundary conditions as we encircle the bolt, that is, $\lambda_{H}(\psi+2\pi)=\lambda_H(\psi)$ and $\lambda_{\pm}(\psi+2\pi)=\lambda_{\pm}(\psi)$.
The angular momentum operators $\hat{\cal J}_z$ and $\hat{\cal J}_\pm$ act on the Wigner
functions $\lvert j,m',m\rangle$ \cite{Sakurai:2011zz} on the $S^3/\mathbb Z_2$ orbits.  A separated ansatz is
\begin{align}
 \lambda_{1A}
 &=g_{1A}(r)\lvert j,m'=j,m\rangle,
 \nonumber\\
 \lambda_{2A}
 &=g_{2A}(r)\lvert j,m'=-j,m\rangle,
 \qquad |m|\leq j.
 \label{su2intgen}
\end{align}
Regularity at the bolt and square integrability at infinity exclude all
$j>0$ solutions in the $Q_{(2)}=1/2$ sector.  For $j=0$, the angular-momentum
operators in \eqref{eq:app-SU2-Weyl-plus}--
\eqref{eq:app-SU2-Weyl-Cartan} vanish, and the remaining radial system has two
independent solutions.

Writing their Grassmann coefficients as $\eta_1$ and $\eta_2$, the result is
\begin{align}
 \lambda_1
 &=\mathcal F_3\frac{\tau_3}{2}\eta_1
 +\mathcal F_{\pm}\frac{\tau_-}{2}\eta_2,
 \nonumber\\
 \lambda_2
 &=-\mathcal F_3\frac{\tau_3}{2}\eta_2
 +\mathcal F_{\pm}\frac{\tau_+}{2}\eta_1,
\end{align}
where
\begin{equation}
 \mathcal F_3
 =\frac{\sqrt{(t^4-a^4)(r^4-a^4)}}{r(r^2+t^2)^2},
 \qquad
 \mathcal F_{\pm}
 =\frac{2(t^2r^2+a^4)}{r(r^2+t^2)^2}.
 \label{eq:app-SU2-final-radial-sizes}
\end{equation}
Direct substitution verifies the radial equations.  At the bolt,
$\mathcal F_3=O(\sqrt{r-a})$ and $\mathcal F_{\pm}$ is finite, while both
profiles fall as $r^{-3}$ at infinity.  They are therefore regular and
square integrable. 

Define
\begin{equation}
 \rho_\lambda(r,t)
 \equiv2\mathcal F_3^2(r,t)+\mathcal F_{\pm}^2(r,t).
 \label{eq:app-SU2-zero-mode-density}
\end{equation}
Using
$\int_{S^3/\mathbb Z_2}\sigma_x\wedge\sigma_y\wedge\sigma_z=-\pi^2$,
one finds the radial normalization identity
\begin{equation}
 \int_{\rm EH}d^4x\sqrt g\,\rho_\lambda(r,t)
 =\pi^2\int_a^\infty dr\,r^3\rho_\lambda(r,t)
 =\pi^2t^2.
 \label{eq:app-SU2-radial-normalization}
\end{equation}
Finally, the generator algebra gives
\begin{equation}
 \operatorname{tr}_{\mathbf 2}(\lambda\lambda)
 =-2\operatorname{tr}_{\mathbf 2}(\lambda_1\lambda_2)
 =\frac12\rho_\lambda(r,t)\eta_1\eta_2,
 \label{eq:app-SU2-bilinear}
\end{equation}
as quoted in the main text.  Together with the index
\eqref{eq:app-SU2-APS-index}, these two explicit profiles exhaust the
$L^2$ kernel.

\subsection{Bosonic zero modes of the $Q_{(2)}=\frac12$ instanton}
\label{sec:bosonic-zero-modes-Q-half}

The minimal $SU(2)$ instanton on EH has adjoint Dirac index
${\cal I}_{\rm adj}=2$.  As shown in
Sec.~\ref{sec:gauge-zero-modes}, the corresponding self-dual deformation
complex has four real bosonic zero modes.  These consist of one scale
modulus and three global color-orientation moduli.  In this section, we
construct these modes explicitly and determine the induced metric and
measure on the bosonic moduli space.

We recall the radial functions appearing in the fermion and bosonic
zero modes:
\begin{equation}
 {\cal F}_3(r,t)
 =
 \frac{
 \sqrt{(t^4-a^4)(r^4-a^4)}
 }{
 r(r^2+t^2)^2
 },
 \qquad
 {\cal F}_{\pm}(r,t)
 =
 \frac{
 2(t^2r^2+a^4)
 }{
 r(r^2+t^2)^2
 }.
 \label{eq:bosonic-mode-radial-functions}
\end{equation}
They obey the useful normalization identity
\begin{equation}
 \int_{\rm EH}d^4x\sqrt{g_{\rm EH}}\,
 \left(
 {\cal F}_{\pm}^2+2{\cal F}_3^2
 \right)
 =
 \pi^2t^2.
 \label{eq:bosonic-radial-normalization-integral}
\end{equation}

\paragraph{The scale zero mode.}

The parameter $t\in[a,\infty)$ appearing in the nonabelian solution is
the radial-size modulus.  A general representative of the associated
deformation is
\begin{equation}
 \delta_{(0)}A_\mu
 =
 \frac{\partial A_\mu^{\rm B}}{\partial t}
 +D_\mu^{\rm B}\Lambda_{(0)},
 \label{eq:general-scale-zero-mode}
\end{equation}
where $\Lambda_{(0)}$ is chosen so that
\begin{equation}
 D_{\rm B}^\mu\delta_{(0)}A_\mu=0.
\end{equation}
For the gauge representative used in
\eqref{generalansatzB}, direct substitution shows that
\begin{equation}
 D_{\rm B}^\mu
 \frac{\partial A_\mu^{\rm B}}{\partial t}=0.
\end{equation}
We may therefore take $\Lambda_{(0)}=0$.

Using
\begin{equation}
 A^{\rm B}
 =
 G(r,t)\sigma_z\frac{\tau_3}{2}
 +
 F(r,t)
 \left(
 \sigma_x\frac{\tau_1}{2}
 +\sigma_y\frac{\tau_2}{2}
 \right),
\end{equation}
together with
\begin{equation}
 \sigma_x=\frac{e^1}{r},
 \qquad
 \sigma_y=\frac{e^2}{r},
 \qquad
 \sigma_z=\frac{f\,e^3}{r},
\end{equation}
we find
\begin{align}
 \delta_{(0)}A_\mu
 ={}&
 \frac{2t}{\sqrt{t^4-a^4}}
 \biggl\{
 {\cal F}_{\pm}
 \left[
 (e^1{}_\mu-ie^2{}_\mu)\frac{\tau_+}{2}
 +(e^1{}_\mu+ie^2{}_\mu)\frac{\tau_-}{2}
 \right]
 +2{\cal F}_3e^3{}_\mu\frac{\tau_3}{2}
 \biggr\}.
 \label{eq:explicit-scale-zero-mode}
\end{align}

\paragraph{Global color-orientation zero modes.}

A rigid color rotation acts on the background according to
\begin{equation}
 A^{\rm B}\longrightarrow
 A_U^{\rm B}=U^\dagger A^{\rm B}U.
\end{equation}
Gauge transformations that approach the identity at infinity are
redundancies of the framed problem.  By contrast, transformations
approaching a nontrivial constant at infinity generate physical color
orientations.  Their background-gauge representatives can be written
as
\begin{equation}
 \delta_{(a)}A_\mu
 =
 D_\mu^{\rm B}\Phi_a,
 \qquad
 D_{\rm B}^\mu D_\mu^{\rm B}\Phi_a=0,
 \qquad
 a=1,2,3,
 \label{eq:global-orientation-zero-modes}
\end{equation}
where the functions $\Phi_a$ approach constant generators at infinity.

For the present instanton, the required solutions are
\begin{align}
 \Phi_1
 &=
 \frac{\sqrt{r^4-a^4}}{r^2+t^2}\,\tau_1,
 &
 \Phi_2
 &=
 \frac{\sqrt{r^4-a^4}}{r^2+t^2}\,\tau_2,
 \nonumber\\
 \Phi_3
 &=
 \frac{t^2r^2+a^4}{t^2(t^2+r^2)}\,\tau_3.
 \label{eq:orientation-gauge-functions}
\end{align}
They satisfy
\begin{equation}
 \Phi_a\longrightarrow\tau_a
 \qquad\text{as}\qquad r\longrightarrow\infty,
\end{equation}
and hence generate global rather than small gauge transformations.
Direct substitution verifies both their normalizability and the
background gauge condition in
\eqref{eq:global-orientation-zero-modes}.

The resulting three zero modes are
\begin{align}
 \delta_{(1)}A_\mu
 ={}&
 4{\cal F}_3e^2{}_\mu\frac{\tau_3}{2}
 +2{\cal F}_{\pm}
 \left[
 (e^0{}_\mu+ie^3{}_\mu)\frac{\tau_+}{2}
 +(e^0{}_\mu-ie^3{}_\mu)\frac{\tau_-}{2}
 \right],
 \label{eq:orientation-zero-mode-1}\\
 \delta_{(2)}A_\mu
 ={}&
 -4{\cal F}_3e^1{}_\mu\frac{\tau_3}{2}
 +2{\cal F}_{\pm}
 \left[
 (-ie^0{}_\mu+e^3{}_\mu)\frac{\tau_+}{2}
 +(ie^0{}_\mu+e^3{}_\mu)\frac{\tau_-}{2}
 \right],
 \label{eq:orientation-zero-mode-2}\\
 \delta_{(3)}A_\mu
 ={}&
 \frac{\sqrt{t^4-a^4}}{t^2}
 \biggl\{
 4{\cal F}_3e^0{}_\mu\frac{\tau_3}{2}
 \nonumber\\
 &\hspace{18mm}
 +2{\cal F}_{\pm}
 \left[
 (-e^2{}_\mu-ie^1{}_\mu)\frac{\tau_+}{2}
 +(-e^2{}_\mu+ie^1{}_\mu)\frac{\tau_-}{2}
 \right]
 \biggr\}.
 \label{eq:orientation-zero-mode-3}
\end{align}

\paragraph{Metric on the bosonic moduli space.}

The zero-mode inner product defines the moduli-space metric
\begin{equation}
 {\cal G}_{IJ}
 =
 \frac{2}{g^2}
 \int_{\rm EH}d^4x\sqrt{g_{\rm EH}}\,
 g^{\mu\nu}
 \operatorname{tr}_{\mathbf 2}
 \left(
 \delta_{(I)}A_\mu\delta_{(J)}A_\nu
 \right),
 \qquad
 I,J=0,1,2,3.
 \label{eq:bosonic-moduli-space-metric2}
\end{equation}
Using
\begin{equation}
 g^{\mu\nu}e^a{}_\mu e^b{}_\nu=\delta^{ab}
\end{equation}
and \eqref{eq:bosonic-radial-normalization-integral}, we find that the
four modes are mutually orthogonal and
\begin{align}
 {\cal G}_{00}
 &=
 \frac{8\pi^2}{g^2}
 \frac{t^4}{t^4-a^4},
 \nonumber\\
 {\cal G}_{11}
 &=
 {\cal G}_{22}
 =
 \frac{8\pi^2}{g^2}t^2,
 \nonumber\\
 {\cal G}_{33}
 &=
 \frac{8\pi^2}{g^2}
 \frac{t^4-a^4}{t^2}.
 \label{eq:bosonic-moduli-metric-components}
\end{align}

To display the global geometry, let $\Sigma_a$ denote the
left-invariant one-forms on the color-orientation group, normalized by
\begin{equation}
 U^\dagger dU=i\tau_a\Sigma_a.
 \label{eq:orientation-Maurer-Cartan-forms}
\end{equation}
The complete moduli-space line element is then
\begin{equation}
 ds_{\cal M}^2
 =
 \frac{8\pi^2}{g^2}
 \left[
 \frac{t^4}{t^4-a^4}\,dt^2
 +t^2\left(\Sigma_1^2+\Sigma_2^2\right)
 +\frac{t^4-a^4}{t^2}\Sigma_3^2
 \right].
 \label{eq:bosonic-moduli-space-EH-metric}
\end{equation}
Up to the overall factor $8\pi^2/g^2$, this is precisely the
EH metric with radial coordinate $t$ and resolution
parameter $a$.  The bosonic moduli space is therefore hyper-Kähler and
isometric to EH space.

The center element $-I_2$ acts trivially on the adjoint background.
Consequently, the generic color-orientation orbit is
\begin{equation}
 \frac{SU(2)}{\mathbb Z_2}\simeq SO(3),
\end{equation}
rather than $SU(2)$.  At $t=a$, the circle generated by $\Sigma_3$
collapses, leaving the two-sphere bolt.  Smoothness requires this circle
to have period $2\pi$, exactly as in the original EH geometry.

A convenient Euler-angle parametrization of the orientation orbit has
the ranges
\begin{equation}
 0\leq\vartheta\leq\pi,
 \qquad
 0\leq\varphi<2\pi,
 \qquad
 0\leq\psi<2\pi.
 \label{eq:SO3-orientation-angle-ranges}
\end{equation}
With the orientation convention used here, the positive Haar measure
is
\begin{equation}
 dV_{SO(3)}
 =
 -\Sigma_1\wedge\Sigma_2\wedge\Sigma_3,
 \qquad
 \int_{SO(3)}dV_{SO(3)}=\pi^2.
 \label{eq:SO3-Haar-volume}
\end{equation}

\paragraph{Bosonic collective-coordinate measure.}

In the orthonormal moduli-space coframe
$(dt,\Sigma_1,\Sigma_2,\Sigma_3)$, the determinant of the metric is
\begin{equation}
 \sqrt{\det{\cal G}}
 =
 \left(\frac{8\pi^2}{g^2}\right)^2t^3
 =
 \left(\frac{2}{g^2}\right)^2
 (4\pi^2)^2t^3.
 \label{eq:bosonic-moduli-metric-determinant}
\end{equation}
Using the standard factor $(2\pi)^{-1/2}$ for each real bosonic zero
mode, the collective-coordinate measure is
\begin{align}
 d{\cal M}_{\rm b}
 &=
 \frac{\sqrt{\det{\cal G}}}{(2\pi)^2}\,
 dt\,dV_{SO(3)}
 \nonumber\\
 &=
 \frac{16\pi^2}{g^4}\,
 t^3dt\,dV_{SO(3)}.
 \label{eq:bosonic-collective-coordinate-measure}
\end{align}

\subsection*{Matching the fermion zero modes to the bosonic moduli}
\label{subsec:matching-fermionic-bosonic-zero-modes}

We now determine explicitly how the two adjoint fermion zero-mode
wavefunctions generate the four real bosonic zero modes associated
with the scale and color-orientation moduli.  For the minimal
$Q_{(2)}=1/2$ instanton, we choose the commuting fermion zero-mode
basis
\begin{equation}
 \lambda_\beta^{(1)}
 =
 \begin{pmatrix}
  {\cal F}_3\,\dfrac{\tau_3}{2}\\[2mm]
  {\cal F}_{\pm}\,\dfrac{\tau_+}{2}
 \end{pmatrix},
 \qquad
 \lambda_\beta^{(2)}
 =
 \begin{pmatrix}
  {\cal F}_{\pm}\,\dfrac{\tau_-}{2}\\[2mm]
  -{\cal F}_3\,\dfrac{\tau_3}{2}
 \end{pmatrix}.
 \label{eq:explicit-adjoint-zero-mode-basis}
\end{equation}
The Grassmann collective coordinates that multiply these
wavefunctions in the gaugino field are not included in
\eqref{eq:explicit-adjoint-zero-mode-basis}.

Before imposing Hermiticity, the most general solution of the
complexified bosonic deformation equation can be written as
\begin{align}
 \delta A_{\beta\dot1}
 &=
 a\,\lambda_\beta^{(1)}
 +b\,\lambda_\beta^{(2)},
 \nonumber\\
 \delta A_{\beta\dot2}
 &=
 c\,\lambda_\beta^{(1)}
 +d\,\lambda_\beta^{(2)}.
 \label{eq:general-complex-bosonic-expansion}
\end{align}
The four coefficients $a,b,c,d$ are not independent for a physical
gauge fluctuation.  Requiring
\begin{equation}
 \delta A_\mu^\dagger=\delta A_\mu
\end{equation}
gives
\begin{equation}
 c=b^*,
 \qquad
 d=-a^*.
 \label{eq:coefficient-matrix-Hermiticity}
\end{equation}
It is therefore convenient to introduce two independent complex
coefficients,
\begin{equation}
 c_1\equiv a,
 \qquad
 c_2\equiv b.
\end{equation}
The Hermitian bosonic bispinor then takes the form
\begin{align}
 \delta A_{\beta\dot1}
 &=
 c_1\lambda_\beta^{(1)}
 +c_2\lambda_\beta^{(2)},
 \nonumber\\
 \delta A_{\beta\dot2}
 &=
 c_2^*\lambda_\beta^{(1)}
 -c_1^*\lambda_\beta^{(2)}.
 \label{eq:Hermitian-bosonic-expansion}
\end{align}
Equivalently, the matrix of coefficients multiplying the two fermion
wavefunctions is
\begin{equation}
 {\cal C}
 =
 \begin{pmatrix}
  c_1 & c_2^*\\
  c_2 & -c_1^*
 \end{pmatrix}.
 \label{eq:bosonic-coefficient-quaternionic-matrix}
\end{equation}
Thus, $c_1$ and $c_2$ contain precisely four independent real
parameters, as required by the index theorem.

The corresponding vector fluctuation is reconstructed according to
\begin{align}
 \delta A_\mu
 =\frac12\bigl[
 &(e^0{}_\mu-ie^3{}_\mu)\delta A_{1\dot1}
 +(-ie^1{}_\mu+e^2{}_\mu)\delta A_{1\dot2}
 \nonumber\\
 &+(-ie^1{}_\mu-e^2{}_\mu)\delta A_{2\dot1}
 +(e^0{}_\mu+ie^3{}_\mu)\delta A_{2\dot2}
 \bigr].
 \label{eq:bosonic-bispinor-vector-reconstruction}
\end{align}
Substituting \eqref{eq:Hermitian-bosonic-expansion} into
\eqref{eq:bosonic-bispinor-vector-reconstruction} and comparing with
the explicit scale and orientation modes
$\delta_{(I)}A_\mu$, $I=0,1,2,3$, gives
\begin{equation}
\begin{array}{c|cc}
 \text{bosonic zero mode}
 & c_1 & c_2 \\ \hline
 \delta_{(0)}A_\mu
 &
 \displaystyle\frac{4it}{\sqrt{t^4-a^4}}
 &0
 \\[3mm]
 \delta_{(1)}A_\mu
 &0&4
 \\[1mm]
 \delta_{(2)}A_\mu
 &0&4i
 \\[1mm]
 \delta_{(3)}A_\mu
 &
 \displaystyle\frac{4\sqrt{t^4-a^4}}{t^2}
 &0
\end{array}
\label{eq:explicit-c1-c2-bosonic-modes}
\end{equation}
in the normalization and phase convention of
\eqref{eq:explicit-adjoint-zero-mode-basis}.

Consequently, the imaginary part of $c_1$ generates the scale
deformation, while its real part generates the third color
orientation.  Similarly, the real and imaginary parts of $c_2$
generate the first and second color orientations:
\begin{equation}
 \operatorname{Im}c_1
 \longleftrightarrow \delta_{(0)}A_\mu,
 \qquad
 \operatorname{Re}c_1
 \longleftrightarrow \delta_{(3)}A_\mu,
 \qquad
 \operatorname{Re}c_2
 \longleftrightarrow \delta_{(1)}A_\mu,
 \qquad
 \operatorname{Im}c_2
 \longleftrightarrow \delta_{(2)}A_\mu.
\end{equation}
This establishes explicitly the correspondence between the two
complex adjoint fermion zero modes and the four real tangent
directions of the bosonic instanton moduli space.

The numerical entries in
\eqref{eq:explicit-c1-c2-bosonic-modes} depend on the normalization
and phases chosen for $\lambda^{(1)}$ and $\lambda^{(2)}$.  Rephasing
or rescaling the fermion basis changes these coefficients but leaves
the physical bosonic modes and their moduli-space metric unchanged.

\subsection{$SU(2)$ fermionic measure}
\label{Fermionic measure}

For the Grassmann coordinates in \eqref{fzeromodest}, define
\begin{equation}
 {\cal U}_F^{ij}
 =\frac{2}{g^2}\int_{\rm EH}d^4x\sqrt g\,
 \operatorname{tr}_{\mathbf 2}\left[
 \epsilon^{\alpha\beta}
 \frac{\partial\lambda_\alpha}{\partial\eta_i}
 \frac{\partial\lambda_\beta}{\partial\eta_j}\right].
\end{equation}
Equation \eqref{eq:app-SU2-radial-normalization} gives
\begin{equation}
 {\cal U}_F^{12}=-{\cal U}_F^{21}
 =-\frac{\pi^2t^2}{2g^2},
 \qquad
 \sqrt{\det{\cal U}_F}=\frac{\pi^2t^2}{2g^2}.
\end{equation}
Because Grassmann Jacobians transform inversely, the fermionic measure is
\begin{equation}
 d{\cal M}_{\rm f}
 =\frac{d\eta_2d\eta_1}{\sqrt{\det{\cal U}_F}}
 =\frac{2g^2}{\pi^2t^2}\,d\eta_2d\eta_1.
 \label{eq:app-SU2-fermionic-measure}
\end{equation}

\section{The $8$-fermion condensate in $SU(4)$}
\label{app:SU4-eight-fermion-details}

In this appendix, we give the details of the computations of the $8$-fermion condensate in the $SU(4)$ case in the background of a unit topological charge $Q_{(4)}=1$ and the central holonomy $h_c=-I_4$.

\subsection{The rationale for computing the moduli space measure and the issue of orthogonality in $SU(4)$}
\label{sec:SU4-moduli-orthogonality}

Let $A(t)$ denote the $Q_{(2)}=1/2$ connection, with radial-size modulus 
$t\geq a$. A subset of the charge-one $SU(4)$ moduli space
is represented by
\begin{equation}
 A^{\rm B}(t_1,t_2,U)
 =
 U^\dagger
 \begin{pmatrix}
  A^{\rm B}(t_1)&0\\
  0&A^{\rm B}(t_2)
 \end{pmatrix}
 U,
 \qquad
 U\in SU(4),
 \qquad
 t_1,t_2\geq a.
 \label{eq:SU4-two-block-family}
\end{equation}
For $t_1\neq t_2$, the stabilizer is\footnote{The stabilizer $H$ is obtained as follows.  Writing a generic $SU(4)$ transformation $H$ in $2\times2$ block form and imposing
$H^\dagger\,\operatorname{diag}\!\left(A(t_1),A(t_2)\right)H
=\operatorname{diag}\!\left(A(t_1),A(t_2)\right)$, one finds that, for $t_1\neq t_2$, the off-diagonal blocks of $H$ must vanish because they would otherwise define constant intertwiners between the inequivalent connections $A(t_1)$ and $A(t_2)$.  Since each $A(t_i)$ is irreducible for generic $t_i>a$, Schur's lemma then requires the diagonal blocks to be proportional to the identity: $H=\operatorname{diag}(z_1I_2,z_2I_2)$, with $z_1,z_2\in U(1)$.  The condition $\det H=1$ gives $(z_1z_2)^2=1$.  Hence
$H=\{\operatorname{diag}(z_1I_2,z_2I_2)\mid z_1,z_2\in U(1),\ (z_1z_2)^2=1\}\simeq U(1)\times\mathbb Z_2$. The matrix $H$ can be parametrized as
\begin{equation}
 H_\epsilon(\alpha)
 =
 \operatorname{diag}
 \left(
 e^{i\alpha}I_2,
 \epsilon e^{-i\alpha}I_2
 \right),
 \qquad
 \epsilon=\pm1,
 \qquad
 0\leq\alpha<2\pi.
 \label{eq:SU4-stabilizer-componentsfoot}
\end{equation}  At the coincident locus $t_1=t_2$, the two blocks become equivalent, and the stabilizer is enhanced.}
\begin{equation}
 H
 =
 \left\{
 \operatorname{diag}(z_1 I_2,z_2 I_2):
 z_1,z_2\in U(1),\ (z_1z_2)^2=1
 \right\}\simeq U(1)\times\mathbb Z_2.
 \label{eq:SU4-generic-stabilizer}
\end{equation}
It has one continuous generator and two connected components. Hence
$\dim_{\mathbb R}SU(4)/H=14$; together with $t_1,t_2$, this gives the
required sixteen real bosonic moduli.   The parametrization
\eqref{eq:SU4-two-block-family} is invariant under simultaneous exchange
of the two blocks and $t_1\leftrightarrow t_2$, so its measure includes
an additional factor $1/2!$. The locus $t_1=t_2$, where the stabilizer
is enhanced, has zero measure in the coordinates used below.

The bosonic collective coordinates consist of the two scale parameters
$t_1,t_2$, six color orientations acting within the two diagonal
$SU(2)$ blocks, and eight color orientations mixing the two blocks. In the following, we shall identify the collective coordinates and compute the measure over this $16$-dimensional moduli space. To tackle such a complex problem, we shall factorize the collective-coordinate measure. 
To justify this factorization, one
must show that imposing background gauge does not mix the
block-diagonal and off-diagonal tangent directions.  This does not
follow merely from the matrix structure of the undressed generators;
the boundary conditions imposed on the gauge-fixing fields must also
be specified.

\subsection*{The block-parity involution}

Introduce the involution
\begin{equation}
 \Gamma
 =
 \begin{pmatrix}
  I_2 & 0 \\
  0 & -I_2
 \end{pmatrix},
 \qquad
 {\cal P}(X)=\Gamma X\Gamma .
 \label{eq:SU4-block-parity}
\end{equation}
Since $\Gamma^2=I_4$, one has ${\cal P}^2=1$.  The Lie algebra therefore
decomposes into the even and odd eigenspaces
\begin{equation}
 \mathfrak{su}(4)=\mathfrak{k}\oplus\mathfrak{m},
 \qquad
 {\cal P}|_{\mathfrak{k}}=+1,
 \qquad
 {\cal P}|_{\mathfrak{m}}=-1,
 \label{eq:SU4-block-parity-decomposition}
\end{equation}
where
\begin{align}
 \mathfrak{k}
 &=
 \left\{
 \begin{pmatrix}
  P & 0 \\
  0 & Q
 \end{pmatrix}
 :
 \operatorname{tr}P+\operatorname{tr}Q=0
 \right\},
 \nonumber\\
 \mathfrak{m}
 &=
 \left\{
 \begin{pmatrix}
  0 & R \\
  R^\dagger & 0
 \end{pmatrix}
 \right\}.
 \label{eq:SU4-block-parity-subspaces}
\end{align}
Thus $\mathfrak{k}$ is the block-diagonal sector and $\mathfrak{m}$ is
the off-diagonal sector.  The background in
\eqref{eq:SU4-two-block-family} is even under ${\cal P}$.  It
follows that
\begin{equation}
 [D_\mu,{\cal P}]=0,
 \qquad
 [D^2,{\cal P}]=0,
 \label{eq:SU4-covariant-derivative-preserves-parity}
\end{equation}
because
\begin{align}
 {\cal P}(D_\mu X)
 &=
 \Gamma\left(
 \partial_\mu X+i[A_\mu^{\rm B},X]
 \right)\Gamma
 \nonumber\\
 &=
 \partial_\mu{\cal P}(X)
 +i[A_\mu^{\rm B},{\cal P}(X)]
 =
 D_\mu{\cal P}(X).
 \label{eq:SU4-explicit-parity-covariance}
\end{align}
The linearized self-duality operator preserves the same decomposition,
since it is constructed from $D_\mu$ and the spacetime Hodge star,
whereas ${\cal P}$ acts only on color indices.

\subsection*{Boundary conditions and background gauge}

It is important to distinguish gauge redundancies from physical color
rotations at infinity.  In the framed problem, the gauge transformations
that are divided out form the based gauge group
\begin{equation}
 {\cal G}_0
 =
 \left\{
 g(x):g(\infty)=I_4
 \right\}.
 \label{eq:SU4-based-gauge-group}
\end{equation}
Infinitesimally, their parameters satisfy
\begin{equation}
 \epsilon(\infty)=0.
 \label{eq:SU4-based-gauge-parameter}
\end{equation}
By contrast, a transformation approaching a nonzero generator $X$ at
infinity is retained as a physical global color rotation whenever it
preserves the specified boundary sector.  For the two-block solution
with central boundary holonomy $h_\infty=-I_4$, every $X\in
\mathfrak{su}(4)$ commutes with $h_\infty$ and is therefore admissible.

\subsection*{Radial-size moduli}

For a radial-size modulus $t_i$, begin with the naive tangent vector
\begin{equation}
 a_{\mu,0}^{(t_i)}
 =
 \partial_{t_i}A_\mu^{\rm B}.
 \label{eq:SU4-naive-scale-mode}
\end{equation}
Its background-gauge representative is
\begin{equation}
 a_\mu^{(t_i)}
 =
 \partial_{t_i}A_\mu^{\rm B}
 +D_\mu\Omega_{t_i},
 \qquad
 D^\mu a_\mu^{(t_i)}=0,
 \label{eq:SU4-gauge-fixed-scale-mode}
\end{equation}
where
\begin{equation}
 D^2\Omega_{t_i}
 =
 -D^\mu\partial_{t_i}A_\mu^{\rm B}.
 \label{eq:SU4-scale-compensator-equation}
\end{equation}
Varying $t_i$ does not change the prescribed boundary framing.
Consequently, the compensating transformation in
\eqref{eq:SU4-gauge-fixed-scale-mode} is chosen from ${\cal G}_0$ and
obeys
\begin{equation}
 \Omega_{t_i}(\infty)=0.
 \label{eq:SU4-scale-compensator-boundary-condition}
\end{equation}
This boundary condition is part of the definition of the horizontal
scale tangent.  It does not follow from
\eqref{eq:SU4-scale-compensator-equation} alone.  Allowing
$\Omega_{t_i}$ to approach a nonzero generator would add an independent
global color rotation to the radial-size mode and would therefore amount to
using a mixed basis of collective coordinates.

The source in \eqref{eq:SU4-scale-compensator-equation} is block
diagonal.  Decomposing
\begin{equation}
 \Omega_{t_i}
 =
 \Omega_{t_i,+}+\Omega_{t_i,-},
 \qquad
 {\cal P}(\Omega_{t_i,\pm})=\pm\Omega_{t_i,\pm},
\end{equation}
the odd part satisfies
\begin{equation}
 D^2\Omega_{t_i,-}=0,
 \qquad
 \Omega_{t_i,-}(\infty)=0.
 \label{eq:SU4-odd-scale-compensator}
\end{equation}
Regularity at the bolt and integration by parts give
\begin{equation}
 \int_{\rm EH}d^4x\sqrt{g}\,
 \operatorname{tr}_{\mathbf 4}
 \left(
 D_\mu\Omega_{t_i,-}D^\mu\Omega_{t_i,-}
 \right)
 =
 -\int_{\rm EH}d^4x\sqrt{g}\,
 \operatorname{tr}_{\mathbf 4}
 \left(
 \Omega_{t_i,-}D^2\Omega_{t_i,-}
 \right)
 =0.
 \label{eq:SU4-scale-compensator-positivity}
\end{equation}
There is no contribution from infinity because
$\Omega_{t_i,-}$ vanishes there, while the bolt is a regular interior
submanifold rather than a boundary.  Hence
$D_\mu\Omega_{t_i,-}=0$, and its vanishing boundary value implies
\begin{equation}
 \Omega_{t_i,-}=0.
 \label{eq:SU4-scale-compensator-odd-part-vanishes}
\end{equation}
The complete gauge-fixed radial-size mode is therefore block diagonal.

\subsection*{Color-orientation modes and their harmonic dressing}

The boundary condition is different for a physical color orientation.
Let $X$ be an admissible constant generator at infinity.  Since
$\partial_\mu X=0$, the undressed infinitesimal rotation is
\begin{equation}
 a_{\mu,0}^{(X)}
 =
 D_\mu X
 =
 i[A_\mu^{\rm B},X].
 \label{eq:SU4-undressed-color-rotation}
\end{equation}
To impose background gauge, add a based gauge correction
$D_\mu\omega_X$:
\begin{equation}
 a_\mu^{(X)}
 =
 D_\mu X+D_\mu\omega_X,
 \qquad
 \omega_X(\infty)=0.
 \label{eq:SU4-based-orientation-correction}
\end{equation}
It is convenient to define the full harmonic dressing
\begin{equation}
 \Theta_X
 =
 X+\omega_X.
 \label{eq:SU4-full-harmonic-dressing}
\end{equation}
Then the gauge-fixed orientation mode takes the form
\begin{equation}
 a_\mu^{(X)}=D_\mu\Theta_X,
 \qquad
 D^2\Theta_X=0,
 \qquad
 \Theta_X(\infty)=X.
 \label{eq:SU4-harmonic-orientation-mode}
\end{equation}
Equivalently, the based correction satisfies
\begin{equation}
 D^2\omega_X=-D^2X,
 \qquad
 \omega_X(\infty)=0.
 \label{eq:SU4-based-orientation-correction-equation}
\end{equation}
Equations \eqref{eq:SU4-based-orientation-correction} and
\eqref{eq:SU4-harmonic-orientation-mode} make the terminology precise:
the genuine gauge-fixing correction $\omega_X$ vanishes at infinity,
whereas the complete dressing $\Theta_X$ does not.  Its nonzero
boundary value is exactly the physical global color rotation.  Calling
$\Theta_X$ itself a ``compensator'' can obscure this distinction. However, in the following treatment, we shall be sloppy about this distinction and shall use $X$ and $\Theta_X$ interchangeably.  

Because $D^2$ commutes with ${\cal P}$, the solution of
\eqref{eq:SU4-harmonic-orientation-mode} has the same block parity as
its boundary generator.  If $X_{\rm diag}\in\mathfrak{k}$, the odd part
of $\Theta_{X_{\rm diag}}$ obeys a homogeneous equation with zero
boundary value and therefore vanishes.  Thus
\begin{equation}
 X_{\rm diag}\in\mathfrak{k}
 \quad\Longrightarrow\quad
 \Theta_{X_{\rm diag}}\in\mathfrak{k},
 \qquad
 D_\mu\Theta_{X_{\rm diag}}\in\mathfrak{k}.
 \label{eq:SU4-diagonal-dressing-parity}
\end{equation}
Similarly, if $X_{\rm off}\in\mathfrak{m}$, the even part has zero
boundary value and vanishes:
\begin{equation}
 X_{\rm off}\in\mathfrak{m}
 \quad\Longrightarrow\quad
 \Theta_{X_{\rm off}}\in\mathfrak{m},
 \qquad
 D_\mu\Theta_{X_{\rm off}}\in\mathfrak{m}.
 \label{eq:SU4-offdiagonal-dressing-parity}
\end{equation}
The radial functions $h_s(r)$ used for the eight cross-block
orientations are precisely the components of
$\Theta_{X_{\rm off}}$.  Their normalization
$h_s(\infty)=1$ implements
$\Theta_{X_{\rm off}}(\infty)=X_{\rm off}$; it is not in conflict with
the requirement $\omega_{X_{\rm off}}(\infty)=0$.

\subsection*{Vanishing of the mixed Gram components}

The Yang--Mills inner product on gauge-field fluctuations is
\begin{equation}
 \left\langle a,b\right\rangle_{\rm YM}
 =
 \frac{2}{g^2}
 \int_{\rm EH}d^4x\sqrt{g}\,
 \operatorname{tr}_{\mathbf 4}
 \left(
 a_\mu b^\mu
 \right).
 \label{eq:SU4-YM-zero-mode-inner-product}
\end{equation}
It is invariant under ${\cal P}$.  Consequently, if $a_+$ is
block diagonal and $b_-$ is off diagonal, then
\begin{align}
 \left\langle a_+,b_-\right\rangle_{\rm YM}
 &=
 \left\langle{\cal P}a_+,{\cal P}b_-\right\rangle_{\rm YM}
 \nonumber\\
 &=
 -\left\langle a_+,b_-\right\rangle_{\rm YM},
\end{align}
and hence
\begin{equation}
 \left\langle a_+,b_-\right\rangle_{\rm YM}=0.
 \label{eq:SU4-even-odd-orthogonality}
\end{equation}
The same conclusion is visible pointwise:
\begin{equation}
 \operatorname{tr}_{\mathbf 4}
 \left[
 \begin{pmatrix}
  P & 0 \\
  0 & Q
 \end{pmatrix}
 \begin{pmatrix}
  0 & R \\
  S & 0
 \end{pmatrix}
 \right]
 =0.
 \label{eq:SU4-pointwise-block-orthogonality}
\end{equation}
Therefore, the radial-size moduli and the six internal block rotations are
orthogonal to all eight off-diagonal rotations after background gauge
has been imposed.

Let $q^I=(t_1,t_2)$ denote the scale coordinates, let $\alpha^a$
denote the six block-diagonal orientation coordinates, and let
$\beta^m$ denote the eight off-diagonal orientation coordinates.  In a
block-parity-adapted moving frame, the bosonic Gram matrix has the
structure
\begin{equation}
 {\cal G}_{\rm b}
 =
 \begin{pmatrix}
  {\cal G}_{qq} & {\cal G}_{q\alpha} & 0 \\
  {\cal G}_{\alpha q} & {\cal G}_{\alpha\alpha} & 0 \\
  0 & 0 & {\cal G}_{\beta\beta}
 \end{pmatrix}.
 \label{eq:SU4-bosonic-Gram-block-structure}
\end{equation}
Thus
\begin{equation}
 \det{\cal G}_{\rm b}
 =
 \det{\cal G}_{\rm block\text{-}diag}\,
 \det{\cal G}_{\rm off}.
 \label{eq:SU4-bosonic-Gram-factorization}
\end{equation}
The parity argument proves the vanishing of every entry connecting the
off-diagonal sector to the scale and internal-rotation sectors.  It
does not, by itself, imply
$ {\cal G}_{q\alpha}=0$.  Orthogonality between radial-size modes and
block-diagonal color rotations was established in the $SU(2)$ case.

\subsection{Off-diagonal color orientations}
\label{subsec:SU4-offdiagonal-orientations}

Consider the generic two-block background
\begin{equation}
 A^{\rm B}(t_1,t_2)
 =
 \begin{pmatrix}
  A_1^{\rm B}&0\\
  0&A_2^{\rm B}
 \end{pmatrix},
 \qquad
 A_i^{\rm B}\equiv A_{\rm EH}^{\rm B}(t_i),
 \qquad
 t_1\neq t_2.
 \label{eq:SU4-offdiag-background}
\end{equation}
The two radial-size moduli $t_1$ and $t_2$ determine the gauge-invariant radial
profile of the solution.  The remaining moduli are color orientations.
Six of them are generated by independent rotations within the two
diagonal $SU(2)$ blocks.  In this subsection we explain how the other
eight, which mix the two two-dimensional color subspaces, are constructed
from bounded adjoint harmonic scalars.

\paragraph{Off-diagonal generators.}
The off-diagonal complement  of the block-diagonal  subalgebra of $\mathfrak{su}(4)$ consists of
matrices of the form (we emphasize that $X(\Phi)$ are off-diagonal hermitian generators of $\mathfrak{su}(4)$ algebra)
\begin{equation}
 X(\Phi)
 =
 \begin{pmatrix}
  0&\Phi\\
  \Phi^\dagger&0
 \end{pmatrix},
 \qquad
 \Phi\in\operatorname{Mat}_{2\times2}(\mathbb C).
 \label{eq:SU4-offdiag-general-generator}
\end{equation}
Since a complex $2\times2$ matrix contains eight real parameters, this
space has real dimension eight.  A convenient complex basis is
\begin{equation}
 M_0=I_2,
 \qquad
 M_i=\tau_i,
 \qquad
 i=1,2,3,
 \qquad
 \operatorname{tr}_{\mathbf2}(M_sM_t)=2\delta_{st}\,, \quad s,t=0,1,2,3\,.
 \label{eq:SU4-offdiag-M-basis}
\end{equation}
For every $M_s$ there are two Hermitian generators,
\begin{align}
 X_s^{\rm R}
 &=
 \frac{1}{2\sqrt{2}}
 \begin{pmatrix}
  0&M_s\\
  M_s&0
 \end{pmatrix},
 &
 X_s^{\rm I}
 &=
 \frac{1}{2\sqrt{2}}
 \begin{pmatrix}
  0&-iM_s\\
  iM_s&0
 \end{pmatrix}.
 \label{eq:SU4-offdiag-real-generators}
\end{align}
They obey
\begin{equation}
 \operatorname{tr}_{\mathbf4}
 \left(X_s^{A}X_t^{B}\right)
 =
 \frac12\delta_{st}\delta^{AB},
 \qquad
 A,B\in\{{\rm R},{\rm I}\}.
 \label{eq:SU4-offdiag-generator-normalization}
\end{equation}
Thus the four channels $s=0,1,2,3$, together with their real and
imaginary generators, account for the eight off-diagonal color
orientations.

\paragraph{A radial dressing is required.}
Let $X$ be one of the constant generators in
\eqref{eq:SU4-offdiag-real-generators}.  A constant color rotation
$U=\exp(i\epsilon X)$ produces (since $X$ is constant, we can add the derivative $\partial_\mu X=0$ for an immediate convenience)
\begin{equation}
 A^{\rm B}
 \longrightarrow
 U^\dagger A^{\rm B}U,
 \qquad
 \delta_XA_\mu
 =
 i[A_\mu^{\rm B},X]
 =
 D_\mu X,
 \label{eq:SU4-offdiag-naive-rotation}
\end{equation}
where
\begin{equation}
 D_\mu X=\partial_\mu X+i[A_\mu^{\rm B},X].
\end{equation}
The deformation \eqref{eq:SU4-offdiag-naive-rotation} is tangent to a
family of self-dual solutions, but it does not in general satisfy the
background-gauge condition.  It must therefore be supplemented by a
small gauge transformation whose parameter vanishes at infinity.
Writing the resulting deformation as
\begin{equation}
 \delta A_\mu=D_\mu\Theta,
 \qquad
 \Theta\longrightarrow X
 \quad\text{as}\quad r\longrightarrow\infty,
 \label{eq:SU4-offdiag-dressed-mode}
\end{equation}
the background-gauge condition becomes
\begin{equation}
 D^\mu\delta A_\mu
 =
 D^2\Theta
 =
 0.
 \label{eq:SU4-offdiag-harmonic-equation}
\end{equation}
The nonzero boundary value is essential: it identifies the deformation
as a true color-orientation modulus rather than a small gauge
transformation.

\paragraph{The full $4\times4$ harmonic scalars.}
The notation $\Theta_s=h_sM_s$, sometimes used for brevity, refers only
to the upper-right $2\times2$ block.  The full Hermitian, traceless
$SU(4)$ scalars are
\begin{align}
 \Theta_s^{\rm R}(r)
 &=
 \frac{h_s(r)}{2\sqrt{2}}
 \begin{pmatrix}
  0&M_s\\
  M_s&0
 \end{pmatrix},
 &
 \Theta_s^{\rm I}(r)
 &=
 \frac{h_s(r)}{2\sqrt{2}}
 \begin{pmatrix}
  0&-iM_s\\
  iM_s&0
 \end{pmatrix},
 \label{eq:SU4-offdiag-full-harmonic-scalars}
\end{align}
where the functions $h_s(r)$ are real and satisfy
\begin{equation}
 h_s(\infty)=1.
 \label{eq:SU4-offdiag-h-boundary-condition}
\end{equation}
The associated bosonic zero modes are
\begin{equation}
 \delta_s^{A}A_\mu
 =
 D_\mu\Theta_s^{A},
 \qquad
 s=0,1,2,3,
 \qquad
 A\in\{{\rm R},{\rm I}\}.
 \label{eq:SU4-offdiag-eight-zero-modes}
\end{equation}

It is sufficient to solve the equation in the upper-right block.  For a
general upper-right matrix $\Phi$, the covariant derivative induced by
\eqref{eq:SU4-offdiag-background} is
\begin{equation}
 D_\mu^{12}\Phi
 =
 \partial_\mu\Phi
 +iA_{1\mu}\Phi
 -i\Phi A_{2\mu}.
 \label{eq:SU4-offdiag-bifundamental-derivative}
\end{equation}
The lower-left equation is its Hermitian conjugate.  Thus solving
\begin{equation}
 \left(D^{12}\right)^2\left(h_sM_s\right)=0
 \label{eq:SU4-offdiag-upper-block-equation}
\end{equation}
automatically constructs the full bosonic modes in
\eqref{eq:SU4-offdiag-eight-zero-modes}.

\subsection*{Detailed solution of the off-diagonal harmonic equations}
\label{subsec:SU4-offdiag-radial-solutions-detailed}

In this subsection we solve explicitly the radial equations obeyed by the
bounded adjoint harmonic scalars that generate the eight off-diagonal
$SU(4)$ color orientations.  The two diagonal blocks of the background
connection are written as
\begin{equation}
 A_i
 =
 F_i(r)
 \left(
  \sigma_x\frac{\tau_1}{2}
  +\sigma_y\frac{\tau_2}{2}
 \right)
 +G_i(r)\sigma_z\frac{\tau_3}{2},
 \qquad
 i=1,2,
 \label{eq:SU4-radial-detail-block-connections}
\end{equation}
where
\begin{equation}
 F_i(r)
 =
 \frac{2\sqrt{t_i^4-a^4}}{r^2+t_i^2},
 \qquad
 G_i(r)
 =
 \frac{2(t_i^2r^2+a^4)}
 {r^2(r^2+t_i^2)},
 \qquad
 f(r)=\left(1-\frac{a^4}{r^4}\right)^{-1/2}.
 \label{eq:SU4-radial-detail-profile-functions}
\end{equation}
In the upper-right $2\times2$ block, the four complex color channels are
\begin{equation}
 \Phi_s(r)=h_s(r)M_s,
 \qquad
 M_0=I_2,
 \qquad
 M_i=\tau_i,
 \qquad
 i=1,2,3.
 \label{eq:SU4-radial-detail-channel-ansatz}
\end{equation}
The lower-left block is the Hermitian conjugate of
\eqref{eq:SU4-radial-detail-channel-ansatz}.  The radial functions are
required to be covariantly regular at the bolt and normalized by
\begin{equation}
 h_s(r)\longrightarrow1
 \qquad\text{as}\qquad
 r\longrightarrow\infty.
 \label{eq:SU4-radial-detail-normalization}
\end{equation}

\subsubsection*{Reduction of the Laplace equation and its solution}

The upper-right block $\Phi$ transforms in the bifundamental
representation of the two diagonal $SU(2)$ factors.  Its covariant
derivative is therefore
\begin{equation}
 D_\mu^{12}\Phi
 =
 \partial_\mu\Phi
 +iA_{1\mu}\Phi
 -i\Phi A_{2\mu}.
 \label{eq:SU4-bifundamental-covariant-derivative2}
\end{equation}
Although $\Phi$ is a spacetime scalar, $D_\mu^{12}\Phi$ carries a
covariant spacetime index.  The second covariant derivative must
therefore include the Levi--Civita connection.  Equivalently,
\begin{equation}
 \left(D^{12}\right)^2\Phi
 =
 \nabla^\mu\left(D_\mu^{12}\Phi\right)
 +iA_1^\mu D_\mu^{12}\Phi
 -i\left(D_\mu^{12}\Phi\right)A_2^\mu.
 \label{eq:SU4-bifundamental-Laplacian-definition}
\end{equation}
Expanding this expression gives
\begin{align}
 \left(D^{12}\right)^2\Phi
 ={}&
 \nabla^2\Phi
 +2iA_1^\mu\partial_\mu\Phi
 -2i\left(\partial_\mu\Phi\right)A_2^\mu
 \nonumber\\
 &+
 i\left(\nabla^\mu A_{1\mu}\right)\Phi
 -i\Phi\left(\nabla^\mu A_{2\mu}\right)
 \nonumber\\
 &-
 A_1^\mu A_{1\mu}\Phi
 -\Phi A_2^\mu A_{2\mu}
 +2A_1^\mu\Phi A_{2\mu}.
 \label{eq:SU4-bifundamental-Laplacian-expanded}
\end{align}
The divergence appearing here is the curved-space divergence
$
 \nabla^\mu A_{i\mu}
 =
 \frac{1}{\sqrt{g}}
 \partial_\mu\left(\sqrt{g}\,A_i^\mu\right).
 \label{eq:SU4-gauge-field-divergence}
$
For the cohomogeneity-one EH connections considered here, the gauge
fields have no radial component and their angular components are
divergence-free.  Hence
\begin{equation}
 A_i^r=0,
 \qquad
 A_i^\mu\partial_\mu\Phi_s=0,
 \qquad
 \nabla^\mu A_{i\mu}=0,
 \qquad i=1,2.
 \label{eq:SU4-vanishing-linear-terms}
\end{equation}
For the radial ansatz
\begin{equation}
 \Phi_s(r)=h_s(r)M_s,
\end{equation}
the ordinary scalar Laplacian is
\begin{align}
 \nabla^2\Phi_s
 &=
 \frac{1}{\sqrt{g}}
 \partial_r\left(
 \sqrt{g}\,g^{rr}\partial_r\Phi_s
 \right)
 \nonumber\\
 &=
 \frac{1}{r^3}\frac{d}{dr}
 \left(
 r^3f^{-2}\frac{dh_s}{dr}
 \right)M_s,
 \label{eq:SU4-radial-scalar-Laplacian}
\end{align}
where $\sqrt{g}=r^3\sin\theta/8$ and $g^{rr}=f^{-2}$.

It follows that the covariant Laplacian reduces to
\begin{align}
 \left(D^{12}\right)^2\Phi_s
 ={}&
 \frac{1}{r^3}\frac{d}{dr}
 \left(
 r^3f^{-2}\frac{dh_s}{dr}
 \right)M_s
 \nonumber\\
 &-
 A_1^\mu A_{1\mu}\Phi_s
 -\Phi_s A_2^\mu A_{2\mu}
 +2A_1^\mu\Phi_s A_{2\mu}.
 \label{eq:SU4-Laplacian-after-linear-cancellation}
\end{align}
Defining
\begin{equation}
 -A_1^\mu A_{1\mu}\Phi_s
 -\Phi_s A_2^\mu A_{2\mu}
 +2A_1^\mu\Phi_s A_{2\mu}
 \equiv
 -V_s(r)\Phi_s,
 \label{eq:SU4-algebraic-term-as-potential}
\end{equation}
the harmonic equation
$\left(D^{12}\right)^2\Phi_s=0$ becomes
\begin{equation}
 \frac{1}{r^3}\frac{d}{dr}
 \left(
 r^3f^{-2}\frac{dh_s}{dr}
 \right)
 -V_s(r)h_s(r)=0.
 \label{eq:SU4-final-harmonic-radial-equation}
\end{equation}

To find the explicit expressions of $V_s$, we use the orthogonality of the invariant
one-forms $\alpha$ and $\beta$,
\begin{eqnarray}
 \langle\alpha,\beta\rangle_g=g^{\mu\nu}\alpha_\mu\beta_\nu\,,
\end{eqnarray}
and hence, we obtain
\begin{equation}
 \langle\sigma_x,\sigma_x\rangle_g
 =
 \langle\sigma_y,\sigma_y\rangle_g
 =
 \frac{1}{r^2},
 \qquad
 \langle\sigma_z,\sigma_z\rangle_g
 =
 \frac{f^2}{r^2},
 \qquad
 \langle\sigma_a,\sigma_b\rangle_g=0
 \quad (a\neq b).
\end{equation}
For $\Phi_s=h_sM_s$, the Pauli-matrix algebra can be summarized as
\begin{equation}
 T_iM_sT_i
 =
 \frac{\varepsilon_{is}}{4}M_s,
 \qquad
 \varepsilon_{is}
 =
 \begin{cases}
 +1, & s=0\ \text{or}\ s=i,\\
 -1, & s\neq0\ \text{and}\ s\neq i.
 \end{cases}
 \label{eq:SU4-Pauli-epsilon-identity}
\end{equation}
Consequently, the quadratic connection term becomes
\begin{align}
 &-A_1^\mu A_{1\mu}\Phi_s
 -\Phi_s A_2^\mu A_{2\mu}
 +2A_1^\mu\Phi_s A_{2\mu}
 \nonumber\\
 &\hspace{10mm}
 =
 -\frac{h_s}{4r^2}
 \bigg[
 2(F_1^2+F_2^2)
 -2F_1F_2(\varepsilon_{1s}+\varepsilon_{2s})
 \nonumber\\
 &\hspace{48mm}
 +f^2\left(
 G_1^2+G_2^2-2\varepsilon_{3s}G_1G_2
 \right)
 \bigg]M_s.
 \label{eq:SU4-quadratic-term-epsilon-form}
\end{align}
Comparing with
\begin{equation}
 -A_1^\mu A_{1\mu}\Phi_s
 -\Phi_s A_2^\mu A_{2\mu}
 +2A_1^\mu\Phi_s A_{2\mu}
 =
 -V_s(r)\Phi_s,
\end{equation}
we find
\begin{align}
 V_0(r)
 &=
 \frac{
 2(F_1-F_2)^2+f^2(G_1-G_2)^2
 }{4r^2},
 \nonumber\\
 V_1(r)=V_2(r)
 &=
 \frac{
 2(F_1^2+F_2^2)+f^2(G_1+G_2)^2
 }{4r^2},
 \nonumber\\
 V_3(r)
 &=
 \frac{
 2(F_1+F_2)^2+f^2(G_1-G_2)^2
 }{4r^2}.
 \label{eq:SU4-radial-potentials-derived}
\end{align}
The equality $V_1=V_2$ follows because the two channels differ only
by interchanging the $\sigma_xT_1$ and $\sigma_yT_2$ contributions.

At the bolt, $F_i(a)=2\kappa_i$ and $G_i(a)=2$, where
$\kappa_i=\sqrt{(t_i^2-a^2)/(t_i^2+a^2)}$.  Since
$f^2(G_1-G_2)^2\rightarrow0$ as $r\rightarrow a^+$, the potentials in
the $s=0,3$ channels remain finite,
\begin{equation}
 V_0(a)=\frac{2}{a^2}(\kappa_1-\kappa_2)^2,
 \qquad
 V_3(a)=\frac{2}{a^2}(\kappa_1+\kappa_2)^2,
\end{equation}
whereas
\begin{equation}
 V_1(r)=V_2(r)=\frac{1}{a(r-a)}+O(a^{-2})
\end{equation}
diverge with the regular centrifugal behavior associated with the
shrinking circle at the bolt.  At large radius,
$F_i=O(r^{-2})$, $G_i=O(r^{-2})$, and $f^2=1+O(r^{-4})$; hence all four
potentials vanish as
\begin{equation}
 V_s(r)=O(r^{-6}),
 \qquad s=0,1,2,3,
 \qquad r\rightarrow\infty.
\end{equation}

To solve (\ref{eq:SU4-final-harmonic-radial-equation}), we introduce 
\begin{equation}
 x=\frac{r^2}{a^2},
 \qquad
 p_i=\frac{t_i^2}{a^2},
 \qquad
 q_i=\sqrt{p_i^2-1},
 \qquad
 x\in[1,\infty),
 \label{eq:SU4-radial-detail-x-p-q}
\end{equation}
so that
\begin{equation}
 F_i(x)=\frac{2q_i}{x+p_i},
 \qquad
 G_i(x)=\frac{2(p_ix+1)}{x(x+p_i)},
 \qquad
 f^2(x)=\frac{x^2}{x^2-1}.
 \label{eq:SU4-radial-detail-dimensionless-profiles}
\end{equation}
The radial differential operator transforms according to
\begin{align}
 \frac{dh_s}{dr}
 &=
 \frac{2r}{a^2}\frac{dh_s}{dx},
 \nonumber\\
 r^3f^{-2}\frac{dh_s}{dr}
 &=
 2a^2(x^2-1)\frac{dh_s}{dx},
 \nonumber\\
 \frac{1}{r^3}\frac{d}{dr}
 \left(
  r^3f^{-2}\frac{dh_s}{dr}
 \right)
 &=
 \frac{4}{a^2x}
 \frac{d}{dx}
 \left[
  (x^2-1)\frac{dh_s}{dx}
 \right].
 \label{eq:SU4-radial-detail-operator-conversion}
\end{align}
It follows that \eqref{eq:SU4-final-harmonic-radial-equation} is equivalent to
\begin{equation}
 {\cal L}_s h_s
 \equiv
 \frac{d}{dx}
 \left[
  (x^2-1)\frac{dh_s}{dx}
 \right]
 -{\cal V}_s(x)h_s
 =0,
 \label{eq:SU4-radial-detail-x-equation}
\end{equation}
where
\begin{equation}
 {\cal V}_s(x)=\frac{a^2x}{4}V_s(r).
 \label{eq:SU4-radial-detail-V-definition}
\end{equation}

For later use, define
\begin{align}
 {\cal D}(x)
 &=(x+p_1)(x+p_2),
 \nonumber\\
 {\cal A}_{-}(x)
 &=(q_1-q_2)x+q_1p_2-q_2p_1,
 \nonumber\\
 {\cal A}_{+}(x)
 &=(q_1+q_2)x+q_1p_2+q_2p_1,
 \nonumber\\
 {\cal B}_{+}(x)
 &=(p_1+p_2)(x^2+1)+2(p_1p_2+1)x.
 \label{eq:SU4-radial-detail-auxiliary-polynomials}
\end{align}
Direct algebra gives
\begin{align}
 F_1-F_2
 &=
 \frac{2{\cal A}_{-}}{{\cal D}},
 &
 F_1+F_2
 &=
 \frac{2{\cal A}_{+}}{{\cal D}},
 \nonumber\\
 G_1-G_2
 &=
 \frac{
  2(p_1-p_2)(x^2-1)
 }{
  x{\cal D}
 },
 &
 G_1+G_2
 &=
 \frac{2{\cal B}_{+}}{x{\cal D}}.
 \label{eq:SU4-radial-detail-profile-combinations}
\end{align}
Substitution into \eqref{eq:SU4-radial-detail-V-definition} yields the
fully rational dimensionless potentials
\begin{align}
 {\cal V}_0
 &=
 \frac{
  2{\cal A}_{-}^2
  +(p_1-p_2)^2(x^2-1)
 }{
  4{\cal D}^2
 },
 \nonumber\\
 {\cal V}_\perp
 &=
 \frac12
 \left[
  \frac{q_1^2}{(x+p_1)^2}
  +\frac{q_2^2}{(x+p_2)^2}
 \right]
 +
 \frac{
  {\cal B}_{+}^2
 }{
  4(x^2-1){\cal D}^2
 },
 \nonumber\\
 {\cal V}_3
 &=
 \frac{
  2{\cal A}_{+}^2
  +(p_1-p_2)^2(x^2-1)
 }{
  4{\cal D}^2
 }.
 \label{eq:SU4-radial-detail-rational-potentials}
\end{align}
Equation \eqref{eq:SU4-radial-detail-x-equation}, together with
\eqref{eq:SU4-radial-detail-rational-potentials}, is the form in which
the solutions can be verified most directly.

\subsubsection*{Useful parametrization of the radial-size moduli}

It is convenient to introduce
\begin{equation}
 u_i=p_i+q_i
 =p_i+\sqrt{p_i^2-1},
 \qquad
 u_i\geq1.
 \label{eq:SU4-radial-detail-u-definition}
\end{equation}
Since
\begin{equation}
 (p_i+q_i)(p_i-q_i)=1,
 \label{eq:SU4-radial-detail-u-inverse}
\end{equation}
one has
\begin{equation}
 p_i=\frac12\left(u_i+\frac{1}{u_i}\right),
 \qquad
 q_i=\frac12\left(u_i-\frac{1}{u_i}\right).
 \label{eq:SU4-radial-detail-p-q-in-u}
\end{equation}
Equivalently, one may write
\begin{equation}
 p_i=\cosh\alpha_i,
 \qquad
 q_i=\sinh\alpha_i,
 \qquad
 u_i=e^{\alpha_i},
 \qquad
 \alpha_i\geq0.
 \label{eq:SU4-radial-detail-hyperbolic-parametrization}
\end{equation}
The latter parametrization makes the positivity and regularity of all
constants below manifest.

\subsubsection*{The $s=0$ channel}

Motivated by the common denominator in
\eqref{eq:SU4-radial-detail-rational-potentials}, consider
\begin{equation}
 h_c(x)=\frac{x+c}{\sqrt{{\cal D}(x)}}.
 \label{eq:SU4-radial-detail-linear-ansatz}
\end{equation}
Its logarithmic derivative is
\begin{equation}
 \frac{h_c'}{h_c}
 =
 \frac{1}{x+c}
 -\frac12
 \left(
  \frac{1}{x+p_1}
  +\frac{1}{x+p_2}
 \right).
 \label{eq:SU4-radial-detail-hc-log-derivative}
\end{equation}
Using \eqref{eq:SU4-radial-detail-p-q-in-u}, direct substitution into the
$s=0$ equation gives
\begin{equation}
 {\cal L}_0h_c
 =
 \frac{
  \left[
   u_1u_2+1-c(u_1+u_2)
  \right]
  \left[
   (u_1u_2+1)x+u_1+u_2
  \right]
 }{
  2u_1u_2\,{\cal D}^{3/2}
 }.
 \label{eq:SU4-radial-detail-L0-residual}
\end{equation}
The second factor in the numerator is strictly positive for
$x\geq1$ and $u_i\geq1$.  Therefore the residual vanishes identically
if and only if
\begin{equation}
 c=c_0
 \equiv
 \frac{u_1u_2+1}{u_1+u_2}.
 \label{eq:SU4-radial-detail-c0}
\end{equation}
The required solution is consequently
\begin{equation}
 h_0(x)
 =
 \frac{x+c_0}
 {\sqrt{(x+p_1)(x+p_2)}}.
 \label{eq:SU4-radial-detail-h0}
\end{equation}

For completeness, the constant $c_0$ may also be written as
\begin{equation}
 c_0
 =
 \frac{
  \cosh\left[(\alpha_1+\alpha_2)/2\right]
 }{
  \cosh\left[(\alpha_1-\alpha_2)/2\right]
 }.
 \label{eq:SU4-radial-detail-c0-hyperbolic}
\end{equation}
It is finite and positive throughout the radial chamber.  At the bolt,
\begin{equation}
 h_0(1)
 =
 \frac{1+c_0}
 {\sqrt{(1+p_1)(1+p_2)}},
 \label{eq:SU4-radial-detail-h0-bolt}
\end{equation}
so the solution is regular.  At infinity,
\begin{equation}
 h_0(x)
 =
 1-
 \frac{
  (p_1+p_2)/2-c_0
 }{x}
 +O(x^{-2}),
 \label{eq:SU4-radial-detail-h0-infinity}
\end{equation}
which verifies the normalization
\eqref{eq:SU4-radial-detail-normalization}.

\subsubsection*{The $s=1,2$ channels}

The $s=1$ and $s=2$ equations are identical.  In these channels the
potential has a simple pole at the bolt.  Since
\begin{equation}
 G_i(1)=2,
 \qquad
 f^2(x)=\frac{1}{2(x-1)}+O(1),
 \label{eq:SU4-radial-detail-perp-near-bolt-data}
\end{equation}
one obtains
\begin{equation}
 {\cal V}_\perp(x)
 =
 \frac{1}{2(x-1)}+O(1).
 \label{eq:SU4-radial-detail-perp-potential-bolt}
\end{equation}
Putting $h_\perp\sim(x-1)^\nu$ in
\eqref{eq:SU4-radial-detail-x-equation} gives the indicial equation
\begin{equation}
 2\nu^2-\frac12=0,
 \qquad
 \nu=\pm\frac12.
 \label{eq:SU4-radial-detail-perp-indicial-equation}
\end{equation}
Covariant regularity selects the positive branch.  The negative branch
diverges at the bolt.

The regular solution is
\begin{equation}
 h_\perp(x)
 \equiv h_1(x)=h_2(x)
 =
 \frac{\sqrt{x^2-1}}
 {\sqrt{{\cal D}(x)}}.
 \label{eq:SU4-radial-detail-hperp}
\end{equation}
Its logarithmic derivative is
\begin{equation}
 \frac{h_\perp'}{h_\perp}
 =
 \frac{x}{x^2-1}
 -\frac12
 \left(
  \frac{1}{x+p_1}
  +\frac{1}{x+p_2}
 \right).
 \label{eq:SU4-radial-detail-hperp-log-derivative}
\end{equation}
Substitution of \eqref{eq:SU4-radial-detail-hperp-log-derivative} and
\eqref{eq:SU4-radial-detail-rational-potentials} into
\eqref{eq:SU4-radial-detail-x-equation} gives
\begin{equation}
 {\cal L}_\perp h_\perp=0,
 \label{eq:SU4-radial-detail-hperp-verification}
\end{equation}
without imposing any additional algebraic condition on $p_1$ and
$p_2$.

Near the bolt,
\begin{equation}
 h_\perp(x)
 =
 \sqrt{
  \frac{2(x-1)}
  {(1+p_1)(1+p_2)}
 }
 +O\left((x-1)^{3/2}\right).
 \label{eq:SU4-radial-detail-hperp-bolt}
\end{equation}
The proper distance from the bolt is proportional to
$\sqrt{x-1}$, so this is the covariantly regular behavior of these
charged color channels.  At infinity,
\begin{equation}
 h_\perp(x)
 =
 1-\frac{p_1+p_2}{2x}
 +O(x^{-2}),
 \label{eq:SU4-radial-detail-hperp-infinity}
\end{equation}
and the solution again satisfies
\eqref{eq:SU4-radial-detail-normalization}.

\subsubsection*{The $s=3$ channel}

The $s=3$ solution has the same functional form as the $s=0$ solution.
Substituting the ansatz \eqref{eq:SU4-radial-detail-linear-ansatz} into
the $s=3$ equation gives
\begin{equation}
 {\cal L}_3h_c
 =
 \frac{
  \left[
   u_1+u_2-c(u_1u_2+1)
  \right]
  \left[
   (u_1+u_2)x+u_1u_2+1
  \right]
 }{
  2u_1u_2\,{\cal D}^{3/2}
 }.
 \label{eq:SU4-radial-detail-L3-residual}
\end{equation}
The residual vanishes identically precisely when
\begin{equation}
 c=c_3
 \equiv
 \frac{u_1+u_2}{u_1u_2+1}.
 \label{eq:SU4-radial-detail-c3}
\end{equation}
Therefore
\begin{equation}
 h_3(x)
 =
 \frac{x+c_3}
 {\sqrt{(x+p_1)(x+p_2)}}.
 \label{eq:SU4-radial-detail-h3}
\end{equation}
The two constants satisfy
\begin{equation}
 c_0c_3=1,
 \label{eq:SU4-radial-detail-c-product}
\end{equation}
and
\begin{equation}
 c_3
 =
 \frac{
  \cosh\left[(\alpha_1-\alpha_2)/2\right]
 }{
  \cosh\left[(\alpha_1+\alpha_2)/2\right]
 }.
 \label{eq:SU4-radial-detail-c3-hyperbolic}
\end{equation}
The bolt and asymptotic expansions are
\begin{align}
 h_3(1)
 &=
 \frac{1+c_3}
 {\sqrt{(1+p_1)(1+p_2)}},
 \nonumber\\
 h_3(x)
 &=
 1-
 \frac{
  (p_1+p_2)/2-c_3
 }{x}
 +O(x^{-2}),
 \label{eq:SU4-radial-detail-h3-limits}
\end{align}
so $h_3$ is regular and has the required normalization.

\subsubsection*{Uniqueness of the regular normalized solutions}

Once one solution $h_s^{(1)}$ of
\eqref{eq:SU4-radial-detail-x-equation} is known, a second linearly
independent solution is obtained by the reduction of order method:
\begin{equation}
 h_s^{(2)}(x)
 =
 h_s^{(1)}(x)
 \int^x
 \frac{dz}
 {(z^2-1)\left[h_s^{(1)}(z)\right]^2}.
 \label{eq:SU4-radial-detail-second-solution}
\end{equation}
For $s=0,3$, the regular solution approaches a nonzero constant at the
bolt.  The integrand in \eqref{eq:SU4-radial-detail-second-solution}
then behaves as $(x-1)^{-1}$, and the second solution has a logarithmic
singularity:
\begin{equation}
 h_{0,3}^{(2)}(x)\sim\log(x-1).
 \label{eq:SU4-radial-detail-second-solution-log}
\end{equation}
For $s=1,2$, the regular solution behaves as $(x-1)^{1/2}$, so the
second solution behaves as
\begin{equation}
 h_{\perp}^{(2)}(x)\sim(x-1)^{-1/2}.
 \label{eq:SU4-radial-detail-second-solution-power}
\end{equation}
It is again singular.  Bolt regularity therefore selects a
one-dimensional solution space in every channel.  The condition
$h_s(\infty)=1$ fixes its remaining overall normalization uniquely.

\subsubsection*{Final result}

Collecting the three channels, the unique covariantly regular and
asymptotically normalized solutions are
\begin{align}
 h_0(x)
 &=
 \frac{x+c_0}
 {\sqrt{(x+p_1)(x+p_2)}},
 &
 c_0
 &=
 \frac{u_1u_2+1}{u_1+u_2},
 \nonumber\\
 h_1(x)=h_2(x)
 &=
 \frac{\sqrt{x^2-1}}
 {\sqrt{(x+p_1)(x+p_2)}},
 \nonumber\\
 h_3(x)
 &=
 \frac{x+c_3}
 {\sqrt{(x+p_1)(x+p_2)}},
 &
 c_3
 &=
 \frac{u_1+u_2}{u_1u_2+1}.
 \label{eq:SU4-radial-detail-final-solutions}
\end{align}

\subsection{Off-diagonal part of the moduli-space metric}

Let $\vartheta_s^A$ be local orbit coordinates associated with the
off-diagonal generators $X_s^A$, where
$s=0,1,2,3$ and $A\in\{{\rm R},{\rm I}\}$.  Infinitesimally, the
off-diagonal part of the Maurer--Cartan form may be written as
\begin{equation}
 U^{-1}dU
 =
 i\sum_{s=0}^{3}\sum_{A={\rm R},{\rm I}}
 X_s^A\,d\vartheta_s^A
 +\cdots,
 \label{eq:SU4-offdiag-Maurer-Cartan-X-frame}
\end{equation}
where the omitted terms lie in the block-diagonal directions.  After
imposing the background-gauge condition, the corresponding variation
of the gauge field is
\begin{equation}
 \delta A_\mu^{\rm off}
 =
 \sum_{s=0}^{3}\sum_{A={\rm R},{\rm I}}
 D_\mu\Theta_s^A\,d\vartheta_s^A,
 \qquad
 \Theta_s^A(r)\longrightarrow X_s^A
 \quad\text{as}\quad r\longrightarrow\infty.
 \label{eq:SU4-offdiag-general-moduli-variation}
\end{equation}
The off-diagonal contribution to the moduli-space line element is
therefore
\begin{align}
 ds_{\mathcal M,{\rm off}}^2
 &=
 \frac{2}{g^2}
 \int_{\rm EH}d^4x\sqrt{g}\,
 \operatorname{tr}_{\mathbf4}
 \left(
  \delta A_\mu^{\rm off}\delta A^{{\rm off}\,\mu}
 \right)
 \nonumber\\
 &=
 \sum_{s,t=0}^{3}
 \sum_{A,B={\rm R},{\rm I}}
 {\cal G}^{(X)}_{sA,tB}\,
 d\vartheta_s^A d\vartheta_t^B,
 \label{eq:SU4-offdiag-moduli-line-element}
\end{align}
with
\begin{equation}
 {\cal G}^{(X)}_{sA,tB}
 =
 \frac{2}{g^2}
 \int_{\rm EH}d^4x\sqrt{g}\,
 \operatorname{tr}_{\mathbf4}
 \left(
  D_\mu\Theta_s^A D^\mu\Theta_t^B
 \right).
 \label{eq:SU4-offdiag-metric-definition}
\end{equation}

The color channels are mutually orthogonal, as are the real and
imaginary generators; see Eq. (\ref{eq:SU4-offdiag-generator-normalization}).  Consequently, the metric is diagonal in the
indices $(s,A)$.  To determine its diagonal entries, write
\begin{equation}
 h_s(r)
 =
 1-\frac{d_s}{r^2}+O(r^{-4})
 \label{eq:SU4-offdiag-h-asymptotics}
\end{equation}
at large $r$.  Since the dressed generators obey
$D^2\Theta_s^A=0$, integration by parts gives
\begin{align}
 &\int_{\rm EH}d^4x\sqrt{g}\,
 \operatorname{tr}_{\mathbf4}
 \left(
  D_\mu\Theta_s^A D^\mu\Theta_t^B
 \right)
 \nonumber\\
 &\qquad
 =
 \lim_{R\rightarrow\infty}
 \int_{r=R}d^3x\sqrt{\gamma}\,
 \operatorname{tr}_{\mathbf4}
 \left(
  \Theta_s^A D_n\Theta_t^B
 \right),
 \label{eq:SU4-offdiag-norm-boundary-term}
\end{align}
where the bolt is a regular interior submanifold and produces no
additional boundary term.

At large $R$, the outward unit normal and the induced volume are
\begin{equation}
 D_n=f^{-1}\partial_r,
 \qquad
 \int_{r=R}d^3x\sqrt{\gamma}
 =
 \pi^2R^3f^{-1}(R),
 \label{eq:SU4-offdiag-asymptotic-boundary-data}
\end{equation}
where $\pi^2$ is the volume of the unit $\mathbb{RP}^3$.
Using
\begin{equation}
 \partial_r h_s
 =
 \frac{2d_s}{r^3}+O(r^{-5}),
 \qquad
 \operatorname{tr}_{\mathbf4}
 \left(X_s^AX_t^B\right)
 =
 \frac12\delta_{st}\delta^{AB},
\end{equation}
the boundary integral becomes
\begin{equation}
 \int_{\rm EH}d^4x\sqrt{g}\,
 \operatorname{tr}_{\mathbf4}
 \left(
  D_\mu\Theta_s^A D^\mu\Theta_t^B
 \right)
 =
 \pi^2d_s\,\delta_{st}\delta^{AB}.
 \label{eq:SU4-offdiag-X-frame-norm}
\end{equation}
Thus, in the coordinates $\vartheta_s^A$ associated directly with the
generators $X_s^A$, the metric components are
\begin{equation}
 {\cal G}^{(X)}_{sA,tB}
 =
 \frac{2\pi^2}{g^2}d_s\,
 \delta_{st}\delta^{AB}.
 \label{eq:SU4-offdiag-X-frame-metric}
\end{equation}

The coefficients obtained from the asymptotic expansions of the
regular radial solutions are
\begin{align}
 d_0
 &=
 a^2
 \frac{
  (u_1-u_2)^2(u_1u_2+1)
 }{
  4u_1u_2(u_1+u_2)
 },
 \nonumber\\
 d_1=d_2
 &=
 a^2\frac{p_1+p_2}{2},
 \nonumber\\
 d_3
 &=
 a^2
 \frac{
  (u_1+u_2)(u_1u_2-1)^2
 }{
  4u_1u_2(u_1u_2+1)
 }.
 \label{eq:SU4-offdiag-d-coefficients}
\end{align}

We note that the asymptotic holonomy of the two-block solution is central,
\begin{equation}
 h_\infty=-I_4,
 \label{eq:SU4-offdiag-central-holonomy}
\end{equation}
and it commutes with every off-diagonal generator, and hence all eight
rotations preserve the selected boundary sector.

\subsection{The measure over the bosonic moduli space}

The color-orientation part of the collective-coordinate measure
contains three logically distinct ingredients:
\begin{enumerate}
 \item the Haar measure on the group $SU(4)$;
 \item the quotient by the stabilizer $H$, which produces an invariant
 measure on $SU(4)/H$;
 \item the Gram determinant of the physical gauge zero modes, which
 supplies a nontrivial $t_1,t_2$-dependent weight.
\end{enumerate}
The Haar measure determines how the angular orbit is integrated.  It
does not, by itself, determine the physical metric on the instanton
moduli space.  The latter is obtained from the Yang--Mills norm of the
zero modes.

\subsection*{Haar measure from the Maurer--Cartan form}

Choose the bi-invariant reference metric on $SU(4)$
\begin{equation}
 ds_0^2
 =
 -\frac{1}{2}
 \operatorname{tr}_{\mathbf 4}
 \left(U^{-1}dU\right)^2.
 \label{eq:SU4-Haar-reference-metric}
\end{equation}
Let $E_A$ be Hermitian generators normalized by
\begin{equation}
 \operatorname{tr}_{\mathbf 4}(E_AE_B)
 =
 2\delta_{AB}.
 \label{eq:SU4-Haar-generator-normalization}
\end{equation}
Expanding the Maurer--Cartan form as
\begin{equation}
 U^{-1}dU=iE_A\omega^A
 \label{eq:SU4-Haar-Maurer-Cartan-expansion}
\end{equation}
gives
\begin{equation}
 ds_0^2=\delta_{AB}\omega^A\omega^B.
 \label{eq:SU4-Haar-orthonormal-coframe}
\end{equation}
Thus the $\omega^A$ are an orthonormal moving coframe.  The
corresponding Riemannian volume form is
\begin{equation}
 dV_{SU(4)}
 =
 \left|
 \omega^1\wedge\cdots\wedge\omega^{15}
 \right|.
 \label{eq:SU4-geometric-Haar-volume-form}
\end{equation}
Because the metric \eqref{eq:SU4-Haar-reference-metric} is
bi-invariant, this volume form is invariant under both left and right
multiplication.  It is therefore a Haar measure, with a definite
geometric normalization.

\subsection*{From the group to the physical color orbit}

Let $A^{\rm B}$ denote the two-block instanton background.  A group
element $h$ belongs to the stabilizer if
\begin{equation}
 h^\dagger A^{\rm B}h=A^{\rm B}.
 \label{eq:SU4-background-stabilizer-definition}
\end{equation}
The rotations $U$ and $Uh$, with $h\in H$, therefore describe the same
background.  The physical color orbit is
\begin{equation}
 {\cal O}_{A^{\rm B}}
 \simeq
 \frac{SU(4)}{H}.
 \label{eq:SU4-physical-color-orbit}
\end{equation}
At a generic point $t_1\neq t_2$, the stabilizer is
\begin{equation}
 H
 =
 \left\{
 \operatorname{diag}(z_1I_2,z_2I_2):
 z_1,z_2\in U(1),\ (z_1z_2)^2=1
 \right\}.
 \label{eq:SU4-generic-stabilizer2}
\end{equation}
It has one continuous dimension and two connected components.  Hence
\begin{equation}
 \dim_{\mathbb R}\frac{SU(4)}{H}
 =
 15-1
 =
 14.
 \label{eq:SU4-color-orbit-dimension}
\end{equation}

Locally, decompose the Maurer--Cartan form into horizontal and vertical
directions:
\begin{equation}
 U^{-1}dU
 =
 iE_a\omega^a+iH_\alpha\eta^\alpha.
 \label{eq:SU4-horizontal-vertical-Maurer-Cartan}
\end{equation}
Here $E_a$ span a complement of $\mathfrak h$ in
$\mathfrak{su}(4)$, whereas $H_\alpha$ span the stabilizer algebra.
The one-form $\eta^\alpha$ moves along the redundant $H$ fiber.  The
fourteen one-forms $\omega^a$ describe physical color orientations.
The induced quotient measure is
\begin{equation}
 dV_{SU(4)/H}
 =
 \left|
 \omega^1\wedge\cdots\wedge\omega^{14}
 \right|.
 \label{eq:SU4-invariant-quotient-measure}
\end{equation}
Strictly speaking, Haar measure is defined on a group.  The expression
\eqref{eq:SU4-invariant-quotient-measure} is the $SU(4)$-invariant
quotient measure induced from Haar measure because $H$ is compact.

\subsection*{The quotient volume}

When the group and subgroup carry the metrics induced by the same
normalization, the projection
\begin{equation}
 SU(4)\longrightarrow SU(4)/H
 \label{eq:SU4-Riemannian-submersion}
\end{equation}
is a Riemannian submersion, and
\begin{equation}
 \operatorname{Vol}\left(\frac{SU(4)}{H}\right)
 =
 \frac{\operatorname{Vol}(SU(4))}
 {\operatorname{Vol}(H)}.
 \label{eq:SU4-quotient-volume-ratio}
\end{equation}
For the reference metric \eqref{eq:SU4-Haar-reference-metric},
\begin{equation}
 \operatorname{Vol}(SU(4))
 =
 \frac{\sqrt{2}\,\pi^9}{3}.
 \label{eq:SU4-reference-volume}
\end{equation}
The two components of $H$ can be parametrized as
\begin{equation}
 h_\epsilon(\alpha)
 =
 \operatorname{diag}
 \left(
 e^{i\alpha}I_2,
 \epsilon e^{-i\alpha}I_2
 \right),
 \qquad
 \epsilon=\pm1,
 \qquad
 0\leq\alpha<2\pi.
 \label{eq:SU4-stabilizer-components}
\end{equation}
On either component,
\begin{equation}
 ds_H^2=2d\alpha^2.
 \label{eq:SU4-stabilizer-line-element}
\end{equation}
Each component is therefore a circle of circumference
$2\pi\sqrt{2}$.  Including both components gives
\begin{equation}
 \operatorname{Vol}(H)=4\pi\sqrt{2}.
 \label{eq:SU4-stabilizer-volume}
\end{equation}
It follows that
\begin{equation}
 \operatorname{Vol}\left(\frac{SU(4)}{H}\right)
 =
 \frac{\pi^8}{12}.
 \label{eq:SU4-final-quotient-volume}
\end{equation}
The disconnected component of $H$ must be included.  Omitting it would
make the quotient volume too large by a factor of two.

\subsection*{The physical zero-mode metric}

The invariant quotient measure is only a reference angular measure.
The physical moduli-space metric is determined by the zero-mode Gram
matrix:
\begin{equation}
 {\cal G}_{ab}(t_1,t_2)
 =
 \frac{2}{g^2}
 \int_{\rm EH}d^4x\sqrt{g}\,
 \operatorname{tr}_{\mathbf 4}
 \left(
 \delta_aA_\mu\delta_bA^\mu
 \right).
 \label{eq:SU4-physical-zero-mode-Gram-matrix}
\end{equation}
In the orthonormal Haar frame, the angular line element is
\begin{equation}
 ds_{{\cal M},{\rm ang}}^2
 =
 {\cal G}_{ab}(t_1,t_2)\omega^a\omega^b.
 \label{eq:SU4-physical-angular-line-element}
\end{equation}
At fixed $t_1,t_2$, gauge invariance makes this a left
$SU(4)$-invariant metric on the homogeneous orbit.  In a frame that
diagonalizes the Gram matrix,
\begin{equation}
 ds_{{\cal M},{\rm ang}}^2
 =
 \sum_{a=1}^{14}
 \lambda_a(t_1,t_2)(\omega^a)^2.
 \label{eq:SU4-diagonal-angular-metric}
\end{equation}
Its volume form is
\begin{equation}
 dV_{{\cal M},{\rm ang}}
 =
 \left[
 \prod_{a=1}^{14}
 \sqrt{\lambda_a(t_1,t_2)}
 \right]
 dV_{SU(4)/H}.
 \label{eq:SU4-physical-angular-volume-form}
\end{equation}
This equation summarizes the division of labor:
\begin{equation}
 \text{physical angular measure}
 =
 \sqrt{\det{\cal G}_{\rm ang}}
 \times
 \text{invariant quotient measure}.
 \label{eq:SU4-angular-measure-division-of-labor}
\end{equation}

\subsection*{Physics-normalized and orthonormal Haar frames}

The off-diagonal generators introduced in the zero-mode construction
obey
\begin{equation}
 \operatorname{tr}_{\mathbf 4}
 \left(
 X_s^AX_t^B
 \right)
 =
 \frac{1}{2}\delta_{st}\delta^{AB}.
 \label{eq:SU4-physics-normalized-offdiagonal-generators}
\end{equation}
These are the standard physics-normalized generators, but they are not
orthonormal under \eqref{eq:SU4-Haar-reference-metric}.  The
orthonormal generators are
\begin{equation}
 E_s^A=2X_s^A,
 \qquad
 \operatorname{tr}_{\mathbf 4}
 \left(
 E_s^AE_t^B
 \right)
 =
 2\delta_{st}\delta^{AB}.
 \label{eq:SU4-orthonormal-Haar-generators}
\end{equation}
Suppose that the same Maurer--Cartan form is expanded in the two
frames:
\begin{equation}
 U^{-1}dU
 =
 iX_s^Ad\vartheta_s^A
 =
 iE_s^A\omega_s^A.
 \label{eq:SU4-two-Maurer-Cartan-frames}
\end{equation}
Then
\begin{equation}
 d\vartheta_s^A=2\omega_s^A.
 \label{eq:SU4-frame-coordinate-transformation}
\end{equation}
For the dressed harmonic scalar associated with $X_s^A$, the boundary
calculation gives
\begin{equation}
 {\cal G}_s^{(X)}
 =
 \frac{2\pi^2}{g^2}d_s.
 \label{eq:SU4-X-frame-zero-mode-coefficient}
\end{equation}
Therefore,
\begin{align}
 {\cal G}_s^{(X)}(d\vartheta_s^A)^2
 &=
 \frac{2\pi^2}{g^2}d_s(2\omega_s^A)^2
 \nonumber\\
 &=
 \frac{8\pi^2}{g^2}d_s(\omega_s^A)^2.
 \label{eq:SU4-frame-coefficient-transformation}
\end{align}
Thus the eigenvalue in the orthonormal Haar frame is
\begin{equation}
 \lambda_{sA}
 =
 \frac{8\pi^2}{g^2}d_s.
 \label{eq:SU4-Haar-frame-eigenvalue}
\end{equation}
The two descriptions are the same metric in different frames.

\subsection*{Off-diagonal determinant and the complete measure}

For $A\in\{{\rm R},{\rm I}\}$ and $s=0,1,2,3$, the eight
off-diagonal eigenvalues are
\begin{equation}
 Cd_0,\ Cd_0,\ Cd_\perp,\ Cd_\perp,\ Cd_\perp,\ Cd_\perp,\
 Cd_3,\ Cd_3,
 \qquad
 C=\frac{8\pi^2}{g^2},
 \label{eq:SU4-eight-offdiagonal-eigenvalues}
\end{equation}
where $d_\perp=d_1=d_2$.  Hence
\begin{equation}
 \sqrt{\det{\cal G}_{\rm off}}
 =
 C^4d_0d_\perp^2d_3.
 \label{eq:SU4-offdiagonal-Gram-determinant}
\end{equation}
Including the two radial-size modes and the six diagonal orientation modes,
the complete sixteen-dimensional bosonic Gram determinant is
\begin{equation}
 \sqrt{\det{\cal G}_{\rm b}}
 =
 C^8t_1^3t_2^3d_0d_\perp^2d_3\,.
 \label{eq:SU4-complete-bosonic-Gram-determinant}
\end{equation}
The Gaussian normalization of sixteen real bosonic zero modes is
$(2\pi)^{-8}$.  The complete collective-coordinate measure is
therefore
\begin{equation}
 d{\cal M}_{\rm b}
 =
 \frac{1}{2!}
 \frac{C^8}{(2\pi)^8}
 t_1^3t_2^3d_0d_\perp^2d_3\,
 dt_1dt_2\,dV_{SU(4)/H}\,,  \qquad
 C=\frac{8\pi^2}{g^2}.
 \label{eq:SU4-complete-bosonic-collective-coordinate-measure}
\end{equation}
The factor $1/2!$ removes the separate exchange $t_1\leftrightarrow
t_2$.  This exchange is not contained in the generic stabilizer $H$
and is therefore not already accounted for by the quotient volume.

\subsection{Fermion zero modes and their normalization}

The two diagonal blocks supply four zero modes.  For block $i=1,2$, define
\begin{equation}
 q_i(r)
 =
 {\cal F}_{H}(r,t_i)^2
 +\frac12{\cal F}_{\pm}(r,t_i)^2.
 \label{eq:SU4-check-block-q}
\end{equation}
Then, from our experience with $SU(2)$, the
diagonal blocks contribution is
\begin{equation}
 {\cal O}_{i}(x)
 =
 q_i(r)\eta_{i1}\eta_{i2},
 \qquad
 d{\cal M}_{{\rm f},i}
 =
 \frac{2g^2}{\pi^2t_i^2}\,
 d\eta_{i2}d\eta_{i1}.
 \label{eq:SU4-check-block-fermion-data}
\end{equation}

The origin of the off-diagonal fermion zero modes can be understood by
exploiting the block-diagonal form of the background,
\begin{equation}
 A_\mu^{\rm B}
 =
 \begin{pmatrix}
  A_{1\mu} & 0\\
  0 & A_{2\mu}
 \end{pmatrix}.
 \label{eq:SU4-block-background-fermions}
\end{equation}
Accordingly, an adjoint Weyl fermion decomposes as
\begin{equation}
 \lambda_\alpha
 =
 \begin{pmatrix}
  \lambda_{1\alpha} & \psi_\alpha\\
  \widetilde\psi_\alpha & \lambda_{2\alpha}
 \end{pmatrix}.
 \label{eq:SU4-adjoint-fermion-block-decomposition}
\end{equation}
The upper-right block transforms in the bifundamental representation of
the two diagonal $SU(2)$ factors.  Its covariant derivative is therefore
\begin{equation}
 D_a^{12}\Phi
 =
 E_a{}^\mu
 \left(
  \partial_\mu\Phi
  +iA_{1\mu}\Phi
  -i\Phi A_{2\mu}
 \right),
 \label{eq:SU4-bifundamental-covariant-derivative}
\end{equation}
whereas the lower-left block is governed by $D_a^{21}$, obtained by
interchanging $A_1$ and $A_2$.

The EH metric admits two covariantly constant spinors of the chirality
appropriate to the self-dual gauge background.  This implies a useful
map from covariantly harmonic bifundamental scalars to Weyl zero modes.
Let $\Phi$ satisfy
\begin{equation}
 \left(D^{12}\right)^2\Phi=0.
 \label{eq:SU4-bifundamental-harmonic-equation}
\end{equation}
Then the bispinor
\begin{equation}
 \Psi_\alpha{}^{\dot\beta}
 =
 \hat\sigma^a_{\alpha}{}^{\dot\beta}
 D_a^{12}\Phi
 \label{eq:SU4-scalar-to-fermion-map}
\end{equation}
obeys the Weyl equation.  Indeed,
\begin{align}
 \bar{\hat\sigma}^{\,b}D_b^{12}
 \left(
  \hat\sigma^aD_a^{12}\Phi
 \right)
 &=
 \left(D^{12}\right)^2\Phi
 +i\bar{\hat\sigma}^{\,ba}
 \left(
  F_{1ba}\Phi-\Phi F_{2ba}
 \right)/2.
 \label{eq:SU4-factorized-Weyl-operator}
\end{align}
The first term vanishes by
\eqref{eq:SU4-bifundamental-harmonic-equation}.  The matrices
$\bar{\hat\sigma}^{\,ab}$ have duality opposite to that of the self-dual
field strengths $F_1$ and $F_2$, and hence the second term vanishes as
well.  Thus
\begin{equation}
 \bar{\hat\sigma}^{\,a}D_a^{12}\Psi=0.
\end{equation}
The free dotted index $\dot\beta=1,2$ labels the two covariantly constant
spinors and therefore produces two independent Weyl zero modes from a
single harmonic scalar.

For the upper-right block, the required scalar is
\begin{equation}
 \Phi_0(r)=h_0(r)I_2,
 \qquad
 \left(D^{12}\right)^2\Phi_0=0,
 \qquad
 h_0(\infty)=1.
 \label{eq:SU4-cross-harmonic-seed}
\end{equation}
Introducing
\begin{equation}
 x=\frac{r^2}{a^2},
 \qquad
 p_i=\frac{t_i^2}{a^2},
 \qquad
 u_i=p_i+\sqrt{p_i^2-1},
 \qquad
 c_0=\frac{u_1u_2+1}{u_1+u_2},
\end{equation}
the solution regular at the bolt is
\begin{equation}
 h_0(x)
 =
 \frac{x+c_0}
 {\sqrt{(x+p_1)(x+p_2)}}.
 \label{eq:SU4-cross-h0-solution}
\end{equation}
Its radial equation is
\begin{equation}
 \frac{1}{r^3}\frac{d}{dr}
 \left(
  r^3f^{-2}\frac{dh_0}{dr}
 \right)
 -
 \frac{
  2(F_1-F_2)^2+f^2(G_1-G_2)^2
 }{4r^2}\,h_0
 =
 0.
 \label{eq:SU4-cross-h0-radial-equation}
\end{equation}
Although $h_0$ approaches a nonzero constant at infinity, the fermion
mode depends only on $D_a^{12}(h_0I_2)$, which is normalizable.

For convenience, define
\begin{equation}
 A_0=\frac{h_0'}{f},
 \qquad
 B_0=\frac{h_0(F_1-F_2)}{2r},
 \qquad
 C_0=\frac{h_0f(G_1-G_2)}{2r},
 \qquad
 {\cal P}_0=A_0^2+2B_0^2+C_0^2.
 \label{eq:SU4-cross-ABC-definitions}
\end{equation}
The tangent-frame derivatives of the harmonic scalar are then
\begin{equation}
 D_0^{12}\Phi_0=A_0I_2,
 \qquad
 D_1^{12}\Phi_0=iB_0\tau_1,
 \qquad
 D_2^{12}\Phi_0=iB_0\tau_2,
 \qquad
 D_3^{12}\Phi_0=iC_0\tau_3.
 \label{eq:SU4-cross-h0-frame-derivatives}
\end{equation}
Using $\hat\sigma^a=(I_2,i\vec\sigma)$ in
\eqref{eq:SU4-scalar-to-fermion-map}, the two upper-right solutions are
\begin{align}
 \Psi^{+\dot1}_1
 &=
 \begin{pmatrix}
  A_0-C_0&0\\
  0&A_0+C_0
 \end{pmatrix},
 &
 \Psi^{+\dot1}_2
 &=
 \begin{pmatrix}
  0&-2B_0\\
  0&0
 \end{pmatrix},
 \nonumber\\
 \Psi^{+\dot2}_1
 &=
 \begin{pmatrix}
  0&0\\
  -2B_0&0
 \end{pmatrix},
 &
 \Psi^{+\dot2}_2
 &=
 \begin{pmatrix}
  A_0+C_0&0\\
  0&A_0-C_0
 \end{pmatrix}.
 \label{eq:SU4-upper-right-fermion-zero-modes}
\end{align}
Here the subscript $\alpha=1,2$ is the undotted Weyl index, while each
displayed matrix acts on the two color indices of the upper-right block.

The same harmonic scalar also solves the lower-left equation with the
two connections interchanged.  Since this reverses the signs of
$B_0$ and $C_0$ but not that of $A_0$, the two lower-left solutions are
\begin{align}
 \Psi^{-\dot1}_1
 &=
 \begin{pmatrix}
  A_0+C_0&0\\
  0&A_0-C_0
 \end{pmatrix},
 &
 \Psi^{-\dot1}_2
 &=
 \begin{pmatrix}
  0&2B_0\\
  0&0
 \end{pmatrix},
 \nonumber\\
 \Psi^{-\dot2}_1
 &=
 \begin{pmatrix}
  0&0\\
  2B_0&0
 \end{pmatrix},
 &
 \Psi^{-\dot2}_2
 &=
 \begin{pmatrix}
  A_0-C_0&0\\
  0&A_0+C_0
 \end{pmatrix}.
 \label{eq:SU4-lower-left-fermion-zero-modes}
\end{align}
The corresponding $SU(4)$ zero modes are
\begin{equation}
 \lambda^{+\dot\beta}_\alpha
 =
 \begin{pmatrix}
  0&\Psi^{+\dot\beta}_\alpha\\
  0&0
 \end{pmatrix},
 \qquad
 \lambda^{-\dot\beta}_\alpha
 =
 \begin{pmatrix}
  0&0\\
  \Psi^{-\dot\beta}_\alpha&0
 \end{pmatrix},
 \qquad
 \dot\beta=1,2.
 \label{eq:SU4-four-cross-fermion-zero-modes}
\end{equation}
Thus the upper-right block supplies two modes and the lower-left block
supplies two more, giving four independent off-diagonal Weyl zero modes.
The two blocks are independent because a Euclidean chiral fermion takes
values in the complexified adjoint bundle; the lower-left Grassmann field
is not required to be the Hermitian conjugate of the upper-right one.

The other harmonic dressings $h_sM_s$, where
$M_s=(I_2,\tau_1,\tau_2,\tau_3)$, do not generate additional independent
fermion zero modes.  Their Dirac images satisfy exact intertwining
relations of the form
\begin{equation}
 \hat\sigma^aD_a^{12}(h_sM_s)
 =
 \kappa_s
 \left[
  \hat\sigma^aD_a^{12}(h_0I_2)
 \right]M_s,
 \label{eq:SU4-cross-fermion-intertwining}
\end{equation}
for constants $ \kappa_s$, where the matrix $M_s$ on the right-hand side acts on the free dotted
index.  They therefore span the same two-dimensional upper-right kernel.
The analogous statement holds in the lower-left block.  Together with
the four block-diagonal modes, the four solutions in
\eqref{eq:SU4-four-cross-fermion-zero-modes} account for the eight
adjoint zero modes required by the index.

It remains to normalize these solutions.  Each upper-right mode obeys
\begin{equation}
 \sum_{\alpha=1}^{2}
 \operatorname{tr}_{\mathbf 2}
 \left[
  \left(\Psi^{+\dot\beta}_\alpha\right)^\dagger
  \Psi^{+\dot\beta}_\alpha
 \right]
 =
 2{\cal P}_0,
 \label{eq:SU4-cross-pointwise-fermion-norm}
\end{equation}
and the same identity holds for the lower-left modes.  Moreover,
\begin{equation}
 2{\cal P}_0
 =
 \operatorname{tr}_{\mathbf 2}
 \left[
  \left(D_\mu^{12}\Phi_0\right)^\dagger
  D^{12\,\mu}\Phi_0
 \right].
 \label{eq:SU4-cross-P0-covariant-norm}
\end{equation}
Let ${\rm EH}_R$ denote the EH space truncated at $r=R$, whose
asymptotic boundary is
\begin{equation}
 Y_R=\partial{\rm EH}_R\simeq\mathbb{RP}^3.
\end{equation}
Using the covariant harmonic equation
\begin{equation}
 \left(D^{12}\right)^2\Phi_0=0,
\end{equation}
covariant integration by parts converts the bulk norm into a
three-dimensional boundary integral:
\begin{align}
 \nu_\times^2
 &\equiv
 \int_{\rm EH}d^4x\sqrt{g}\,2{\cal P}_0
 \nonumber\\
 &=
 \lim_{R\rightarrow\infty}
 \int_{Y_R}d^3x\sqrt{\gamma}\,
 \operatorname{tr}_{\mathbf 2}
 \left(
  \Phi_0^\dagger D_n^{12}\Phi_0
 \right).
 \label{eq:SU4-cross-norm-boundary-integral}
\end{align}
The bolt is a regular interior submanifold and is not a boundary of
the EH space.  Equivalently, if a small tubular neighborhood of the
bolt is temporarily excised, its inner boundary contribution vanishes
when the size of the neighborhood is taken to zero.

At $r=R$, the outward normal derivative and the induced boundary
volume are
\begin{equation}
 D_n^{12}=f^{-1}\partial_r,
 \qquad
 \int_{Y_R}d^3x\sqrt{\gamma}
 =
 \pi^2R^3f^{-1}(R).
 \label{eq:SU4-cross-asymptotic-boundary-data}
\end{equation}
Since $\Phi_0=h_0I_2$, one has
\begin{equation}
 \operatorname{tr}_{\mathbf 2}
 \left(
  \Phi_0^\dagger D_n^{12}\Phi_0
 \right)
 =
 2h_0f^{-1}h_0'.
\end{equation}
It follows that
\begin{align}
 \nu_\times^2
 &=
 2\pi^2
 \lim_{R\rightarrow\infty}
 R^3f^{-2}(R)h_0(R)h_0'(R).
 \label{eq:SU4-cross-norm-asymptotic-limit}
\end{align}
Using
\begin{equation}
 h_0(r)
 =
 1-\frac{d_0}{r^2}+O(r^{-4}),
 \qquad
 h_0'(r)
 =
 \frac{2d_0}{r^3}+O(r^{-5}),
\end{equation}
together with $f^{-2}(r)\rightarrow1$ as $r\rightarrow\infty$, we
finally obtain
\begin{equation}
\nu_\times^2=4\pi^2d_0.
 \label{eq:SU4-cross-fermion-norm}
\end{equation}

Finally, with the fermionic action normalized as
\begin{equation}
 S_{\rm f}
 =
 \frac{2}{g^2}
 \int_{\rm EH}d^4x\sqrt{g}\,
 \operatorname{tr}_{\mathbf4}
 \left(
  \bar\lambda\,
  \bar{\hat\sigma}^{\,a}D_a\lambda
 \right),
\end{equation}
a zero mode of norm $\nu$ contributes $g/(\sqrt{2}\nu)$ to the
Grassmann measure.  The four off-diagonal modes therefore give
\begin{equation}
 d{\cal M}_{{\rm f},\times}
 =
 \frac{g^4}{64\pi^4d_0^2}\,
 d\xi_{-\dot2}\,d\xi_{-\dot1}\,
 d\xi_{+\dot2}\,d\xi_{+\dot1}.
 \label{eq:SU4-cross-fermion-measure-corrected}
\end{equation}
Combining this result with the two diagonal pairs gives
\begin{equation}
 d{\cal M}_{\rm f}
 =
 \mu^{12}
 \frac{g^8}{16\pi^8t_1^2t_2^2d_0^2}\,
 d\eta_{1B}d\eta_{1A}\,
 d\eta_{2B}d\eta_{2A}\,
 d\xi_{-\dot2}d\xi_{-\dot1}\,
 d\xi_{+\dot2}d\xi_{+\dot1}.
 \label{eq:SU4-complete-fermion-measure-corrected}
\end{equation}

\subsection{Integration over the collective-coordinate space: the $8$-fermion condensate}
\label{subsec:SU4-four-bilinear-integration}

We now evaluate the charge-one contribution to the correlator of four
gaugino bilinears,
\begin{equation}
 {\cal G}_4(x_1,x_2,x_3,x_4)
 \equiv
\left\langle
 \left[
 \displaystyle\prod_{\ell=1}^{4}{\cal O}(x_\ell)
 \right]\right\rangle,
 \qquad
 {\cal O}(x)
 =
 \operatorname{tr}_{\mathbf4}
 \bigl(\lambda\lambda\bigr)(x).
 \label{eq:SU4-four-bilinear-correlator}
\end{equation}
 The
two-block background has
\begin{equation}
 Q_{(4)}=1,
 \qquad
 S_{\rm cl}
 =
 \frac{8\pi^2}{g^2}-i\theta,
\end{equation}
and supports eight adjoint Weyl zero modes.  Consequently, four
bilinear insertions are required to saturate the fermionic measure.
Assuming the supersymmetric cancellation of the nonzero-mode
determinants, the leading semiclassical contribution is
\begin{equation}
 {\cal G}_4^{\rm B}
 =
 e^{-S_{\rm cl}}
 \int d{\cal M}_{\rm b}\,d{\cal M}_{\rm f}\,
 {\cal O}^4.
 \label{eq:SU4-four-bilinear-collective-coordinate-integral}
\end{equation}

\paragraph{Bosonic collective-coordinate measure.}
The moduli space contains the two scale coordinates $t_1,t_2$ and
fourteen color orientations.  With
\begin{equation}
 C=\frac{8\pi^2}{g^2},
\end{equation}
the bosonic measure is
\begin{equation}
 d{\cal M}_{\rm b}
 =
 \frac{1}{2!}
 \frac{C^8}{(2\pi)^8}
 t_1^3t_2^3d_0d_\perp^2d_3\,
 dt_1dt_2\,dV_{SU(4)/H}.
 \label{eq:SU4-four-bilinear-bosonic-measure}
\end{equation}
The factor $1/2!$ removes the redundancy under exchange of the two
identical $SU(2)$ blocks.  The functions appearing in the measure are
\begin{align}
 d_0
 &=
 a^2
 \frac{
 (u_1-u_2)^2(u_1u_2+1)
 }{
 4u_1u_2(u_1+u_2)
 },
 \nonumber\\
 d_\perp
 &=
 a^2\frac{p_1+p_2}{2},
 \nonumber\\
 d_3
 &=
 a^2
 \frac{
 (u_1+u_2)(u_1u_2-1)^2
 }{
 4u_1u_2(u_1u_2+1)
 },
 \label{eq:SU4-four-bilinear-d-functions}
\end{align}
where
\begin{equation}
 p_i=\frac{t_i^2}{a^2},
 \qquad
 u_i=p_i+\sqrt{p_i^2-1}.
\end{equation}
For the group metric
\begin{equation}
 ds_0^2
 =
 -\frac12
 \operatorname{tr}_{\mathbf4}
 \left(U^{-1}dU\right)^2,
\end{equation}
the quotient volume is
\begin{equation}
 \operatorname{Vol}\left(\frac{SU(4)}{H}\right)
 =
 \frac{\pi^8}{12}.
 \label{eq:SU4-four-bilinear-orbit-volume}
\end{equation}

\paragraph{Fermionic measure.}
The four block-diagonal modes and the four off-diagonal modes give the
fermionic collective-coordinate measure
\begin{equation}
 d{\cal M}_{\rm f}
 =
 \mu^{12}
 \frac{g^8}
 {16\pi^8t_1^2t_2^2d_0^2}\,
 d^4\eta\,d^4\xi.
 \label{eq:SU4-four-bilinear-fermionic-measure}
\end{equation}
Here $\eta_{i1},\eta_{i2}$ are the two Grassmann coordinates associated
with block $i$, while $\xi_1,\ldots,\xi_4$ multiply the four
off-diagonal modes.  The factor $\mu^{12}$ is the Pauli--Villars
factor appropriate to $SU(4)$ with $Q_{(4)}=1$.

\paragraph{Saturation of the eight fermion zero modes.}
We use the convention
\begin{equation}
 {\cal O}(x)
 \equiv
 \operatorname{tr}_{\mathbf4}\bigl(\lambda\lambda\bigr)(x)
 =
 \operatorname{tr}_{\mathbf4}
 \left(
  \epsilon^{\alpha\beta}\lambda_\beta\lambda_\alpha
 \right)
 =
 -2\operatorname{tr}_{\mathbf4}
 \left(\lambda_1\lambda_2\right),
 \qquad
 \epsilon^{12}=+1.
 \label{eq:SU4-bilinear-convention}
\end{equation}
Decompose the adjoint fermion into its two block-diagonal parts and
its off-diagonal part:
\begin{equation}
 \lambda_\alpha
 =
 \lambda_\alpha^{(1)}
 +
 \lambda_\alpha^{(2)}
 +
 \lambda_{\times,\alpha}.
 \label{eq:SU4-fermion-block-decomposition}
\end{equation}
Products of a block-diagonal matrix with an off-diagonal matrix have
vanishing color trace.  Moreover, the two diagonal blocks have
disjoint support.  Consequently,
\begin{equation}
 {\cal O}
 =
 {\cal O}_1+{\cal O}_2+{\cal O}_\times.
 \label{eq:SU4-bilinear-block-decomposition}
\end{equation}

\subparagraph{Diagonal contributions.}
In the $i$th diagonal $SU(2)$ block, the two zero modes can be written
as
\begin{align}
 \lambda^{(i)}_1
 &=
 {\cal F}_{3i}\frac{\tau_3}{2}\eta_{i1}
 +
 {\cal F}_{\pm i}\frac{\tau_-}{2}\eta_{i2},
 \nonumber\\
 \lambda^{(i)}_2
 &=
 -{\cal F}_{3i}\frac{\tau_3}{2}\eta_{i2}
 +
 {\cal F}_{\pm i}\frac{\tau_+}{2}\eta_{i1}.
 \label{eq:SU4-diagonal-zero-mode-components}
\end{align}
Using
\begin{equation}
 \operatorname{tr}_{\mathbf2}
 \left[
  \left(\frac{\tau_3}{2}\right)^2
 \right]
 =
 \frac12,
 \qquad
 \operatorname{tr}_{\mathbf2}
 \left(
  \frac{\tau_-}{2}\frac{\tau_+}{2}
 \right)
 =
 \frac14,
 \label{eq:SU4-Pauli-traces-for-bilinear}
\end{equation}
together with
\begin{equation}
 \eta_{i2}\eta_{i1}
 =
 -\eta_{i1}\eta_{i2},
\end{equation}
one obtains
\begin{align}
 \operatorname{tr}_{\mathbf2}
 \left(
  \lambda^{(i)}_1\lambda^{(i)}_2
 \right)
 &=
 -\left(
  \frac12{\cal F}_{3i}^2
  +\frac14{\cal F}_{\pm i}^2
 \right)
 \eta_{i1}\eta_{i2}.
 \label{eq:SU4-diagonal-lambda1-lambda2}
\end{align}
Therefore,
\begin{align}
 {\cal O}_i
 &=
 -2\operatorname{tr}_{\mathbf2}
 \left(
  \lambda^{(i)}_1\lambda^{(i)}_2
 \right)
 \nonumber\\
 &=
 \left(
  {\cal F}_{3i}^2
  +\frac12{\cal F}_{\pm i}^2
 \right)
 \eta_{i1}\eta_{i2}
 \nonumber\\
 &\equiv
 q_i\eta_{i1}\eta_{i2},
 \label{eq:SU4-diagonal-bilinear-explicit}
\end{align}
where
\begin{equation}
 q_i(r)
 =
 {\cal F}_{3i}(r)^2
 +\frac12{\cal F}_{\pm i}(r)^2.
 \label{eq:SU4-qi-definition}
\end{equation}

\subparagraph{Off-diagonal contribution.}
We label the four off-diagonal Grassmann coordinates by
\begin{equation}
 \xi_1=\xi_{+\dot1},
 \qquad
 \xi_2=\xi_{+\dot2},
 \qquad
 \xi_3=\xi_{-\dot1},
 \qquad
 \xi_4=\xi_{-\dot2},
 \label{eq:SU4-cross-Grassmann-labels}
\end{equation}
so that
\begin{equation}
 \lambda_{\times,\alpha}
 =
 \xi_1\lambda^{+\dot1}_\alpha
 +
 \xi_2\lambda^{+\dot2}_\alpha
 +
 \xi_3\lambda^{-\dot1}_\alpha
 +
 \xi_4\lambda^{-\dot2}_\alpha.
 \label{eq:SU4-cross-zero-mode-expansion}
\end{equation}
Define the antisymmetric pairings
\begin{equation}
 {\cal B}_{IJ}
 =
 \operatorname{tr}_{\mathbf4}
 \left(
  \lambda^I_1\lambda^J_2
  -
  \lambda^I_2\lambda^J_1
 \right),
 \qquad
 {\cal B}_{JI}=-{\cal B}_{IJ}.
 \label{eq:SU4-cross-bilinear-pairing}
\end{equation}
Because the product of two upper-right blocks or two lower-left
blocks vanishes, only an upper-right mode paired with a lower-left
mode can contribute.

Using the explicit zero-mode matrices gives
\begin{align}
 {\cal B}_{14}
 &=
 \operatorname{tr}_{\mathbf2}
 \left(
  \Psi^{+\dot1}_1\Psi^{-\dot2}_2
  -
  \Psi^{+\dot1}_2\Psi^{-\dot2}_1
 \right)
 \nonumber\\
 &=
 \left[
  (A_0-C_0)^2+(A_0+C_0)^2
 \right]
 -
 \left(-4B_0^2\right)
 \nonumber\\
 &=
 2A_0^2+2C_0^2+4B_0^2
 =
 2{\cal P}_0,
 \label{eq:SU4-cross-pairing-14}
\end{align}
where
\begin{equation}
 {\cal P}_0=A_0^2+2B_0^2+C_0^2.
 \label{eq:SU4-P0-definition}
\end{equation}
Similarly,
\begin{align}
 {\cal B}_{23}
 &=
 \operatorname{tr}_{\mathbf2}
 \left(
  \Psi^{+\dot2}_1\Psi^{-\dot1}_2
  -
  \Psi^{+\dot2}_2\Psi^{-\dot1}_1
 \right)
 \nonumber\\
 &=
 -4B_0^2
 -
 \left[
  (A_0+C_0)^2+(A_0-C_0)^2
 \right]
 \nonumber\\
 &=
 -2{\cal P}_0.
 \label{eq:SU4-cross-pairing-23}
\end{align}
All other independent pairings vanish.  It follows from
\eqref{eq:SU4-bilinear-convention} that
\begin{align}
 {\cal O}_\times
 &=
 -2\sum_{I<J}\xi_I\xi_J{\cal B}_{IJ}
 \nonumber\\
 &=
 -4{\cal P}_0
 \left(
  \xi_1\xi_4-\xi_2\xi_3
 \right).
 \label{eq:SU4-cross-bilinear-explicit}
\end{align}
The overall sign in this equation depends on the phases chosen for
the lower-left zero-mode basis.  Redefining
\begin{equation}
 \xi_3\longrightarrow-\xi_3,
 \qquad
 \xi_4\longrightarrow-\xi_4
\end{equation}
gives the equivalent convention
\begin{equation}
 {\cal O}_\times
 =
 4{\cal P}_0
 \left(
  \xi_1\xi_4-\xi_2\xi_3
 \right).
 \label{eq:SU4-cross-bilinear-positive-convention}
\end{equation}
The square of ${\cal O}_\times$ is independent of this overall sign.

To evaluate it, note that
\begin{equation}
 (\xi_1\xi_4)^2=(\xi_2\xi_3)^2=0,
 \qquad
 (\xi_1\xi_4)(\xi_2\xi_3)
 =
 (\xi_2\xi_3)(\xi_1\xi_4),
\end{equation}
because both factors have even Grassmann degree.  Hence
\begin{align}
 \left(
  \xi_1\xi_4-\xi_2\xi_3
 \right)^2
 &=
 -2\xi_1\xi_4\xi_2\xi_3
 \nonumber\\
 &=
 -2\xi_1\xi_2\xi_3\xi_4,
 \label{eq:SU4-cross-Grassmann-square}
\end{align}
where moving $\xi_4$ past $\xi_2$ and $\xi_3$ requires two
interchanges and therefore produces no additional sign.  We thus find
\begin{equation}
 {\cal O}_\times^2
 =
 -32{\cal P}_0^2\,
 \xi_1\xi_2\xi_3\xi_4.
 \label{eq:SU4-cross-bilinear-square}
\end{equation}

\subparagraph{Saturation in the fourth power.}
Introduce the three Grassmann-even quantities
\begin{equation}
 X=q_1\eta_{11}\eta_{12},
 \qquad
 Y=q_2\eta_{21}\eta_{22},
 \qquad
 Z={\cal O}_\times.
\end{equation}
They commute with one another because they all have even Grassmann
degree.  They also obey
\begin{equation}
 X^2=Y^2=0,
 \qquad
 Z^3=0.
\end{equation}
To saturate the four diagonal and four off-diagonal Grassmann
coordinates in
\begin{equation}
 {\cal O}^4=(X+Y+Z)^4,
\end{equation}
one must select one factor of $X$, one factor of $Y$, and two factors
of $Z$.  The number of such terms is
\begin{equation}
 \frac{4!}{1!\,1!\,2!}=12.
\end{equation}
Consequently,
\begin{align}
 \left.{\cal O}^4\right|_{\eta^4\xi^4}
 &=
 12XYZ^2
 \nonumber\\
 &=
 12
 \left(q_1\eta_{11}\eta_{12}\right)
 \left(q_2\eta_{21}\eta_{22}\right)
 \left(
  -32{\cal P}_0^2\xi_1\xi_2\xi_3\xi_4
 \right)
 \nonumber\\
 &=
 -384q_1q_2{\cal P}_0^2\,
 \eta_{11}\eta_{12}
 \eta_{21}\eta_{22}
 \xi_1\xi_2\xi_3\xi_4
 .
 \label{eq:SU4-four-bilinear-Grassmann-saturation}
\end{align}
The sign of the displayed Grassmann monomial depends on the chosen
orientation of the fermionic collective-coordinate measure.  The
absolute coefficient $384$ is independent of this convention.

\paragraph{Reduction to the radial-size moduli integral.}
The chiral Ward identity implies that the correlator is independent of
the insertion points, provided that no boundary terms arise at spatial
infinity or at the ends of the instanton moduli space.  Under this
assumption, the four insertions may be brought to the bolt, where the integral is easier to perform. We have also checked that the result is independent of $r$.  At $r=a$,
\begin{equation}
 q_i(a)
 =
 \frac{2a^2}{(t_i^2+a^2)^2},
 \label{eq:SU4-four-bilinear-q-bolt}
\end{equation}
and
\begin{equation}
 {\cal P}_0(a)
 =
 \frac{
 32u_1u_2(u_1-u_2)^2
 }{
 a^2(u_1+u_2)^2
 (u_1+1)^2(u_2+1)^2
 }.
 \label{eq:SU4-four-bilinear-P-bolt}
\end{equation}
After performing the Grassmann integrations and combining the powers
of $t_i$ and $d_0$ in the bosonic and fermionic measures, the remaining
scale integral is
\begin{equation}
 {\cal J}_4
 =
 \int_a^\infty dt_1
 \int_a^\infty dt_2\,
 t_1t_2
 \frac{d_3d_\perp^2}{d_0}\,
 q_1(a)q_2(a){\cal P}_0(a)^2.
 \label{eq:SU4-four-bilinear-scale-integral}
\end{equation}
Using
\begin{equation}
 t_i\,dt_i
 =
 \frac{a^2}{4}
 \frac{u_i^2-1}{u_i^2}\,du_i,
\end{equation}
one obtains
\begin{equation}
 {\cal J}_4
 =
 256
 \int_1^\infty du_1
 \int_1^\infty du_2\,
 \frac{
 (u_1-1)(u_2-1)(u_1-u_2)^2
 (u_1u_2-1)^2
 }{
 (u_1+1)^7(u_2+1)^7
 }.
 \label{eq:SU4-four-bilinear-u-integral}
\end{equation}
All dependence on the bolt radius $a$ cancels.  Introducing
\begin{equation}
 y_i=\frac{u_i-1}{u_i+1},
 \qquad
 0\leq y_i<1,
\end{equation}
reduces the integral to
\begin{align}
 {\cal J}_4
 &=
 4
 \int_0^1dy_1
 \int_0^1dy_2\,
 y_1y_2(y_1-y_2)^2(y_1+y_2)^2
 \nonumber\\
 &=
 4
 \left(
 \frac1{12}-\frac18+\frac1{12}
 \right)
 =
 \frac16.
 \label{eq:SU4-four-bilinear-J-result}
\end{align}
The apparent singularity at $t_1=t_2$ is canceled because
\begin{equation}
 \frac{{\cal P}_0(a)^2}{d_0}
 =
 O\left((u_1-u_2)^2\right).
\end{equation}
Thus the coincident-scale locus is integrable.

\paragraph{Final result.}
Collecting the bosonic and fermionic normalization factors gives
\begin{equation}
 \frac{C^8}{(2\pi)^8}
 \frac{g^8}{16\pi^8}
 =
 \frac{2^{12}}{g^8}.
 \label{eq:SU4-four-bilinear-Jacobian-constant}
\end{equation}
The remaining numerical factors are the Grassmann saturation
coefficient, the block-exchange factor, the scale integral, and the
color-orbit volume:
\begin{align}
 384
 \times
 \frac{2^{12}}{g^8}
 \times
 \frac{1}{2!}
 \times
 \frac16
 \times
 \frac{\pi^8}{12}
 &=
 \frac{2^{15}\pi^8}{3g^8}.
 \label{eq:SU4-four-bilinear-final-arithmetic}
\end{align}
It follows that
\begin{equation}
 {\cal G}_4^{\rm B}
 =
 \frac{2^{15}\pi^8}{3}
 \frac{\mu^{12}}{g^8}
 \exp\left(
 -\frac{8\pi^2}{g^2}+i\theta
 \right).
 \label{eq:SU4-four-bilinear-PV-result}
\end{equation}
Defining the Pauli--Villars strong scale by
\begin{equation}
 \Lambda_{\rm PV}^{12}
 =
 \frac{\mu^{12}}{g^8}
 \exp\left(
 -\frac{8\pi^2}{g^2}+i\theta
 \right),
 \label{eq:SU4-PV-scale-definition}
\end{equation}
the two-block charge-one contribution becomes
\begin{equation}
 {\cal G}_4^{\rm B}
 =
 \frac16
 \left(
 16\pi^2\Lambda_{\rm PV}^3
 \right)^4
.
 \label{eq:SU4-four-bilinear-final-result}
\end{equation}

  \bibliography{RefEHcondensates.bib}
  
  \bibliographystyle{JHEP}
  \end{document}